\expandafter\edef\csname hypers@fe\endcsname{\catcode
                                             `\noexpand @=\the\catcode`\@}%
\catcode`\@=11
%
%
\ifx\hyperd@ne\hyper@ndefined
 \global\let\hyperd@ne=\relax
\else
 \errhelp{hyperbasics.tex needs to be included only once outside
          of any {...} or \begingroup...\endgroup. You have tried to
          include it more than once. If the previous include was indeed
          outside any groupings, continue and all will be well.}%
 \errmessage{Input this file only once!}%
  
\fi
%
%
\def\hyperv@rsion{8}%
%
%
\newread\hyperf@le
\def\hyperf@lename{\jobname.hrf}%
\immediate\openin\hyperf@le\hyperf@lename\relax
\ifeof\hyperf@le\relax
 \immediate\closein\hyperf@le\relax
\else
 \immediate\closein\hyperf@le\relax
 \input \hyperf@lename
\fi
%
%
\newwrite\hyperf@le
\immediate\openout\hyperf@le\hyperf@lename
%
%
\newtoks\hypert@ks
%
%
\edef\hypert@mp{\catcode`\noexpand\#=\the\catcode`\#}%
\catcode`\#=12
\def\hyperh@sh{#}%
\hypert@mp
\let\hypert@mp=\relax
\let\hyper@nd=\relax
\def\hyperstr@pquote"#1"#2\hyper@nd{\ifx\hyper@ndefined#2\hyper@ndefined#1\else
                                    \ifx\hyper@ndefined#1\hyper@ndefined
                                    \hyperstr@pquote#2"\hyper@nd\else
                                    #1\hyperstr@pquote"#2"\hyper@nd\fi\fi}%
\def\hyperstr@pblank" #1 #2\hyper@nd"{\ifx\hyper@ndefined#2\hyper@ndefined#1\else
                                    \ifx\hyper@ndefined#1\hyper@ndefined
                                    \hyperstr@pblank"#2 \hyper@nd"\else
                                    #1\hyperstr@pblank" #2 \hyper@nd"\fi\fi}
\long\def\hyper@nchor#1#2{\edef\hyperm@cro{html:<A #1>}%
                          \special\expandafter{\hyperm@cro}%
                          {#2}}%
\def\hyper@atm@ning#1->#2\hyper@nd{#2}
\def\hyperlink#1{\edef\hypert@mp{#1}%
               \edef\hypert@mp{\expandafter\hyper@atm@ning\meaning\hypert@mp
                               \hyper@nd}%
               \edef\hypert@mp"{ \expandafter\hyperstr@pquote\expandafter"%
                               \hypert@mp"\hyper@nd}%
               \edef\hypert@mp{\expandafter\hyperstr@pblank\expandafter%
                               "\hypert@mp" \hyper@nd"}%
               \hyper@nchor{href=\expandafter"\hypert@mp"}}%
\def\hypertarget#1{\edef\hypert@mp{#1}%
               \edef\hypert@mp{\expandafter\hyper@atm@ning\meaning\hypert@mp
                               \hyper@nd}%
               \edef\hypert@mp"{ \expandafter\hyperstr@pquote\expandafter"%
                               \hypert@mp"\hyper@nd}%
               \edef\hypert@mp{\expandafter\hyperstr@pblank\expandafter%
                               "\hypert@mp" \hyper@nd"}%
               \hyper@nchor{name=\expandafter"\hypert@mp"}}%
\def\hyperref{\afterassignment\hyperr@f\let\hyperp@ram}
\def\hyperr@f{\ifx\hyperp@ram{\iffalse}\fi
               \expandafter\expandafter\expandafter\hyperr@@
               \expandafter{%
              \else
               \iffalse}\fi
               \ifx\hyperp@ram\hyper@ndefined
                 \message{Undefined reference}%
                 \def\hyperp@r@m{{}{undefined}{}}%
               \else
                 \edef\hyperp@r@m{\hyperp@ram}%
               \fi
               \expandafter\expandafter\expandafter\hyperr@@
               \expandafter\hyperp@r@m
              \fi}%
\def\hyperr@@#1#2#3{\ifx\hyper@ndefined#1\hyper@ndefined
                    \hypert@ks\expandafter{\hyperh@sh#2.#3}%
                    \else
                     \ifx\hyper@ndefined#2#3\hyper@ndefined
                      \hypert@ks{#1}%
                     \else
                      \def\hypert@mp{#1}%
                      \hypert@ks\expandafter\expandafter\expandafter
                      {\expandafter\hypert@mp\hyperh@sh#2.#3}%
                     \fi
                    \fi
                    \expandafter\hyperlink\expandafter{\the\hypert@ks}}%
\def\hyperdef#1#2#3{{\global\escapechar=`\\\relax
                     \edef\hypert@mp{\hyperstr@pquote"#2.#3"\hyper@nd}%
                     \expandafter\ifx\csname hyperd@\meaning\hypert@mp
                     \endcsname
                     \relax
                     \expandafter\gdef\csname hyperd@\meaning\hypert@mp
                     \endcsname{}%
                     \gdef#1{{}{\hyperstr@pquote"#2"\hyper@nd}%
                               {\hyperstr@pquote"#3"\hyper@nd}}%
                     \immediate\write\hyperf@le{\def\noexpand#1{#1}}%
                     \xdef\hypert@mp{\global\let\noexpand\hypert@mp=\relax
                                     \noexpand\hypertarget{\hypert@mp}}%
                     \global\hypert@ks={\hypert@mp}%
                     \else
                     \message\expandafter{'\hypert@mp' duplicate}%
                     \global\let\hypert@mp=\relax
                     \global\hypert@ks={\hyperdef{#1}{#2}{#3@}}%
                     \fi}\the\hypert@ks}%

\def\hyper@nique#1#2#3#4{\global\escapechar=`\\\relax
                     \edef\hypert@mp{\hyperstr@pquote"#2.#3"\hyper@nd}%
                     \expandafter\ifx\csname hyperd@\meaning\hypert@mp
                     \endcsname
                     \relax
                     \gdef#1{{}{\hyperstr@pquote"#2"\hyper@nd}%
                               {\hyperstr@pquote"#3"\hyper@nd}}%
                     \global\let\hypert@mp=\relax
                     #4%
                     \else
                     \global\let\hypert@mp=\relax
                     \hyper@nique{#1}{#2}{#3@}{#4}%
                     \fi
                     }%

\let\hyper@@@@=\relax
\def\hyper@@{\let\hyper@@@=\relax}%
\hyper@@
\def\hyper@{\relax\let\hyper@@@\noexpand\hyper@\noexpand}%
\def\hyperpr@ref{\hyper@@\hyperref}
\def\hyperpr@def{\hyper@@\hyperdef}

\let\href\hyperlink

%
%
\hypers@fe

\input epsf

\catcode`\@=11
\def\unredoffs{\voffset=13mm \hoffset=6.5truemm}
\def\redoffs{\voffset=-12.truemm\hoffset=-3truemm}
\def\speclscape{}
\newbox\leftpage \newdimen\fullhsize \newdimen\hstitle \newdimen\hsbody
\newdimen\hdim
\hfuzz=1pt
\ifx\hyperdef\UNd@FiNeD\def\hyperdef#1#2#3#4{#4}\def\hyperref#1#2#3#4{#4}\fi
\def\newans{y }
\def\answb{y }
\ifx\answb\newans\message{(This uses normal fonts.)}%
\def\bigans{b }
\def\answ{b }
\ifx\answ\bigans\message{(Format simple colonne 12pts.}
\magnification=1200 \unredoffs\hsize=147truemm\vsize=219truemm
\hsbody=\hsize \hstitle=\hsize 
\else\message{(Format double colonne, 10pts.} \let\l@r=L
\magnification=1000 \vsize=182.5truemm
\redoffs%
\hstitle=122.5truemm\hsbody=122.5truemm\fullhsize=258truemm\hsize=\hsbody
\output={
  \almostshipout{\leftline{\vbox{\makeheadline\pagebody\makefootline}}}
\advancepageno%
}
\def\almostshipout#1{\if L\l@r \count1=1 \message{[\the\count0.\the\count1]}
      \global\setbox\leftpage=#1 \global\let\l@r=R
 \else \count1=2
  \shipout\vbox{\speclscape{\hsize\fullhsize}
      \hbox to\fullhsize{\box\leftpage\hfil#1}}  \global\let\l@r=L\fi}
\fi

\def\sla#1{\mkern-1.5mu\raise0.4pt\hbox{$\not$}\mkern1.2mu #1\mkern 0.7mu}
\def\Dbar{\mkern-1.5mu\raise0.4pt\hbox{$\not$}\mkern-.1mu {\rm D}\mkern.1mu}
\def\Abar{\mkern1.mu\raise0.4pt\hbox{$\not$}\mkern-1.3mu A\mkern.1mu}
\def\Bbar{\mkern-0.mu\raise0.4pt\hbox{$\not$}\mkern-.3mu B\mkern 0.6mu}
\newskip\tableskipamount \tableskipamount=8pt plus 3pt minus 3pt

  
\newdimen\chapskip
 \font\ssbx=cmssbx10
 \font\caprm=cmr9 \font\capit=cmti9
\font\capbf=cmbx9 \font\capsl=cmsl9 \font\capmi=cmmi9
\font\capex=cmex9 \font\capsy=cmsy9 \chapskip=17.5mm
\def\makeheadline{\vbox to 0pt{\vskip-22.5pt
\line{\vbox to8.5pt{}\the\headline}\vss}\nointerlineskip}
\font\tenbi=cmmib10 \font\ninebi=cmmib9 \font\sevenbi=cmmib7
\font\fivebi=cmmib5 \textfont4=\tenbi \scriptfont4=\sevenbi
\scriptscriptfont4=\fivebi \font\headrm=cmr10 

\font\sixrm=cmr6 \font\fiverm=cmr5 \font\sixmi=cmmi6
\font\fivemi=cmmi5 \font\sixsy=cmsy6 \font\fivesy=cmsy5
\font\sixbf=cmbx6 \font\fivebf=cmbx5 \skewchar\capmi='177
\skewchar\sixmi='177 \skewchar\fivemi='177 \skewchar\capsy='60
\skewchar\sixsy='60 \skewchar\fivesy='60

\def\elevenpoint{
\textfont0=\caprm \scriptfont0=\sixrm \scriptscriptfont0=\fiverm
\def\rm{\fam0\caprm}
\textfont1=\capmi \scriptfont1=\sixmi \scriptscriptfont1=\fivemi
\textfont2=\capsy \scriptfont2=\sixsy \scriptscriptfont2=\fivesy
\textfont3=\capex \scriptfont3=\capex \scriptscriptfont3=\capex
\textfont\itfam=\capit \def\it{\fam\itfam\capit} 
\textfont\slfam=\capsl  \def\sl{\fam\slfam\capsl} 
\textfont\bffam=\capbf \scriptfont\bffam=\sixbf
\scriptscriptfont\bffam=\fivebf
\def\bf{\fam\bffam\capbf} 
\textfont4=\ninebi \scriptfont4=\sevenbi
\scriptscriptfont4=\fivebi \abovedisplayskip=11pt plus 3pt minus
8pt \belowdisplayskip=\abovedisplayskip
\smallskipamount=2.7pt plus 1pt minus 1pt
\medskipamount=5.4pt plus 2pt minus 2pt
\bigskipamount=10.8pt plus 3.6pt minus 3.6pt
\normalbaselineskip=11pt \setbox\strutbox=\hbox{\vrule height7.8pt
depth3.2pt width0pt} \normalbaselines \rm}


\catcode`\@=11

\font\tenmsa=msam10 \font\sevenmsa=msam7 \font\fivemsa=msam5
\font\tenmsb=msbm10 \font\sevenmsb=msbm7 \font\fivemsb=msbm5
\newfam\msafam
\newfam\msbfam
\textfont\msafam=\tenmsa  \scriptfont\msafam=\sevenmsa
  \scriptscriptfont\msafam=\fivemsa
\textfont\msbfam=\tenmsb  \scriptfont\msbfam=\sevenmsb
  \scriptscriptfont\msbfam=\fivemsb

\def\hexnumber@#1{\ifcase#1 0\or1\or2\or3\or4\or5\or6\or7\or8\or9\or
    A\or B\or C\or D\or E\or F\fi }

\font\teneuf=eufm10 \font\seveneuf=eufm7 \font\fiveeuf=eufm5
\newfam\euffam
\textfont\euffam=\teneuf \scriptfont\euffam=\seveneuf
\scriptscriptfont\euffam=\fiveeuf
\def\frak{\ifmmode\let\next\frak@\else
 \def\next{\Err@{Use \string\frak\space only in math mode}}\fi\next}
\def\goth{\ifmmode\let\next\frak@\else
 \def\next{\Err@{Use \string\goth\space only in math mode}}\fi\next}
\def\frak@#1{{\frak@@{#1}}}
\def\frak@@#1{\fam\euffam#1}

\edef\msa@{\hexnumber@\msafam} \edef\msb@{\hexnumber@\msbfam}

\mathchardef\boxdot="2\msa@00 \mathchardef\boxplus="2\msa@01
\mathchardef\boxtimes="2\msa@02 \mathchardef\square="0\msa@03
\mathchardef\blacksquare="0\msa@04
\mathchardef\centerdot="2\msa@05 \mathchardef\lozenge="0\msa@06
\mathchardef\blacklozenge="0\msa@07
\mathchardef\circlearrowright="3\msa@08
\mathchardef\circlearrowleft="3\msa@09
\mathchardef\rightleftharpoons="3\msa@0A
\mathchardef\leftrightharpoons="3\msa@0B
\mathchardef\boxminus="2\msa@0C \mathchardef\Vdash="3\msa@0D
\mathchardef\Vvdash="3\msa@0E \mathchardef\vDash="3\msa@0F
\mathchardef\twoheadrightarrow="3\msa@10
\mathchardef\twoheadleftarrow="3\msa@11
\mathchardef\leftleftarrows="3\msa@12
\mathchardef\rightrightarrows="3\msa@13
\mathchardef\upuparrows="3\msa@14
\mathchardef\downdownarrows="3\msa@15
\mathchardef\upharpoonright="3\msa@16

\mathchardef\downharpoonright="3\msa@17
\mathchardef\upharpoonleft="3\msa@18
\mathchardef\downharpoonleft="3\msa@19
\mathchardef\rightarrowtail="3\msa@1A
\mathchardef\leftarrowtail="3\msa@1B
\mathchardef\leftrightarrows="3\msa@1C
\mathchardef\rightleftarrows="3\msa@1D \mathchardef\Lsh="3\msa@1E
\mathchardef\Rsh="3\msa@1F \mathchardef\rightsquigarrow="3\msa@20
\mathchardef\leftrightsquigarrow="3\msa@21
\mathchardef\looparrowleft="3\msa@22
\mathchardef\looparrowright="3\msa@23
\mathchardef\circeq="3\msa@24 \mathchardef\succsim="3\msa@25
\mathchardef\gtrsim="3\msa@26 \mathchardef\gtrapprox="3\msa@27
\mathchardef\multimap="3\msa@28 \mathchardef\therefore="3\msa@29
\mathchardef\because="3\msa@2A \mathchardef\doteqdot="3\msa@2B

\mathchardef\triangleq="3\msa@2C \mathchardef\precsim="3\msa@2D
\mathchardef\lesssim="3\msa@2E \mathchardef\lessapprox="3\msa@2F
\mathchardef\eqslantless="3\msa@30
\mathchardef\eqslantgtr="3\msa@31
\mathchardef\curlyeqprec="3\msa@32
\mathchardef\curlyeqsucc="3\msa@33
\mathchardef\preccurlyeq="3\msa@34 \mathchardef\leqq="3\msa@35
\mathchardef\leqslant="3\msa@36 \mathchardef\lessgtr="3\msa@37
\mathchardef\backprime="0\msa@38
\mathchardef\risingdotseq="3\msa@3A
\mathchardef\fallingdotseq="3\msa@3B
\mathchardef\succcurlyeq="3\msa@3C \mathchardef\geqq="3\msa@3D
\mathchardef\geqslant="3\msa@3E \mathchardef\gtrless="3\msa@3F
\mathchardef\sqsubset="3\msa@40 \mathchardef\sqsupset="3\msa@41
\mathchardef\vartriangleright="3\msa@42
\mathchardef\vartriangleleft="3\msa@43
\mathchardef\trianglerighteq="3\msa@44
\mathchardef\trianglelefteq="3\msa@45
\mathchardef\bigstar="0\msa@46 \mathchardef\between="3\msa@47
\mathchardef\blacktriangledown="0\msa@48
\mathchardef\blacktriangleright="3\msa@49
\mathchardef\blacktriangleleft="3\msa@4A
\mathchardef\vartriangle="0\msa@4D
\mathchardef\blacktriangle="0\msa@4E
\mathchardef\triangledown="0\msa@4F \mathchardef\eqcirc="3\msa@50
\mathchardef\lesseqgtr="3\msa@51 \mathchardef\gtreqless="3\msa@52
\mathchardef\lesseqqgtr="3\msa@53
\mathchardef\gtreqqless="3\msa@54
\mathchardef\Rrightarrow="3\msa@56
\mathchardef\Lleftarrow="3\msa@57 \mathchardef\veebar="2\msa@59
\mathchardef\barwedge="2\msa@5A
\mathchardef\doublebarwedge="2\msa@5B \mathchardef\angle="0\msa@5C
\mathchardef\measuredangle="0\msa@5D
\mathchardef\sphericalangle="0\msa@5E
\mathchardef\varpropto="3\msa@5F \mathchardef\smallsmile="3\msa@60
\mathchardef\smallfrown="3\msa@61 \mathchardef\Subset="3\msa@62
\mathchardef\Supset="3\msa@63 \mathchardef\Cup="2\msa@64

\mathchardef\Cap="2\msa@65

\mathchardef\curlywedge="2\msa@66 \mathchardef\curlyvee="2\msa@67
\mathchardef\leftthreetimes="2\msa@68
\mathchardef\rightthreetimes="2\msa@69
\mathchardef\subseteqq="3\msa@6A \mathchardef\supseteqq="3\msa@6B
\mathchardef\bumpeq="3\msa@6C \mathchardef\Bumpeq="3\msa@6D
\mathchardef\lll="3\msa@6E

\mathchardef\ggg="3\msa@6F

\mathchardef\circledS="0\msa@73 \mathchardef\pitchfork="3\msa@74
\mathchardef\dotplus="2\msa@75 \mathchardef\backsim="3\msa@76
\mathchardef\backsimeq="3\msa@77 \mathchardef\complement="0\msa@7B
\mathchardef\intercal="2\msa@7C \mathchardef\circledcirc="2\msa@7D
\mathchardef\circledast="2\msa@7E
\mathchardef\circleddash="2\msa@7F
\def\ulcorner{\delimiter"4\msa@70\msa@70 }
\def\urcorner{\delimiter"5\msa@71\msa@71 }
\def\llcorner{\delimiter"4\msa@78\msa@78 }
\def\lrcorner{\delimiter"5\msa@79\msa@79 }
\def\yen{\mathhexbox\msa@55 }
\def\checkmark{\mathhexbox\msa@58 }
\def\circledR{\mathhexbox\msa@72 }
\def\maltese{\mathhexbox\msa@7A }
\mathchardef\lvertneqq="3\msb@00 \mathchardef\gvertneqq="3\msb@01
\mathchardef\nleq="3\msb@02 \mathchardef\ngeq="3\msb@03
\mathchardef\nless="3\msb@04 \mathchardef\ngtr="3\msb@05
\mathchardef\nprec="3\msb@06 \mathchardef\nsucc="3\msb@07
\mathchardef\lneqq="3\msb@08 \mathchardef\gneqq="3\msb@09
\mathchardef\nleqslant="3\msb@0A \mathchardef\ngeqslant="3\msb@0B
\mathchardef\lneq="3\msb@0C \mathchardef\gneq="3\msb@0D
\mathchardef\npreceq="3\msb@0E \mathchardef\nsucceq="3\msb@0F
\mathchardef\precnsim="3\msb@10 \mathchardef\succnsim="3\msb@11
\mathchardef\lnsim="3\msb@12 \mathchardef\gnsim="3\msb@13
\mathchardef\nleqq="3\msb@14 \mathchardef\ngeqq="3\msb@15
\mathchardef\precneqq="3\msb@16 \mathchardef\succneqq="3\msb@17
\mathchardef\precnapprox="3\msb@18
\mathchardef\succnapprox="3\msb@19 \mathchardef\lnapprox="3\msb@1A
\mathchardef\gnapprox="3\msb@1B \mathchardef\nsim="3\msb@1C
\mathchardef\ncong="3\msb@1D

\mathchardef\varsubsetneq="3\msb@20
\mathchardef\varsupsetneq="3\msb@21
\mathchardef\nsubseteqq="3\msb@22
\mathchardef\nsupseteqq="3\msb@23
\mathchardef\subsetneqq="3\msb@24
\mathchardef\supsetneqq="3\msb@25
\mathchardef\varsubsetneqq="3\msb@26
\mathchardef\varsupsetneqq="3\msb@27
\mathchardef\subsetneq="3\msb@28 \mathchardef\supsetneq="3\msb@29
\mathchardef\nsubseteq="3\msb@2A \mathchardef\nsupseteq="3\msb@2B
\mathchardef\nparallel="3\msb@2C \mathchardef\nmid="3\msb@2D
\mathchardef\nshortmid="3\msb@2E
\mathchardef\nshortparallel="3\msb@2F
\mathchardef\nvdash="3\msb@30 \mathchardef\nVdash="3\msb@31
\mathchardef\nvDash="3\msb@32 \mathchardef\nVDash="3\msb@33
\mathchardef\ntrianglerighteq="3\msb@34
\mathchardef\ntrianglelefteq="3\msb@35
\mathchardef\ntriangleleft="3\msb@36
\mathchardef\ntriangleright="3\msb@37
\mathchardef\nleftarrow="3\msb@38
\mathchardef\nrightarrow="3\msb@39
\mathchardef\nLeftarrow="3\msb@3A
\mathchardef\nRightarrow="3\msb@3B
\mathchardef\nLeftrightarrow="3\msb@3C
\mathchardef\nleftrightarrow="3\msb@3D
\mathchardef\divideontimes="2\msb@3E
\mathchardef\varnothing="0\msb@3F \mathchardef\nexists="0\msb@40
\mathchardef\mho="0\msb@66 \mathchardef\eth="0\msb@67
\mathchardef\eqsim="3\msb@68 \mathchardef\beth="0\msb@69
\mathchardef\gimel="0\msb@6A \mathchardef\daleth="0\msb@6B
\mathchardef\lessdot="3\msb@6C \mathchardef\gtrdot="3\msb@6D
\mathchardef\ltimes="2\msb@6E \mathchardef\rtimes="2\msb@6F
\mathchardef\shortmid="3\msb@70
\mathchardef\shortparallel="3\msb@71
\mathchardef\smallsetminus="2\msb@72
\mathchardef\thicksim="3\msb@73 \mathchardef\thickapprox="3\msb@74
\mathchardef\approxeq="3\msb@75 \mathchardef\succapprox="3\msb@76
\mathchardef\precapprox="3\msb@77
\mathchardef\curvearrowleft="3\msb@78
\mathchardef\curvearrowright="3\msb@79
\mathchardef\digamma="0\msb@7A \mathchardef\varkappa="0\msb@7B
\mathchardef\hslash="0\msb@7D \mathchardef\hbar="0\msb@7E
\mathchardef\backepsilon="3\msb@7F
\def\Bbb{\ifmmode\let\next\Bbb@\else
 \def\next{\errmessage{Use \string\Bbb\space only in math mode}}\fi\next}
\def\Bbb@#1{{\Bbb@@{#1}}}
\def\Bbb@@#1{\fam\msbfam#1}
 \catcode`\@=12
\def\sla#1{\mkern-1.5mu\raise0.4pt\hbox{$\not$}\mkern1.2mu #1\mkern 0.7mu}
\def\Dbar{\mkern-1.5mu\raise0.4pt\hbox{$\not$}\mkern-.1mu {\rm D}\mkern.1mu}
\def\Abar{\mkern1.mu\raise0.4pt\hbox{$\not$}\mkern-1.3mu A\mkern.1mu}
\nopagenumbers
\headline={\ifnum\pageno=1\hfill\else\draftdate\hfil{\headrm\folio}%
\hfil\hphantom{\draftdate}\fi} \else\message{(This uses pseudo
12pts fonts.} \hoffset=8mm \voffset=16mm
\input lfont12 

\def\sla#1{\mkern-1.5mu\raise0.5pt\hbox{$\not$}\mkern1.2mu #1\mkern 0.7mu}
\def\Dbar{\mkern-1.5mu\raise0.5pt\hbox{$\not$}\mkern-.1mu {\rm D}\mkern.1mu}
\def\Abar{\mkern1.mu\raise0.5pt\hbox{$\not$}\mkern-1.3mu A\mkern.1mu}
\fi


\newcount\yearltd\yearltd=\year\advance\yearltd by -2000
\newif\ifdraftmode
\draftmodefalse
\def\draft{\draftmodetrue{\count255=\time\divide\count255 by 60
\xdef\hourmin{\number\count255}
  \multiply\count255 by-60\advance\count255 by\time
  \xdef\hourmin{\hourmin:\ifnum\count255<10 0\fi\the\count255}}}
\def\draftdate{\ifdraftmode{\headrm\quad (\jobname,\ le
\number\day/\number\month/\number\yearltd\ \ \hourmin)}\else{}\fi}
\newif\iffrancmode
\francmodefalse
\def\e{\mathop{\rm e}\nolimits}
\def\sgn{\mathop{\rm sgn}\nolimits}

\def\d{{\rm d}}
\def\ud{{\textstyle{1\over 2}}}
\def\half{\ud}
\def\tr{\mathop{\rm tr}\nolimits}
\def\det{\mathop{\rm det}\nolimits}
\def\del{\partial}
\def\dd{\d^d\hskip-1pt}

\chardef\sigmat=27

\def\rf{\par\item{}}

\def\frac#1#2{{\textstyle{#1\over#2}}}

\def\leaderfill{\leaders\hbox to 1em{\hss.\hss}\hfill}
\catcode`\@=11
\def\deqalignno#1{\displ@y\tabskip\centering \halign to
\displaywidth{\hfil$\displaystyle{##}$\tabskip0pt&$\displaystyle{{}##}$
\hfil\tabskip0pt &\quad
\hfil$\displaystyle{##}$\tabskip0pt&$\displaystyle{{}##}$
\hfil\tabskip\centering& \llap{$##$}\tabskip0pt \crcr #1 \crcr}}
\def\deqalign#1{\null\,\vcenter{\openup\jot\m@th\ialign{
\strut\hfil$\displaystyle{##}$&$\displaystyle{{}##}$\hfil
&&\quad\strut\hfil$\displaystyle{##}$&$\displaystyle{{}##}$
\hfil\crcr#1\crcr}}\,}
\def\xlabel#1{\expandafter\xl@bel#1}\def\xl@bel#1{#1}
\def\label#1{\l@bel #1{\hbox{}}}
\def\l@bel#1{\ifx\UNd@FiNeD#1\message{label \string#1 is undefined.}%
\xdef#1{?.? }\fi{\let\hyperref=\relax\xdef\next{#1}}%
\ifx\next\em@rk\def\next{}%
\else\def\next{#1}\fi\next}
\def\DefWarn#1{\ifx\UNd@FiNeD#1\else
\immediate\write16{*** WARNING: the label \string#1 is already defined%
***}\fi}%
\newread\ch@ckfile
\def\cinput#1{\def\filen@me{#1 }
\immediate\openin\ch@ckfile=\filen@me
\ifeof\ch@ckfile\message{<< (\filen@me\ DOES NOT EXIST in this pass)>>}\else%
\closein \ch@ckfile\input\filen@me\fi}
\ifx\UNd@FiNeD\prefix\def\prefix{}\fi 
\newread\ch@ckfile
\immediate\openin\ch@ckfile=\jobname.def
\ifeof\ch@ckfile\message{<< (\jobname.def DOES NOT EXIST in this
pass) >>} \else
\def\DefWarn#1{}%
\closein \ch@ckfile%
\input\jobname.def\fi
\def\listcontent{
\immediate\openin\ch@ckfile=\jobname.tab 
\ifeof\ch@ckfile\message{no file \jobname.tab, no table of
contents this
pass}%
\else\closein\ch@ckfile\centerline{\bf\iffrancmode Table des
mati\`eres \else Contents\fi}\nobreak\medskip%
{\baselineskip=12pt\parskip=0pt\catcode`\@=11\input\jobname.tab
\catcode`\@=12\bigbreak\bigskip}\fi}
\newcount\nosection
\newcount\nosubsection
\newcount\neqno
\newcount\notenumber
\newcount\nofigure
\newcount\notable
\newcount\noexerc
\newif\ifappmode
\def\equation{\jobname.equ}
\newwrite\equa

\newdimen\hulp
\def\maketitle#1{
\edef\oneliner##1{\centerline{##1}}
\edef\twoliner##1{\vbox{\parindent=0pt\leftskip=0pt plus
1fill\rightskip=0pt plus 1fill
                     \parfillskip=0pt\relax##1}}
\setbox0=\vbox{#1}\hulp=0.5\hsize
                 \ifdim\wd0<\hulp\oneliner{#1}\else
                 \twoliner{#1}\fi}
\def\preprint#1{\ifdraftmode\gdef\prepname{\jobname/#1}\else%
\gdef\prepname{#1}\fi\hfill{
\expandafter{\prepname}}\vskip20mm}
\def\title#1\par{\gdef\titlename{#1}
\maketitle{\ssbx\uppercase\expandafter{\titlename}} \vskip20truemm
\nosection=0 \neqno=0 \notenumber=0 \nofigure=0 \notable=0
\def\prefix{}
\appmodefalse \mark{\the\nosection} \message{#1}
\immediate\openout\equa=\equation }
\def\abstract{\vskip8mm\iffrancmode\centerline{R\'ESUM\'E}\else%
\centerline{ABSTRACT}\fi \smallskip \begingroup\narrower
\elevenpoint\baselineskip10pt}
\def\endabstract{\par\endgroup \bigskip}
\def\section#1\par{\vskip0pt plus.1\vsize\penalty-100\vskip0pt plus-.1
\vsize\bigskip\vskip\parskip
\ifnum\nosection=0\ifappmode\relax\else\writetoc
\fi\fi
\advance\nosection by 1\global\nosubsection=0\global\neqno=0
\vbox{\noindent\bf{\hyperdef\hypernoname{section}{\prefix\the\nosection}%
{\prefix\the\nosection}\ #1}}
\writetoca{{\string\hyperref{}{section}{\prefix\the\nosection}%
{\prefix\the\nosection}} {#1}} \message{\the\nosection\ #1}
\mark{\the\nosection}\bigskip\noindent }
\def\appendix#1#2\par{\bigbreak \nosection=0 \appmodetrue \notenumber=0 \neqno=0
\def\prefix{A}
\mark{\the\nosection} \message{APPENDICES} {\centerline{\bf
Appendices}
\hyperdef\hypernoname{appendix}{\prefix}{
\leftline{\uppercase\expandafter{#1}}
\leftline{\uppercase\expandafter{#2}}}}
\writetoca{\string\hyperref{}{appendix}{\prefix}{Appendices}.\ #1 \ #2}%
}
\def\subsection#1\par {\vskip0pt plus.05\vsize\penalty-100\vskip0pt
plus-.05\vsize\bigskip\vskip\parskip\advance\nosubsection by 1
\vbox{\noindent\it{\hyperdef\hypernoname{subsection}{\prefix\the\nosection.%
\the\nosubsection}{\prefix\the\nosection.\the\nosubsection\ #1}}}%
\smallskip\noindent
\writetoca{{\string\hyperref{}{subsection}{\prefix\the\nosection.%
\the\nosubsection}{\prefix\the\nosection.\the\nosubsection}} {#1}}
\message{\the\nosection.\the\nosubsection\ #1} }
\def\note #1{\advance\notenumber by 1
\footnote{$^{\the\notenumber}$}{\sevenrm #1}}

\parindent=1em
\newinsert\margin
\dimen\margin=\maxdimen \count\margin=0 \skip\margin=0pt
\def\sslbl#1{\DefWarn#1%
\ifdraftmode{\hfill\escapechar-1{\rlap{\hskip-1mm%
\sevenrm\string#1}}}\fi%
\ifnum\nosection=0\if\prefix{}\xdef#1{}%
\edef\ewrite{\write\equa{{\string#1}}%
\write\equa{}}\ewrite%
\else
\xdef#1{\noexpand\hyperref{}{appendix}{\prefix}{\prefix}}%
\edef\ewrite{\write\equa{{\string#1},\prefix}%
\write\equa{}}\ewrite%
\writedef{#1\leftbracket#1} \fi
\else%
\ifnum\nosubsection=0%
\xdef#1{\noexpand\hyperref{}{section}{\prefix\the\nosection}%
{\prefix\the\nosection}}%
\edef\ewrite{\write\equa{{\string#1},\prefix\the\nosection}%
\write\equa{}}\ewrite%
\writedef{#1\leftbracket#1}
\else%
\xdef#1{\noexpand\hyperref{}{subsection}{\prefix\the\nosection.%
\the\nosubsection}{\prefix\the\nosection.\the\nosubsection}}%
\writedef{#1\leftbracket#1}
\edef\ewrite{\write\equa{{\string#1},\prefix\the\nosection%
.\the\nosubsection}\write\equa{}}\ewrite\fi\fi}%
\newwrite\tfile \def\writetoca#1{}
\def\writetoc{\immediate\openout\tfile=\jobname.tab
\def\writetoca##1{{\edef\next{\write\tfile{\noindent ##1 \string\leaderfill%
\noexpand\number\pageno\par}}\next}}}

%
\def\nolabels{\def\wrlabeL##1{}\def\eqlabeL##1{}\def\reflabeL##1{}}
\def\writelabels{\def\wrlabeL##1{\leavevmode\vadjust{\rlap{\smash%
{\line{{\escapechar=` \hfill\rlap{\sevenrm\hskip.03in\string##1}}}}}}}%
\def\eqlabeL##1{{\escapechar-1\rlap{\sevenrm\hskip.05in\string##1}}}%
\def\reflabeL##1{\noexpand\llap{\noexpand\sevenrm\string\string\string##1}}}
\nolabels

\global\newcount\refno \global\refno=1
\newwrite\rfile
\def\ref{[\hyperref{}{reference}{\the\refno}{\the\refno}]\nref}
\def\nref#1{\DefWarn#1%
\xdef#1{[\noexpand\hyperref{}{reference}{\the\refno}{\the\refno}]}%
\writedef{#1\leftbracket#1}%
\ifnum\refno=1\immediate\openout\rfile=\jobname.ref\fi
\chardef\wfile=\rfile\immediate\write\rfile{\noexpand\item{[\noexpand\hyperdef%
\noexpand\hypernoname{reference}{\the\refno}{\the\refno}]\ }%
\reflabeL{#1\hskip.31in}\pctsign}\global\advance\refno
by1\findarg}
\def\findarg#1#{\begingroup\obeylines\newlinechar=`\^^M\pass@rg}
{\obeylines\gdef\pass@rg#1{\writ@line\relax #1^^M\hbox{}^^M}%
\gdef\writ@line#1^^M{\expandafter\toks0\expandafter{\striprel@x #1}%
\edef\next{\the\toks0}\ifx\next\em@rk\let\next=\endgroup\else\ifx\next\empty%
\else\immediate\write\wfile{\the\toks0}\fi\let\next=\writ@line\fi\next\relax}}
\def\striprel@x#1{} \def\em@rk{\hbox{}}
\def\lref{\begingroup\obeylines\lr@f}
\def\lr@f#1#2{\DefWarn#1\gdef#1{\let#1=\UNd@FiNeD\ref#1{#2}}\endgroup\unskip}

\def\addref#1{\immediate\write\rfile{\noexpand\item{}#1}} 
\def\listrefs{{}\vfill\supereject\immediate\closeout\rfile\writestoppt
\baselineskip=14pt\centerline{{\bf\iffrancmode R\'eferences\else References%
\fi}}
\bigskip{\parindent=20pt%
\frenchspacing\escapechar=` \input
\jobname.ref\vfill\eject}\nonfrenchspacing}
\def\startrefs#1{\immediate\openout\rfile=\jobname.ref\refno=#1}
\def\xref{\expandafter\xr@f}\def\xr@f[#1]{#1}
\def\refs#1{\count255=1[\r@fs #1{\hbox{}}]}
\def\r@fs#1{\ifx\UNd@FiNeD#1\message{reflabel \string#1 is undefined.}%
\nref#1{need to supply reference \string#1.}\fi%
\vphantom{\hphantom{#1}}{\let\hyperref=\relax\xdef\next{#1}}%
\ifx\next\em@rk\def\next{}%
\else\ifx\next#1\ifodd\count255\relax\xref#1\count255=0\fi%
\else#1\count255=1\fi\let\next=\r@fs\fi\next}
\newwrite\lfile
{\escapechar-1\xdef\pctsign{\string\%}\xdef\leftbracket{\string\{}
\xdef\rightbracket{\string\}}\xdef\numbersign{\string\#}}
\def\writedefs{\immediate\openout\lfile=\jobname.def \def\writedef##1{%
{\let\hyperref=\relax\let\hyperdef=\relax\let\hypernoname=\relax
 \immediate\write\lfile{\string\def\string##1\rightbracket}}}}%
\def\writestop{\def\writestoppt{\immediate\write\lfile{\string\pageno%
\the\pageno\string\startrefs\leftbracket\the\refno\rightbracket%
\string\def\string\secsym\leftbracket\secsym\rightbracket%
\string\secno\the\secno\string\meqno\the\meqno}\immediate\closeout\lfile}}
\def\writestoppt{}\def\writedef#1{}
\writedefs
\def\biblio\par{\vskip0pt plus.1\vsize\penalty-100\vskip0pt plus-.1
\vsize\bigskip\vskip\parskip \message{Bibliographie}
{\leftline{\bf \hyperdef\hypernoname{biblio}{bib}{Bibliographical
Notes}}} \nobreak\medskip\noindent\frenchspacing
\writetoca{\string\hyperref{}{biblio}{bib}{Bibliographical Notes}}}%

\def\biblionote{\iffrancmode Notes Bibliographiques\else Bibliographical Notes
\fi}
\def\beginbib\par{\vskip0pt plus.1\vsize\penalty-100\vskip0pt plus-.1
\vsize\bigskip\vskip\parskip \message{Bibliographie}
{\leftline{\bf \hyperdef\hypernoname{biblio}{\the\nosection}%
{\biblionote}}} \nobreak\medskip\noindent\frenchspacing
\writetoca{\string\hyperref{}{biblio}{\the\nosection}%
{\biblionote}}}%

\def\Exercises{\iffrancmode Exercices\else Exercises
\fi}
\def\exerc\par{\vskip0pt plus.1\vsize\penalty-100\vskip0pt plus-.1
\vsize\bigskip\vskip\parskip\global\noexerc=0 \message{Exercises}
{\leftline{\bf
\hyperdef\hypernoname{exercise}{\the\nosection}{\Exercises}}}
\nobreak\medskip\noindent\frenchspacing
\writetoca{\string\hyperref{}{exercise}{\the\nosection}{\Exercises}}
}
\def\esubsec{\ifnum\noexerc=0\vskip-12pt\else\vskip0pt plus.05\vsize%
\penalty-100\vskip0pt plus-.05\vsize\bigskip\vskip\parskip\fi%
\global\advance\noexerc by 1
\hyperdef\hypernoname{exercise}{\the\nosection.\the\noexerc}{}%
\vbox{\noindent\it \iffrancmode Exercice\else Exercise\fi\
\the\nosection.\the\noexerc}\smallskip\noindent}
\def\exelbl#1{\ifdraftmode{\hfill\escapechar-1{\rlap{\hskip-1mm%
\sevenrm\string#1}}}\fi%
{\xdef#1{\noexpand\hyperref{}{exercise}{\the\nosection.\the\noexerc}%
{\the\nosection.\the\noexerc}}}%
\edef\ewrite{\write\equa{{\string#1}\the\nosection.\the\noexerc}%
\write\equa{}}\ewrite%
\writedef{#1\leftbracket#1}}
\def\eqnn{\global\advance\neqno by 1 \ifinner\relax\else%
\eqno\fi(\prefix\the\nosection.\the\neqno)}
%
\def\eqnd#1{\DefWarn#1%
\global\advance\neqno by 1
{\xdef#1{($\noexpand\hyperref{}{equation}{\prefix\the\nosection.\the\neqno}%
{\prefix\the\nosection.\the\neqno}$)}}
\ifinner\relax\else\eqno\fi(\hyperdef\hypernoname{equation}{\prefix\the%
\nosection.\the\neqno}{\prefix\the\nosection.\the\neqno})
\writedef{#1\leftbracket#1}
\ifdraftmode{\escapechar-1{\rlap{\hskip.2mm\sevenrm\string#1}}}\fi
\edef\ewrite{\write\equa{{\string#1},(\prefix\the\nosection.\the\neqno)
{\noexpand\number\pageno}}\write\equa{}}\ewrite}
%
\def\checkm@de#1#2{\ifmmode{\def\f@rst##1{##1}\hyperdef\hypernoname{equation}%
{#1}{#2}}\else\hyperref{}{equation}{#1}{#2}\fi}
\def\f@rst#1{\c@t#1a\em@ark}\def\c@t#1#2\em@ark{#1}
\def\eqna#1{\DefWarn#1%
\global\advance\neqno by1\ifdraftmode{\hfill%
\escapechar-1{\rlap{\sevenrm\string#1}}}\fi%
\xdef #1##1{(\noexpand\relax\noexpand%
\checkm@de{\prefix\the\nosection.\the\neqno\noexpand\f@rst{##1}1}%
{\hbox{$\prefix\the\nosection.\the\neqno##1$}})}
\writedef{#1\numbersign1\leftbracket#1{\numbersign1}}%
}
%

%
\def\em@rk{\hbox{}}
\def\xeqn{\expandafter\xe@n}\def\xe@n(#1){#1}
\def\xeqna#1{\expandafter\xe@na#1}\def\xe@na\hbox#1{\xe@nap #1}
\def\xe@nap$(#1)${\hbox{$#1$}}
\def\eqns#1{(\e@ns #1{\hbox{}})}
\def\e@ns#1{\ifx\UNd@FiNeD#1\message{eqnlabel \string#1 is undefined.}%
\xdef#1{(?.?)}\fi{\let\hyperref=\relax\xdef\next{#1}}%
\ifx\next\em@rk\def\next{}%
\else\ifx\next#1\xeqn#1\else\def\n@xt{#1}\ifx\n@xt\next#1\else\xeqna#1\fi
\fi\let\next=\e@ns\fi\next}
\def\figure#1#2{\global\advance\nofigure by 1 \vglue#1%
\hyperdef\hypernoname{figure}{\the\nofigure}{}%
{\elevenpoint \setbox1=\hbox{#2}
\ifdim\wd1=0pt\centerline{Fig.\ \the\nofigure\hskip0.5mm}%
\else\def\caption{Fig.\ \the\nofigure\quad#2\hskip0mm}
\setbox0=\hbox{\caption} \ifdim\wd0>\hsize\noindent Fig.\
\the\nofigure\quad#2\else
                 \centerline{\caption}\fi\fi}\par}
\def\lfigure#1#2{\global\advance\nofigure by
1\vglue#1%
\hyperdef\hypernoname{figure}{\the\nofigure}{}%
\leftline{\elevenpoint\hskip10truemm  Fig.\ \the\nofigure\quad
#2}}
\def\figlbl#1{\ifdraftmode{\hfill\escapechar-1{\rlap{\hskip-1mm%
\sevenrm\string#1}}}\fi%
{\xdef#1{\noexpand\hyperref{}{figure}{\the\nofigure}%
{\the\nofigure}}}%
\edef\ewrite{\write\equa{{\string#1}\the\nofigure}%
\write\equa{}}\ewrite%
\writedef{#1\leftbracket#1}}
\def\tablbl#1{\global\advance\notable by
1\ifdraftmode{\hfill\escapechar-1{\rlap{\hskip-1mm%
\sevenrm\string#1}}}\fi%
{\xdef#1{\noexpand\hyperref{}{table}{\the\notable}%
{\the\notable}}}%
\hyperdef\hypernoname{table}{\the\notable}{}%
\edef\ewrite{\write\equa{{\string#1}\the\notable}%
\write\equa{}}\ewrite%
\writedef{#1\leftbracket#1}}

\catcode`@=12




\francmodefalse

\def\*{********** new, please check **********}

\def\slam#1{\mkern-1.5mu\raise-0.2pt\hbox{$\scriptstyle{\not}$}\mkern1.2mu
#1\mkern 0.7mu}
\def\r{{\rm r}}

\def\D{{\rm D}}
\def\r{{\rm r}}
\def\psib{\psi}
\def\phib{\phi}
\def\pib{\pi}

\def\biggclb{\mathopen{\vcenter{\hbox{\tenex\char'32}}}}

\def\biggcrb{\mathclose{\vcenter{\hbox{\tenex\char'33}}}}
\def\bigbra{\mathopen{\vcenter{\hbox{\tenex\char'12}}}}
\def\bigket{\mathclose{\vcenter{\hbox{\tenex\char'13}}}}


\def\half{\frac{1}{2}}
\def\lb{\hfil\break}
\def\del{\partial}

\def\delslash{\mathord{\not\mathrel{\partial}}}
\def\psibar{\bar\psi}

\def\phivec{\vec \phi}

\def\ma{m_\varphi}
\def\mpsi{m_\psi}
\def\del{\partial}

\def\mpone{m_{\psi_1}}
\def\mptwo{m_{\psi_2}}

\def\sqr#1#2{{\vcenter{\hrule height.#2pt
     \hbox{\vrule width.#2pt height#1pt \kern#1pt
          \vrule width.#2pt}
       \hrule height.#2pt}}}

\def\to{\rightarrow}
\def\half{\frac{1}{2}}
\def\del{\partial}
\def\phib{{\phi}}
\def\Vc{\Omega_c}
\newskip\tableskipamount \tableskipamount=10pt plus 4pt minus 4pt

\def\upar{\uparrow\kern-9.\exec1pt\lower.2pt \hbox{$\uparrow$}}



\preprint{May 2003}

\title{Quantum Field Theory in the Large $N$ Limit: a review}

\centerline{{\bf Moshe~Moshe}${}^{1 ,a,c}$ and {\bf
Jean~Zinn-Justin}${}^{2,b}$}
\bigskip
{\baselineskip14pt
\centerline{${}^1$\it Department of Physics, Technion - Israel
Institute of Technology,} \centerline{\it Haifa, 32000 ISRAEL}
\smallskip
\centerline{${}^2$\it  Service de Physique Th\'eorique* CEA/Saclay}
\centerline{\it F-91191 Gif-sur-Yvette C\'edex, FRANCE}
\centerline{\it and} \centerline{\it Institut de Math\'ematiques
de Jussieu--Chevaleret,} \centerline{\it Universit\'e de Paris
VII} }
\footnote{}{(a)~email: moshe@physics.technion.ac.il}
\footnote{}{(b)~email: zinn@spht.saclay.cea.fr} \footnote{}{(c)
~Supported in part by the Israel Science Foundation grant number
193-00}
 \footnote{}{${^*}$Laboratoire de la Direction des Sciences
de la Mati\`ere du Commissariat \`a \indent \ l'Energie Atomique
Unit\'e de recherche associ\'ee au CNRS} \vskip 3mm


\abstract We review the solutions of $O(N)$ and $U(N)$ quantum field theories  in
the large $N$ limit and as $1/N$ expansions, in the case of vector
representations. Since invariant composite fields have small
fluctuations for large $N$, the method relies on constructing
effective field theories for composite fields after integration
over the original degrees of freedom. We first solve  a general
scalar $U(\phib^2)$ field theory for $N$ large and discuss various
non-perturbative physical issues such as critical behaviour.  We
show how large $N$ results can also be obtained from variational
calculations.We illustrate these ideas by showing that the large
$N$ expansion allows to relate the $(\phib^2)^2$ theory and the
non-linear $\sigma$-model, models which are renormalizable in
different dimensions. Similarly, a  relation
between $CP(N-1)$ and abelian Higgs models is exhibited.
Large $N$
techniques also allow solving self-interacting fermion models.
A relation between  the Gross--Neveu,   a theory with a four-fermi
self-interaction, and a Yukawa-type theory renormalizable in four
dimensions then follows. We discuss dissipative dynamics,
which is relevant to the approach to equilibrium, and which in
some formulation exhibits quantum mechanics supersymmetry. This
also serves as an introduction to the study of the 3D
supersymmetric quantum  field theory. Large
$N$ methods are useful in problems that involve a crossover between different dimensions. We thus  briefly discuss  finite
size effects, finite temperature scalar and supersymmetric field
theories. We also use large $N$ methods to investigate the weakly interacting Bose gas. The solution of the general scalar $U(\phib^2)$ field theory is then applied to other  issues like tricritical behaviour and double
scaling limit.
\endabstract
\vfill\eject
\listcontent
\vfill\eject
\section Introduction

Studies of the non-perturbative features of quantum field
theories are at the forefront of theoretical physics research.
 Though remarkable progress
has been achieved in recent years, still, some of the
more fundamental questions have only a descriptive answer,
whereas non-perturbative calculable schemes are seldom at hand.
The absence of calculable
dynamics in realistic models is often supplemented
by simpler models in which the essence of the dynamics
is revealed. Such a calculable framework for exploring theoretical
ideas is given by large $N$ quantum field theories. \par
Very early quantum field  theorists have looked for methods to solve
field theory beyond perturbation theory and obtain confirmation of
perturbative results. Moreover, some important physical questions are often intrinsically non-perturbative. Let us mention, for
illustration, the problem of fermion-pair condensation. A number of
similar schemes were proposed, all of mean-field theory nature,
variational methods, self-consistent approximations, all reducing the
interacting theory to a free fermion theory with self-consistently
determined parameters. For example, the quartic fermion self-interaction
$(\bar\psi\psi)^2$ would be replaced by a term  proportional to
$\langle \bar\psi\psi\rangle \bar\psi\psi$, where
$\langle \bar\psi\psi\rangle $ is the free field average. However,
all these methods have several drawbacks:  it is unclear how to
improve the results systematically, the domain of validity of the
approximations are often unknown,  in fact there is no obvious small
parameter.  To return to the fermion example, one realizes that the
approximation would be justified if for some reasons the fluctuations
of the composite field $\bar\psi\psi$ were much smaller than the
fluctuations of the fermion field itself.
\par
Large $N$ techniques solve this problem in the spirit of the
central limit theorem of the theory of probabilities. If the field
has $N$ components, in the large $N$ limit scalar (in the group
sense) composite fields are sums of many terms and therefore may
have small fluctuations (at least if the different terms are sufficiently
uncorrelated). Therefore, if we are able to construct an effective
field theory for the scalars, integrating out the original degrees
of freedom, we can solve the field theory not only in the large
$N$ limit, but also in a systematic $1/N$ expansion. On the
technical level, one notes that in vector representations the
number of independent scalars is finite and independent of $N$,
unlike what happens for matrix representations. This explains why
vector models have been solved much more generally than matrix
models.
\par
 In this review we describe a few applications of large $N$
techniques to quantum field theories (QFT) with $O(N)$ or $U(N)$
symmetries, where the fields are in the {\it vector}\/
representation  \ref\rZJTai{Some sections are
directly inspired from  J. Zinn-Justin, lectures given at {\it 11$^{\rm th}$ Taiwan Spring School}, Taipei 1997, hep-th/9810198.}. A summary of results are presented here  in the
study of the phase structure of quantum field theories. It is
demonstrated that large $N$ results nicely complement results
obtained from more conventional perturbative renormalization group
(RG) \ref\rbook{For a general
background with analogous notation see  J. Zinn-Justin,
{\it Quantum Field Theory and Critical Phenomena}, Clarendon
Press (Oxford 1989, fourth ed.~2002).}. Indeed, the shortcoming of the latter method is that it
mainly applies to gaussian or near gaussian fixed points. This
restricts space dimension to dimensions in which the corresponding
effective QFT is renormalizable, or after dimensional
continuation, to the neighbourhood of such dimensions. In some
cases, large $N$ techniques allow a study in generic dimensions.
In this review we will, in particular, stress two points: first,
it is always necessary to check that the $1/N$ expansion is both
IR finite and renormalizable.  This is essential for the stability of the large
$N$ results and the existence of a $1/N$ expansion. Second, the large $N$
expansion is just a technique, with its own (often unknown)
limitations and it should not be discussed in isolation. Instead,
as we shall do in the following examples, it should be combined, when possible,
with other perturbative techniques and the reliability of the
$1/N$ expansion should be inferred from the general consistency of
all results.\par
Second-order phase transitions in classical
statistical physics provide a first illustration of
the usefulness of the large $N$ expansion. Due to the divergence
of the correlation length at the critical temperature, one finds
that near $T_c$, system share universal properties which can be
described by effective continuum quantum field theories. The
$N$-vector model that we discuss in {\bf Sections \label{\scfivN}} and {\bf \label{\ssNsymspace}} is the
simplest example but it has many applications since it allows to
describe the critical properties of systems like vapour--liquid,
binary mixtures, superfluid Helium or ferromagnetic transitions as
well as the statistical properties of polymers. Before showing
what kind of information can be provided by large $N$ techniques,
we will first shortly recall what can be learned from perturbative
RG methods.
Long distance properties can be described by a
$(\phib^2)^2$ field theory in which analytic calculations can be performed
only in an $\varepsilon=4-d$ expansion. From lattice model considerations, one expects that
the same properties can also be derived from a different
 QFT,  the $O(N)$ non-linear $\sigma$ model, which, however, can be solved only
as an $\varepsilon=d-2$ expansion. It is somewhat surprising
that the same statistical model can be described by two different
theories. Since the results derived in this  way are valid {\it a
priori}\/ only for $\varepsilon$ small, there is no overlap to
test the consistency. The large $N$ expansion  enables one  to
discuss generic dimensions and thus to understand the relation
between both field theories. \par
Similar large $N$ techniques can also be applied to other non-linear models
and we briefly examine the example of the $CP(N-1)$ model. \par
Four-fermi interactions have been
proposed to generate a composite Higgs particle in four
dimensions, as an alternative to a Yukawa-type theory, as one
finds in the Standard Model. Again, in the specific example of
the Gross--Neveu model, in {\bf Section \label{\ssNfermions}}  we will use large $N$
techniques to clarify the relations between these two approaches.
We will finally briefly investigate other
models with chiral properties, like massless QED or the Thirring
model.\par
Preceding the
discussion of supersymmetric models, we study  in {\bf Section \label{\ssCrDY}}
critical dynamics of purely dissipative systems, which are known
to provide a simple field theory extension of supersymmetric quantum mechanics. We
then study in {\bf Section \label{\ssNSUSYsc}} two SUSY models in two and three dimensions: the
supersymmetric $\phi^4$ field theory, which is shown to have a
peculiar phase structure in the large $N$ limit, and a
supersymmetric non-linear $\sigma$ model. \par
Other applications of the large $N$ expansion include finite size effects ({\bf Section \label{\ssNFSS}}) and finite temperature field theory to which we devote {\bf Section \label{\scFTQFT}} for non-supersymmetric theories and  {\bf Section \label{\ssNSUSYT}} for the supersymmetric theories of  {\bf Section \label{\ssNSUSYsc}}. In these situations a dimensional crossover occurs
between the large size or zero temperature situation, where the
infinite volume theory is relevant, to a dimensionally reduced
theory in the small volume or high temperature limit. Both
effective field theories being renormalizable in different
dimensions, perturbative RG cannot describe correctly both
situations. Again, large $N$ techniques help    understanding
the crossover. We often compare in this review the large $N$
results with those obtained by variational methods, since it is
often possible to set up variational calculations which parallel
the large $N$ limit calculations.\par
The effect of weak  interactions on Bose gases at the Bose--Einstein condensation temperature are analyzed in {\bf Section \label{\scBEC}} where large $N$ techniques are
employed for the non-perturbative calculations of physical
quantities.
\par
We then return  in {\bf Section \label{\ssdblescal}} to general scalar boson field theories,
and examine multi-critical points (where the large $N$ technique
will show some obvious limitations), and the double scaling limit,
a toy model for discussing problems encountered in matrix models
of 2D quantum gravity.
\par
In  this review, we also discuss the breaking of scale
invariance. In most quantum field theories spontaneous and
explicit breaking of scale invariance  occur simultaneously and
thus the breaking of scale symmetry is not normally accompanied by
the appearance of a massless Nambu--Goldstone boson. Spontaneous
breaking of scale invariance, unaccompanied by explicit breaking,
associated with a non-zero fixed point and the creation of a
massless bound state is demonstrated in {\bf Section \label{\ssNscaleinv}}.
In some cases the dynamics by which scale invariance is broken in
a theory which has no trace anomalies in perturbation theory is
directly related to the breaking of internal symmetry. This
appears in the phase structure of $O(N)\times O(N)$
symmetric models  for $N$ large, where the breaking of scale invariance is
directly related to the breaking of the internal symmetry.
The spontaneous breaking of scale invariance in a supersymmetric,
$O(N)$ symmetric vector model in three dimensions was also studied
in {\bf Section \label{\ssNSUSYsc}},
were one finds the creation of a massless fermionic bound state as
the supersymmetric partner of the massless boson in the
supersymmetric ground state.
 \par

\vfill\eject
\def\vv{v}
\def\tu{\tau}
\def\ro{\rho}
\nref\rBKStan{As shown by   H.E. Stanley, {\it Phys. Rev.} 176
(1968) 718, the large $N$-limit of the classical $N$-vector model
coincides with the spherical model solved in ref. \rBerlin. }
\nref\rBerlin{ T.H. Berlin and M. Kac, {\it Phys. Rev.} 86 (1952)
821.} \nref\rWWFis{The modern formulation of the RG ideas is due
to: \rf K.G. Wilson, {\it Phys. Rev.} B4 (1971) 3174, {\it
ibidem}\/ 3184; \rf
  and presented in an expanded form in \rWKogut. }
\nref\rWKogut{K.G. Wilson and J. Kogut, {\it Phys. Rep.} 12C
(1974) 75. }
 \nref\rWFisher{The idea of the $\varepsilon$-expansion
is due to \rf K.G. Wilson and M.E. Fisher, {\it Phys. Rev.  Lett.}
28 (1972) 240. } \nref\rAMSDSW{Early work on calculating critical
properties for large $N$ includes \rf R. Abe, {\it Prog. Theor.
Phys.} 48 (1972) 1414; 49 (1973) 113, 1074, 1877 and refs.~\refs{\rMa{--}\rWilson}.} \nref\rMa{S.K. Ma, {\it Phys. Rev.
Lett.} 29 (1972) 1311 ; {\it Phys. Rev.} A7 (1973) 2172 .}
\nref\rSuzuki{M. Suzuki, {\it Phys. Lett.} 42A (1972) 5; {\it
Prog. Theor. Phys.} 49 (1973) 424, 1106, 1440 .} \nref\rFerrel{
R.A. Ferrel and D.J. Scalapino, {\it Phys. Rev. Lett.} 29 (1972)
413.} \nref\rWilson{K.G. Wilson, {\it Phys. Rev.} D7 (1973) 2911.}
\nref \rFAAAH{The spin--spin correlation in zero field is obtained
in \rf M.E. Fisher and A. Aharony, {\it Phys. Rev. Lett.} 31
(1973) 1238 and refs.~\refs{\rAharony,\rAbe }.}\nref\rAharony{ A.
Aharony, {\it Phys. Rev.} B10 (1974) 2834.}\nref\rAbe{ R. Abe and
S. Hikami, {\it Prog. Theor. Phys.} 51 (1974) 1041.}

\nref\rBrWa{The contribution of order $1/N$ to the equation of
state is given in \rf E. Br\'ezin and D.J. Wallace, {\it Phys.
Rev.} B7 (1973) 1967.} \nref\rSKMa{The exponent $\omega$ has been
calculated to order $1/N$ by S.K. Ma, {\it Phys. Rev.} A10 (1974)
1818.} \nref\rMaDGvi{See also the review of S.K. Ma  in {\it Phase
Transitions and Critical Phenomena} vol. 6, C. Domb and M.S. Green
eds. (Academic Press, London 1976).}
\nref\rBLGZJDBvi{The study of the large $N$ limit by the steepest
descent method is explained in \rf
 E. Br\'ezin, J.C. Le Guillou and J. Zinn-Justin, contribution to {\it Phase
Transitions and Critical Phenomena} vol. 6, C. Domb and M.S. Green
eds. (Academic Press, London 1976). It is applied to more general
scalar field theories in \rHalpern .}\nref\rHalpern{
 M.B. Halpern, {\it Nucl. Phys.} B173 (1980) 504.}
\nref\rBarMos{For a Hartree--Fock variational approach to large
$N$ theories and large $N$  QFT at finite temperature see  \rf
W.A. Bardeen and M. Moshe, {\it Phys. Rev.} D28 (1983) 1372 and
{\it Phys. Rev.} D34 (1986) 1229.} \nref\rBrZJsig{The non-linear
$\sigma $-model is discussed in the spirit of this review in \rf
E. Br\'ezin and  J. Zinn-Justin, {\it Phys. Rev. Lett.} 36 (1976)
691; {\it Phys. Rev.} B14 (1976) 3110.} \nref\rnlslat{Lattice
calculations of the non-linear $\sigma $ model with the large $N$
expansion are reported in \rf M. Campostrini, P. Rossi, {\it
Phys. Lett.} B242 (1990) 81 and in \rBiscari .}\nref\rBiscari{P.
Biscari, M. Campostrini, P. Rossi, {\it Phys. Lett.} B242 (1990)
225.} \nref\rCPN{The $CP(N-1)$ model is discussed in two
dimensions with the large $N$ expansion in \rf M. L\"uscher, {\it
Phys. Lett.} 78B (1978) 465 and in refs.~\refs{\rDadda{--}\rCampos}.}\nref\rDadda{ A. D'Adda, P. Di
Vecchia and M. L\"uscher, {\it Nucl. Phys.} B146 (1978) 63; B152
(1979) 125. }\nref\rMunster{ G. M\"unster, {\it Nucl. Phys.} B218
(1983) 1.} \nref\rDiVecch{P. Di Vecchia, R. Musto, F. Nicodemi,
R. Pettorino and P. Rossi, {\it Nucl. Phys.} B235 (1984)
478.}\nref\rCampos{ M. Campostrini, P. Rossi,
    {\it Phys. Lett.} B272 (1991) 305;
    {\it Phys. Rev.} D45 (1992) 618; Erratum-{\it ibid.} D46 (1992) 2741.}
\nref\rBrCaPe{Some finite size calculations are reported in \rf E.
Br\'ezin, {\it J. Physique (Paris)} 43 (1982) 15 and in refs.~\refs{\rSingh {--}\rCarac}. } \nref\rSingh{ S. Singh and
R.K.~Pathria, {\it Phys. Rev.} B31 (1985) 4483.} \nref\rCompost{
M. Campostrini, P. Rossi, {\it Phys. Lett.} B255 (1991) 89.
}\nref\rCarac{S. Caracciolo and A. Pelissetto, {\it Phys. Rev.}
D58 (1998) 105007, hep-lat/9804001.} \nref\rparamet{The parametric
representation has been introduced in \rf P. Schofield, J.D.
Litster and J.T. Ho, {\it Phys. Rev. Lett.} 23 (1969) 1098  and in
\rJoseph. } \nref\rJoseph{B.D. Josephson, {\it J. Phys. C} 2
(1969) 1113.} \nref\rSymanza{K.~Symanzik, Carg\`ese lectures, DESY
preprint 73/58, Dec 1973; {\it Lett. Nuovo Cim.} 8 (1973) 771;
  {\it    Kyoto 1975, Proceedings, Lecture Notes In Physics}, Berlin 1975, 102-106.
  \rf
A similar analysis in the case of theories renormalizable beyond
perturbation theory (like the non-linear $\sigma $ or Thirring
models) is found in \rParisi .} \nref\rParisi{G. Parisi, {\it
Nucl. Phys.} B100 (1975) 368.}
 \nref\rTriv{On the triviality of $\phi^4$ theory in $d=4$ dimensions
see reference 8 in  K.G. Wilson, {\it Phys. Rev.}   D6 (1972) 419
and the review \rCalla .} \nref\rCalla{ D.J.E. Callaway, {\it
Phys. Rep.} 167 (1988) 241.} \nref\rUVrenorm{B. Lautrup, {\it
Phys. Lett.} B69 (1977) 109 and refs.~\refs{\rtHooft,\rPari}.
{}For more references see also the reprint volume \refs\rLeGuill
.} \nref\rtHooft{ G. 't Hooft, {\it Erice Lectures 1977}.}
\nref\rPari{ G. Parisi, {\it Phys. Lett.} 76B (1978) 65. }
\nref\rLeGuill { {\it Large Order Behaviour of Perturbation
Theory}, J.C. Le Guillou and J. Zinn-Justin eds. (North Holland,
Elsevier Science Pub., Amsterdam 1989).} \nref\rDavid{F. David,
{\it Nucl. Phys.} B209 (1982) 433.} \nref\rHiggsbound{The bound on
the Higgs mass is discussed in \rf R.F. Dashen, H. Neuberger, {\it
Phys. Rev. Lett.} 50 (1983) 1897 and in refs.~\refs{\rHasen,\rLusche}.} \nref\rHasen{ A. Hasenfratz, K. Jansen,
C.B. Lang, T. Neuhaus, H. Yoneyama, {\it Phys. Lett.} B199( 1987)
531. } \nref\rLusche{M. L\"uscher, P. Weisz, {\it Nucl. Phys.}
B290 (1987) 25, B295 (9188) 65, B318 (1989) 705.}
\nref\rSymanzb{K. Symanzik, DESY preprint 77/05, Jan. 1977.}
\nref\rFRG{Functional renormalization group has been discussed in
the large $N$ limit in \rf M. Reuter, N. Tetradis, C. Wetterich,
{\it Nucl. Phys.} B401 (1993) 567 and in refs.~\refs{\rAttan,\rBerges}.}\nref\rAttan{ M. D'Attanasio, T.R.
Morris, {\it Phys. Lett.} B409 (1997) 363, hep-th/9704094.}
\nref\rBerges{J. Berges, N. Tetradis, C. Wetterich, {\it Phys.
Rept.} 363 (2002) 223, hep-ph/0005122.}

\nref \rPeRoVi{Large-$N$ critical behaviour of $O(N)\times O(m)$
spin models is considered in \rf
 A. Pelissetto, P. Rossi and E.
Vicari, {\it Nucl. Phys.}   B607 (2001) 605 and in
\rGracey.}\nref\rGracey{J.A. Gracey, {\it Nucl. Phys.} B644 (2002)
433, hep-th/0209053.}

\nref\rJaStr {see e.g.~R. Jackiw and A. Strominger, {\it Phys.
Lett.} 99B  (1981)133.} \nref\rNsigamiiD{The non-linear $\sigma $
model two-point functions and mass gap in dimension 2 have been
studied by large $N$ techniques in \rf V. F. M\"uller, T Raddatz
and W. R\"uhl, {\it Nucl. Phys.} B251 (1985) 212, {\it Erratum
ibid.}\/ B259 (1985) 745. } \nref\rANPVN{The consistency of the
$1/N$ expansion to all orders has been proven in \rf I. Ya
Aref'eva, E.R. Nissimov and S.J. Pacheva, {\it Commun. Math.
Phys.} 71 (1980) 213, see also ref. \rVasilev. The method presented
here is taken from ref.~\rbook.}\nref\rVasilev{ A.N. Vasil'ev and
M.Yu. Nalimov, {\it Teor. Mat. Fiz.} 55 (1983) 163.}

\nref\rKTOOVPH{At present the longest $1/N$ series for exponents
and amplitudes are found in \rf I. Kondor and T. Temesvari, {\it
J. Physique Lett. (Paris)} 39 (1978) L99 and refs.~\refs{\rOkabe{--}\rVasilevB}. } \nref\rOkabe{Y. Okabe and M. Oku,
{\it Prog. Theor. Phys.} 60 (1978) 1277, 1287; 61 (1979)
443.}\nref\rVasilevA{ A.N. Vasil'ev, Yu.M. Pis'mak and Yu.R.
Honkonen, {\it Teor. Mat. Fiz.} 46 (1981) 157 {\it ibid.} 47
(1981) 465; {\it ibid.} 50 (1982) 195.}\nref\rVasilevB{
A.~N.~Vasilev, M.~Y.~Nalimov and Y.~R.~Khonkonen, {\it Theor.\
Math.\ Phys.} 58 (1984) 111.} \nref \rKTH{See also I. Kondor, T.
Temesvari and L. Herenyi, {\it Phys. Rev.} B22 (1980) 1451.}
\nref\rPeVi{Results concerning the $\beta$-function at order $1/N$
in the massive theory renormalized at zero momentum have been
reported in \rf A. Pelissetto and E. Vicari, {\it Nucl. Phys.}
B519 (1998) 626, cond-mat/9711078.} \nref\rDerMa{In particular, a
calculation of the dimensions of composite operators are reported
and the consequences for the stability of the fixed point of the
non-linear $\sigma$ model discussed in \rf S. E. Derkachov, A. N.
Manashov, {\it Nucl. Phys.} B522 (1998) 301,  hep-th/9710015; {\it
Phys. Rev. Lett.} 79 (1997) 1423, hep-th/9705020.}
\nref\rCamros{See also M. Campostrini, P. Rossi, {\it  Phys.
Lett.} B242 (1990) 81.} \nref\rGraci{The crossover exponent in
$O(N)$ $\phi^4$ theory at $O(1/N^2)$ is given in \rf J.A. Gracey,
{\it Phys. Rev.} E66 (2002) 027102, cond-mat/0206098.}
\nref\rBrHi{S.~Hikami ,  E.~Br\'ezin, {\it J.~Phys.} A 12, 759-770
(1979) [reprinted in {\it Currents Physics - Sources and
Comments,} CPSC 7,  Le Guillou J.C., Zinn-Justin J. eds.
(North-Holland, 1990)].} \nref\rVegAv{H.J. de Vega, {\it Phys.
Lett.} 98B (1981) 280; J. Avan and H.J. de Vega, {\it Phys. Rev.}
D29 (1984) 2891; {\it ibidem} 2904 [all reprinted in {\it Currents
Physics - Sources and Comments,} CPSC 7,  Le Guillou J.C.,
Zinn-Justin J. eds. (North-Holland, 1990)].} \nref\rBZJFSS{E.
Br\'ezin, J. Zinn-Justin, {\it Nucl. Phys.} B257 (1985) 867.}
\nref\rLaRue{Renormalization of operators is discussed in \rf K.
Lang and W. R\"uhl, {\it Nucl. Phys.} B400 (1993) 597; {\it Z.
Phys.} C61 (1994) 459.} \nref\rMalong{ The case of long range
forces has been discussed in \rf S.K. Ma, {\it Phys. Rev.} A7
(1973) 2172.}
\section{Scalar field theory for  $N$ large: general formalism and applications}

In this section we present a general formalism that allows
studying $O(N)$ symmetric scalar field theories in the large $N$
limit and, more generally, order by order in a large $ N
$-expansion. \par Of particular interest is the $(\phib^2)^2$
statistical field theory that describes the universal properties
of a number of phase transitions. The  study of phase transitions
and critical phenomena in statistical physics has actually been
one of the early applications of large $N$ techniques. It was
realized that an $N$-component spin-model (the spherical model)
could be solved exactly in the large $N$ limit
\refs{\rBKStan,\rBerlin}, and the solutions revealed scaling laws
and non-trivial (i.e.~non-gaussian or mean-field like) critical
behaviour . Later, following Wilson (and Wilson--Fisher)
\refs{\rWWFis{--}\rWFisher} it was discovered that universal
properties of critical systems  could be derived  from the
$(\phib^2)^2$ field theory within the framework of the formal $
\varepsilon =4-d$ expansion, by a combination of perturbation
theory and renormalization group (RG). The peculiarity of this
scheme, whose reliability in the physical dimension $d=3$, and
thus $\varepsilon=1$, could only be guessed, demanded some
independent confirmation. This was provided in particular by
developing a scheme to solve the $N$-component $(\phib^2)^2$ field
theory in the form of an $1/N$ expansion, whose leading order
yields the results of the spherical spin-model
\refs{\rAMSDSW{--}\rMaDGvi}. \sslbl\scfivN
\par
Here, we first solve more general $O(N)$ symmetric field theories
in the large $N$ limit, reducing the problem to a steepest descent
calculation \refs{\rBLGZJDBvi,\rHalpern}. The $( \phib^2)^2$ field
theory is then discussed more thoroughly from the point of view of
phase transitions and critical phenomena. We also stress the
relation between the large $N$ limit and  variational principles
\refs{\rBarMos}. Moreover, some other issues relevant to particle
physics like triviality, renormalons or Higgs mass are examined in
the large $N$ limit. \par A remarkable implication of the large
$N$ analysis is that two different field theories, the
$(\phib^2)^2$ theory and the non-linear $\sigma $ model, describe
the same critical phenomena, a results that holds to all orders in
the $1/N$ expansion \refs{\rBrZJsig{--}\rBiscari}. This result has
several generalizations, leading for instance to a relation
between the $CP(N-1)$  \refs{\rCPN{--}\rCampos} and the abelian
Higgs models.
\par Finally, large $N$ techniques are well adapted to the
analysis of finite size effects in critical systems \rBrCaPe, a
question we investigate in the more convenient formalism of the
non-linear $\sigma $ model in section \label{\ssNFSS}.

\subsection Scalar field theory: the large $N$ formalism

We consider an $O(N)$ symmetric euclidean  action (or classical hamiltonian) for an $N$-component scalar field $\phib$: \sslbl\ssNbosgen
$$ {\cal S} ( \phib )= \int \left[\ud \left[
\partial_{\mu} \phib (x) \right]^2+NU\bigl(\phib^2(x)/N\bigr)  \right] \d^{d}x\,,
\eqnd\eactONgen $$
where $U(\rho )$ is a general polynomial, and the explicit $N$ dependence
has been chosen to lead to a large $N$ limit.
The corresponding partition function is  given by a functional integral:
$$ {\cal Z}= \int \left[ \d \phib (x) \right] \exp \left[-{\cal
S}(\phib)\right] . \eqnd\eONpart $$
To render the perturbative expansion finite, a cut-off $ \Lambda $ consistent with the symmetry  is implied. \par
The solution of the model in the large $N$ limit is based on an idea of
mean field  type: it can be expected that, for $N$ large,
$O(N)$ invariant quantities like
$$\phib^2(x)=\sum_{i=1}^N \phi_i^2(x) $$
self-average and therefore have small
fluctuations (as for the central limit theorem this relies on the assumption that the components
$\phi_i$ are somehow  weakly correlated). Thus, for example,
$$\left< \phib^2(x)\phib^2(y)\right>
\mathop{\sim}_{N
\rightarrow \infty}\left< \phib^2(x)\right>  \left<
\phib^2(y)\right> .$$
This observation suggests to take $\phib^2(x)$ as a dynamical variable, rather than $\phib(x)$ itself.
For this purpose, we introduce two additional fields $\lambda$ and $\rho$ and impose the
constraint $\rho(x)=\phib^2(x)/N$ by an integral over $\lambda$. For each point
of space $x$, we use the identity
$$1=N\int\d\rho\,\delta(\phib^2-N\rho)=
{N\over 4i\pi}\int\d\rho\d\lambda\e^{\lambda(\phib^2-N\rho)/2} ,\eqnd\eNbasicid $$
where the $\lambda$ integration contour runs parallel to the imaginary axis.
The insertion of the identity into the integral \eONpart\ yields a new representation of the partition function:
$$ {\cal Z}= \int [ \d \phib][\d\rho][\d\lambda] \exp \left[-{\cal
S}(\phib,\rho,\lambda)\right]   \eqnd\eONpartgen $$
with
$${\cal S}(\phib,\rho,\lambda)= \int \left[\ud \left[
\partial_{\mu} \phib (x) \right]^2+NU\bigl(\rho(x)\bigr) +\ud\lambda(x)
\bigl(\phib^2(x)-N\rho(x)\bigr) \right]
\d^{d}x\,. \eqnd\eactONgenii $$
The  functional integral \eONpartgen\ is then gaussian in $\phib $, the integral
over the field $ \phib $ can be performed and the dependence on $N$ of
the partition function becomes explicit. Actually, it is convenient
to separate the components of $ \phib $ into one component $\sigma
$, and $ N-1 $ components $ \pib $, and integrate over $ \pib $ only
(for $T<T_c$ it may even be convenient to integrate over only $N-2$
components). This does not affect the large $N$ limit but only the $1/N$ corrections. To generate
$\sigma$-correlation functions, we  add also a source $ H(x)$ (a space-dependent magnetic field in the ferromagnetic language) to the action. The partition function then becomes
$$ {\cal Z} (H)= \int[\d\sigma][\d\rho] [\d \lambda] \exp \left[ -{\cal S}_N
(\sigma,\rho,\lambda)+ \int \d^{d}x\, H (x)\sigma (x) \right]
\eqnd{\eZeff} $$
with
$$ \vbox{\eqalignno{ {\cal S}_N (\sigma,\rho,\lambda)= &
\int \left[\ud \left(\partial_{\mu}\sigma \right)^2
+NU(\rho)+ \ud\lambda (x) (\sigma^2 (x)-N\rho(x)) \right] \d^{d}x & \cr
&\quad +\ud (N-1)\tr\ln \left[ - \nabla^2  +\lambda   \right] . &
\eqnd\eactONef \cr}} $$
\smallskip
{\it $\phib^2$-field correlation functions.} In this formalism it is
natural to consider also correlation functions involving the
$\rho$-field which by construction is proportional to the $\phib^2$ composite field. In the framework of phase transitions, near the critical temperature, the $\phib^2$ field plays the role of  the energy operator.
\smallskip
{\it Remark.} One can wonder how much one can still generalize this formalism
(with only one vector field). Actually, one can solve also the most general
$O(N)$ symmetric field theory with two derivatives. Indeed, this involves adding
the two terms
$$Z(\phib^2/N)(\partial _\mu\phib)^2,\quad V(\phib^2/N)(\partial _\mu \phib\cdot \phib)^2/N , $$
where $Z$ and $V$ are two arbitrary functions. These terms can be rewritten
$$Z(\rho)(\partial _\mu\phib)^2, \quad NV(\rho)(\partial _\mu \rho )^2, $$
in such a way that the $\phib$ integral remains gaussian and can be performed.
This reduces again the study of the large $N$ limit to the steepest descent method.
\subsection Large $ N $ limit: saddle points and phase transitions

We now study the large $ N $ limit, the function $U(\rho)$ being
considered as $N$ independent. If we take $\sigma=O(N^{1/2})$,
$\rho=O(1)$, $\lambda=O(1)$ all terms in ${\cal S}_N$ are of order
$N$ and the functional integral can be calculated for $N$ large by the steepest descent method  \rBLGZJDBvi. \sslbl\ssNUfige
\smallskip
{\it Saddle points.}  We look for a uniform saddle point ($\sigma(x),\rho(x),\lambda(x)$ are space-independent because we look for the ground state, thus excluding instantons or solitons)
$$ \sigma (x)=\sigma\, ,\quad\rho(x)=\rho \quad{\rm and}\quad \lambda (x)=m^2 \eqnd\eNsaddpts $$
because the $\lambda $ saddle point value must be positive. \par
The action density $\cal E$ in zero field $H$ then becomes
$${\cal E}=NU(\rho)+\ud m^2 (\sigma ^2-N\rho)+{ N\over2}\int{\d^d k
\over(2\pi)^d}\,\ln[(k^2+m^2 )/k^2].\eqnd\eNEner $$
Differentiating then ${\cal E}$ with respect to $\sigma$, $\rho$ and
$m^2$, we obtain the saddle point equations
\eqna\esaddleN
$$ \eqalignno{ m^2\sigma & = 0\,, & \esaddleN{a} \cr
\ud m^2&=U'(\rho) , &\esaddleN{b} \cr
\sigma^2/N-\rho+{1 \over  (2\pi)^{d}} \int^\Lambda { \d^{d}k \over k^2+m^2} & =
0\,. &
\esaddleN{c}  \cr} $$
\smallskip
{\it Regularization and large cut-off expansion.}
In the last equation we have now introduced a cut-off $\Lambda$ explicitly. This means,
more precisely, that we have replaced the propagator by some regularized form
$${1\over k^2+m^2}\ \mapsto  \tilde\Delta _\Lambda (k)= {1\over k^2
D(k^2/\Lambda^2)+m^2}\quad {\rm with}\ D(z)=1+O(z ), \eqnd\epropreg $$
where the function $D(z)$ is a function strictly positive for $z> 0$, analytic in
the neighbourhood of the real positive semi-axis, and increasing faster than
$z^{(d-2)/2}$ for $z\to+\infty $. \par
We  set
$${1\over (2\pi)^d }\int^\Lambda {\d^d k \over k^2
+m^2}\equiv {1\over (2\pi)^d }\int \d^d k\, \tilde\Delta _\Lambda (k) \equiv   \Omega_d(m)\equiv \Lambda^{d-2} \omega_d  (m/\Lambda). \eqnd\etadepole $$
Below we need the first terms of the expansion of $\Omega_d(m)$ for
$m^2\to 0$. One finds for $z\to 0$ and $d>2$ an expansion which we
parametrize as  (for details see appendix \label{\apiloop})
$$\omega_d(z)= \omega _d(0)-K(d)z^{d-2}+a(d)z^2 +O\left(  z^4, z^d\right).
\eqnd\etadepolii $$
The constant $K(d)$ is universal, that is independent of the cut-off procedure:
\eqna\etadpolexp
$$ \eqalignno{ K(d)&=-{1\over
(4\pi)^{d/2}}\Gamma(1-d/2)=- { \pi \over2\sin(\pi d/2)}N_d \, , & \etadpolexp{a} \cr
N_d &={2 \over(4\pi)^{d/2} \Gamma(d/2) }\,,  & \etadpolexp{b}
\cr }
$$
where we have introduced for later purpose the usual loop factor
$N_d$.  \par
The constant $a(d)$, in contrast,
depends explicitly on the regularization, that is on the way large momenta are cut,
$$a(d)=N_d \times\cases{ \displaystyle \int_0^\infty k^{d-5}\d k\left(1-{1\over D^2(k^2)} \right)  & for $d<4$,
\cr \displaystyle - \int_0^\infty {k^{d-5} \d k \over D^2(k^2)}  & for $d>4$, \cr}
 \eqnd\eadef $$
but  for
$\varepsilon=4-d\to 0$ satisfies
$$ a(d)\mathop{\sim}_{\varepsilon=4-d\to 0} 1/ (8\pi^2\varepsilon). \eqnd \eaespzero $$
Integrating $\Omega_d(m)$ over $m^2$, we then obtain a finite
expression for the regularized integral arising from the $\phib$
integration and given in \eNEner:
$$  {1\over(2\pi)^d}\int^\Lambda \d^d k\,\ln[(k^2+m^2 )/k^2] =\int_0^{m  }2s\d s\, \Omega _d(s) .
 \eqnd\eNtrlnreg  $$
>From the expansion \etadepolii,  we infer
$$\eqalignno{
\int_0^{m  }2s\d s\, \Omega _d(s) &=-2{K(d)\over d}m^d +\Omega _d(0)m^2+{a(d)\over2}m^4\Lambda^{d-4}\cr&\quad +O(m^{6}\Lambda^{d-6},m^{d+2}\Lambda^{ -2} ).& \eqnd\eNtrlnexp
\cr} $$
Finally, for $d=4$  these expressions have to be modified because a logarithmic contribution appears:
$$ \omega _d(z)- \omega _d(0)\sim {1\over8\pi^2}z^2\ln z \,. \eqnd \etadepoliv $$
\medskip
{\it Phase transitions.}
Eq.~\esaddleN{a} implies either $\sigma=0 $ or $m=0$. We see from the $\tr\ln$ term in expression \eactONef~that
$m$, at this order,  is also the  mass of the  $\pib$ field.
When $\sigma \ne 0$,   the $O(N)$ symmetry is spontaneously broken, $m$ vanishes and the massless $\pib$-field corresponds to the expected $N-1$ Goldstone modes.  If, instead, $\sigma=0$  the $O(N)$  symmetry is
unbroken and the $N$ $\phib$-field components have the same mass $m$. \par
We then note from Eq.~\esaddleN{c} that
the solution $m=0$ can exist only for $d>2$, because at $d=2$ the
integral is IR divergent. This result is consistent with the Mermin--Wagner--Coleman
theorem: in a system with only short range forces a continuous symmetry cannot
be broken for $ d\leq 2$, in the sense that the expectation value $\sigma$ of the order
parameter must necessarily vanish. The potential Goldstone modes are
responsible for this property: being massless, as we expect from
general arguments and verify here, they induce an IR instability for $d\le 2$.
Therefore, we discuss below only the dimensions $d>2$; the dimension
$d=2$ will be examined separately in the more appropriate formalism
of the non-linear $\sigma $ model in section \label{\ssLTsN}.\par
Moreover, we assume now that the polynomial $U(\rho )$ has for $\rho \ge 0$
a unique minimum at a strictly positive value of $\rho $ where
$U''(\rho)$ does not vanish,
otherwise the critical point would turn out to be a multicritical
point,  a situation that  will be studied in section  \label{\ssdblescal}.
 \smallskip
(i) {\it Broken phase.}
When $m=0$, the saddle point Eqs.~\esaddleN{} reduce to
$$U'(\rho)=0 \,,\quad \sigma ^2/N-\rho+{1 \over  (2\pi)^{d}} \int^\Lambda { \d^{d}k \over
k^2} =0\,.$$
The first equation implies that $\rho $ is given by the minimum of $U$ and the second equation then determines the
field expectation value. Clearly a solution can be found only if
$$\rho> \rho_c={1 \over  (2\pi)^{d}} \int^\Lambda { \d^{d}k \over k^2}
=\Omega_d(0),\eqnd\eONrhoc  $$
where the definition \etadepole~has been used, and then
$$\sigma =\sqrt{N(\rho-\rho_c)}. \eqnd\esponmag $$
  \smallskip
(ii) {\it The symmetric phase.} When $\sigma =0$, the saddle point equations \esaddleN{} can be written as
$$ \rho- \rho_c=  \Omega_d(m)-\Omega_d(0) ,\quad m^2=2 U'(\rho)\,. \eqnd\eNsaddsym $$
The first equation \eNsaddsym\ implies $\rho \le \rho _c$. At the value $\rho =\rho _c$ a  transition
takes place between an ordered phase $\rho >\rho _c$ and a symmetric phase
$\rho \le \rho _c$. The condition
$$U'(\rho _c)=0 \eqnd\eNUcrit $$
determines critical potentials. \par

In expression \eactONef\ we see that the
$\sigma$-propagator then becomes  \rFAAAH
$$\Delta_\sigma\mathop{\sim}_{|p|,m\ll \Lambda }{1\over p^2+m^2}\,.\eqnd\eDeltasigN$$
Therefore, $m $ is at this order  the physical mass or  the
inverse of the correlation length $\xi $ of the field $\sigma $ (and thus of all components of the $\phib$-field).
\par
The condition
$m\ll \Lambda $, or equivalently $\xi\gg 1/\Lambda $  defines the critical domain. The first equation \eNsaddsym\ then implies that $\rho-\rho _c $  is small in the critical domain. From the second equation \eNsaddsym\ follows that $U'(\rho )$ is small and thus $\rho $ is close  to the minimum of $U(\rho )$. We can then expand $U(\rho )$ around $\rho _c$:
$$U(\rho)=U'(\rho _c) (\rho -\rho _c)+ \ud U''_c(\rho -\rho _c)^2
+O\bigl( (\rho -\rho _c)^3\bigr), \eqnd\eNUparam $$
and it is convenient to set
$$U'(\rho _c)=\ud \tau \,,\quad |\tau |\ll  \Lambda ^2\,.$$
With this parametrization
$$m^2= 2U''_c(\rho -\rho _c)+ \tau  +O\bigl( (\rho -\rho _c)^2\bigr).$$
With our assumptions  $U''_c $  is strictly positive (the sign ensures that the extremum is a minimum). Then $\tau $ is positive in the symmetric phase, while $\tau <0$ corresponds to the broken phase. \par
At this point we realize that, in the case of a generic critical point, $U(\rho)$ can be approximated by a quadratic polynomial. The problem then reduces to a discussion of the $ (\phib^2 )^2$ field theory
that indeed, in the framework of the $\varepsilon=4-d$ expansion, describes critical phenomena. Therefore, we postpone a more detailed analysis of
the solutions of the saddle point equations and first summarize a few
properties of the $\phi^4$ field theory from the point of view of perturbative RG.
\subsection $ (\phib^2 )^2$ field theory, renormalization group, universality and large $N$ limit

>From now on, the discussion in this section will be specific to the
$(\phib^2)^2$ field theory. In terms of the initial $N$-component scalar field $\phib$,  we write
the action  as \sslbl\ssfivNi
$$ {\cal S} ( \phib )= \int \left\{{ 1 \over 2} \left[
\partial_{\mu} \phib (x) \right]^2+{1 \over 2}r
\phib^2 (x)+{1\over 4!} {u \over  N} \left[ \phib^2(x) \right]^2 \right\} \d ^{d}x\,.
\eqnd\eactON $$
>From the point of view of classical statistical physics, this model has the interpretation of an effective field theory that encodes large distance properties
of various statistical models near a second order phase transition. In this framework $r$ is a regular function of the temperature $T$ near the critical temperature $T_c$.  To the critical temperature corresponds a value $r_c$ of the parameter $r$ at which the correlation length $\xi$ (the inverse of the
physical mass in field theory language) diverges. For $r$ close to its critical value $r_c$, $\xi(r)\Lambda\gg 1$ and  a
continuum limit can be defined.
\par
We denote by $ \Gamma^{(\ell,n)}$ the vertex or 1PI functions of $\ell$
$\phib^2$ and $n$ $\phib$ fields (the coefficients of the expansion of the
thermodynamic potential) in Fourier representation. We set
$$  r=r_c+ \tu  \eqnd\eTmTc $$
and  $\tu$ thus characterizes the deviation from the critical value $r_c$.
In the symmetric phase ($\tau \ge0$) in zero field, the $\phib$ and $\phib^2$ correlation functions then satisfy, as functions of $\Lambda$, the dimensionless
coupling constant $g=u\Lambda^{4-d}/N$ and the deviation $\tau \ll \Lambda^2$, RG equations:
$$ \left[ \Lambda{ \partial \over \partial
\Lambda} +\beta(g){\partial \over \partial g}-{n \over 2}\eta(g)-\left(\tau{\partial \over\partial  \tau}+\ell\right)\eta_2
(g) \right] \Gamma^{(\ell,n)} =\delta_{n0}\delta_{\ell2} \Lambda^{d-4}B(g),
\eqnd\eRGgamln $$
where $\beta(g)$ is associated with the flow of the coupling constant $g$,
$\eta(g)$ and $\eta_2(g)$ to the anomalous dimensions of the fields $\phib$
and $\phib^2$, and $B(g)$ is associated to the additive renormalization of the $\phib^2$
two-point function. \par
The RG $\beta$-function in dimension $d=4-\varepsilon$,
$$\beta(g,\varepsilon)=-\varepsilon g+{N+8 \over 48\pi^2}g^2+O(g^3), $$
has for $d<4$ a non-trivial   zero
$$ g^*={48\pi^2\varepsilon\over N+8}+O(\varepsilon^2) , $$
which IR attractive since
$$\beta'(g^*)\equiv \omega =\varepsilon+O(\varepsilon^2)>
0\,.\eqnd\edefomega $$
This property is the starting point of the determination of the universal critical properties of the
model within the framework of  the so-called $\varepsilon$-expansion. \par
We explain here, instead, how
the model can be solved  in the large $N$ limit. In this way we will be
able to verify at fixed dimension, in some limit, many results obtained by perturbative methods.
\medskip
{\it Large $N$ limit.}
To the action \eactON~ corresponds the function
$$U(\rho)={1\over2} r\rho+{ u\over4!}\rho^2.\eqnd\eUONii $$
The large $N$ limit is taken at $U(\rho)$ fixed and this
implies with our conventions that $u$, the coefficient of $\phi^4/N$, is fixed. \par
The integral over $\rho$ in \eZeff\ is then gaussian. The integration results in simply replacing $\rho (x)$  by the solution of
$$\frac{1}{6}u\rho (x)+ r=\lambda (x).\eqnd\eONrhola $$
and one obtains the action
$$\eqalignno{{\cal S}_N ( \sigma ,\lambda) &  = {1 \over 2}\int \d ^{d}x \left[
\left(\partial_{\mu}\sigma \right)^2+ \lambda  \sigma^2  - {3N\over u}\lambda^2 +{6 Nr\over u}  \lambda \right] \cr &\quad +{ (N-1
 ) \over 2}  \tr\ln \left[ -\nabla^2 +\lambda (\bullet) \right]
. &\eqnd  \eactONsigla  \cr}$$
Note, however, that the field $\rho$ has a more direct physical interpretation
than the field $\lambda(x) $.
\smallskip
{\it Diagrammatic interpretation.} In  the $(\phib^2)^2$
field theory, the leading perturbative contributions  in the large $N$
limit come from chains of ``bubble'' diagrams of the form displayed in
figure \label{\figfivNi}. These
diagrams asymptotically form a geometric series, which  the
algebraic techniques explained in this section allow to sum.
\midinsert
\epsfxsize=102mm
\epsfysize=12mm
\centerline{\epsfbox{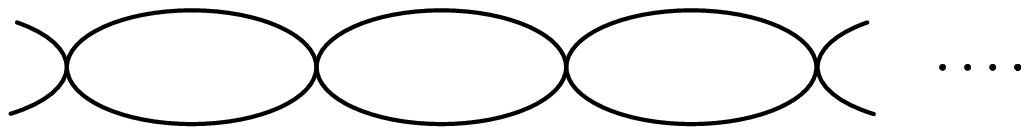}}
\figure{4.mm}{The dominant diagrams in the large $N$ limit.}
\figlbl\figfivNi
\endinsert
\medskip
{\it The low temperature phase.} We first assume that $ \sigma $, the expectation value of the
field, does not vanish, and thus the $O(N)$
symmetry is spontaneously broken. The constant $\rho$ is then given by Eq.~\esaddleN{b} which reduces to $U'(\rho )=0$. The solution must satisfy
$\rho>\rho_c$  and Eq.~\esponmag\ then yields
$\sigma=\sqrt{N(\rho-\rho_c)}$ (we recall $\rho_c=\Omega_d(0)$, Eq.~\eONrhoc).
\par
The condition \eNUcrit\ determines the critical potential $U$:
$$U'(\rho_c)=0\ \Rightarrow \ r=r_c=-u\rho _c/6\,.$$
The expectation value of the field vanishes for $r=r_c$, which
thus corresponds to the critical temperature $T_c$. Then  ($r-r_c=\tau$),
$$U'(\rho)=0\ \Rightarrow\ \rho-\rho_c =-(6/u) \tau\,.\eqnn $$
The $O(N)$ symmetry is broken for $\tu<0$,  that is at low temperature, and we
can rewrite Eq.~\esponmag\ as
$$ \sigma^2=- (6/u)\tu\propto  (-\tu)^{2\beta}\quad {\rm with} \
\beta=\ud\, .\eqnd{\expbeta} $$
We find that, for $ N $ large, the exponent $\beta $ remains mean-field like or quasi-gaussian in all dimensions.
\medskip
{\it The high temperature phase.} For $\tu>0$,  that is above $ T_c $, $ \sigma $
vanishes.
Using Eqs.~\eqns{\eONrhoc,\eTmTc} in Eqs.~\esaddleN{b} and \esaddleN{c}, we then find
\eqna\ecorleng
$$\eqalignno{m^2&=(u/6)(\rho-\rho_c)+\tu\,,& \ecorleng{a} \cr
\rho-\rho_c&= \Omega_d(m)-\Omega_d(0) . &\ecorleng{b}\cr} $$
\smallskip
(i) For $ d>4 $,  the expansion \etadepolii\ implies that the leading
contribution to  $\rho-\rho_c$  is proportional to $m^2$,
as the l.h.s.~of Eq.~\ecorleng{a}, and thus,  at leading order,
$$ m^2=\xi^{-2}\sim \tu^{2\nu} \quad  {\rm with} \quad \nu =\ud\,,
\eqnn $$
which is the mean-field or gaussian result for the correlation exponent $\nu$. \par
(ii) For $ 2<d<4 $,   the leading term is now of order $m^{d-2}$:
$$\rho-\rho_c\sim -K(d)m^{d-2}. $$
In Eq.~\ecorleng{a} the leading $m$-dependent contribution  for $m\to 0$
now comes from $\rho-\rho_c$. Keeping only the
leading term in  \etadepolii, we obtain ($\varepsilon=4-d$)
$$ m=\xi^{-1}\sim \tu^{1/(2-\varepsilon)}, \eqnd{\ecorlenb} $$ which
shows that the exponent $ \nu $ is no longer gaussian (or mean-field like):
$$ \nu ={1 \over 2-\varepsilon} ={1 \over d-2}\cdot \eqnd{\enuNlim}
$$
\par
(iii) For $d=4$, the leading $m$-dependent contribution in Eq.~\ecorleng{a} still comes from $\rho-\rho_c$:
$$m^2 \sim {48\pi^2\over u} {\tau \over \ln(\Lambda /m)}.\eqnd\eNfivdivm $$
The correlation length has no longer  a power law behaviour but, instead, the behaviour of the gaussian model  modified by a logarithm. This is typical of a situation
where the gaussian fixed point is stable, in the presence of a marginal operator. \par
(iv) Examining Eq.~\esaddleN{c} for $\sigma=0$ and $d=2$,  we find
that the correlation length becomes large only for $r\to-\infty$.
This peculiar situation will be discussed in the framework of the
non-linear  $\sigma$-model.
\smallskip
{\it Critical limit $\tu=0$}. At $\tu=0$, $m$ vanishes and
from the form \eDeltasigN\ of the $\sigma$-propagator, we find that
the critical exponent $\eta$ remains gaussian for all $d$:
$$\eta=0\ \Rightarrow \ d_\phi=\ud(d-2)\,.\eqnd\eetaNg$$
We verify that for $d\le 4$, the exponents $\beta,\nu,\eta$ satisfy the scaling relation
proven within the framework of the $\varepsilon$-expansion:
$$\beta=\nu d_\phi=\ud\nu(d-2+\eta).$$
\medskip
{\it Singular free energy and scaling equation of state.} In a
constant magnetic field $H$ in the $\sigma$ direction, the free
energy density $W(H) $ (defined here as the opposite of the action density $\cal E$ when the saddle point equations are used)   is given by  \rBrWa
$$\eqalign{W(H) &=\ln {\cal Z} /\Omega =- {\cal E}\cr &  =N\left[{3\over 2u}m^4-{3r\over
u}m^2-{1\over2} m^2\sigma^2/N+H\sigma/N- \int_0^{m  }s\d s\,
\Omega _d(s) \right],\cr} $$ where $\Omega$ is the  $d$
dimensional space volume and $\rho$ has been eliminated using
Eq.~\esaddleN{c}. The saddle point values $m^2,\sigma$  are given
by Eq.~\esaddleN{b} and the modified saddle point
Eq.~\esaddleN{a}:
$$m^2 \sigma = H\,. \eqnd{\esaddfld} $$
The magnetization $M$, expectation value of $\phib$, is
$$M= {\partial W\over\partial H}=\sigma\,,\eqnd\eNfivmag $$
because
partial derivatives of $W$ with respect to $m^2 $ and $\sigma$ vanish as a consequence of the saddle point equations.  The thermodynamic potential density ${\cal G}(M)$, Legendre transform of $W(H)  $,
follows:
$$\eqalignno{{\cal G}(M) &=HM-W(H) &\eqnd\eNfivGam \cr
&=N\left[-{3\over 2u}m^4+{3r\over u}m^2+{1\over2} m^2 M^2/N +
\int_0^{m  }s\d s\, \Omega _d(s) \right]. \cr}$$
As a property of
the Legendre transformation, the saddle point equation for $m^2$ is now
obtained by expressing that the derivative of ${\cal G}$ with
respect to $m^2$ vanishes.
\par
The expansion for large $\Lambda$ of  the $\tr\ln$ has been given  in Eq.~\eNtrlnexp.
Introducing $r_c$, one obtains
$${\cal G}(M)/N={3\over 2}\left({1\over u^*}-{1\over u}\right)m^4
+{3(r-r_c)\over u}m^2+{1\over2}m^2 M^2/N-{K(d)\over
d} m^d ,
\eqnn $$
where we have defined
$$u^*={6\over a(d)}\Lambda^\varepsilon.\eqnd\eustari $$
Note that for $d<4$ the term proportional to $m^4$ is
negligible for $m $ small with respect to the singular term $m^d$. Thus, at
leading order in the critical domain,
$${\cal G}(M)/N= {3\over u}\tu m^2+{1\over2}m^2
M^2/N-{K(d)\over d}m^d , \eqnn $$ where $\tu$ has been
defined in \eTmTc.\par
Expressing that the derivative with respect to $m^2$ vanishes,
$$(6/u)\tu + M ^2/N-K(d)m^{d-2}=0\, , $$
we obtain
$$m =\left[{1\over  K(d)}\left((6/u)\tu
+M^2/N\right)\right]^{1/(d-2)}. $$ It follows that the leading

contribution to the thermodynamic
potential, in the critical domain, is given by
$$ {\cal G}(M)/N \sim {(d-2)\over 2d}{1\over\bigl(K(d)
\bigr)^{2/(d-2)}} \left[(6/u)\tu +M^2/N\right]^{d/(d-2)}.\eqnd\ethermscN $$
>From ${\cal G}(M)$ can be derived various other quantities  like the equation
of state, which is obtained by differentiating with respect to $M$. It can be cast into the scaling form
$$H={\partial {\cal G}\over \partial M}= h _0
M^{\delta}f\left(a_0 \tu/M^2 \right), \eqnd \estatNli  $$
where $h_0$ and $a_0$ are normalization constants. The exponent $\delta$ is given by
$$\delta= {d+2 \over d-2}\,, \eqnn $$ in agreement with the general
scaling relation $\delta=d/ d_\phi-1$, and the function
$f(x)$ by
$$f(x)= (1+ x)^{2/(d-2)}. \eqnd{\estatNgb} $$
The asymptotic form of
$f(x)$ for $x$ large implies $\gamma=2/(d-2)$ again in agreement
with the scaling relation $\gamma=\nu(2-\eta)$. Taking into account
the values of the critical exponents $\gamma$ and $\beta$, it is then
easy to verify that the function $f$ satisfies all required
properties like for example Griffith's analyticity. In particular, the equation
of state can be cast into the parametric form
 \rparamet
$$\eqalign{M & =(a_0)^{1/2} R^{1/2}\theta\,,\cr
\tu & =3R\left(1-\theta^2\right),\cr
H& =h_0 R^{\delta/2}\theta\left(3-2\theta^2\right)^{2/(d-2)}.\cr}$$
\medskip
{\it Leading corrections to scaling.} The $m^4$ term yields the
leading corrections to scaling. It is subleading by a power of
$\tu$:
$$m^4/m^d=O(\tu^{(4-d)/(d-2)}).$$
The exponent governing the leading corrections to scaling in the temperature variable is  $\omega \nu $ ($\omega $ is defined in Eq.~\edefomega) and thus \rSKMa
$$\omega\nu=(4-d)/(d-2)\ \Rightarrow\ \omega=4-d\,.\eqnd\eNomega $$
Note that for the special value $u=u^*$ this
correction vanishes.
\medskip
{\it Specific heat exponent. Amplitude ratios.}  Differentiating
twice ${\cal G}(M)$ with respect to $\tu$, we obtain the specific heat
at fixed magnetization
$$C_H\propto   \left[(6/u)\tu
+M^2/N\right]^{(4-d)/(d-2)}.\eqnd\esphCH $$
For $M=0$, we identify the
specific exponent
$$\alpha={4-d\over d-2}, \eqnn $$
which indeed is equal to $2-d\nu$,
as predicted by scaling relations. Among the universal ratios of amplitudes, one can
calculate for example $R^+_\xi$ and $R_c$ (for definitions, see
chapter 29 of ref.~\rbook)
$$(R^+_\xi)^d={4N\over(d-2)^3}{\Gamma(3-d/2)\over(4\pi)^{d/2}},\quad
R_c={4-d\over(d-2)^2}. \eqnn $$
\midinsert
\epsfxsize=50.mm
\epsfysize=16.mm
\centerline{\epsfbox{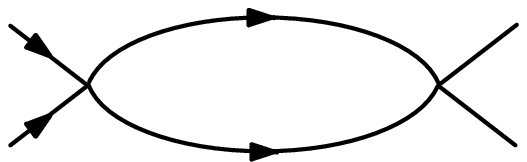}}
\vskip-15.8mm
\centerline{$q$}
\vskip4.6mm
\centerline{$p-q$}
\figure{3.mm}{The ``bubble" diagram $B_\Lambda(p,m)$.}
\figlbl\figbubii
\endinsert
\medskip
{\it The $\lambda$ and $ \phib ^2$ two-point functions.}
In the high temperature phase, differentiating twice the action \eactONef\ with
respect to $\lambda(x),\rho(x)$ and replacing  the field $\lambda(x)$ by its
expectation value $m^2$, we find the $ \lambda $-propagator
 $$ \Delta_\lambda(p)=-{2 \over N} \left[ {6\over
u}+B_\Lambda(p,m)\right]^{-1} \,, \eqnn $$ where $B_\Lambda(p,m)$
is the bubble diagram of figure \figbubii:
$$B_\Lambda(p,m)={1 \over  (2\pi )^{d}} \int^\Lambda { \d
^{d}q
\over \left(q^2+m^2 \right) \left[ \left(p-q \right)^2+m^2
\right]} , \eqnd\ediagbul $$
and the cut-off symbol $\Lambda $ means calculated with a regularized propagator
as in \epropreg.\par
The $\lambda$-propagator is negative because the $\lambda$-field is imaginary. Using the relation \eONrhola, we obtain the $\rho$ two-point function (in the Fourier representation the constant shift only produces a $\delta $-function at $p=0$) and thus as noted in \ssfivNi,  the
$\phib^2$ two-point function
$$\left<\phib^2 \phib^2\right>=N^2\left<\rho\rho\right>=-{12N/u \over
1+(u/6)B_\Lambda(p,m)}.
\eqnn $$
At zero momentum we recover the specific heat. The small $m$ expansion of
$B_\Lambda(0,m)$ can be derived from the expansion \etadepolii.
One finds
$$\eqalignno{B_\Lambda(0,m)&={1\over(2\pi)^d} \int^\Lambda {\d^{d}q \over
\left(q^2+m^2 \right)^2} \cr &=-{\partial \over \partial m^2}
\Omega_d(m) \mathop{=}_{m\ll\Lambda}
 (d/2-1)K(d) m^{-\varepsilon}-a(d)\Lambda^{-\varepsilon}
+\cdots\hskip8mm &\eqnd\eBLamze \cr}$$
The singular part of the specific heat thus
vanishes as $m^{\varepsilon}$, in agreement with Eq.~\esphCH\ for $M=0$.
\par
In the critical theory ($ m=0 $ at this order) for $2\le d< 4$, the
denominator is also dominated at low momentum by the integral
$$B_\Lambda(p,0)= {1 \over \left(2\pi \right)^{d}} \int^\Lambda {
\d^{d}q \over q^2(p-q)^2} \mathop{=}_{2<d<4} b(d)p^{-\varepsilon}
-a(d)\Lambda^{-\varepsilon}
+O\left(\Lambda^{d-6}p^ 2 \right),\eqnd\ebullecrit  $$
where
$$ b (d)=-{\pi \over\sin(\pi d/2)}
{\Gamma^2 (d/ 2) \over \Gamma (d-1)}N_d \,, \eqnd\econstb $$
and thus
$$ \Delta_{\lambda}(p)\mathop{\sim}_{p\to 0} -{2 \over N
b(d)} p^{\varepsilon}. \eqnd{\eprocrit} $$  We again
verify consistency with scaling relations. In particular, we note
that in the large $N$ limit the {\it dimension  of the
field $\lambda$}\/ is
$$[\lambda]=[\rho]=[\phib^2]=d-1/\nu=\ud(d+\varepsilon)=2\,, \eqnn $$
a result important for the $1/N$ perturbation theory.
\smallskip
{\it Remarks.} \par
(i) For $d=4$,  the integral  has a logarithmic behaviour:
$$B_\Lambda(p,0)\mathop{\sim}_{p\ll \Lambda} {1\over 8\pi^2}\ln (\Lambda/p)+\
{\rm const.}\,, \eqnd\eBivasym $$
and still gives {\it the leading contribution}\/ to the inverse propagator
$\Delta_\lambda\propto 1/\ln(\Lambda/p)$.\par
(ii) Note, therefore,
that for $d\le 4$ the contributions generated by the term
proportional to $\lambda^2(x)$ in \eactONef\ are always negligible in the critical domain.
\medskip
{\it The $\left<\sigma \sigma \right>$ two-point function at low temperature.}
In the phase of broken symmetry the action, after translation of
expectation  values, includes a term proportional to $\sigma\lambda$ and thus the
propagators of the fields $\sigma$ and $\lambda$ are elements of a  $2\times2$ matrix $\bf M$:
$${\bf M^{-1}}(p)=\pmatrix{p^2 & \sigma \cr \sigma & -3N/u-\ud N B_\Lambda(p,0)
\cr} \ ,\eqnd\eNsigiipt $$
where $\sigma= \left<\sigma(x)\right>$ and  $B_\Lambda$ is given by Eq.~\ebullecrit.
For $d<4$ at leading order for $|p|\ll \Lambda $, the determinant is given by
$$1/\det{\bf M}(p)\sim-N\left[b(d)p^{d-2}+6\tau /u\right],$$
where the relation \expbeta\ has been used.
For $|r-r_c|\ll \Lambda ^2$, this expression defines a crossover mass scale
$$m_{\rm cr}=(-\tau /u)^{1/(d-2)}\propto \Lambda \bigl((r_c-r) /\Lambda ^2\big)^{1/(d-2)}= \Lambda\bigl((r_c-r) /\Lambda ^2\bigr)^ \nu  , \eqnd\eNmcrossa $$
at which a crossover between Goldstone behaviour ($N-1$ massless free particles) and critical behaviour  ($N$ massless interacting particles) occurs. \par
At $d=4$, the form \eBivasym\ becomes relevant and
$$ m_{\rm cr}^2\propto {r_c-r \over \ln[\Lambda ^2/(r_c-r)]} .\eqnd\eNmcrossb $$
Finally, for $d>4$, $B_\Lambda(p,0)$ has a limit for $p=0$ and therefore
$$ m_{\rm cr}\propto \sqrt{r_c-r }\,.\eqnd\eNmcrossc $$
In all dimensions $m_{\rm cr}$    scales near $r_c$ as the physical mass above $r_c$.
\subsection  RG functions and leading corrections to scaling

 {\it The RG functions.} For a more detailed verification of the
consistency between the large $N$ results and RG predictions, we now
calculate RG functions at leading order for $N\to\infty $.
We set (Eq.~\eustari)  \sslbl\sssEGRN
$$ u=Ng\Lambda^{\varepsilon},\quad g^*=u^*\Lambda^{-\varepsilon}/N=
6/(Na)\,,\eqnd\eustar $$
where the constant $ a(d) $ has been defined in
\etadepolii\ and behaves for $\varepsilon=4-d\to 0$ like
$ a(d)\sim 1/ (8\pi^2\varepsilon) $ (Eq.~\eaespzero).
\par
One then verifies
that $m$ solution of
Eqs.~\ecorleng{} satisfies asymptotically for $\Lambda$ large an equation that expresses that it is
RG invariant:
$$ \left( \Lambda{ \partial \over \partial \Lambda} +\beta (g){\partial \over \partial g}-\eta_2(g)\tu{\partial
\over \partial \tu} \right) m(\tu,g,\Lambda)=0\,, \eqnn $$
where in the r.h.s.\ contributions of order $1/\Lambda^2$ have been neglected.
The RG functions $\beta(g)$ and $\eta_2(g)$ are then given by
$$ \eqalignno{ \beta (g) & = -\varepsilon g(1- g/g^*) , &
\eqnd{\ebetaN} \cr
\nu^{-1}(g)= 2+\eta_2(g)& = 2-\varepsilon g/g^*. & \eqnn \cr} $$
When $a(d)$ is positive (but this not true for all regularizations,
see the discussion below), one finds an IR fixed point $g^*$, as
well as exponents  $\omega=\varepsilon$, and $\nu^{-1}=d-2$, in
agreement with Eqs.~\eqns{\eNomega,\enuNlim}. In the framework of the
$\varepsilon$-expansion, $\omega$ is associated with the leading
corrections to scaling. In the large $N$ limit $\omega$ remains
smaller than 2 for $\varepsilon<2$, and this extends the property established near $d=4$ to all dimensions $2\le d\le 4$. \par
Finally, applying the RG
equations to the propagator \eDeltasigN, one finds
$$\eta(g)=0\ , \eqnn $$
a result consistent with the value \eetaNg\ found for $\eta=\eta (g^*)$.
\medskip
{\it Leading corrections to scaling.} From the general RG analysis, we
expect the leading corrections to scaling to vanish for $u=u^*$.
This property has already been verified for the free energy. Let us now
consider the correlation length or mass $m$
given by Eq.~\ecorleng{}. If we keep the leading correction to
the integral for $m$ small (Eq.~\etadepolii), we find
$$ 1- {u\over u^*} +(u/6) K(d) m^{-\varepsilon}
+O\left({m^{2-\varepsilon} \Lambda^{-2}}
\right) ={\tu\over m^2}\,, \eqnd\ecorlengb $$
where Eq.~\eustar~has been used.  We see that the leading
correction again vanishes for $u=u^*$. Actually, all correction terms
suppressed by powers of order $\varepsilon$ for $d\to 4$ vanish
simultaneously as expected from the RG analysis of the $\phi^4$
field theory. Moreover, one verifies that the leading correction is
proportional to $(u-u^*)\tu^{\varepsilon/(2-\varepsilon)}$, which
leads to $\omega\nu=\varepsilon/(2-\varepsilon)$, in agreement with
Eqs.~\eqns{\eNomega,\enuNlim}.
\par
In the same way, if we keep the leading correction to the
$\lambda$-propagator in the critical theory (equation
\ebullecrit), we find
$$ \Delta_{\lambda} (p )=-{2 \over N} \left[{6\over u}-
{6\over u^*}+ b(d)p^{-\varepsilon}
\right]^{-1}, \eqnd{\epropNli}  $$
where terms of order $ \Lambda^{-2} $ and $1/N$ have been neglected.
The leading corrections to scaling again  cancel for $
u=u^{\ast} $ exactly, as expected.
\smallskip
{\it Discussion.} \par
 (i) One can show that a perturbation due to
irrelevant operators is equivalent, at leading order in the critical
region, to a modification of the $(\phib^2)^2$ coupling.  This can
be explicitly  verified here. The amplitude of the leading
correction to scaling has been found to be proportional to
$6/u-a(d)\Lambda^{-\varepsilon}$, where the value of $a(d)$ depends
on the cut-off procedure and thus on contributions of irrelevant
operators. Let us call $u'$ the $(\phib^2)^2$ coupling constant in
another scheme where $a$ is replaced by $a'$. Identifying the
leading correction to scaling, we find the  relation
$${6\Lambda^{\varepsilon} \over u}-a(d)={6\Lambda^{\varepsilon}
\over u'}-a'(d),$$
a homographic relation that is consistent with the special form
\ebetaN\ of the $\beta$-function.\par
 (ii) {\it The sign of
$a(d)$.} It is generally assumed that $a(d )$ is positive for $2<d<4$.  This is indeed what one finds in the simplest regularization schemes, for example
when the function $D(k^2)$ in \epropreg\ is an increasing function of $k^2$.
Moreover, $a(d)$ is always positive near four
dimensions where it diverges like
$$a(d)\mathop{\sim}_{d\to 4} {1\over 8\pi^2\varepsilon}.$$
Then, for $2<d<4$ there exists an IR fixed point, corresponding to a non-trivial zero $u^*$ of the
$\beta$-function. For the value $u=u^*$ the leading corrections to
scaling vanish.\par
However, for $d<4$ fixed  that positivity of $a(d)$ is not assured. For example, in the case of simple lattice
regularizations it has been shown that in $d=3$ the sign is
arbitrary. \par
When $a(d)$ is negative, the RG method for
large $N$ (at least in the perturbative framework) is confronted
with a serious difficulty.  Indeed, the coupling flows in the IR
limit to large values where the large $N$ expansion is no longer
reliable. It is not known whether this signals a real pathology of the model in the RG sense, or is just an artifact of the large $N$ limit. \par
Another way of viewing the problem is to examine directly the
relation between bare and renormalized coupling constant. Calling
$g_{\rm r} m^{4-d}$ the renormalized four-point function at zero
momentum, we find
$$m^{4-d}g_{\rm r}={\Lambda^{4-d}g\over 1+\Lambda^{4-d}g N
B_\Lambda(0,m)/6} . \eqnd\egrenor$$ In the limit $m\ll\Lambda$, the
relation can be written as
$${1\over g_\r}= {1\over g_\r^*}+\left(m\over\Lambda\right)^{4-d}
\left({1\over g}-{N a(d)\over6}\right),\quad {1\over g_\r^*}={(d-2) N K(d)\over
12}.\eqnd\egrenor $$
We see that when $a(d)<0$, the limiting value $g_\r=g_\r^* $ for $m/ \Lambda=0 $
 cannot be reached by varying $g$ when $m/\Lambda$ is small but finite  (since $g>0$). In
the same way, leading corrections to scaling can no longer be
cancelled.
\subsection Small coupling constant and large momentum
expansions for $d<4$

Section \label{\sssfivNRT} is devoted to a systematic discussion of the $1/N$
expansion. However, the $1/N$ correction to
the two-point function will help us to investigate  immediately
the following problem: the perturbative expansion of the massless $\phi^4$ field theory has IR divergences for any dimension $d<4$, although we believe the critical theory to exist.
In the framework of the $1/N$ expansion, instead, the critical theory ($T=T_c, m^2=0$) is defined for any dimension $d<4$. This
implies that the coefficients of the $1/N$ expansion cannot be expanded in
a Taylor series of the coupling constant. \par
To understand the phenomenon, we consider the
$\left<\sigma\sigma\right>$
correlation function at order $1/N$.  At this order only one diagram
contributes (figure \label{\figbubiii}), containing two $\lambda^2\sigma$ vertices.
After mass renormalization,   in the  large cut-off limit, we find \sslbl\ssNplarge \par
$$\Gamma^{(2)}_{\sigma\sigma}(p)= p^2 +{2\over N (2\pi)^{d}}\int
{\d^{d}q \over(6/u)+b(d)q^{-\varepsilon}}\left({1 \over
(p+q)^2} -{1 \over q^2}\right) +O\left({1 \over N^2}\right)
. \eqnd{\eONpropi} $$
We now expand $\Gamma^{(2)}_{\sigma\sigma}$ for $u\to0$.
Note that since the gaussian fixed point is an UV fixed
point, the small coupling expansion is also a large momentum
expansion. \par
An analytic study  then reveals that the integral has an expansion of the form
$$\sum_{k\ge 1} \alpha_k u^k p^{2-k\varepsilon}+\beta_k
u^{(2+2k)/\varepsilon} p^{-2k} .\eqnn $$
The coefficients $\alpha_k,\beta_k$ can be obtained by performing a Mellin
transformation over $u$ on the integral. Indeed, if a function $f(u)$
behaves like $u^t$ for $u$ small, then the Mellin transform
$$M(s)=\int_0^\infty\d u\,u^{-1-s}f(u), $$
has a pole at $s=t$. Applying the transformation to the
integral  and inverting $q$ and $u$ integrations, we have to calculate
the integral
$$\int_0^\infty\d
u\,{u^{-1-s}\over(6/u)+b(d)q^{-\varepsilon}}
={1\over 6}\left(b(d)q^{-\varepsilon}\over
6\right)^{1-s}{\pi
\over \sin\pi s}\,\cdot $$
Then, the value of the remaining $q$ integral follows from the generic
result \eqns{\eintmunu}.\par
The terms with integer powers of $u$ correspond to the formal
perturbative expansion where each integral is calculated for
$\varepsilon$ small enough. $\alpha_k$
has poles at $\varepsilon=(2l+2)/k$ for which  the corresponding power of
$p^2$ is $-l$, that is an integer. One verifies that $\beta_l$ has a pole at
the same value of $\varepsilon$ and that the singular contributions cancel in
the sum \rSymanza. For these dimensions logarithms of $u$ appear in the small $u$ expansion.
\midinsert
\epsfxsize=50.mm
\epsfysize=16.mm
\centerline{\epsfbox{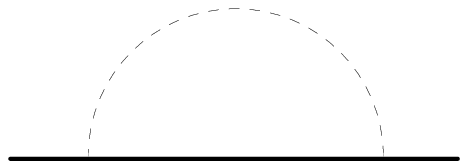}}
\vskip-17.mm
\centerline{$\lambda$}
\vskip7.mm
\centerline{$\sigma$}
\figure{3.mm}{The diagram contributing to $\Gamma^{(2)}_{\sigma\sigma}$ at
order  $1/N$.}
\figlbl\figbubiii
\endinsert
\subsection Dimension four: triviality, renormalons, Higgs mass

A number of issues concerning the physics of the $(\phib^2)^2$  theory
in four dimensions can be addressed within the framework of the large $N$
expansion. For  simplicity reasons, we consider here only the critical
(i.e.~massless) theory.\sslbl \sstrivia
\medskip
{\it Triviality and UV renormalons.} One
verifies that the renormalized coupling constant $g_\r$,
defined as the  value of the vertex $\left<\sigma\sigma\sigma\sigma\right>$ at
momenta of order $\mu\ll\Lambda$, is given by
$$g_\r={g\over1+\frac{1}{6}N g B_\Lambda(\mu,0)}\,,\eqnd\egfivivren $$
where  $B_\Lambda(p,0)$, which  corresponds to the bubble diagram  (figure \figbubii), is given by Eq.~\eBivasym:
$$B_\Lambda(p,0)\mathop{\sim}_{p\ll \Lambda} {1\over 8\pi^2}\ln (\Lambda/p)+\
{\rm const.}\,.  $$
We see that when the ratio $\mu/\Lambda$ goes to zero, the renormalized
coupling constant vanishes,  independently of the value of $g$ (here $g$ is physical, that is $g>0$). 
This is the so-called {\it triviality}\/ property. In the traditional presentation of quantum field field, one usually insists in taking the
infinite cut-off $\Lambda$ limit. Here, one finds then only a free field
theory.
Another way of formulating the problem is the following: it seems impossible to
construct in four dimensions a $\phi^4 $ field  theory consistent (in the
sense of satisfying all usual physical requirements) on all scales for non
zero coupling. Of course, in the logic of {\it effective} field theories this
is no longer an issue. The triviality property just implies that the
renormalized or effective charge is logarithmically small as indicated by
Eqs.~\eqns{\egfivivren,\eBivasym}. Note that if $g$ is generic (not too
small) and $\Lambda/\mu$ large, $g_r$ is essentially independent of the
initial coupling constant. The renormalized coupling remains an adjustable, though bounded, quantity only when the bare coupling is small enough and the RG flow is thus very slow.\par
If we proceed formally and, ignoring the problem,
express the leading contribution to the four-point function in terms of the
renormalized constant,
$${g\over 1+{N\over 48\pi^2}g\ln (\Lambda/p)}={g_\r\over 1+{N\over
48\pi^2}g_\r\ln (\mu/p)} \,,$$
we then find that the function has a pole for
$$p=\mu\e^{48\pi^2/(Ng_\r)}.$$
This unphysical pole (called sometimes Landau's ghost) is generated because $g=0$ is an IR fixed point. If we calculate contributions of
higher orders, for example to the two-point function, this pole makes
the loop integrals  diverge. In an expansion in powers of
$g_\r$, each term is instead calculable
but one finds, after renormalization, UV contributions of the type
$$\int^\infty{\d^4 q\over q^6}\left(-{Ng_\r\over48\pi^2}\ln(\mu/q)\right)^k
\mathop{\propto}_{k\to\infty}\left({Ng_\r\over96\pi^2}\right)^k k!\,.$$
The perturbative manifestation of  Landau's ghost is the appearance of
contributions to the perturbation series which are not Borel summable.
This effect is called  UV renormalon effect \rUVrenorm. By contrast the contributions due to the finite momentum region, which can be
evaluated by a semi-classical analysis, are Borel summable, but invisible for
$N$ large.  Note,
finally,  that this UV problem is independent of the mass of the field $\phib$,
which we have taken zero only for simplicity.
\medskip
{\it IR renormalons.}
We now illustrate the problem of IR renormalons with the same example
of the massless  $(\phib^2)^2$ theory (but now zero mass is essential), in
four dimensions, in the large $N$ limit \rDavid. We calculate the contribution
of the small momentum region to the mass renormalization, at
cut-off $\Lambda$ fixed. In the large $N$ limit, the mass renormalization
then  is proportional to (see Eq.~\eONpropi)
$$\int^\Lambda{\d^4 q\over q^2\bigl(1+\frac{1}{6}NgB_\Lambda(q)\bigr)}
\sim\int{\d^4 q\over q^2\bigl(1+{N\over 48\pi^2}g\ln (\Lambda/q) \bigr)}\,.$$
It is easy to expand this expression in powers of the coupling constant $g$.
The term of order $k$ in the limit $k\to\infty$  behaves as $(-1)^k
(N/ 96\pi^2)^k k!$. This contribution has the alternating sign of the
semi-classical contribution. Note that more generally for $N$ finite
on finds $(-\beta_2/2)^k k!$. IR singularities are responsible for
additional, Borel summable, contributions to the large order behaviour.\par
In a theory asymptotically free for large momentum, clearly the roles
of IR and UV singularities are interchanged.
\medskip
{\it The mass of the $\sigma$ field in the phase of broken  symmetry}.
In four dimensions the $\phi^4$ theory is an ingredient of the Standard Model, and the field
$\sigma$ then represents the Higgs field. 
With some reasonable assumptions, it is possible to establish for finite $N$
a  semi-quantitative bound on the Higgs mass. Let us examine here this question for $N$ large.\par
In the phase of broken symmetry the action, after translation of
expectation  values, includes a term proportional to $\sigma\lambda$ and thus the
propagators of the fields $\sigma$ and $\lambda$ are elements of the  $2\times2$
matrix $\bf M$  defined in \eNsigiipt,
 where $B_\Lambda(p,0)$ is given by Eq.~\eBivasym. \par
It is convenient to introduce a  RG invariant mass scale $M$, that we  define by
$${48\pi^2\over Ng}+8\pi^2 B_\Lambda(p,0)= \ln(M/p).$$
Then,
$$M\propto \e^{48\pi^2/Ng}\Lambda\,.$$
Poles of the propagator correspond to zeros of  $\det{\bf M} $. Solving
the equation $\det{\bf M} =0$,
one finds for the mass $m_\sigma$ of the field $\sigma$ at this order
$$p^2 \ln (M/p)=-(16\pi^2/N)\sigma^2\ \Rightarrow\ m_\sigma^2\ln (i M/
m_\sigma) =(16\pi^2/N)\sigma^2.$$
The solution of the equation is complex, because the particle
$\sigma$ can  decay into massless Goldstone bosons. At $\sigma$ fixed, the
mass decreases when the cut-off increases or when the coupling  constant goes
to zero. Expressing  that the mass must be smaller than the cut-off, one
obtains an upper-bound on $m_\sigma$ (but which  depends somewhat on the precise regularization) \rHiggsbound. 
%
\subsection Other methods. General vector field theories

The large $N$ limit can be obtained by several other algebraic methods.
Without being exhaustive, let us list a few. Schwinger--Dyson equations
for $N$ large lead to a self-consistent equation for the two-point function
\rSymanzb.
Some versions of the Hartree--Fock approximation or variational methods
also yield the large $N$
result as we show in section \label{\ssNVarfiv}. The functional (also called exact) RG takes the form of partial differential equations in the large $N$ limit \rFRG.
\sslbl\ssfivNge \par
>From the point of view of stochastic quantization or critical dynamics the
Langevin equation also becomes linear and self-consistent for $N$ large,
because the fluctuations of $\phib^2(x,t)$ are small. As a byproduct the
large $N$ expansion of the equilibrium equal-time correlation functions
is recovered. This is a topic we study in section \label{\ssCrDY}.
\par
\medskip
{\it General vector field theories.} We have shown how the large $N$ expansion can be generated for a general function $NU(\phib^2/N)$.
This method will be applied
in section \label{\ssdblescal} to the study of multicritical points and
double scaling limit.
\par
We now briefly explain how the algebraic
method of section \ssfivNi~can be generalized to $O(N)$ symmetric actions that depend on several vector fields. Again, the composite fields which are expected to have small
fluctuations, are the $O(N)$ scalars constructed from all $O(N)$ vectors.  One thus introduces pairs of fields and Lagrange multipliers for all independent
$O(N)$ invariant scalar products constructed from the many-component fields \rPeRoVi.
\par
Let us illustrate the idea with the example of two fields $\phib_1$ and $\phib_2$, and  a symmetric interaction  arbitrary function of the three scalar  invariants $\rho_{12}=\phib_{1}\cdot\phib_2/N$, $\rho_{11}=\phib^2_{1}/N$ and $\rho_{22}=\phib^2_2/N$:
$${\cal S}(\phib_1, \phib_2)= \int\d^d x \left\{\ud \left[ \partial_{\mu} \phib_1 (x) \right]^2
+\ud \left[ \partial_{\mu} \phib_2 (x) \right]^2+
NU (\rho_{11},\rho_{12},\rho_{22}) \right\} .\eqnd{\eactONg}$$
We then introduce three pairs of fields $\rho_{ij}(x)$ and $\lambda_{ij}(x)$ and use the identity
$$\eqalignno{&\exp\left[-\int \d^{d}x\,NU (\phib^2_{1}/N, \phib_{1}\cdot\phib_2/N,\phib^2_2/N)\right] \propto \int
\left[\d\rho_{ij}(x)\,\d\lambda_{ij}(x) \right] \cr & \quad \times \exp\left\{-\int
\d^{d}x\left[\sum_{ij}\ud\lambda_{ij}\left(\phib_i\cdot \phib_j-N\rho_{ij}\right) +
NU(\rho_{11},\rho_{12},\rho_{22})\right]\right\}. \hskip7mm &\eqnd\egeniden \cr}$$
The identity \egeniden\ transforms the action into a quadratic form in $\phib_i$,
the integration over $\phib$ can thus be performed.
The large $N$ limit is then again obtained by the steepest descent method.
In the special case in which $U(\ro)$ is a quadratic function, the
integral over all $\rho$'s can also be performed.
If the action is a general $O(N)$ invariant function of $p$ fields $\phib_i$,
it is necessary to introduce $p(p-1)/2$ pairs of $\rho$ and $\lambda $ fields.
\subsection{Variational calculations in large $N$ quantum field theory}

A possible extra insight into the large $N$ limit and its non-perturbative nature
is revealed when  variational methods are employed \rBarMos.
The gaussian variational wave functional reproduces for $N$ large
the saddle point or gap equations of the $1/N$ expansion but also yields a clear
picture of the end-point contribution in the variational
parameter phase space.  It is  applicable for finite values of $N$ though it simplifies for $N$ large. However, as usual with variational methods, it is difficult to improve results systematically, while the large $N$ limit is the leading term in an  $1/N$ expansion.\sslbl\ssNVarfiv
\par
To explain the variational method, we  use again the concrete
example of the $O(N)$ symmetric scalar field theory \eONpart\
with the euclidean action   \eactONgen:
$$ {\cal S}  ( \phib )= \int \left[\ud \left[
\partial_{\mu} \phib (x) \right]^2+NU\bigl(\phib^2(x)/N\bigr)  \right] \d^{d}x\,.
 $$
In section \label{\scFTQFT} the more general situation
of a  finite euclidean time interval $\beta$  with periodic
boundary conditions will be discussed. The functional integral \eONpart\
then represents the quantum partition function
$${\cal Z}(\beta)=\tr\e^{-\beta H},$$
where $H$ is the quantum hamiltonian corresponding to the euclidean
action, and $\beta =1/T$ the inverse temperature. \par
For any   action  ${\cal S}$ and auxiliary action ${\cal S}_0$,
we can write the identity
$$ {\cal Z} = \int [\d \phib] \e^{-{\cal S}(\phib)}
= {\cal Z}_0 \left< \e^{-({\cal S}-{\cal S}_0)} \right>_0\,,\eqnn $$
where $\left<\bullet\right>_0$ means expectation value with respect
to $\e^{-{\cal S}_0}$.
The variational principle then relies on the general (convexity) inequality
$$\bigbra \e^{-({\cal S}-{\cal S}_0)} \bigket_{\!0}~~\geq
~~\e^{-\bigbra {\cal S}-{\cal S}_0 \bigket_{\!0}}. \eqnd\evarineq $$
The trial action ${\cal S}_0$  then is chosen to maximize the r.h.s., a general strategy that also leads to mean field approximations.\par
In the limit of infinite $d$-dimensional volume, or equivalently
$d-1$-dimensional volume $V_{d-1}$ and zero-temperature ($\beta\to\infty $)
$$\lim_{\beta\to\infty }-{1\over\beta V_{d-1}}\ln{\cal Z}= {\cal E} \,,\quad
\lim_{\beta\to\infty }-{1\over\beta V_{d-1}}\ln{\cal Z}_0= {\cal E}_0 \,,$$
where  ${\cal E}$  and  ${\cal E}_0$ are the ground state energy (or action) densities  corresponding to ${\cal S}$ and ${\cal S}_0$ respectively.
As a consequence  one obtains the  inequality
$$  {\cal E}
  \le {\cal E}_{\rm var.}={\cal E}_0 +{1\over V_{d-1} \beta  } \left<  {\cal S}-{\cal S}_0\right>_0 ,$$
One verifies that $V_{d-1} {\cal E}_{\rm var.}$ is also the expectation value
of the hamiltonian corresponding to ${\cal S}$ in the ground state of the
hamiltonian corresponding to ${\cal S}_0$.
In the zero temperature limit   the choice of a trial action ${\cal
S}_0$  thus becomes equivalent to the choice of a trial wave
functional (the ground state) in a Schr\"odinger representation.\par
For trial action  ${\cal S}_0$, we take a free field action with a mass
$m$ that is a variational parameter:
$$  {\cal S}_0={1\over2}\int \d^ d x \,
 \left( {(\partial_{\mu} \phib)}^2 +
 m^2   \phib^2 ~\right)  . $$
This corresponds, in the Schr\"odinger representation, to choosing a gaussian wave functional as a variational state:
$$\Psi(\phi)=\exp\left[-{1\over2}\int\d^{d-1}k\,  \tilde \phib(k)  \sqrt{k^2+m^2} \tilde\phib(-k)\right],$$
where $\tilde \phib(k)$ are the Fourier components of $\phib(x)$  or
more generally,
$$\Psi(\phi)=\exp\left[-{1\over2}\int\d^{d-1}x \,\int\d^{d-1}y\, \phib(x) K _m(x-y)\phib(y)\right],$$
where $K _m$ is determined by minimizing ${\cal E}_{\rm var.}$
\rJaStr .\par

The corresponding energy density is 
$$ {\cal E}_0 = {N\over 2 (2\pi)^d}\int \d^d k\, \ln[(k^2+m^2 )/k^2] ={N \over
{(2\pi)}^{d-1} }
\int\d^ {d-1}\!k\left[\ud \omega(k)\right],    $$
where $\omega(k)=\sqrt{k^2+m^2}$.\par
The variational energy  then becomes
$$ {\cal E}_{\rm var.}= {N\over 2 (2\pi)^d}\int \d^d k\, \ln[(k^2+m^2 )/k^2]
+N\left<U\bigl(\phib^2(x)/N\bigr)
-\ud m^2 \phib^2(x)/N\right>_0\,.\eqnd\evarTzero $$
The r.h.s.~contains gaussian expectation  values that can be calculated explicitly.
We introduce the parameter
$$\rho=\left<\phib^2(x)/N\right>_0={1\over (2\pi)^d}\int{\d^d k\over k^2+m^2}=\Omega _d(m),
\eqnd\esaddleNc $$
where $\Omega _d(m)$ is defined by Eq.~\etadepole.
The form of ${\cal S}_0$ implies
$${\partial {\cal E}_0\over \partial m^2}=\ud \left<\phib^2\right>=\ud
N\rho\,.$$
The important remark, which explains the role of the large $N$ limit,
is (see also appendix \label{\appnormprod})
that  for  $N\to\infty $ the r.h.s.~in Eq.~\evarTzero\ simplifies:
$$\left< U(\phib^2 /N) \right>  = U(\left< \phib^2 /N\right>)
\left[1 + O({1 / N})\right] .\eqnd\eNvarav $$
We then find
$${\cal E}_{\rm var.}/N\mathop{\to}_{N\to \infty } {1\over 2 (2\pi)^d}
\int \d^d k\, \ln[(k^2+m^2 )/k^2] +
 U(\rho)-\ud m^2 \rho \,,$$
an expression identical (for $\sigma =0$) to \eNEner. Note, however, that here Eq.~\esaddleN{c}
is automatically satisfied, since it defines the parameter $\rho $ (Eq.~\esaddleNc).
The quantity $\cal E$ can now be minimized with respect to the free parameter $m^2$, while
without the constraint \esaddleNc\ it has no minimum.
\par
In the broken phase, previous equations have no solution
and one must take as a trial action a free action with a shifted
field:
$${\cal S}_0={1\over2}\int \d^ d x \,
 \left( {(\partial_{\mu} \phib(x))}^2 +
 m^2   (\phib(x)-\phib_0)^2 ~\right)  , $$
where $\phib_0^2=\sigma ^2$ is an additional variational parameter.
Eq.~\esaddleNc~is replaced by
$$\rho=(1/N)\left<\phib^2(x)\right>_0= \Omega _d(m) +\sigma ^2/N\,.\eqnd\esaddleNcii $$
Then,
$${\cal E}_{\rm var.}/N\mathop{=}_{N\to \infty }  {1\over 2 (2\pi)^d}
\int \d^d k\, \ln[(k^2+m^2 )/k^2] +U(\rho)-\ud m^2( \rho-\sigma
^2/N).$$
We  note that again ${\cal E}_{\rm var}$ is identical to
\eNEner~when Eq.~\esaddleN{c} is used. \par
Differentiating the
variational energy with respect to  $m^2$, we first notice
$${\partial \over \partial m^2}{\cal E}_0=\ud\left<(\phib-\phib_0)^2\right>_0=
\ud(N\rho-\sigma ^2)$$
and thus find
$$\eqalign{ {\partial {\cal E}_{\rm var.}\over \partial m^2}&=
\ud N(\rho-\sigma ^2/N)+N {\partial \over \partial
m^2}\left[U(\rho)-\ud m^2( \rho-\sigma ^2/N)\right] \cr
&=N{\partial \rho \over \partial m^2}\left[U'(\rho)-\ud m^2\right].\cr} $$
Since ${\partial \rho / \partial m^2}$ is strictly negative, the
derivative vanishes only when
$$U'(\rho)-\ud m^2=0\,, $$
which, combined with Eq.~\esaddleNcii, yields the set of equations
\esaddleN{b} and \esaddleN{c}.
Differentiating with respect to $\sigma $, we recover Eq.~\esaddleN{a}.
\par
Therefore, in the large $N$ limit the saddle point equations and the
variational equations coincide.
It should, however, be remembered that since we perform a variational
calculation the lowest energy eigenstate is not necessarily the
state determined by the solution of Eqs.~\esaddleN{}.
The end-points in the range of variation of $m^2$ and ${\phib_0}^2$
have to be considered as well and the values of ${\cal E}_{\rm var}$ at these
points have to be examined and compared to the extremum values \rBarMos.
\medskip
{\it The $(\phib^2)^2$ field theory at negative coupling for $d=4$.}
We again consider the large $N$ self interacting scalar field in $d=4$
dimensions  with ($\rho\equiv \phib^2/N$)
 $$U(\rho)={1\over2} r\rho+{u\over4!}\rho^2.        \eqnd \phiPotential $$
The large $N$ limit is defined as $N \to \infty$ holding
$u $, $r$  and the necessary
ultraviolet cutoff, $\Lambda$, fixed.
We have already discussed the triviality of the $(\phib^2)^2$ field theory
in $d=4$ for  positive coupling in section \label{\sstrivia}. Here, we add some comments about the situation for negative coupling. Even though the initial functional integral is defined only for $u>0$, the large $N$ expansion seems to have a well-defined continuation.\par
The only divergence appears in
$$ \rho(m^2)=\left<\phib^2\right>/N=\rho_c- {m^2 \over 8 \pi^2} \ln(\Lambda /m)+\sigma ^2/N \,,\quad
\rho_c= \Omega _4(0)\,,     \eqnn   $$
where a numerical, regularization dependent, constant has been cancelled
by adjusting the definition of the cut-off $\Lambda $.\par
We then obtain the energy density
$$\eqalignno{ { 1 \over N} {\cal E}_{\rm var}( \sigma ,m^2)&=U(\rho)-\ud \int_0^{m^2}
s{\partial \rho(s)\over \partial s}\d s=
U(\rho_c) +{1\over 32\pi^2} m^4( \ln( \Lambda/m)- \frac{1}{4}) \cr
& +{u\over 24}\biggclb \sigma^2/N +6 {r-r_c \over u} - {1\over 8\pi^2}    m^2 \ln(\Lambda/ m)  \biggcrb^2 -{3\over 2} {(r-r_c)^2 \over
u}\,.\hskip8mm &\eqnd\WfourD  }$$
The gap equation   can be expressed in terms of the
renormalized parameters (see Eq.~\egfivivren),
$$ {1 \over u_\r} ={1\over u}  +{1 \over  48\pi^2} \ln(\Lambda/\mu)\,,  \quad
{M^2 \over u_\r}  =  {r-r_c  \over u}  \,,  \eqnd \ecoupmassRe   $$
where $\mu$ is the renormalization scale and $M$ a renormalized mass parameter, as
$${m^2\over u_\r}=-{m^2\ln(\mu/m) \over 48\pi^2}+{M^2\over u_\r}+{\sigma ^2\over 6N}\,.$$
In the  case $u < 0 $ the renormalized coupling constant $u_\r$ does
not necessarily approach zero as the cutoff is removed (if $u
\rightarrow 0^-$ as $\Lambda \to \infty$).
One also finds that the solutions of the gap equation reproduce the past
results in the literature.
However, the ground state energy is lower
at the end point $m^2=0$ value in which the $O(N)$ symmetry is broken
down to $O(N-1)$. One finds in Eq.~\WfourD~that with $u < 0$,
the ground state energy density,
${\cal E}_{\rm var} \rightarrow -\infty$ as
$ \sigma $ is becoming larger. Since ${\cal E}_{\rm var}$, in this case, is
the upper limit to the exact ground state energy density,
one concludes that the theory
is inconsistent for $u< 0$.

One finds, however, that when $u<0$ the $m\ne 0$ solution of
 \esaddleN{} 
corresponds to an $O(N)$ symmetric metastable state whose life-time
can be calculated. One finds
that this metastable state is stabilized by a large
tunnelling barrier; its decay rate
is proportional to $\e^{-N}$
and thus, in our large $N$ limit, is an acceptable
ground state for the theory.
However, this metastable state is not a good
candidate for the vacuum of a realistic model because of its
peculiar behaviour at finite temperature
(see  section \label{\ssNvarfivT}).
At first, it may sound strange that an $O(N)$ symmetric state will
become unstable as the temperature is raised. Indeed, one may expect that
the usual behaviour will appear here,
in which, as the temperature increases
the $O(N)$ symmetric phase is stabilized, whereas
the $O(N)$ broken phase,
(which here is the end point $m^2=0$ phase) will be destabilized.
This expectation is based on the fact that at finite temperature,
thermal fluctuations add up to the quantum fluctuations of the field
operator $\phib^2$ .
Thus, one usually expects that a possible
negative ``mass" at low temperature
will turn positive as the temperature
increases, and the possible broken
symmetry will be restored. This, however, is true as long as one
discusses a stable  theory $u>0$. The effect of the temperature
is reversed in a metastable situation (here, $u<0$) since thermal fluctuations help to overcome potential barriers.

The triviality of $ (\phib^2)^2$ for $u > 0 $
and its inconsistency for $u< 0$
have been derived here for the large $N$ limit of
the theory. It seems very likely that these results
persist also at finite $N$.
Interesting suggestions came
from different view points
that the
positively coupled theory may have non-trivial
implications on practical physics
issues. In the standard Weak-Electromagnetic theory,
bounds on the Higgs mass can be
derived (see for example section \sstrivia).

\vfill\eject

\section Models on symmetric spaces in the large $N$ limit

A whole class of geometric models involving only scalar fields, for which the non-linear $\sigma $ model is the simplest example, shares  various classical and quantum  properties: The field belongs to a symmetric space, associated to a coset $G/H$, where $G$ is a Lie group and $H$ a compact maximal subgroup. The action is unique, up to a multiplicative factor which is the coupling constant, and takes the general form \sslbl\ssNsymspace
$${\cal S}(\varphi)={1\over 2T}\int\d^d x\, \partial _\mu \varphi^i(x)g_{ij}(\varphi)  \partial _\mu \varphi^j(x), $$
where $g_{ij}$ is the metric tensor on the corresponding manifold $G/H$.
This action leads to an infinite number of classical conservation laws in two dimensions.
The corresponding quantum models are renormalizable in two dimensions, and then UV asymptotically free. In the classical limit the fields are massless, and correspond to the Goldstone bosons of the $G$ symmetry broken down to $H$. Since continuous symmetries cannot be broken in two  dimensions, the
 spectrum of such theories is non-perturbative. In dimension $2+\varepsilon$, one finds a critical coupling constant $T_c$ where a phase transition occurs. \par
To go beyond the $\varepsilon=d-2$ expansion, one can consider using large $N$ techniques.
However, only a subclass can be studied in this way. To this subclass belong the $O(N)$ non-linear
$\sigma $-model and the $CP(N-1)$ model that are discussed below. A number of other models based on
Grassmannian manifolds could also be  investigated, like $O(N)/O(N-p)\times O(p)$ or similarly
$U(N)/ U(N-p)\times U(p)$. More general homogeneous spaces with several coupling constants could also be considered.
\subsection The non-linear $\sigma$-model in the large $N$ limit

In reference \rBrZJsig\ it was first shown that universal
quantities could be determined in the form of an $\varepsilon=d-2$ expansion (at least for $N>2$). Again, as for the $\varepsilon=4-d$ expansion, it is somewhat reassuring  that at least in the
limiting case $N\to\infty $,  the results
obtained in this way remained valid even when $ \varepsilon$
is no longer infinitesimal \refs{\rDerMa,\rCamros}. \par
Moreover, the $1/N$ expansion allows exhibiting a remarkable relation between  the non-linear $\sigma$-model  and the  $(\phib^2)^2$ field theory,  a relation expected on physical grounds \rBrZJsig. \par
Finally, large $N$ techniques are well adapted to the analysis of finite size effects in critical systems, a feature we illustrate with a system in a finite volume with periodic boundary conditions.
\sslbl\ssLTsN
\medskip
{\it Non-linear $\sigma $-model and $(\phib^2)^2 $ field theory.}
We have noticed that the term proportional to $\int\d^d x\,\lambda^2(x)$ has dimension $4-d$ for $N\to \infty $. It is thus irrelevant in
the critical domain for  all dimensions $d<4$ and can be omitted at leading order (this
also applies to $d=4$ where it is marginal but yields only logarithmic
corrections).
Actually, the constant part in the inverse propagator as written in equation
\epropNli~plays the role of a large momentum cut-off. We thus omit the $\lambda^2$ term in the action \eactONsigla\ after shifting
 the field
$\lambda(x)$ by its expectation value $m^2$  (Eq.~\eNsaddpts),
$\lambda(x) \mapsto m^2+\lambda(x)$:\sslbl\sssfivNRT 
$$\eqalignno{{\cal S}_N ( \sigma ,\lambda) &  = {1 \over 2}\int \d ^{d}x \left[
\left(\partial_{\mu}\sigma \right)^2+ m^2\sigma
^2+ \lambda  \sigma^2  - {3N\over u}\lambda^2  -{6 N\over u} \left(m^2-r\right) \lambda \right] \cr &\quad +{ (N-1
 ) \over 2}  \tr\ln \left[ -\nabla^2 +m^2+\lambda (\bullet) \right]
. &\eqnd  \eacteffb  \cr}$$
If we then work backwards, reintroduce the initial field $\phib$ and
integrate over $ \lambda (x) $, we find
$$ {\cal Z}= \int \left[ \d  \phib (x)\right]\delta
\left[ \phib^2 (x)/N- 6   (m^2-r  )/u
\right] \exp\left[- { 1 \over 2}\int\left(\partial_{\mu} \phib(x)\right)^2
\d ^{d}x\right]. \eqnd{\epartsig} $$
Under this form we recognize the partition function of the $ O(N) $ symmetric
non-linear $ \sigma $-model in an unconventional normalization. We have,
therefore, discovered a remarkable correspondence,  to all orders in an $ 1/N $
expansion, between the non-linear $ \sigma $-model and the $ ( \phib^2 )^2 $ field theory. \par
Actually, reviewing carefully the arguments, one verifies that the identity between the two models has been derived here under the implicit
condition $u_{\phi^4}\propto \Lambda ^{4-d}\gg(m_{\phi})^{4-d}$, which is the generic situation if
the $\phi^4$ interaction is an effective long distance interaction generated
 by some microscopic model. If the condition is not satisfied the situation
is less clear.
\medskip
{\it The large $N$ limit.} We now  write the partition function of the non-linear $\sigma $ model, with slightly more usual notation, as
$${\cal Z}= \int \left[\d\phi(x)\d\lambda(x)\right]
\exp\left[-{\cal S}(\phib,\lambda)\right]  \eqnn  $$
with
$${\cal S}(\phib,\lambda) = {1 \over 2T}\int \d^{d}x \left[ \left(
\partial_{\mu}\phib \right)^2 + \lambda \left(\phib^2 -N \right)\right].
\eqnd\eactsigla $$
By solving  in the large $N$ limit the $\sigma$-model directly, we are able to exhibit more explicitly the correspondence
between the different set of parameters used in the two models.
\par
The field $\phib$ represents a classical spin of size $\sqrt{N}$. The
coupling constant $T$, which is a loop expansion parameter, represents
the temperature of the classical spin model. Eventually, it will be
useful to introduce a dimensionless coupling constant $t$, setting
$$T=\Lambda^{2-d} Nt\,. \eqnd\eNsigtdef $$
\par
We now separate the field $\phib$ into $N-1$ components,
which we call $\pib$ in what follows, and over which we
integrate as we did in section \ssNbosgen,  and a remaining component $\sigma$. We obtain
$${\cal Z}= \int \left[\d\sigma(x)\d\lambda(x)\right]
\exp\left[-{\cal S}_N(\sigma,\lambda)\right]  \eqnn $$
with, for $N\gg 1$,
$${\cal S}_N  (\sigma,\lambda  )=
 {1 \over 2T}\int \left[ \left(\partial_{\mu}\sigma \right)^2+
\left(\sigma^2 (x)-N\right) \lambda (x)
\right] \d^{d}x +{N \over 2}   \tr\ln \left[
-\nabla^2 +\lambda (\cdot) \right] .  \eqnn $$
The large $N$ limit is  taken here at $T $ fixed. The saddle point equations,
analogous to Eqs.~\esaddleN{}, are
\eqna\emgNsig
$$\eqalignno{m^2\sigma &=0\, ,& \emgNsig{a} \cr
\sigma^2/N& = 1 -   \Omega _d(m)T\,,& \emgNsig{b}\cr} $$
where we have set $\left<\lambda(x)\right>=m^2$ and introduced the function \etadepole. At low temperature and $d>2$, $\sigma $
is different from zero and thus $m$, which is the mass of the $\pi$-field,
vanishes. Eq.~\emgNsig{b} yields the spontaneous magnetization:
$$\sigma^ 2 /N = 1 -  \Omega _d(0)T   .\eqnd\emagNsig $$
Setting
$$  T_c = 1/ \Omega _d(0) ,\eqnn $$
we can write Eq.~\emagNsig\ as
$$\sigma^2/N = 1 - T/T_c\, . \eqnd\emgTNsig $$
Thus, $T_c$ is the critical temperature where $\sigma$ vanishes.\par
Above $T_{c}$, $\sigma$ instead vanishes and $m$, which is now the
common mass of the $\pi$- and $\sigma$-field, is for $d>2$
given by
$$ {1 \over  T_c}- {1 \over  T} = \Omega_d(0)-\Omega_d(m)
  .\eqnd\emasNsig $$
The physical mass $m$ is solution of an equation quite similar to \ecorleng{}. In particular, for $d<4$, we recover  the scaling form \ecorlenb~of the correlation length:
$$\Omega_d(0)-\Omega_d(m)
=K(d)m^{d-2} +O(\Lambda^{d-4}m^2) \ \Rightarrow \ \xi=1/ m\propto (T-T_c)^{-1/(d-2)}.$$
\par
In terms of the cut-off $\Lambda$ and the dimensionless coupling $t$
(Eq.~\eNsigtdef), vertex and correlation functions of the non-linear $\sigma$
model satisfy for $d<4$ RG equations:
$$\eqalign{\left(\Lambda {\partial \over \partial\Lambda}+\beta(t){\partial \over
\partial t}- {n\over 2} \zeta(t)\right)\Gamma^{(n)}(p,\Lambda,t)&=0\,, \cr
\left(\Lambda {\partial \over \partial\Lambda}+\beta(t){\partial \over
\partial t}+{n \over2}\zeta(t)\right)W^{(n)}(p,\Lambda,t)&=0\,. \cr} $$
Applying the second equation with $n=1$ to Eq.~\emgTNsig~and
expressing in Eq.~\emasNsig~that $m$ is RG invariant, we determine the RG
functions at leading order for $N$ large:
$$ \beta (t) = (d-2) t(1-t/t_c)\, ,\qquad
 \zeta (t) =(d-2) t/t_c\quad {\rm with} \ t_c=\Lambda ^{d-2}T_c/N\,. \eqnn  $$
If we add to the action \eactsigla~the contribution due to a magnetic field  in
the $\sigma $ direction,
$${\cal S}(\phib,\lambda) \mapsto {\cal S}(\phib,\lambda) -{H\over T}
\int\d^d x\, \sigma (x),$$
we can calculate the free energy density $W(H)=T\ln {\cal Z}(H)/\Omega $ ($\Omega $ is the volume) and  its Legendre
transform (Eqs.~\eqns{\eNfivmag,\eNfivGam}), the thermodynamic potential density  function of the magnetization $M$:
$$ {\cal G}(M)=N{d-2\over2d}{T^{2/(2-d)}\over\bigl(K(d)\bigr)^{2/(d-2)}}
(M^2/N-1+T/T_c)^{d/(d-2)},\eqnn $$
a result that extends the scaling form \ethermscN~to all temperatures below
$T_c$. The calculation of other physical quantities and the
expansion in powers of $1/N$ follow from the considerations of previous sections
and section \sssfivNRT.
\medskip
{\it Two dimensions and the question of Borel summability.} For $d=2$, the
critical  temperature vanishes and the parameter $m$ has the form
$$m \sim \Lambda_0 \e^{-2\pi / T}, \eqnn $$
where (with the definition \epropreg)
$$\ln(\Lambda _0/\Lambda )={ 1\over4\pi}\int_0^\infty {\d s\over s}\left(
{1\over D(s)}-\theta (1-s)\right), $$
in agreement with RG predictions. Note that the field inverse two-point
function in the large $N$-limit is given by
$$\tilde\Gamma^{(2)}_{\sigma\sigma}(p)=p^2 + m^2\,. \eqnd\eNsigmii $$
The mass term vanishes to all orders in the expansion in powers of
the coupling constant $t$, preventing a perturbative calculation of
the mass of the $\sigma$-field. The perturbation series is
trivially not Borel summable. Most likely, this property remains true for the
model at finite $N$. On the other hand, if the $O(N)$ symmetry is broken by
the addition of a term proportional to $ \int\d x\,\sigma(x)$ to the action (a magnetic field), the
physical mass becomes calculable in perturbation theory.
\medskip
{\it Corrections to scaling and the dimension 4.} In Eq.~\emasNsig\
we have neglected corrections to scaling. If we take into account the leading
correction, which becomes increasingly important when $d$ approaches 4 from below,  we get instead
$$ {1\over T_c}-{1\over  T}=K(d)m^{d-2} -a(d)\Lambda ^{d-4}m^2   +O\left(\Lambda ^{d-6}m^4, \Lambda ^{-2}m^d\right),$$
where $a(d)$, as we have already discussed, is a constant that explicitly
depends on the cut-off procedure and can thus be varied by changing
contributions from irrelevant operators.\par
We can compare this result with the solution of Eqs.~\ecorleng{} expanded
at the same order:
$${6\over u}(r-r_c-m^2)=K(d)m^{d-2} -\tilde a(d)\Lambda ^{d-4}m^2   +O\left(\Lambda ^{d-6}m^4\right),$$
where the constant $\tilde a(d)$ has the same formal expression as $a(d)$ but corresponds to a different regularization. Eliminating the leading contribution, we find
$$ {1\over  T_c}-{1\over  T}-{6\over u}(r-r_c)=\Lambda ^{d-4}m^2
\left(\tilde a(d)-{6\over Ng}-a(d)\right).$$
We note that it is thus possible to find a regularization of the non-linear $\sigma $-model that reproduces the effect of the $\phi^4$ coupling constant.
\par
More generally, by comparing with the results of section \sssEGRN, we discover that,
although the non-linear $\sigma$-model superficially depends on one parameter
less than the corresponding $\phib^4$ field theory, actually this parameter is
hidden in the cut-off function. This remark becomes important in the four
dimensional limit where most leading contributions come from the leading
corrections to scaling. For example, for $d=4$ Eq.~\emasNsig\ takes a
different form, the dominant term in the r.h.s.\ is proportional to
$m^2\ln m$. We recognize in the factor $\ln m$ the effective
$\phi^4$ coupling at mass scale $m$. However,
to describe with perturbation theory and RG the physics
of the non-linear $\sigma$ model beyond the $1/N$ expansion,  it is necessary  to return to the $\phi^4$ field theory. This involves addding to the action
the operator $\int\d^d x\,\lambda^2(x)$, which irrelevant for $d<4$,
becomes marginal in four dimensions.
\subsection $1/N$-expansion and renormalization group: an alternative formulation

{\it Preliminary remarks. Power counting.}
Higher order terms in the steepest descent calculation of the
functional integral \eZeff\ generate a systematic $ 1/N $ expansion.

We now analyze the terms in the action \eacteffb~from the
point of view of large $N$ power counting. The
dimension of the field $ \sigma (x) $ is $ (d-2)/2 $. From the critical
behaviour \eprocrit\ of the $\lambda$-propagator, we inferred the
canonical dimension  $[\lambda]$ of the field $ \lambda (x) $:
$$ 2 \left[ \lambda \right] -\varepsilon =d\,, \qquad {\rm i.e.} \quad \left[
\lambda \right] =2\,. $$
As noted above, $\lambda^2$ has dimension $4>d$ and thus is  irrelevant below  four dimensions.
The interaction term $ \int \lambda(x)\sigma^2 (x)\d^d x $ has
dimension zero. It is easy to verify that the non-local interactions
involving the {$ \lambda$-field},  coming from the expansion of the $ \tr\ln $,
have all also  canonical dimension zero:
$$ \left[ \tr \left[ \lambda (x) \left(-\nabla^2 +m^2 \right)^{-1}
\right]^{k} \right] =k \left[ \lambda \right] -2k=0\,. $$
This power counting  has the following implication: in contrast with
usual perturbation theory,  the $ 1/N $ expansion is exactly renormalizable and thus generates only logarithmic
corrections to the leading long distance behaviour for any fixed dimension
$d$, $ 2<d\leq 4$. A similar behaviour is found in  the $ \varepsilon $-expansion (at the IR fixed point) and, thus, one expects here also to be able to
calculate universal quantities like critical exponents for example as power
series in $ 1/N$. However, because the interactions are non-local, it
is not obvious that the general results of renormalization theory
apply here. Therefore, we now construct
an alternative quasi-local field theory, for which the standard RG
analysis is valid, and which reduces to the large $N$ field theory in
some limit \rANPVN.
\medskip
{\it An alternative field theory.}
To be able to use the standard results of renormalization theory, we
reformulate the critical theory to deal with the non-local
interactions. Neglecting corrections to scaling, we start from the
non-linear $\sigma$-model in the form \eactsigla:
$$ \eqalignno{ {\cal Z} & = \int \left[ \d  \lambda (x)
\right] \left[ \d  \phib (x) \right]
\exp\left[-{\cal S} (\phib ,\lambda )\right] , & \eqnn
\cr {\cal S} ( \phib ,\lambda) & = {1 \over 2T}\int \d^{d}x \left[ \left(
\partial_{\mu}\phib \right)^2 + \lambda \left(\phib^2 -N \right)\right].
  & \eqnd{\eactphla}   \cr} $$
The difficulty arises from the $\lambda$-propagator, absent in the
perturbative formulation, and generated by the large $N$ summation.
We thus add to the action \eactphla\ a term quadratic in $\lambda$
that, in the tree approximation of standard perturbation theory, generates a
$\lambda$-propagator of the form \eprocrit.
We thus consider the modified action
$$ {\cal S}_{\vv}  (  \phib ,\lambda ) = {1 \over 2}\int
\d^{d}x \left\{
{1\over T}\left[\left( \partial_{\mu}\phib \right)^2 + \lambda
\left(\phib^2 -N\right)\right]-
{1\over\vv^2}\lambda(-\partial^2)^{-\varepsilon/2}\lambda\right\}
. \eqnd{\eactSg}$$
In the limit where the parameter $\vv$ goes to infinity, the coefficient of
the additional term vanishes and the initial action is recovered. \par
Only the critical theory is discussed  below, and thus the
couplings of all relevant interactions are set to their critical values.
These interactions contain a term linear in $\lambda$ and a polynomial in
$\phib^2$ of degree depending on the dimension. Note that in some discrete
set of dimensions some monomials become just renormalizable. Therefore, we work
in generic dimensions and rely on the property that the quantities we calculate are regular
functions of the dimension. \par
The field theory with the  action \eactSg~can be studied with
standard field theory methods. The peculiar form of the $\lambda$ quadratic
term, which is not strictly local, does not create a problem. Similar terms
are encountered in statistical systems with long range forces. The
main consequence is
that the $\lambda$-field is not renormalized because counter-terms are
always local.\par
It is convenient to rescale $\phib\mapsto \phib\sqrt{T}$, $\lambda \mapsto
\vv\lambda$:
$$ {\cal S}_{\vv} ( \phib ,\lambda ) = {1 \over 2}\int \d^{d}x \left[
\left( \partial_{\mu}\phib \right)^2 + \vv\lambda\phib^2
- \lambda(-\partial^2)^{-\varepsilon/2}\lambda +{\rm relevant\
terms}\right].$$
The renormalized critical action then reads
$$[{\cal S}_\vv]_{\rm ren} =  {1 \over 2}\int \d^{d}x \left[
Z_\phi \left( \partial_{\mu}\phib \right)^2 +\vv_\r Z_{\vv}\lambda\phib^2
- \lambda(-\partial^2)^{-\varepsilon/2}\lambda +{\rm relevant\
terms}\right] . \eqnd\eactSren $$
It follows that the RG equations for vertex functions of $l$
$\lambda$ fields and $n$ $\phib$ fields in the critical theory take the
form
$$\left[\Lambda {\partial \over \partial \Lambda}+
\beta_{\vv^2}(\vv){\partial \over \partial \vv^2}-{n\over 2}\eta(\vv)\right]
\Gamma^{(l,n)}=0\,.\eqnd \eqRGm $$
The solution to the RG equations \eqRGm\ can be written as
$$\Gamma^{(l,n)}(\ell p, \vv,\Lambda)=Z^{-n/2}(\ell)
\ell^{d-2l-n(d-2)/2} \Gamma^{(l,n)}(p, \vv(\ell) \Lambda)
\eqnd\esolRG $$
with the usual definitions
$$\ell{\d \vv^2\over \d \ell}=\beta(\vv(\ell))\,,\quad \ell{\d \ln Z
\over \d \ell}=\eta(\vv(\ell))\, .$$
We can then calculate the RG functions as power series in $1/N$. It
is easy to verify that $\vv^2$ has to be taken of order $1/N$. Therefore,
to generate a $1/N$ expansion, one first has to sum the multiple insertions of
the one-loop $\lambda $ two-point function, contributions that form a
geometric series. The $\lambda$ propagator then becomes
$$\Delta_\lambda(p)=-{2 p^{4-d}\over b(d) D(\vv)}  \eqnn $$
($b(d)$ is given in Eq.~\econstb), where we have defined
$$D(\vv)=2/b(d)+ N\vv^2.$$
We are interested in the neighbourhood of the fixed point $\vv^2=\infty$.
One verifies that the RG function $\eta(\vv)$ approaches the exponent
$\eta$ obtained by direct calculation, and the RG $\beta$-function
behaves like  $\vv^2$. The flow equation
for large coupling constant  becomes
$$\ell{\d \vv^2\over \d \ell}\sim\rho \vv^2  \ \Rightarrow\ \vv^2(\ell)\sim
\ell^{\rho}.\eqnn $$
We then note that to each power of the field $\lambda$ corresponds a power of
$\vv$. It follows that
$$\eqalignno{\Gamma^{(l,n)}(\ell p,\vv,\Lambda)&\propto
v^l(\ell)\ell^{d-2l-n(d-2+\eta)} & \cr
&\propto\ell^{d-(2-\rho/2)l-n(d-2+\eta)} .& \eqnn \cr}
$$
To compare with the result obtained from the perturbative
RG, one has still to take into account that the functions
$\Gamma^{(l,n)}$ defined here are obtained by an additional Legendre
transformation with respect to the source of $\phib^2$. Therefore,
$$ 2-\rho/2=d_{\phib^2}=d-1/\nu \,. \eqnn $$
\midinsert
\epsfxsize=37.6mm
\epsfysize=18.2mm
\centerline{\epsfbox{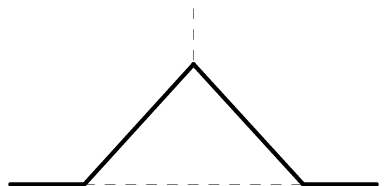}}
\figure{3.mm}{Diagram contributing to $\Gamma^{(3)}_{\sigma\sigma\lambda}$ at
order  $1/N$.}
\figlbl\figNtriangl
\endinsert
\midinsert
\epsfxsize=34mm
\epsfysize=27.2mm
\centerline{\epsfbox{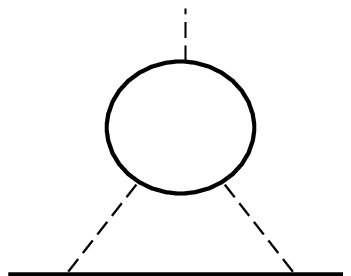}}
\figure{3.mm}{Diagram contributing to $\Gamma^{(3)}_{\sigma\sigma\lambda}$ at
order  $1/N$.}
\figlbl\figNtrianglii
\endinsert

\medskip
{\it RG functions at order $1/N$.}
Most calculations at order $1/N$ rely on the evaluation of the generic
integral
$${1\over(2\pi)^d}\int{\d^d q \over (p+q)^{2\mu}
q^{2\nu}}=p^{d-2\mu-2\nu}{\Gamma(\mu+\nu-d/2)\Gamma(d/2-\mu)\Gamma(d/2-\nu)
\over (4\pi)^{d/2}\Gamma(\mu)\Gamma(\nu)\Gamma(d-\mu-\nu)}\,
. \eqnd\eintmunu $$
For later purpose, it is convenient to set
$$X_1={2 N_d\over b(d)}= {4 \Gamma(d-2)\over
\Gamma(d/2)\Gamma(2-d/2)\Gamma^2(d/2-1)}= 
{4 \sin(\pi\varepsilon/2) \Gamma(2-\varepsilon)\over
\pi\Gamma(1-\varepsilon/2) \Gamma(2-\varepsilon/2)} .\eqnd{\eXone}$$
To compare with fixed dimension results, note that $X_1\sim 2(4-d)$ for $d\to 4$
and $X_1\sim(d-2)$ for $d\to 2$.\par
The calculation of the $\left<\phi\phi\right>$ correlation function
at order $1/N$ involves the evaluation of the diagram
of figure  \figbubiii.
 We want to determine the coefficient of
$p^2\ln\Lambda/ p$. Since we work at one-loop order, we can instead
replace the $\lambda$ propagator $q^{-\varepsilon}$ by $q^{2\nu}$ and send the
cut-off to infinity. We then use the result \eintmunu\ with
$\mu=1$. In the limit $2\nu \to-\varepsilon$, the integral has a
pole. The residue of the pole  yields the coefficient of
$p^2\ln\Lambda$ and the finite part contains the $p^2\ln p$ contribution:
$$\tilde\Gamma^{(2)}_{\sigma\sigma}(p)= p^2 +
{\varepsilon\over 4-\varepsilon} {2 N_d\over b(d) D(\vv)} \vv^2
p^2\ln(\Lambda/p) .$$
Expressing that the function satisfies the RG equation, we obtain the
function $\eta(\vv)$.\par
The second RG function can be deduced from the divergent parts
of the $\left<\phi\phi\lambda\right>$ function:
$$\tilde\Gamma^{(3)}_{\sigma\sigma\lambda}=\vv+A_1 \vv^3 D^{-1}(\vv)\ln\Lambda +A_2
\vv^5 D^{-2}(\vv)\ln \Lambda +\ {\rm finite} $$
with
$$\eqalign{A_1 &=-{2\over b(d) } N_d=-X_1\,, \cr
A_2 &=- {4N\over b^2(d)} (d-3)b(d) N_d=-2N(d-3)X_1\,,
\cr} $$
where  $A_1$ and $A_2$ correspond to the diagrams of figures \figNtriangl~and
\figNtrianglii, respectively. \par
Applying the RG equation, one finds at order $1/N$ the relation
$$\beta_{\vv^2}(\vv)=2\vv^2\eta(\vv)-2A_1 \vv^4 D^{-1}(\vv)-2A_2
\vv^6 D^{-2}(\vv) . \eqnn $$
One thus obtains
$$\eqalignno{\eta(\vv) &={ \varepsilon\vv^2 \over 4-\varepsilon}X_1
D^{-1}(\vv), & \eqnn \cr
\beta_{\vv^2}(\vv) &={8 \vv^4 \over 4-\varepsilon}X_1 D^{-1}(\vv)
+4N(1-\varepsilon) \vv^6 X_1 D^{-2}(\vv), &\eqnn \cr}$$
where the first term in $\beta_{\vv^2}$ comes from $A_1$ and $\eta$
and the second from $A_2$.\par
Extracting the large $\vv^2$ behaviour,  one infers
$$\eqalignno{\eta  & =  { \varepsilon \over N (4-\varepsilon)} X_1 +O(1/N^2)
,& \eqnn \cr
\rho&={4(3-\varepsilon)(2-\varepsilon)\over N(4-\varepsilon)}X_1 > 0\,, \cr}$$
and thus
$${1\over\nu}=d-2 + {2(3-\varepsilon)(2-\varepsilon)\over N(4-\varepsilon)}X_1
+O(1/N^2) .\eqnn $$
\subsection Some  higher order results

The calculations beyond the order $1/N$ are rather technical. The reason
is easy to understand: because the effective
field theory is  renormalizable in all dimensions $2\leq d \leq 4$, the
dimensional regularization, which is so helpful in perturbative calculations,
can no longer be used. Therefore, either one keeps a true cut-off or one
introduces more sophisticated regularization schemes. For
details the reader is referred to the literature \refs{\rKTOOVPH{--}\rGraci}.
\medskip
{\it Generic dimensions}. The exponents $\gamma$ and  $\eta$
are known up to order $1/N^2$ and $1/N^3$, respectively, in arbitrary
dimensions but the expressions are too complicated to be reproduced here.
The expansion of $\gamma$ up to order
$1/N$ can be directly deduced from the results of the preceding
sections:
$$ \gamma  =  { 1 \over 1- \varepsilon / 2}\left(1-{3 \over
2N}X_1 \right)+O \left({1 \over N ^2} \right) .\eqnn $$
The exponents $ \omega $ and $\theta=\omega\nu$, governing the leading
corrections to scaling, can also be calculated, for example, from the
$\left<\lambda^2\lambda\lambda\right>$ function:
$$ \eqalignno{\omega & = \varepsilon\left(1-{2(3-\varepsilon)^2 \over
(4-\varepsilon)N}X_1\right) +O \left({1 \over N^2}
\right), & \eqnn \cr
\theta=\omega \nu &={\varepsilon \over 2-\varepsilon}
\left(1-{2(3-\varepsilon) \over N}X_1\right) +O \left({1 \over N^2}
\right). & \eqnn \cr}$$
Note that the exponents are regular functions of $\varepsilon$ up to
$\varepsilon =2$ and free of renormalon singularities at $\varepsilon
=0$.\par
The equation of state and the spin--spin correlation function in zero field
are also known at order $ 1/N $, but since the expressions are
complicated we again refer the reader to the literature for details.
\medskip
{\it Three dimensional results.} Let us give the expansion of $\eta$  in
three dimensions at the order presently available:
$$ \eta  = {\eta_{1} \over N}+{\eta_2 \over N^2}+ {\eta_{3} \over N^{3}}
+O \left({1 \over N^{4}} \right) $$
with
$$\eta_1 ={\textstyle{8 \over 3\pi^2}}\, ,\quad \eta_2= -{\textstyle {8
\over 3} \eta_1^2}\, ,\quad
\eta_{3} = \eta_1^3 {\textstyle\left[ -{797 \over 18} - { 61 \over 24}\pi^2
+ {27 \over 8}\psi''(1 / 2 ) + {9 \over 2}\pi^2 \ln 2
\right]}, $$
$\psi(x)$ being the logarithmic derivative of the $\Gamma$ function.\par
The exponent $\gamma$ is  known only up to order $1/N^2$:
$$\gamma  = 2 -{\textstyle {24 \over N \pi^2}} + {\textstyle{64 \over
N^2\pi^{4}}\left({44 \over 9} - \pi^2 \right)+O \left({1 \over N^{3}}
\right)}. $$
Note that the $1/N$ expansion seems to be rapidly divergent and certainly a
direct summation of these terms does not provide precise estimates
of critical exponents in three dimensions for relevant values of $N$.
\medskip
{\it The nature of the large $N$ expansion.} The large order
behaviour of the $N$ expansion has been determined explicitly in
zero and one dimension (simple integrals and quantum
mechanics) \rBrHi.
In quantum field theory some estimates are available in
ref.~\rVegAv.
All results seem to indicate that the expansion is  divergent in all
dimensions, but Borel summable for dimensions $d<4$.

\subsection Finite size effects: the non-linear $\sigma $ model

Because finite size effects involve crossover phenomena between different effective dimensions, large $N$ techniques provide convenient tools to study them \rBrCaPe. It is difficult to discuss systematically finite size
effects because the results depend both on the geometry of the system and on
 boundary conditions. In particular, one must discuss separately boundary
conditions whether they break or not translation invariance. In
the first case new effects appear, which are surface effects, and that we do
not consider here.
We study here only periodic conditions, although they are not the only ones  preserving translation invariance. For systems that have a symmetry, one can
glue the boundaries after having made a group transformation. Thus,
here one could
also discuss anti-periodic conditions or, more generally, fields differing on both sides by a
transformation of the $O(N)$ group.\sslbl\ssNFSS \par
Even with periodic boundary conditions, the number of different possible situations remains large, the finite sizes in different directions may differ,
some sizes may be infinite. Note  that
QFT at finite temperature, which we begin studying in section \scFTQFT, can
be considered as another example of finite size effects, since the
functional integral representing the partition function of a quantum
system at finite temperature is also the partition function of a
classical system with finite size  and periodic
boundary conditions in one dimension.
\par
>From the point of view of the RG, finite size effects, which
only affect the IR domain, do not change UV divergences. RG
equations remain the same, only the solutions are modified due to
the existence of new dimensional parameters. Thus, if finite sizes
are characterized by only one length $L$, solutions will be
functions of an additional  argument like $L/\xi$ where $\xi$ is
the correlation length \rBZJFSS. \par Since we want only to
demonstrate that large $N$ techniques are useful in the context of
finite size effects, we discuss here the non-linear $\sigma
$-model in a simple geometry with periodic boundary conditions
(the geometry of the hypercube  or better hypertorus of linear
size $L$).\par A characteristic property of a system of finite
size is the quantization of momenta, the arguments of field
Fourier components. For periodic boundary conditions, if  $L$ is
the size of the system in a direction $\mu$, we find
$$p_\mu=2\pi n_\mu /L \,,\quad n_\mu\in {\Bbb Z}\,.$$
In particular, in a massless theory in a finite volume the zero mode
${\bf p}=0$   corresponds to an isolated pole of the propagator.
This automatically leads to IR divergences in all dimensions.
Therefore, in Eqs.~\esaddleN{} the solution $\sigma\ne 0$ no longer
exists. This is not surprising:   no phase  transition is expected in a
finite volume.
 \medskip
{\it Finite size scaling in the non-linear $\sigma$ model.}
The gap equation \emgNsig{b} becomes ($N\gg1$)
$$G(m,L,\Lambda )\equiv L^{-d}\sum_{n_\mu\in{\cal Z}^d}{1\over
m^2+(2\pi {\bf n}/L)^2}={1\over  T }\,,
\eqnd\esaddNL $$
where the sums are cut by a cut-off $\Lambda$.\par
To write the equation  in a more  manageable form and be able
to define it for continuous dimension, one uses Schwinger's
representation
$${1\over p^2+m^2}=\int_0^\infty\d s\,\e^{-s(p^2+m^2)}.$$
However, in contrast with  the infinite volume limit,  gaussian
integrals over momenta are here replaced by  infinite sum over
integers which can no longer be calculated exactly. One thus
introduces the function
$$\vartheta_0(s)= \sum^{+ \infty}_{n=- \infty} \e^{-\pi sn^2},
\eqnd\eJacobi $$
related to Jacobi's elliptic function $\theta_3$ by
$$\vartheta_0(s)=\theta_3(0,\e^{-\pi s}).$$
Poisson's transformation allows to prove the useful identity
$$\vartheta_0(s) = s^{-1/2} \vartheta_0\left(1/s\right).\eqnd \ePoisson $$
In terms of $\vartheta_0(s)$, the sums can then be written as
$$G(m,L,\Lambda) =L^{-d}\int^\infty\d s\,
\e^{-sm^2}\vartheta_0^d(4s\pi/L^2), $$
where UV convergence is here ensured by a small $s$ cut-off of order
$1/\Lambda^2$. \par
For $d>2$, we can introduce the  critical temperature $T_c$, which with
the same regularization reads
$${1\over T_c}={1\over(4\pi)^{d/2}}\int^\infty s^{-d/2}\d s\,.$$
It follows from the identity \ePoisson~that the difference between
both integrals is UV convergent for $d<4$. After
a change of variables $4\pi s/L^2 \mapsto s$ (and keeping only the leading
contribution in the critical domain), one finds that
the gap equation can be written as
$$L^{d-2}\left({1\over T}-{1\over T_c}\right)= F(mL)  \eqnd\eFSsigN  $$
with
$$F(z)={1\over 4\pi } \int_0^\infty\d s\,
\left(\e^{-s z^2 /(4\pi)}\vartheta_0^d(s)-s^{-d/2}\right). \eqnn $$
For $|T-T_c|\ll \Lambda^{d-2}$, we thus find a scaling form
which is consistent with  RG predictions:
$$L m(T,L)\equiv L/\xi(T,L)=f\bigl(L/\xi(T,L=\infty)\bigr) , $$
where in addition $f$ is a regular function of $T$ at $T_c$.
More precisely, in the large $N$ limit, using Eq.~\emasNsig~we obtain
$$K(d)\bigl[Lm(T,L=\infty)\bigr]^{d-2}=-F\bigl[L m(T,L)\bigr],$$
and we recall that $d-2=1/\nu+O(1/N)$.
Note that the length $\xi$ has the meaning of a correlation length only for
$\xi<L$. Since $\eta=0$ at this order, $m$ is also directly related to
the magnetic susceptibility $\chi$ in zero field, $\chi=1/m^2$. \par
The function $F(z)$ is decreasing. For $z\to\infty$ the integral is
dominated by the small $s$ region, which corresponds to the infinite
volume limit,
$$F(z)\sim \Gamma(1-d/2) {z^{d-2}\over(4\pi)^{d/2}}=-K(d)z^{d-2} .$$
One then verifies that for $T>T_c$ fixed, $L\to\infty$ and thus for
$mL\to\infty$, one
recovers the infinite volume limit. Alternatively, in the low
temperature phase for $T<T_c$ fixed, $L\to\infty$, $mL$ goes to zero.
Thus, the contribution of the zero mode ${\bf n}=0$ dominates  the
sum in equation
\esaddNL. Using the relation \ePoisson, one then finds
$$\eqalign{F(z)&={1\over z^2}+K_1(d)+O\left(z^2\right), \cr
K_1(d)&={1\over 4\pi } \int_0^\infty\d s\,
\left(\vartheta_0^d(s)-s^{-d/2}-1\right)
, \cr}$$
and thus
$$\chi(L,T)/N={1\over m^2}= \left({1\over T}-{1\over T_c} \right)L^d-  L^2
K_1(d)+O\left(L^{4-d}/(T-T_c)\right)   . \eqnn $$
We see that the susceptibility diverges with the volume, a precursor of
the low temperature phase with broken symmetry.\par
Note, finally, that it is instructive to make a similar analysis
for other boundary conditions that have no zero mode.
\par
For $d=2$, the regime where finite size effects can be seen
corresponds to $T\ln(L\Lambda)=O(1)$, that is to a regime of low
temperature. The zero mode dominates for $T\ln(L\Lambda)\ll 1$, and the
susceptibility then is given by
$$\chi(T,L)\sim  L^2\left[1+O(T\ln(L\Lambda))\right] \,.$$
\subsection The $CP(N-1)$ models in the large $N$ limit

We discuss here only one other  family of models based on
symmetric spaces that can be solved in the large $N$ limit, the $CP(N-1)$ models,
because they have been the subject of many studies \rCPN.
In particular, one can show that these models in two dimensions
have instanton solutions.
\par
Again one discovers that, within the framework of the large $N$
expansion,   $CP(N-1)$ models are related to  other models, abelian Higgs models, which are renormalizable in four dimensions.

\def\varphib{\varphi}
\smallskip
{\it The models.} The field $\varphi_\alpha $ is an $N$-component complex vector of unit length:
$$ \bar \varphi \cdot \varphi=N\,. \eqnd \ezzbar  $$
In addition two vectors $ \varphi_{\alpha} $ and $ \varphi'_{\alpha} $ are
equivalent if
$$ \varphi'_{\alpha} (x)=\e^{i\Lambda(x)}\varphi_{\alpha}(x)\,.
\eqnd\egaugequi $$
These conditions characterize the symmetric space  $U(N)/[U(N-1) \times U(1)] $,
a  complex Grassmannian manifold,
 which is isomorphic to the complex projective manifold  $CP(N-1)$.\par
One form of the unique symmetric classical
action is
$$ {\cal S}(\varphib, A_\mu) ={1 \over T} \int \d^2x\,
\overline{ {\rm D}_{\mu}\varphib}\cdot {\rm D}_{\mu}\varphib\,, \eqnn $$
in which $T$ is a coupling constant and $ {\rm D}_{\mu}$   the covariant derivative:
$$ {\rm D}_{\mu}=\partial_{\mu}+i A_\mu\,. \eqnn $$
The field $A_\mu$ is a gauge field for the $U(1)$ transformations
$$ \varphib'(x)=\e^{i\Lambda(x)}\varphib (x)\,, \quad A'_\mu(x)=A_\mu(x)-\partial _\mu \Lambda (x).\eqnd{\egaugequi} $$
The $U(N)$ symmetry of the action is obvious  and the gauge symmetry
implements the equivalence  \egaugequi. \par
In the tree approximation the fields are massless, and the $2N-2$ independent
real components correspond to the Goldstone bosons of the broken symmetry
$U(N)\to U(N-1)$, one Goldstone boson being suppressed by the abelian
gauge symmetry (the Higgs mechanism).\par
Since the action contains no kinetic term for   $A_\mu$, the gauge field
is not a dynamical field but only an auxiliary field
that can be integrated out. The action is quadratic in $A$ and the gaussian  integration  results in  replacing in the action $A_\mu$ by the solution of the $A $-field equation
$$NA_\mu=\ud i\left(\overline{  \varphi}\cdot \partial_\mu  \varphi-
\overline{\partial_\mu  \varphi}\cdot \varphi \right) =i\bar
\varphib \cdot \partial_{\mu}\varphib\,,\eqnd\eCPNAphi $$
where Eq.~\ezzbar\ has been used.
After this substitution, the  composite field $\bar \varphib \cdot \partial_{\mu}\varphib/N$ acts as a gauge field.
In the following, however, we find it more convenient to keep $A_\mu$ as an independent  field.\par
Note that the $CP(1)$ model is locally isomorphic to the $ O (3)$ non-linear
$ \sigma$-model, with the identification
$$ \phi^i=\bar \varphi_{\alpha} \sigma^i_{\alpha \beta}\varphi_{\beta}\,. \eqnn
$$
\medskip
{\it Large $N$ limit.}
As for the non-linear $\sigma $-model, we   introduce a Lagrange multiplier $\lambda (x) $ to implement the constraint \ezzbar, and obtain the action
$$ {\cal S}(\varphi,A_\mu ,\lambda )={1 \over T} \int \d^{d}x\left[
\overline{ {\rm D}_{\mu}\varphi}\cdot {\rm D}_{\mu}\varphi+ \lambda\left( \bar\varphi\cdot \varphi-N\right)\right] . \eqnn $$
The integral over $\varphi$ now is gaussian and can  be performed. Integrating over $N-1$ components, one finds (with $\varphi\equiv \varphi_1$),
$$ {\cal S}_N(\varphi,A_\mu ,\lambda )={1 \over T} \int \d^{d}x\left[
| {\rm D}_{\mu}\varphi|^2+ \lambda\left(  | \varphi|^2
-N\right)\right] +(N-1)\tr\ln(-{\rm D}_\mu^2+\lambda ). \eqnn $$
Of course, the remaining functional integral over $\varphi,A_\mu
,\lambda$ is well-defined only after a choice of gauge.\par The
gauge field plays no role in the saddle point equations, which are
those of the $O(2N)$ non-linear $\sigma $-model. However, the
model embodies the physics of the abelian Higgs model and of the
Landau--Ginzburg theory of superconductivity. In dimensions $d>2$,
in the broken symmetry phase, the gauge field becomes massive and
the number of Goldstone modes is $2N-2$ instead of $2N-1$. In two
dimensions a phase with massless modes is excluded; the model
exhibits a symmetric phase with confinement because the Coulomb
force is linear.
 \medskip
{\it The abelian Higgs model.} The action of the abelian Higgs model can be written as
$${\cal S}(A_\mu, \varphi)=\int\ \d^{d } x\left[{N\over 4
e^2}F_{\mu\nu}^2+  \overline{{\rm D}_\mu\varphi}\cdot{\rm D}_\mu\varphi + NU(\bar\varphi \cdot \varphi/N)\right], \eqnn $$
where the  potential  is quadratic:
$$U(z)=rz +\frac{1}{6} u z^2. \eqnd\eFTUabHig $$
This model with $N$ charged scalars is renormalizable in $d=4$ dimensions. In dimension $d=4-\varepsilon$, the RG $\beta $-functions are
$$\eqalign{\beta_u&=-\varepsilon u+{1\over 24\pi^2 N}\left[(N+4)u^2-18ue^2+54 e^4\right] ,\cr
\beta_{e^2}&=- \varepsilon e^2+ {1\over 24\pi^2} e^4.\cr} $$
For $d=4$, the origin $e^2=g=0$ is a stable IR fixed point only for $N\ge N_c=90+24\sqrt{15}\approx 183$. Correspondingly, the model has a stable IR fixed point in dimension $4-\varepsilon$  for $N\ge N_c= N_c(d=4)+O(\varepsilon)$. This is of course the situation which prevails in the large $N$ limit, and one finds the IR fixed point
$$u^*= e^2{}^*=24\pi^2\varepsilon+O(\varepsilon^2) .$$
We thus expect both effective couplings $g$ and $e$ to run at low momentum to the IR fixed point and the model to depend only on one parameter.
As in the case of the
non-linear $\sigma $ model, the linear $|\varphi|^4$ theory and the
non-linear $CP(N-1)$ model are equivalent. \par
This result can be verified by the large $N$ techniques.
We introduce the two fields $\rho (x)$ and $\lambda (x)$  as in section \ssNbosgen, to implement the constraint $\rho (x)=\bar\varphi (x)\cdot \varphi(x)/N$. We then obtain an action of the form
$$ {\cal S}(\varphi,A_\mu ,\lambda,\rho  )=  \int \d^{d}x\left[{N\over 4
e^2}F_{\mu\nu}^2+
\overline{ {\rm D}_{\mu}\varphi}\cdot {\rm D}_{\mu}\varphi+ \lambda\left( \bar\varphi\cdot \varphi-N\rho \right)+NU(\rho )\right] . \eqnn $$
We again integrate over $N-1$ components of the complex field and find ($\varphi\equiv \varphi_1$)
$$\eqalignno{ {\cal S}_N(\varphi,A_\mu ,\lambda,\rho  )&= \int \d^{d}x\left[{N\over 4
e^2}F_{\mu\nu}^2+ | {\rm D}_{\mu}\varphi|^2+ \lambda\left(  |
\varphi|^2 -N\rho \right)+NU(\rho )\right] \cr &+(N-1)\tr\ln(-{\rm
D}_\mu^2+\lambda ). &\eqnn\cr} $$ For the scalar field, the
arguments that show that after translation of the expectation
value the four-point interaction is negligible are the same as for
the usual $\phi^4$ theory. In the symmetric phase the expansion of
the $\tr\ln$ yields an additional contribution to the gauge field
propagator of the form
$$\Delta _{\mu\nu}(p)=\ud (p^2\delta _{\mu\nu}-p_\mu p_\nu)B_\Lambda (p,0).$$
We  face the same situation as in section \ssLTsN. For all dimensions $d<4$, $B_\Lambda (p,0)$ behaves for $p$ small like $p^{d-4}$ and, therefore, the contribution to the propagator coming from $F_{\mu\nu}^2$ is negligible. At $d=4$, it is subleading by $1/\ln p$. Therefore, $1/e^2$ is the coefficient of an irrelevant contribution, which can be omitted in the action. The gauge field
then is no longer dynamical and we return to the $CP(N_1)$ model.

\vfill\eject
\def\ggn{u}
\def\tt{\tau}
\def\GG{G}
\nref\rNaJoLa{A four-fermion interaction with $U(1)$ chiral
invariance was proposed by Nambu and Jona-Lasinio  as a basic
mechanism to generate nucleon, scalar and pseudo-scalar
$\sigma,\pi $ masses: \rf Y. Nambu and G. Jona-Lasinio, {\it Phys.
Rev.} 122 (1961) 345.} \nref\rWilGN{The difficulties connected
with this approach (approximate treatment of Dyson--Schwin\-ger
equations without small parameter,  non renormalizable theory with
cut-off) have been partially solved, the $1/N$ expansion
introduced and the existence of IR fixed points pointed out  in
\rf K.G. Wilson, {\it Phys. Rev.} D7 (1973) 2911 and in ref.
\rGN.} \nref\rGN{D.J. Gross and A. Neveu, {\it Phys. Rev.} D10
(1974) 3235.} \nref\rGNZJ{J. Zinn-Justin, {\it Nucl. Phys.}  B367
(1991) 105.} \nref\rGNeps{Calculations of RG functions in
dimensions 2 and $2+\varepsilon$ have been reported  in \rf N.A.
Kivel, A.S. Stepanenko, A.N. Vasil'ev, {\it Nucl.Phys.} B424
(1994) 619, hep-th/9308073 and refs. \refs{\rDerk{--}\rGrace}.}
\nref\rDerk{S.E. Derkachov, N.A. Kivel, A.S. Stepanenko, A.N.
Vasiliev, hep-th/9302034.}\nref\rWenzel{ W. Wentzel, {\it Phys.
Lett.} 153B (1985) 297.} \nref\rGrace{ J.A. Gracey, {\it Nucl.
Phys.} B367 (1991) 657.}

\nref\rDHN{The semi-classical spectrum for $d=2$ in the large $N$
limit of the GN model (with discrete chiral invariance) was
obtained from soliton  calculation in \rf R. Dashen, B. Hasslacher
and A. Neveu, {\it Phys. Rev.} D12 (1975) 2443.} \nref\rZZKT{ This
study as well as some additional considerations concerning the
factorization of $S$ matrix elements at order $1/N$ have led to a
conjecture of the exact spectrum at $N$ finite \rf A.B.
Zamolodchikov and Al.B. Zamolodchikov, {\it Phys. Lett.} 72B
(1978) 481 and also in refs. \refs{\rKaro,\rFor}.} \nref\rKaro{ M.
Karowski and H.J. Thun, {\it Nucl. Phys.} B190 (1981)
61.}\nref\rFor{ P. Forgacs, F. Niedermayer and P. Weisz, {\it
Nucl. Phys.} B367 (1991) 123, 144.} \nref\rLow{ The properties of
the NJL model in two dimensions are discussed in \rf J.H.
Lowenstein, {\it Recent Advances in Field Theory and Statistical
Mechanics}, Les Houches 1982, J.B. Zuber and R. Stora eds.,
(Elsevier Science Pub., Amsterdam 1984).} \nref\rKMV{For rigorous
results see \rf C. Kopper, J. Magnen and V. Rivasseau, {\it Comm.
Math. Phys.} 169 (1995) 121.} \nref\rRVW{Approximate functional RG
has also been used in \rf L. Rosa, P. Vitale, C. Wetterich, {\it
Phys. Rev. Lett.} 86 (2001) 958, hep-th/0007093.} \nref\rGNY{The
relation between the GN and GNY models is discussed in \rf A.
Hasenfratz, P. Hasenfratz, K. Jansen, J. Kuti and Y. Shen, {\it
Nucl. Phys.} B365 (1991) 79 and reference \rGNZJ.} \nref\rKLLP{A
comparison in three dimensions between numerical simulations of
the GN model and the expansion to order $\varepsilon^2$ obtained
from the GNY model is reported in \rf L. K\"arkk\"ainen, R.
Lacaze, P.Lacock and B. Petersson, {\it Nucl. Phys.} B415 (1994)
781, Erratum {\it ibidem} B438 (1995) 650 and
\rFocht.}\nref\rFocht{ E. Focht, J. Jerzak and J. Paul, {\it Phys.
Rev.} D53 (1996) 4616.} \nref\rFFJS{ The models are also compared
numerically in dimension two in \rf A.K. De, E. Focht, W. Franski,
J. Jersak and M.A. Stephanow, {\it Phys. Lett.} B308 (1993) 327.}
\nref\rGNtop{The GN and GNY models are related to the physics of
the top quark condensate. For a  review see for instance \rf G.
Cvetic,  {\it Rev. Mod. Phys.} 71 (1999) 513, hep-ph/9702381.}
\nref\rFNRK{Further RG calculations concerning the GN and NJL
models at $d=4$ can be found in \rf P.M. Fishbane and R.E. Norton,
{\it Phys. Rev.} D48 (1993) 4924 and \rRose.}\nref\rRose{ B.
Rosenstein, H.L. Yu and A. Kovner, {\it Phys. Lett.} B314 (1993)
381.} \nref\rGNDiii{The large $N$ expansion of the GN model in
$d=3$ has been discussed in\rf B. Rosenstein, B. Warr and S.H.
Park, {\it Phys. Rev. Lett.} 62 (1989) 1433 and \rGat.} \nref\rGat{
 G. Gat, A. Kovner and B. Rosenstein,
{\it Nucl. Phys.} B385 (1992) 76.} \nref\rGNDiiirev{For an early
review on the large $N$ expansion of the GN model in $d=3$ see B.
Rosenstein, B. Warr and S.H. Park, {\it Phys. Rep.} 205 (1991)
59.} \nref\rGNGracey{A number of $1/N$  calculations concerning
the GN and NJL models have been reported \rf J.A. Gracey, {\it
Phys. Lett.} B342 (1995) 297, hep-th/9410121;
{\it Phys. Rev.} D50 (1994) 2840; {\it Erratum-ibid.} D59 (1999) 109904,  hep-th/9406162;
{\it Int. J. Mod. Phys.} A9 (1994) 727, hep-th/9306107;
{\it Int. J. Mod. Phys.} A9 (1994) 567,  hep-th/9306106;
{\it Phys. Lett.} B308 (1993) 65, hep-th/9305012.}

\nref\rSchwinger{A few references on the Schwinger model and its
relation with the confinement problem: \rf J. Schwinger, {\it
Phys. Rev.} 128 (1962) 2425 and refs. \refs{\rLowen
{--}\rCol}.}\nref\rLowen{ J.H. Lowenstein and J.A. Swieca, {\it
Ann. Phys. (NY)} 68 (1971) 172.} \nref\rCasher{A. Casher, J. Kogut
and L. Susskind, {\it Phys. Rev.} D10 (1974) 732.} \nref\rCJS{ S.
Coleman, R. Jackiw and L. Susskind, {\it Ann. Phys. (NY)} 93
(1975) 267.}\nref\rCol{ S. Coleman, {\it Ann. Phys. (NY)} 101
(1976) 239.} \nref\rQEDRG{ For the calculation of the QED RG
$\beta$ function in the $\overline{\rm MS}$  scheme see \rf S.G.
Gorishny, A.L. Kataev and S.A. Larin, {\it Phys. Lett.} B194
(1987) 429; B256 (1991) 81.} \nref\rNSchwinger{A few references on
Schwinger or QED in the large $N$ limit: \rf J.A. Gracey, {\it
Phys. Lett.} B317 (1993) 415, hep-th/9309092; {\it Nucl. Phys.}
B414 (1994) 614, hep-th/9312055  and refs.
\refs{\rEspriu,\rPalan}.} \nref\rEspriu{ D. Espriu, A.
Palanques-Mestre, P. Pascual and R. Tarrach, {\it Z. Phys} C13
(1982) 153.} \nref\rPalan{A. Palanques-Mestre and P. Pascual, {\it
Comm. Math. Phys.} 95 (1984) 277.} \nref\rThirringeps{An early
calculation for dimension $d=2+ \varepsilon$  in the Thirring
model is found in \rf S. Hikami and T. Muta, {\it Prog. Theor.
Phys.} 57 (1977) 785.} \nref\rNThirring{The Thirring model has
been
 investigated for $N$ large in \rf
 S.J. Hands, {\it Phys. Rev.}
D51 (1995) 5816; hep-th/9411016; hep-lat/9806022.}
\nref\rDuHaMe{Recent simulations concerning the 3D Thirring model
are reported in \rf
 Simon Hands, Biagio Lucini, {\it  Phys. Lett.} B461 (1999) 263,
 hep-lat/9906008 and \rDelDeb.}\nref\rDelDeb{
 L. Del Debbio, S.J. Hands, {\it Nucl. Phys.} B552 (1999) 339, hep-lat/9902014.
}

\section Fermions in the large $N$ limit

We now illustrate the large $N$ techniques that we have started
describing in section \ssNbosgen, with the study of models involving
fermions in the vector representation of the $U(N)$ group.
We first explain how a class of self-interacting fermion models can be solved in the large $N$ limit. When these models have a discrete chiral symmetry
they may exhibit two phases, a symmetric phase with massless fermions, and a massive phase where the symmetry is broken.  The simplest realization is
the Gross--Neveu (GN) model, which we study in more detail \refs{\rNaJoLa,\rWilGN}.
The model  is renormalizable in two dimensions, and
describes in perturbation theory only one phase, the symmetric phase.
We also summarize what can  be learned from RG equations and $d=2-\varepsilon$ expansion.
\par
A  different field theory, the Gross--Neveu--Yukawa (GNY)
model has  the same symmetry, but a different field content since fermions interact through their coupling to a scalar field. The  model is renormalizable in four dimensions and, moreover, allows
a perturbative analysis of the chiral phase transition.
We  recall some properties of the  model  using perturbation theory and
 RG equations at and near four dimensions.  We
then show that additional information can be obtained from a large
$N$ expansion, and that a relation between  the GNY and GN models
follows.\par Finally, we also briefly examine QED and the massless
Thirring model in the large $N$ limit.\par In all examples, one of
the physical issues we  explore is the possibility of spontaneous
chiral symmetry breaking and fermion mass generation.
\sslbl\ssNfermions
\subsection Large $N$ techniques and fermion self-interactions

We consider a model characterized by a $U(\tilde N)$ symmetric
action for a set of $\tilde N$ Dirac fermions $\{\psi^i,
\bar\psi^i \}$. Since fermions have also a spin, we concentrate on
the simple examples where interactions involve only the scalar
combination both in the $U(\tilde N)$ and the spin group sense:
$$\bar\psib\cdot \psib\equiv \sum_{\alpha ,i}\bar\psi_\alpha^i \psi_\alpha^i.$$
The action can then be written as
 \sslbl\ssNferself
$${\cal S} (\bar \psib, \psib )= -\int \d^d x \left[ \bar\psib(x)\cdot
\sla{\partial} \psib(x) +NU\bigl(\bar\psib(x)\cdot \psib(x)/N \bigr)
\right],\eqnd \eNactferg $$
where $U $ is a general polynomial potential. We have introduced the notation  $N =\tilde N\tr{\bf 1}$, the matrix ${\bf 1}$ being the identity in the space
of Dirac $\gamma $ matrices and $N$ thus the total number of $\psi$ components. Moreover, a chiral invariant regularization (see Eq.~\epropreg) is assumed:
$$\sla{\partial} \mapsto \sla{\partial}\sqrt {D(-\nabla ^2/\Lambda ^2)}\,.$$
\par From the general discussion of section \ssNbosgen, it is now quite
clear how to solve such a model in the large $N$ limit and how to
generate a systematic large $N$ expansion. We introduce two scalar
fields $\sigma$ and $\rho$ and impose the
constraint $\rho(x)=\bar\psib(x)\cdot \psib(x) /N $ by an integral over $\sigma$.
The partition function can then be written as
$$ {\cal Z}= \int [ \d\psib\d\bar\psib][\d\rho][\d\sigma] \exp \left[-{\cal
S}(\bar\psib,\psib,\rho,\sigma)\right]  \eqnd\eONpartgen $$
with
$${\cal S}(\bar\psib,\psib,\rho,\sigma)=- \int \left[ \bar\psib\cdot
\sla{\partial} \psib +NU\bigl(\rho(x)\bigr) +\sigma(x)
\bigl((\bar\psib(x)\cdot \psib (x)-N \rho(x)\bigr) \right]
\d^{d}x\,. \eqnd\eNactfergii $$
In the representation \eONpartgen,  the integration over fermion fields is gaussian and can be performed, the $N$-dependence of
the partition function becoming explicit. Again it is convenient
to integrate only over $ \tilde N-1 $ components. The action suited for large
$N$ calculations then takes the form (now $\psi\equiv \psib_1$)
$$\eqalignno{{\cal S}_N(\bar\psi,\psi,\rho,\sigma)&=- \int \left[ \bar\psi
\sla{\partial} \psi +NU\bigl(\rho(x)\bigr) +\sigma(x)
\bigl((\bar\psib(x) \psib (x)-N \rho(x)\bigr) \right]
\d^{d}x \cr &\quad -(\tilde N-1)\tr \ln\left(\sla{\partial}+\sigma
\right).&
 \eqnd\eNactfergiii \cr} $$  For  large $N$, the action is
 proportional to $N$  and can be calculated
by the steepest descent method. The action density $\cal E$ for constant fields
$\rho,\sigma=M$ reduces to
$${\cal E}(M,\rho )/N=-U(\rho)+ M \rho-{1\over2}\int^\Lambda{\d^d q\over(2\pi)^d}
\ln[(q^2+M^2)/q^2].\eqnn $$
At this order $M$, the $\sigma $ expectation value, is also the fermion mass.
In the continuum limit (or in the critical domain) it must satisfy the physical condition $|M|\ll \Lambda $.
\par
The saddle point equations, expressed in terms of the function \etadepole, are
\eqna\eNsadferg
$$\eqalignno{M&=U'(\rho),&\eNsadferg{a} \cr
 \rho&={M\over(2\pi)^d}\int^\Lambda {\d^d q \over
q^2+M^2} = M \Omega_d(M) \,.  &\eNsadferg{b}\cr} $$
In terms of the function $\Omega_d$, the action density then reads
$${\cal E}(M,\rho )/N=-U(\rho)+ M \rho-\int_0^M s\d s
\, \Omega _d(s).\eqnd\eNEfer $$ From
Eq.~\eNsadferg{b}, we infer that $\rho /M$ is positive and
that  $\rho $ and $M$ vanish simultaneously for $d>1$.  Moreover,
the  condition $|M|\ll \Lambda $ implies that  $\rho $ is small in
the natural cut-off scale. We can, therefore, expand $U$ for
$\rho$  small:
$$U( \rho  )={\cal M}\rho +\ud G\rho ^2+O(\rho ^3).\eqnd\eNferUgen $$
Using Eq.~\eNsadferg{b} to eliminate $\rho $, we infer from Eq.~\eNsadferg{a}
$$M={\cal M}+G M \Omega_d(M)+O(M^2 \Omega_d^2).$$
In particular, the fermion mass $M$ vanishes when $\cal M$ goes to zero except
if the equation $ G\Omega_d(M)=1$
has a solution. The latter condition implies $G>0$,
that is attraction between fermions. For a repulsive interaction ($G<0$), at ${\cal M}=0$ the saddle point is $\rho =M=0$ and  the fermion mass always vanishes.\par
The value ${\cal M}=0$ is natural if the model has a discrete symmetry
$U(\rho )=U(-\rho )$, which prevents the addition of an explicit fermion mass term. Such a symmetry can be realized in the fermion representation, in the form of discrete chiral transformations in even dimensions, and in odd dimensions is a consequence of parity symmetry. In the case of attractive interactions, the issue one can then address concerns the possibility of spontaneous fermion mass generation, consequence of the spontaneous breaking of the symmetry. This is the question we now discuss at  large $N$.
\subsection Discrete chiral symmetry and spontaneous mass generation

We now consider models with a discrete symmetry that prevents the addition of a fermion mass (${\cal M}=0$).
In even dimensions it is a discrete chiral symmetry
($\gamma_S \equiv \gamma_{d+1}$) \sslbl\ssNGNgen
$$\psib \mapsto \gamma_S \psib, \quad\bar \psib \mapsto -\bar\psib \gamma_S\,
, \eqnd\echirald $$
while in odd dimensions  it is simply space reflection. Actually, it is possible to find a unique transformation, which corresponds to space reflection in odd dimensions, and makes sense in all dimensions
$${\bf x}=\{x_1  ,\ldots,x_\mu,\ldots, x_{d }\}\mapsto\tilde {\bf
x}=\{x_1 ,\ldots,-x_\mu,\ldots,x_{d }\},
\quad \cases{\psi({\bf x})\mapsto \gamma_\mu \psi(\tilde{\bf x}), \cr \bar\psi({\bf x})\mapsto -\bar\psi(\tilde {\bf x})\gamma_\mu \cr} .\eqnd\echirpar$$
Then, the potential has the expansion
$$U(\rho)=\ud \GG \rho^2 +O(\rho^4)\ \Rightarrow\ \rho \sim
M/\GG\,,\eqnd\eNferUquad $$
where we have assumed an attractive fermion self-interaction  that  excludes multicritical points ($G>0$) . \par
Then, the gap equation  \eNsadferg{b} has two solutions,
a symmetric solution
$M=\rho =0$ and a solution with non-vanishing mass and broken symmetry,
  $|M|  \ll \rho^{1/(d-1)}\ll \Lambda$,  which satisfies
$$1\sim \GG \Omega_d (M)  . \eqnd\eNGNmassgap $$
\smallskip
{\it Two dimensions.} We  first  examine dimension 2,
which is peculiar.   The solution with non-vanishing mass  and broken symmetry leads to
$$1= \GG \Omega_2(M) \ \Rightarrow \
 M\propto \Lambda \e^{-2\pi/ \GG},$$
where $|M|  \ll \Lambda$ implies that the parameter $\GG$  has to be small enough.
The massive solution has always lower energy density, and is therefore realized (see also the variational analysis at the end of the section).
\smallskip
{\it Higher dimensions.} In   dimensions $d>2$,
the massive phase can exist only if the  ratio $\rho/M$ is smaller than
some critical value
$$\rho/M=  \Omega_d(0).$$
Since  $M/\rho \sim \GG$, this implies the existence of  a critical coupling constant
$$G_c=1/ \Omega_d(0). \eqnn $$  For  $\GG<\GG_c $
the symmetry is unbroken and fermions are
massless: $\rho =M=0$.\par
For $\GG>\GG_c$, instead, both the  trivial symmetric solution
$\rho =M=0$ and a massive solution can be found.
In the broken symmetry phase, the gap equation can  be written as
$$ {1\over \GG_c}-{1\over \GG}  = \Omega_d(0)-\Omega_d(M) .$$
A comparison of energies  then shows that the broken phase has a lower energy
than the symmetric phase  (see also the variational
argument at the end of the section).  For
$\GG>\GG_c$, the symmetry is broken and the fermions are
massive.  At the critical value $\GG_c$,
 a phase transition occurs.
\par
Note that while for $d>2$ the massless symmetric phase  $M=\rho=0$ can exist,  for $d=2$ instead the symmetric solution does
not exist anymore.  This phenomenon is associated with the divergence
of momentum integral diverges at $q=0$ in
the saddle point equation  \eNsadferg{b}.
The mechanism is reminiscent of the Goldstone
phenomenon, but with one difference: here it is the symmetric phase that is
massless and therefore does not exist in low dimensions.
\par
Continuum physics in the broken phase is
possible only if $|M|\ll \Lambda$, that is when $\GG$ is close to its critical value, $  \GG  -  \GG_c\ll \Lambda ^{2-d}$.
 \par
Depending on the position of the dimension with respect to 4,
one then finds
$$\eqalign{M\propto \left({1\over \GG_c}-{1\over \GG}\right)^{1/(d-2)}
&{\rm for}\ d<4\,, \cr M\propto{1\over\sqrt{\ln(\Lambda /M)}}
\left({1\over \GG_c}-{1\over \GG}\right)^{1/2} &{\rm for}\ d=4\,,
\cr M\propto   \left({1\over \GG_c}-{1\over \GG}\right)^{1/2}
&{\rm for}\ d>4 \,.\cr}$$  Finally, we note that in the continuum
limit the function $U(\rho)$ can  always be approximated by the
quadratic polynomial \eNferUquad.

\smallskip
{\it The scalar bound state.} In the quadratic approximation relevant to the continuum limit, we can
integrate over $\rho(x)$. The result of the gaussian integration
amounts to replace $\rho(x)$ by the solution of the field equation
$$\rho(x)=\sigma(x)/\GG\,,\eqnd\eNferquadrs $$
and the action becomes
$${\cal S}_N(\bar\psi,\psi,\sigma)=- \int \left[ \bar\psi
\left(\sla{\partial}+\sigma(x)\right) \psi -N\sigma^2(x)/2\GG \right]
\d^{d}x -(\tilde N-1)\tr \ln\left(\sla{\partial}+\sigma
\right).\eqnd\eNactfergiv  $$
It is then instructive to calculate, at leading order, the $\sigma$
two-point function $\Delta_\sigma$ in the massive phase. Differentiating the action twice
with respect to $\sigma(x)$, and then setting $\sigma(x)=M$, one finds
after some algebra
$$\eqalignno{\Delta^{-1}_\sigma(p)&={N \over \GG} - {N  \over (2\pi)^d }
\int^\Lambda{\d^d q \over q^2+M^2}\cr&\quad  + {N  \over 2 (2\pi)^d }
\left(p^2+4M^2 \right)\int^\Lambda{\d^d q \over \left(
q^2+M^2 \right) \left[(p+q)^2 +M^2 \right]} \cr
&= \ud N   \left(p^2+4M^2 \right) B_\Lambda (p,m), &\eqnd\eNfergsig \cr}
$$
where the definition \ediagbul, the  saddle point equation \eNsadferg{b} and the relation
\eNferUquad~have been used.
We see that the inverse propagator vanishes for $ip=2M$ (we use
euclidean conventions), which means that in the broken phase, in the
 quadratic approximation, one finds
a scalar bound state with mass $2M$ (the two-fermion threshold), independently of the dimension $d$ of  space.
\medskip
{\it Variational calculations.} Here, like for the scalar fields in section \ssNVarfiv, it is possible to relate the large $N$ results to results obtained from variational calculations in the large $N$ limit.
One takes as a variational action
$${\cal S}_0  (\bar \psib, \psib )= -\int \d^d x\,\bar\psib\cdot
(\sla{\partial}+M) \psib \,.$$
The arguments follow directly what has been done in the scalar example.
We introduce the parameter
$$\rho = \left<\bar\psib\cdot \psib\right>_0/N={\tilde N \over N} \tr
\int^\Lambda {\d^d k \over (2\pi)^d}  { M-i\sla{k} \over k^2+M^2}= M \Omega _d(M)\,,\eqnd\eNsadferrho $$
where $\langle\bullet\rangle$ means expectation value with respect to $\e^{-{\cal S}_0}$.\par
Then, the variational energy density ${\cal E}_{\rm var}$ is given by
$${\cal E}_{\rm var}/N=-\left<U(\bar\psib\cdot \psib)/N\right>_0+M \left<\bar\psib\cdot \psib\right>_0/N
-\tr\ln(\sla{\partial }+M)/\tr{\bf 1}\, .$$
Again, in the large $N$ limit,
$$\left<U(\bar\psib\cdot \psib)/N\right>_0\sim U(\rho)  $$
and thus
$${\cal E}_{\rm var}/N= -U(\rho)+ M \rho-\int_0^M s\d s \, \Omega _d(s)\,.
\eqnd\eGNeVAR$$ We recognize Eq.~\eNEfer\ but here $\rho $ and $M$
are related by Eq.~\eNsadferrho\ and thus Eq.~\eNsadferg{b} is
automatically satisfied. We have now to look for the minimum of
${\cal E}_{\rm var}$ as a function of $M$. Except for possible
end-point solutions, which are not relevant here, and taking into
account Eq.~\eNEfer, we recover Eq.~\eNsadferg{a}. \par We can now
examine the behaviour of ${\cal E}_{\rm var}$ near the trivial
symmetric solution $M=0$.  From Eq.~\eGNeVAR, one finds
$${1\over N}{\del{\cal E_{\rm var}}\over\del M }=M(1-G\Omega_d(M))
{\del \{ M\Omega_d(M)\}\over \del M}\,.\eqnd\eGNextrem $$
Thus, for $d>2$,
$${\cal E}_{\rm var}/N\mathop{\sim}_{M\to0}  \ud \Omega _d(0)M^2(1-\GG/\GG_c).$$
Therefore, for $\GG<\GG_c$ the massless symmetric phase has lower
energy, while it is  the massive broken phase for $\GG>\GG_c$.\par
\noindent For $d=2$,
$${\cal E}_{\rm var}/N\mathop{\sim}_{M\to0} -\ud G M^2\Omega _2^2(M)$$
and, therefore, the broken phase is always lower.
\subsection The Gross--Neveu model

In the general discussion of the fermion self-interaction in the large
$N$ limit, we have  seen that interesting physics can be
studied by considering only a quartic fermion interaction. Then,\sslbl\ssNGNmod
$${\cal S} (\bar \psib, \psib )= -\int \d^d x \left[ \bar\psib\cdot
\sla{\partial} \psib +\frac{1}{2N}\GG\left(\bar\psib\cdot \psib \right)^2 \right].\eqnd\eGNact $$
This characterizes the Gross--Neveu (GN) model which we now study in more detail.
The model illustrates, for $\GG>0$, the mechanism of spontaneous mass
generation and, in even dimensions, chiral symmetry breaking.
It is renormalizable and asymptotically free in two dimensions.
However, as in the case of the non-linear $\sigma$ model, the perturbative
GN model describes only one phase.  \par
Since in the GN model the symmetry breaking mechanism is non-perturbative,
it will eventually be instructive to compare it with a different model
with the same symmetries, but where the mechanism is perturbative: the
Gross--Neveu--Yukawa model.
\smallskip
{\it RG equations in two and near two dimensions.} The GN model is renormalizable in two dimensions, and in perturbation
theory describes only the massless symmetric phase.
Perturbative calculations in two dimensions can be made with an IR cut-off
of the form of a mass term ${\cal M}\bar\psi\psi$, which breaks softly the chiral
symmetry. It is possible to use dimensional regularization in
practical calculations. \par
Note that in two dimensions the symmetry group
is really $O( N)$, as one verifies after some relabelling of the
fields. Therefore, the $(\bar \psi \psi)^2$ interaction is
multiplicatively renormalized. \par
In generic non-integer dimensions $d>2$, the situation is more complicated because the algebra of $\gamma$ matrices is infinite-dimensional and an infinite number of four-fermion interactions mix under renormalization. The coupling \eqns{\eNGNefg} thus has the interpretation of
a coupling constant that parametrizes the RG flow that joins the gaussian fixed point to the non-trivial UV fixed point.
 Its flow equation is obtained by first eliminating all other couplings.
This remark is important from the point of view of explicit calculations  in a $d-2$ expansion, but because the problem does not appear at leading order, it does not affect the analysis and we  disregard here this subtlety.\par
It is convenient to introduce here a
dimensionless coupling constant
$$\ggn=G\Lambda^{2-d}/N. \eqnd\eNGNefg $$
As  a function of the cut-off $\Lambda$, the bare vertex functions
satisfy the RG equations  \rGNZJ.
$$\left[ \Lambda{ \partial \over \partial
\Lambda} +\beta (\ggn){\partial \over \partial \ggn}-{n \over
2}\eta_{\psi}(\ggn)-\eta_{\cal M}(\ggn){\cal M}{\partial \over \partial {\cal M}} \right]
 \Gamma^{(n)}\left(p_{i};\ggn,{\cal M},\Lambda \right)=0\, .\eqnd{\eRGGN}$$
A direct calculation of the $\beta$-function in $d=2+\varepsilon$
dimension yields \rGNeps
$$\beta(\ggn)=\varepsilon \ggn-(N-2){\ggn^2 \over
2\pi}+(N-2){\ggn^3 \over4\pi^2}+{(N-2)(N-7)\over 32\pi^3}\ggn^4
+O\left(\ggn^5\right),\eqnd \ebetGNii $$ Here $N=\tilde N\tr{\bf
1}$ is the number of fermion degrees of freedom and thus for $d=2$
$N=2 \tilde N$.\par The special case $N=2$, for which the
$\beta$-function vanishes identically in two dimensions,
corresponds to the Thirring model (since for $N=2$ ~,~ $(\bar\psi
\gamma_{\mu}\psi)^2=-2 (\bar \psi\psi )^2$). The latter model is
also equivalent to the sine-Gordon or the $O(2)$ vector model.

\par Finally, the field and mass RG functions, at the presently
available, order are
$$\eqalignno{\eta_{\psi}(\ggn)&={N-1\over 8 \pi^2}\ggn^2-{(N-1)(N-2) \over
32\pi^3} \ggn^3+{(N-1)(N{}^2-7N+7)\over 128\pi^4}\ggn^4 , \hskip10mm &\eqnd\egamii \cr
\eta_ {\cal M} (\ggn)&={N-1\over 2\pi}\ggn-{N-1\over
8\pi^2}\ggn^2-{(2N-3)(N-2)\over 32\pi^3} \ggn^3+O(\ggn^4). & \cr}
$$
\smallskip
{\it Repulsive interactions.} If $u$ is negative, the form of the $\beta $-function shows that the model is IR free for any dimension $d\ge 2$ (at least for $u$ small enough), chiral symmetry is never broken and, for ${\cal M}=0$,
fermions remain massless. In dimension 2, one finds a gaussian behaviour
modified by logarithms. In higher dimensions the theory is gaussian.
Therefore, in what follows we discuss only the
situation of  an attractive self-interaction, that is $u>0$.
\medskip
{\it Attractive interactions.}
As in the case of the non-linear $\sigma$  model, it is then convenient to express the solutions of the RG
equations \eRGGN~in terms of  a RG invariant mass scale $\Lambda(\ggn)$ or its inverse, a length scale $\xi$ of the type of a correlation
length,
$$\xi^{-1}(\ggn)\equiv \Lambda(\ggn) \propto \Lambda\exp\left[-\int^\ggn{\d
\ggn'\over \beta(\ggn')}\right]\,. \eqnd\eferMass $$
We then have to consider separately dimension 2, which is special, and higher dimensions.
\smallskip
{\it Two dimensions} \refs{\rDHN{--}\rKMV}. For $d=2$, the model is UV asymptotically
free. In the chiral limit (${\cal M}=0$) the spectrum, then, is
non-perturbative, and a number of arguments lead to the conclusion
that the chiral symmetry is always broken and a fermion  mass
generated. From the statistical point of view, this corresponds to
a gap in the spectrum of fermion excitations (like in a
superconductor). All masses are proportional to the mass parameter
$\Lambda(\ggn)$ defined in equation \eferMass. For $\ggn$ small
$$\Lambda(\ggn)\propto \Lambda
\ggn^{1/(N-2)}\e^{-2\pi/(N-2)\ggn}\bigl(1+O(\ggn)\bigr)\,. \eqnn $$
We see that the continuum limit, which is reached when the masses are
small compared to the cut-off, corresponds to $\ggn\to 0$.\par
$S$-matrix considerations have then led to the conjecture that, for $N$
finite, the spectrum is
$$m_n=\Lambda(\ggn) {(N-2) \over \pi}\sin\left({n\pi \over
N-2}\right),\quad
n=1,2\ldots <N/2\,,\ N>4\,. $$
The  first values of $N$ are special, the model $N=4$ is
conjectured to be equivalent to two decoupled sine-Gordon models.\par
To each mass value corresponds a representation of the $O( N)$ group.
The nature of the representations leads to the conclusion that $n$ odd
corresponds to fermions and $n$ even to bosons.\par
This result is consistent with the spectrum for $N$ large evaluated by
semi-classical methods. In particular, the ratio of the masses of the
fundamental fermion $\psi$ and the lowest lying boson $\sigma$ is
$${m_{\sigma}\over m_{\psi}}=2\cos\left({\pi \over N-2}\right)=2+O(1/N^2).
\eqnd{\eGNspec}$$
Note that the results about breaking of chiral symmetry, the coupling
 constant dependence of the mass scale, and the ratio of \eGNspec~are
completely consistent with the large $N$ results
found in sections \label{\ssNferself,\ssNGNgen}.
\smallskip
{\it Dimension $d=2+\varepsilon$}  \rGNZJ.
As in the case of the $\sigma$-model, asymptotic freedom implies the
existence of a non-trivial UV fixed point $\ggn_c$  in $2+\varepsilon$
dimension:
$$\ggn_c={2\pi \over N-2}\varepsilon\left(1-{\varepsilon \over N-2}\right)
+O\left(\varepsilon^3\right). $$
$\ggn_c$ is also the critical coupling constant for a transition
between a
phase in which the chiral symmetry is spontaneously broken and a massless
small $\ggn$ phase.  \par
Setting $u=u_c$ in the RG functions, one infers the correlation length exponent $\nu$:
$$\nu^{-1}=-\beta'(\ggn_c)=\varepsilon-{\varepsilon^2 \over N-2}
+O\left(\varepsilon^3\right),\eqnd{\expnuii}$$
and the fermion field dimension $[\psi]$:
$$2[\psi]=d-1+\eta_{\psi}(\ggn_c)= 1+\varepsilon+ {N-1 \over
2(N-2)^2}\varepsilon^2 +O\left(\varepsilon^3\right) .\eqnd\expetaii $$
The dimension of the composite field $\sigma=\bar\psib\psib$ is given by
$$[\sigma]=d-1-\eta_M(\ggn_c)=1-{\varepsilon \over N-2}+O(\varepsilon^2). $$
As for the $\sigma$-model, the existence of a non-trivial UV fixed point
implies that large momentum behaviour is not given by perturbation theory
above two dimensions. This explains why the perturbative result that
indicates that  the model cannot be renormalized in higher dimensions, cannot
be trusted.
However, to investigate whether the $\varepsilon$ expansion makes
sense beyond an infinitesimal neighbourhood of dimension two, other methods are required  \rRVW, like the $1/N$ expansion, which is discussed in sections
\ssNferself, \ssNGNgen, \label{\sssGNYN}. \par
\subsection The Gross--Neveu--Yukawa model

The Gross--Neveu--Yukawa (GNY)  and the GN models have the
same chiral and $U(\tilde N)$ symmetries. The action of the
GNY model is (now $\varepsilon=4-d$) \refs{\rGNY{--}\rFFJS}
$${\cal S} (\bar \psib, \psib,\sigma )=\int \d^d x \left[-
\bar\psib\cdot \left(\sla{\partial}+g\Lambda^{\varepsilon/2}\sigma
\right) \psib +\ud\left(\partial_{\mu}\sigma\right)^2+\ud m^2
\sigma^2+{\lambda \over 4!} \Lambda^{\varepsilon}\sigma^4
\right],\eqnd{\eactGNg} $$ where $\sigma$ is an additional scalar
field, $\Lambda$ the momentum cut-off, and $g,\lambda$
dimensionless ``bare",  that is effective coupling constants at
large momentum scale $\Lambda$.\par The action still has a
reflection symmetry, $\sigma$ transforming into $-\sigma$ when the
fermions transform by \echirald. In contrast with the GN model,
however, the chiral transition can  be discussed here by
perturbative methods. An analogous situation has already been
encountered when comparing the $(\phib^2)^2$ field theory with the
non-linear $\sigma$ model. An additional analogy is provided by
the property that the GN model is renormalizable in dimension 2
and the GNY model in four dimensions.
\medskip
{\it The phase transition.} Examining the action \eactGNg, we see that in the
tree approximation when $m^2$ is negative the chiral symmetry is spontaneously
broken. The $\sigma$ expectation value gives a mass to the fermions, a
mechanism reminiscent of the Standard Model of weak-electromagnetic
interactions,
$$m_\psi =g\Lambda ^{\varepsilon/2}\left<\sigma \right>,\eqnn $$
while the $\sigma$ mass  is then
$$m^2_{\sigma}={\lambda \over 3g^2}m^2_{\psi}\,.\eqnd \eratiobf $$
Interactions modify the transition value $m^2_c$ of the parameter
$m^2$ and thus in what follows we set
$$m^2=m^2_c+\tt \,. \eqnn $$
The new parameter $\tt$ plays, in the language of phase
transitions, the role of the deviation from the critical
temperature.\par In order to study the model beyond the tree
approximation, we discuss now shortly RG equations near four
dimensions.
\subsection RG equations near  dimension 4

The model \eactGNg\ is trivial above four dimensions,
renormalizable in four dimensions and can thus be studied near
dimension 4 by RG techniques. Five renormalization constants are
required, corresponding to the two
field renormalizations, the $\sigma$ mass, and the two coupling
constants. The RG equations thus involve five RG functions. The
vertex functions $\Gamma^{(\ell,n)}$, for $l$ $\psi$ and $n$
$\sigma$ fields, then satisfy
$$\left(\Lambda{\partial\over \partial \Lambda}+\beta_{g^2}{\partial \over
\partial
g^2}+\beta_{\lambda}{\partial \over \partial\lambda} -\ud \ell\eta_{\psi}-\ud
n\eta_{\sigma}-\eta_m \tt{\partial \over \partial \tt}
\right)\Gamma^{(\ell,n)}=0\,.
\eqnd{\eRGiv}$$
The RG functions at one-loop
order are
$$\eqalignno{\beta_{\lambda}&= -\varepsilon \lambda+{1\over 8\pi^2}
\left({3\over 2}\lambda^2+N\lambda g^2-6N g^4\right), &\eqnd \ebetl  \cr
\beta_{g^2}&= -\varepsilon g^2+{ N+6 \over16\pi^2} g^4, &\eqnd \ebetgde
\cr}$$
where $N=\tilde N\tr{\bf 1}$ is the total number of fermion components. In four
dimensions $\tr {\bf 1}=4$ and thus $N=4\tilde N$.
\smallskip
{\it Dimension 4:} In  dimension 4, the origin
$\lambda=g^2=0$ is IR stable. Indeed  Eq.~\ebetgde\ implies that
$g$ goes to zero, and  Eq.~\ebetl \ then implies that also
$\lambda$ goes to zero. As a consequence, if the bare coupling
constants are generic,  that is if the effective couplings at
cut-off scale are of order 1, the effective couplings at scale
$\mu\ll\Lambda$ are small and become asymptotically independent
from the initial bare couplings. One finds
$$g^2(\mu)\sim {16\pi^2 \over ( N+6)\ln(\Lambda/\mu)}\,,\quad
\lambda(\mu)\sim {16\pi^2 \over
\ln(\Lambda/\mu)}R \eqnd\eGNYgrunas $$ 
where we have defined
$$R={24 N\over ( N+6)\left[( N-6)+\sqrt{ N^2+132N+36}\right]}.
\eqnn $$
This observation allows using renormalized perturbation theory to calculate
physical observables. For example, we can evaluate the ratio between the
masses of the scalar and fermion fields. To minimize quantum corrections we take  for $\mu$ a value of order $\langle \sigma\rangle$. A remarkable consequence
follows: the ratio \eratiobf~of scalar and fermion masses is fixed \refs{\rGNZJ,\rGNtop,\rFNRK}:
$${m^2_{\sigma}\over m^2_{\psi}}={\lambda_* \over
3g_*^2}={8N\over ( N-6)+\sqrt{ N^2+132N+36} }, \eqnd\emassratio $$
while in the classical limit it is arbitrary. Note that in the large $N$ limit
$${m _ \sigma \over m _ \psi }=2\left(1-\frac{15}{N}\right) +O(1/N^2).$$
The ratio has the same limit 2 as in two dimensions (Eq.~\eGNspec), a result that will be
explained by the study of the large $N$ limit in section \label{\sssGNYN}.
\par
Of course, if
the initial bare couplings are ``unnaturally'' small, the ratio
$\Lambda/\mu$ may not be large enough for the asymptotic regime
\eGNYgrunas\ to be reached. The renormalized couplings at scale
$\mu$ may then be even smaller than in Eq.~\eGNYgrunas\ and the
ratio will remain arbitrary.
\smallskip
{\it Dimension $d=4-\varepsilon$.}
One then finds a non-trivial IR fixed point
$$g_*^{2}={16\pi^2\varepsilon \over N +6},\quad \lambda_*=16\pi^2
R\varepsilon \,.\eqnd{\estar}$$
The  matrix of derivatives of the $\beta$-functions has two positive
eigenvalues $\omega _1,\omega _2$,
$$0<\omega_1=\varepsilon < \omega_2=\varepsilon
\sqrt{N^2+132N +36}/(N +6),
\eqnn $$
and thus the fixed point is IR stable.
\par
The field renormalization RG functions at the same order are
$$\eta_{\sigma}={N\over 16\pi^2}  g^2,\qquad
\eta_{\psi}= {1 \over 16\pi^2}  g^2 .\eqnn $$
At the fixed point one finds
$$\eta_{\sigma}={N\varepsilon\over N+6},\quad \eta_{\psi}={\varepsilon
\over (N+6)} , \eqnd\expeps $$
and thus the dimensions $d_\psi$ and $d_\sigma$ of the fields:
$$d_\psi={3\over2}-{N+4 \over2(N+6)}\varepsilon\,,\qquad
d_\sigma=1-{3\over N+6}\varepsilon\,. \eqnn $$
The  RG function $\eta_m$ corresponding
to the mass operator is at one-loop order:
$$\eta_m=-{\lambda \over 16\pi^2}-\eta_{\sigma}\,,$$
and the correlation length exponent $\nu$ is given by 
$${1\over\nu}=2+\eta_m =2- \varepsilon R-  {N\varepsilon\over N+6}=  2-\varepsilon{5N+6+\sqrt{N^2+132N+36}
\over6(N+6)}.\eqnn $$Finally, we can evaluate the ratio of masses
\eratiobf\ at the fixed point:
$${m^2_{\sigma}\over m^2_{\psi}}={\lambda_* \over
3g_*^2}={8N\over (N-6)+\sqrt{N^2+132N+36}} .$$
In $d=4$ and $d=4-\varepsilon$, the existence of an IR fixed point has the
same consequence: if we assume that the $\sigma$ expectation value is much
smaller than the cut-off and that the coupling
constants are generic at the cut-off scale, then {\it the ratio  of fermion
and scalar masses is fixed}.
\subsection GNY and GN models in the large $N$ limit

We now show that the GN model plays with respect to the GNY model
\eactGNg~the role the non-linear $\sigma$-model plays with respect
to the $\phi^4$ field theory \rGNZJ. For this purpose we start
from the action \eactGNg~of the GNY model
and integrate over $\tilde N-1$ fermion fields. We also rescale
for convenience $\Lambda^{(4-d)/2}g\sigma$ into $\sigma$, and then
get the large $N$ action \sslbl\sssGNYN
$$\left.\eqalign{ {\cal S}_N (\bar \psi, \psi,\sigma )&=\int
\d^d x \left\{-
\bar\psi \left(\sla{\partial} +\sigma \right) \psi
+\Lambda^{d-4}\left[{1\over2g^2} \bigl(
 \left(\partial_{\mu}\sigma\right)^2+m^2\sigma^2
 \bigr)+{\lambda\sigma^4\over 4!g^4}
 \right]\right\}\cr &\quad -(\tilde N-1)\tr \ln\left(\sla{\partial}+\sigma
\right).\cr}\right.\eqnd\eactefGN  $$
We take the large $N$ limit with $Ng^2$, $N\lambda$ fixed. When
$\sigma$ is of order one, the action is of order $N$ and can be
calculated by the steepest descent method.\par
We denote by ${\cal E}(\sigma)$ the action density for constant field
$\sigma(x)$ and vanishing fermion fields:
$$\eqalignno{{\cal
E}(\sigma)&=\Lambda^{d-4}\left({m^2\over2g^2}\sigma^2+{\lambda\over
4!g^4} \sigma^4 \right) -{\tilde N}\tr
\ln\left(\sla{\partial}+\sigma \right) \cr
&=\Lambda^{d-4}\left({m^2\over2g^2}\sigma^2+{\lambda\over 4!g^4}
\sigma^4 \right) -{ N\over 2}\int^\Lambda{\d^d q\over
(2\pi)^d}\ln[(q^2+\sigma^2)/q^2] .&\eqnn \cr}$$
 The expectation
value of $\sigma$ for $N$ large is a solution to  the {\it gap}\/
equation
$${\cal E}'(\sigma)\Lambda^{4-d}={m^2\over g^2} \sigma+{\lambda \over
6g^4}\sigma^3-N\Lambda^{4-d} \sigma\,  \Omega _d(\sigma )=0\,,\eqnd\evac $$
where, again, we have introduced the function \etadepole.
It is also useful to calculate the second derivative to check
stability of the extrema:
$${\cal E}''(\sigma)\Lambda^{4-d}={m^2\over g^2}+{\lambda \over
2g^4}\sigma^2+N \Lambda^{4-d} \int^\Lambda
{\d^d q\over (2\pi)^d} {\sigma^2-q^2 \over( q^2+\sigma^2)^2}.$$
The solution $\sigma=0$ is stable provided
$${\cal E}''(0)>0\ \Leftrightarrow\ {m^2\over g^2}>N \Lambda^{4-d} \Omega _d(0) .$$
Instead, the non-trivial solution to the gap equation exists only
for
$${m^2\over g^2}<N \Lambda^{4-d}\Omega _d(0) ,$$
but then it is stable. We conclude that the bare mass $m_c$ given by
$${m^2_c\over g^2}=N  \Lambda^{4-d}\Omega _d(0),\eqnd\eTcrit $$
is the critical bare mass (the analogue of the critical temperature for a
classical statistical system) where a phase transition occurs.
The expression  shows that the fermions favour the
chiral transition. In particular when $d$ approaches 2, we observe that
$m^2_c\to +\infty$ which implies that the chiral symmetry is always broken in
two dimensions. Using Eq.~\eTcrit~and setting
$$\tt=\Lambda^{d-4}(m^2-m^2_c)/g^2 , \eqnn $$
we can write the equation for the non-trivial solution as
$$\tt +\Lambda^{d-4}{\lambda \over 6g^4}\sigma^2+N\bigl( \Omega _d(0)-\Omega _d(\sigma )\bigr)=0\,.$$
We now expand $\Omega _d$  for $\sigma$ small (see Eq.~\etadepolii): $$\Omega _d(0)-\Omega _d(\sigma)=K(d)\sigma^{2-\varepsilon}
-a(d)\Lambda^{-\varepsilon}\sigma ^2 + O \left({\sigma^4/
\Lambda^{2+\varepsilon}} \right). \eqnn $$
Keeping only the leading terms for $\tt\to 0$, we obtain for $d<4$ the
scaling behaviour
$$  \sigma\sim \bigl(-\tt/NK(d)\bigr)^{1/(d-2)}.\eqnd{\expbet} $$
Since, at leading order, the fermion mass $m_{\psi}=\sigma$, it
follows  immediately that the exponent $\nu$ is also given by
$$\nu\sim \beta\sim 1/(d-2)\ \Rightarrow\
\eta_{\sigma}=4-d \ \Leftrightarrow \ d_\sigma =\ud(d-2+\eta_{\sigma})=1 \,.\eqnd \eGNNnu $$
At leading order  for $N\to\infty$, $\nu$ has the same value as in the
non-linear $\sigma$-model.\par
At leading order in the scaling limit, the  thermodynamic potential density  then becomes
$${\cal G}(\sigma)= \ud \tt\sigma^2+(N/d)K(d)|\sigma|^d .\eqnd\efpotGN $$
We note that, although in terms of the $\sigma$-field the model
has a simple Ising-like symmetry, the scaling equation of state
for $N$ large is different. This reflects the property that the
fermions become massless at the transition and thus do not
decouple.\par We also read from the large $N$ action that at this
order $\eta_{\psi}=0$. Finally, from the large $N$ action we can
calculate the $\sigma$-propagator at leading order. Quite
generally, using the saddle point equation,  one finds for the
inverse $\sigma$-propagator in the massive phase
$$\eqalignno{\Delta^{-1}_{\sigma}(p)&=\Lambda^{d-4}\left({p^2\over
g^2}+{\lambda \over3g^4 }\sigma^2\right) \cr&\quad + {N
\over 2 (2\pi)^d }\left(p^2+4\sigma^2 \right)\int^\Lambda{\d^d q \over \left(
q^2+\sigma^2 \right) \left[(p+q)^2 +\sigma^2 \right]}.&\eqnd\eGNsigprop \cr}$$
We see that in the scaling limit $p,\sigma\to 0$, the integral yields the
leading contribution. Neglecting corrections to scaling, we find that
the propagator vanishes for $p^2=-4\sigma^2$ which is
just the $\bar\psi\psi$  threshold. Thus, in this limit,  $m_{\sigma}=2
m_{\psi}$ in all dimensions, a result consistent with the $d=2$ and $d=4$ exact
values (Eq.~\eqns{\eGNspec,\emassratio}).\par
At the transition the propagator reduces to
$$\Delta_\sigma\sim {2 \over N b(d)  p^{d-2}} \eqnn
$$
with (Eq.~\econstb)
$$ b(d)=-{\pi \over\sin(\pi d/2)}
{\Gamma^2 (d/ 2) \over \Gamma (d-1)}N_d \,. \eqnn $$ The result is
consistent with the value of $\eta_{\sigma}$ in Eq.~\eGNNnu.
\par Finally, we note that the behaviour of the propagator at the
critical point, $\Delta_\sigma(p) \propto p^{2-d}$, implies that
the field $\sigma$ from the point of view of the large $N$
expansion, for $2\le d\le 4$, has the canonical dimension
$$[\sigma]=1\,. \eqnn $$
\medskip
{\it Corrections to scaling and the IR fixed point.}  The IR fixed point is
determined by demanding the cancellation of the leading corrections to
scaling. In the example of the action density ${\cal E}(\sigma)$, the leading correction to scaling is proportional to
$$\left({\lambda\over4! g^4}-{N a(d)\over 4}\right)\sigma^4 ,$$
($ a(d)\sim{1 / 8\pi^{2}\varepsilon}
$).
We now assume $a(d)>0$, otherwise we are led to problems analogous to those already discussed in section \sssEGRN. Demanding the
cancellation of the coefficient of $\sigma^4$, we obtain a relation between
$\lambda$ and $g^2$,
$$ g_*^4={\lambda_*\over 6N a(d)}
={4\lambda_*\varepsilon \pi^2\over 3N}+O\left(\varepsilon^2\right).$$
In the same way, it is possible to calculate the leading correction to the
$\sigma$-propagator \eGNsigprop. Demanding the cancellation of the leading
correction, we obtain
$${p^2\over g^2_*}+{\lambda_* \over3g_*^4}\sigma^2 -\ud N
\left(p^2+4\sigma^2\right)a(d)=0\,.$$
The coefficient of $\sigma^2$ cancels from the previous relation and
the cancellation of the coefficient of $p^2$ implies
$$g_*^2={2\over Na(d)}
={16\pi^2\varepsilon\over N}
+O\left(\varepsilon^2\right) \ \Rightarrow \
\lambda ^*={192 \pi^2 \varepsilon \over N}\,,$$
in agreement with the $\varepsilon$-expansion for $N$ large.
\medskip
{\it  The relation to the GN model for dimensions $2\le d\le 4$.}
In several examples we have observed that the contributions coming
from the terms $(\partial_{\mu}\sigma)^2$ and $\sigma^4$ in the large
$N$ action could be neglected in the IR critical region for $d\le 4$.
Power counting confirms this property because both terms have a
canonical dimension $4>d$ and are therefore irrelevant.
  We recognize a situation already encountered in
the $(\phib^2)^2$ field theory in the large $N$ limit. In the scaling region
it is possible to omit them and one then finds the action
$${\cal S}_N (\bar \psib, \psib,\sigma )=\int \d^d x \left[-
\bar\psib\cdot \left(\sla{\partial}+\sigma \right) \psib
+\Lambda^{d-4}{ m^2\over2g^2} \sigma^2 \right].\eqnd\eactGN $$
The gaussian integral over the $\sigma$ field can then  be performed explicitly
and yields the action of the GN model
$${\cal S}_N (\bar \psib, \psib )=-\int\d^d x  \left[\bar\psib\cdot
\sla{\partial} \psib +{\Lambda^{4-d} \over
2m^2}g^2\left(\bar\psib\cdot \psib \right)^2 \right].$$ The GN
(with an attractive interaction) and GNY models are thus
equivalent for  large distance physics, that is in the continuum
limit. Again, the arguments above rely on the condition that the
effective coupling constants at cut-off scale are generic. In the
GN model, in the large $N$ limit, the $\sigma$ particle, simply
appears as a $ \bar\psi\psi$ bound state at threshold  \refs{\rGNDiii,\rGNDiiirev}.\par
One may
then wonder whether the corrections to scaling are different.
Indeed superficially it would seem that the GN model depends on a
smaller number of parameters than the GNY model. Again this
problem is only interesting in four dimensions where corrections
to the leading contributions vanish only logarithmically. However,
if we examine the divergences of the term $\tr\ln
\left(\sla{\partial}+\sigma \right)$ in the effective action
\eactefGN\ relevant for the large $N$ limit, we find a local
polynomial in $\sigma$ of the form
$$\int\d^4x\left[A \sigma^2(x)+B\left(\partial_{\mu}\sigma\right)^2 +C
\sigma^4(x)\right].$$ Therefore, the value of the determinant can
be modified by a local polynomial of this form by changing the way
the cut-off is implemented: additional parameters, as in the case
of the non-linear $\sigma$-model, are hidden in the cut-off
procedure. Near two dimensions these operators can be identified
with $(\bar\psi\psi)^2, [\partial_{\mu} (\bar\psi\psi)]^2,
(\bar\psi\psi)^4$. It is clear that by changing the cut-off
procedure, we change the amplitude of higher dimension operators.
These bare operators in the IR limit have a component on all lower
dimensional renormalized operators.  \par Finally, note that we
could have added to the GNY model an explicit breaking term linear
in the $\sigma$ field, which becomes a fermion mass term in the GN
model, and which would have played the role of the magnetic field
of ferromagnets.
\subsection The large $N$ expansion

Using the large $N$ dimension of fields and power counting
arguments, one can then prove that the $1/N$ expansion is
renormalizable with arguments quite similar to those presented in
section \sssfivNRT\ \refs{\rZJTai, \rbook,\rGNDiiirev}.
\medskip
{\it Alternative theory.} To prove that the large $N$
expansion is renormalizable, one proceeds as in the case of the scalar theory
in section \sssfivNRT. One starts from a critical action with an additional
term quadratic in $\sigma$ which generates the large $N$ $\sigma$-propagator
already in perturbation theory:
$${\cal S}(\psi,\bar\psi,\sigma)=\int \d^d x\left[-\bar\psi
(\sla{\partial}+\sigma)\psi +{1\over
2\vv^2}\sigma(-\partial^2)^{d/2-1}\sigma\right] .\eqnn $$
The initial theory  is recovered in the limit $\vv\to\infty$. One then
rescales $\sigma$ in $\vv \sigma$. The model is renormalizable without
$\sigma$ field renormalization because divergences generate only local
counter-terms. The renormalized action then reads
$${\cal S}_\r(\psi,\bar\psi,\sigma)=\int \d^d x\left[-Z_\psi\bar\psi
(\sla{\partial}+\vv_\r Z_\vv\sigma)\psi +{1\over
2}\sigma(-\partial^2)^{d/2-1}\sigma\right] .\eqnn $$
RG equations follow:
$$\left[\Lambda {\partial \over \partial \Lambda}+
\beta_{\vv^2}(\vv){\partial \over \partial \vv^2}-{n\over
2}\eta_\psi(\vv)\right] \Gamma^{(l,n)}=0\,.\eqnn $$
Again, the large $N$ expansion is obtained by first summing the bubble
contributions to the $\sigma$-propagator. We define
$$D(\vv)={2\over b(d)}+N \vv^2 .$$
Then, the $\sigma$ propagator for $N$ large reads
$$\left<\sigma\sigma\right>={2\over b(d)D(\vv) p^{d-2}}.\eqnn $$
The solution to the RG equations can be written as
$$\Gamma^{(l,n)}(\ell p, \vv,\Lambda)=Z^{-n/2}(\ell)
\ell^{d-l-n(d-2)/2} \Gamma^{(l,n)}(p, \vv(\ell),\Lambda)
\eqnn $$
with the usual definitions
$$\ell{\d \vv^2\over \d \ell}=\beta(\vv(\ell))\,,\quad \ell{\d \ln Z
\over \d \ell}=\eta_\psi(\vv(\ell))\, .$$
We are interested in the neighbourhood of the fixed point $\vv^2=\infty$.
Then, the RG function $\eta(\vv)$ approaches the exponent
$\eta$. The flow equation
for the coupling constant becomes
$$\ell{\d \vv^2\over \d \ell}=\rho \vv^2 , \ \Rightarrow\ \vv^2(\ell)\sim
\ell^{\rho}. $$
We again note that a correlation function with $l$ $\sigma$ fields becomes
proportional to $\vv^l$. Therefore,
$$\Gamma^{(l,n)}(\ell p, \vv,\Lambda)\propto
\ell^{d-(1-\rho/2)l-n(d-2+\eta_\psi)/2} .
\eqnn  $$
We conclude
$$ d_\sigma=\ud(d-2+\eta_\sigma)=1-\ud\rho \ \Leftrightarrow\
\eta_\sigma=4-d-\rho \,.\eqnn   $$
\medskip
{\it RG functions at order $1/N$} \rGNGracey.
A new generic integral is useful here:
$${1\over(2\pi)^d}\int{\d^d q (\sla{p}+\sla{q})\over (p+q)^{2\mu} q^{2\nu}}=
\sla{p}
p^{d-2\mu-2\nu}{\Gamma(\mu+\nu-d/2)\Gamma(d/2-\mu+1)\Gamma(d/2-\nu)
\over (4\pi)^{d/2}\Gamma(\mu)\Gamma(\nu)\Gamma(d-\mu-\nu+1)}
\, . \eqnn $$
We first calculate the
$1/N$ contribution to the fermion two-point function at the critical point
(from a diagram similar to diagram \figbubiii)
$$\Gamma^{(2)}_{\bar\psi\psi}(p)=i\sla{p}+{2i \vv^2\over
b(d)D(\vv)(2\pi)^d}
\int^\Lambda{\d^d q (\sla{p}+\sla{q})\over q^{d-2}(p+q)^2}.$$
We need the coefficient of $\sla{p}\ln\Lambda/p$.
Since we work only at one-loop order, we again replace the $\sigma$
propagator $1/q^{d-2}$ by $1/q^{2\nu}$  and send the cut-off to
infinity. The residue of the pole at $2\nu=d-2$ gives the coefficient
of the term $\sla{p} \ln\Lambda$ and the finite part the $\sla{p}\ln p$
contribution. We find
$$\Gamma^{(2)}_{\bar\psi\psi}(p)=i\sla{p}+{2i \vv^2\over
b(d)D(\vv)}N_d \left(d-2\over d\right)
\sla{p}\ln(\Lambda/p)\,, \eqnn $$
where $N_d$ is the loop factor \etadpolexp{b}.
Expressing that the $\left<\bar \psi \psi\right>$ function satisfies
RG equations, we obtain  immediately the RG function
$$\eta_\psi(\vv)={\vv^2\over D(\vv)}{(d-2) \over d}X_1\,, \eqnn $$
where $X_1$ is given by equation \eXone.
We then calculate the function $\left<\sigma\bar\psi\psi\right>$ at
order $1/N$:
$$\Gamma^{(3)}_{\sigma\bar\psi\psi}(p)=\vv+A_1 D^{-1}(\vv)\vv^3 \ln\Lambda
 \,,$$
where $A_1$ corresponds to the diagram of figure \figNtriangl:
$$A_1 =-{2\over b(d) } N_d =-X_1 \,.$$
The diagram
of figure \figNtrianglii\ vanishes because the $\sigma$ three-point function vanishes for symmetry reasons. \par
The $\beta$-function follows:
$$\beta_{\vv^2}(\vv) ={4(d-1) \vv^4 \over d}X_1 D^{-1}(\vv)
 \eqnn $$
and thus
$$\rho={8(d-1)N_d\over d b(d) N}={4(d-1)\over d N}X_1\,.$$
The exponents $\eta_{\psi}$ and $\eta_\sigma$ at order $1/N$  and, thus,
the corresponding dimensions $d_\psi, d_\sigma$ of the fields, follow:
$$\eqalignno{\eta_{\psi}&= {(d-2)\over d}{X_1\over N}= {(d-2)^2 \over
d}{\Gamma(d-1)\over \Gamma^3(d/2)\Gamma(2-d/2)N}.&\eqnd\eetapsii
\cr 2d_\psi&=d-1-{2(d-2)\over d}{X_1\over N}\,. &\eqnn \cr}$$ For
$d=4-\varepsilon$, we find $\eta_{\psi}\sim \varepsilon/N$, result
consistent with \expeps\ for  $N$ large. Whereas for
$d=2+\varepsilon$, one finds $\eta_{\psi}\sim \varepsilon^2/2N$,
consistent with \expetaii. The dimension of the field $\sigma$ is
$$d_\sigma=\ud(d-2+\eta_\sigma)=1-{2(d-1)\over d N}X_1 +O(1/N{}^2)
.\eqnn $$
A similar evaluation of the $\left<\sigma^2\sigma\sigma\right>$ function
allows to determine the exponent $\nu$ to order $1/N$:
$${1\over\nu}=d-2-{2(d-1)(d-2)\over dN}X_1\,. \eqnn $$
Actually all exponents are known to order $1/N^2$, except $\eta_\psi$ which is
known to order $1/N^3$.
%
\smallskip

We discuss now shortly  two other models with chiral fermions, in
which large $N$ techniques can be applied, massless QED and the
$U(N)$ massless Thirring model.

\subsection Massless electrodynamics with $U(\tilde N)\times U(\tilde N)$ symmetry.

We first consider $\tilde N$ charged massless fermion fields
$\psib,\bar \psib$, interacting through an abelian gauge field
$A_{\mu}$ (massless QED with $\tilde N$ flavours)  \rSchwinger:
$$ {\cal S} (\bar \psib ,\psib,A _{\mu} ) = \int \d^{d}x
\left[{\textstyle{ 1 \over 4e^2}}  F ^{2} _{\mu \nu}(x) - \bar
\psib (x) \cdot\left(\sla{\partial} + i\Abar \right) \psib(x)
\right] . \eqnd\eactQEDN $$ This model possesses, in addition to
the $U(1)$ gauge invariance, a chiral  $U(\tilde N)\times U(\tilde
N)$ symmetry since the fermions are massless. Again, an
interesting question is whether the model exhibits in some
dimensions $2\le d\le 4$ a spontaneous breaking of chiral
symmetry. As before we use here $N={\tilde N} \tr{\bf 1}$, where
$N$ is the total number of fermion  components in $\tilde N$
differently flavoured Dirac fields. \sslbl\ssNQED
\smallskip
{\it Dimension $d=4-\varepsilon$.}
In terms of the  coupling constant standard in dimension 4,
$$\alpha\equiv {e^2\over 4\pi}\Lambda ^{-\varepsilon}\,, \eqnn $$
the RG $\beta$-function reads (taking  $\tr {\bf 1}=4$ in the space of
$\gamma$ matrices) \rQEDRG:
$$\eqalignno{\beta(\alpha)&=-\varepsilon \alpha +{2 {\tilde N}\over 3\pi}\alpha^2+
{{\tilde N}\over2\pi^2}\alpha^3-{{\tilde N}(22{\tilde
N}+9)\over144\pi^3}\alpha^4 \cr &-{1\over 64 \pi^4}{\tilde
N}\left[\frac{616}{243}{\tilde N}^2+\left(\frac{416}{9}\zeta(3)
-\frac{380}{27}\right){\tilde N}+23\right]\alpha^5+
O\left(\alpha^6\right). &\eqnd\ebetQED \cr} $$
The model is IR
free  in four dimensions. Therefore no phase  transition is
expected, at least for $e^2$ small enough. A hypothetical  phase
transition would rely on the existence on non-trivial fixed points
outside of the perturbative regime. \par In the perturbative
framework, the model provides an example of the famous triviality
property. For a generic effective coupling constant at cut-off
scale (i.e.~bare coupling), the effective coupling constant at
scale $\mu\ll\Lambda$ is given by
$$\alpha(\mu)\equiv {e^2(\mu)\over 4\pi}\sim {3\pi\over2{\tilde N}\ln(\Lambda/\mu)}.$$
This result can be used to bound the number of charged fields
(the number is not huge).\par
In $4-\varepsilon$ dimension
instead, one finds a non-trivial  IR fixed point corresponding to
a coupling constant
$$e^2_*=24 \pi^2 \varepsilon\Lambda^{\varepsilon}/N\,,$$
($N={\tilde N}\tr{\bf 1}$) and  correlation functions have a
scaling behaviour at large distance. As we have discussed in the
case of the $\phi^4$ field  theory, the effective coupling
constant at large distance becomes close to the IR fixed  point,
except when the  initial coupling constant is very small.
\par
The RG function associated  with the field renormalization is also
known up to order $\alpha^3$:
$$\eta_\psi= \xi {\alpha\over2\pi}-{4{\tilde N}+3\over 16\pi^2}\alpha^2+{40 {\tilde N}^2 +
54{\tilde N}+27 \over 576\pi^3}\alpha^3+O\left(\alpha^4\right),$$
where the gauge is specified by a term $(\partial_\mu
A_\mu)^2/2\xi$,  but it is a non-physical quantity because
it is gauge dependent. The simple dependence of $\eta _\psi$ in $\xi$ reflects the property that
if $\bar\psi(x)\psi(y)$ is not gauge invariant, $\bar\psi(x)\exp[i\oint_y^x A_\mu(s)\d s_\mu]\psi(y) $, instead is.

\medskip
{\it The large $N$ limit} \rNSchwinger.
To solve the model for $N\to \infty $, one first integrates over the
fermion fields and one obtains the large $N$ action
$${\cal S}_N  (A _{\mu} )= \int \d^{d}x
\left[{\textstyle{ 1 \over 4e^2}}  F^{2}_{\mu \nu}(x) -{\tilde
N}\tr\ln \left(\sla{\partial} + i\Abar \right)\right] .
\eqnd{\eactQEDN} $$ The  large  $N$ limit ($N={\tilde N}\tr{\bf
1}$) is taken with $e^2N$ fixed. Therefore, at leading order, only
${\cal S}_{N,{\rm F}} (A_\mu)$, the quadratic term in $A_{\mu}$ in the
expansion of the fermion determinant, contributes. A short
calculation yields
$$\left.\eqalign{{\cal S}_{N,{\rm F}} (A_\mu)&=-N \int \d^d k\, A_{\mu}(k)
\left[k^2\delta_{\mu\nu}-k_{\mu}k_{\nu}\right]K(k) A_{\nu}(-k),
\cr {\rm with}\quad K(k)&={d-2\over4 (d-1)}\left[b(d)k^{d-4}
-a(d)\Lambda^{d-4}\right]+O\left(\Lambda^{-2}\right),\cr}\right.
\eqnd\eAAlN$$ where $b(d)$ is the universal constant \econstb, and
$a(d)$ is the constant \eadef\ that depends on the regularization.
\par For $d<4$, the leading term in the IR region comes from the
integral. The behaviour at small momentum of the vector field is
modified, which confirms the existence of a non-trivial IR fixed
point. The fixed point is found by demanding cancellation of the
leading corrections to scaling coming from $F_{\mu\nu}^2$ and the
divergent part of the loop integral,
$$e_*^2={2(d-1)\over (d-2)a(d)}{\Lambda^{4-d}\over N}.$$
However, there is again no indication of chiral symmetry breaking.
Power counting within the $1/N$ expansion confirms that the IR
singularities have been eliminated, because the large $N$ vector
propagator is less singular than in perturbation theory. Of course,
this result is valid only for $N$ large. Since the long range
forces generated by the gauge coupling have not been totally
eliminated, the problem remains open for $d$ not close to four, or
for $e^2$ not very small and $N$ finite. Some numerical
simulations indeed suggest a  chiral phase transition for $d=4$
and $d=3$, ${\tilde N}\le N_c \sim 3$. \par
The exponents
corresponding to the IR fixed point have been calculated up to
order $1/N^2$. At order $1/N$ ($X_1$ is defined  by Eq.~\eXone)
$$\eqalign{\eta_\psi(\xi=0)&=-{(d-1)^2(4-d)\over
d(d-2)} {X_1\over N}+O\left(1/  N ^2\right), \cr \eta_m&
=-4{(d-1)^2\over d(d-2)} {X_1\over N }+O\left(1/   N ^2\right)
,\cr \beta'(\alpha^*)&= 4-d -{(d-3)(d-6)(d-1)^2(4-d)\over d(d-2)}
{X_1\over N }+O\left(1/  N ^2\right)  . \cr}$$  Finally, note that
in the  $d=2$ limit, the integral generates a contribution
$Ne^2/\pi k^2$ times the propagator of the free gauge field
$$K(k)\mathop{\sim}_{d\to 2}{1\over 4 \pi k^2}. $$
As a direct  analysis of the $d=2$  case confirms, this corresponds
to a massive bound state, of mass squared $Ne^2/\pi$. However, for
generic values of the coupling constant, the mass is of the order of
the cut-off $\Lambda$. Only when $e$ is ``unnaturally" small with respect to the
microscopic scale, as one assumes in conventional renormalized
perturbation theory, does this mass correspond in the continuum limit
to a propagating particle.
\smallskip
{\it Two dimensions.}
We now assume that the dimensional quantity $e^2$ is small
in the microscopic scale. The model then  is a simple
 extension  of the Schwinger model and can be exactly solved by the same
method. For ${\tilde N}=1$, the model exhibits the simplest example
of a chiral anomaly, illustrates  the property of confinement and
spontaneous chiral symmetry breaking. For ${\tilde N}>1$, the
situation is more subtle. The neutral $\bar\psi\psi$ two-point
function decays algebraically,
$$\left< \bar{\psib}(x)\cdot\psib(x)\bar{\psib}(0)\cdot\psib(0) \right>
\propto x^{2/{\tilde N}-2},$$ indicating the presence of a
massless mode and $\left<\bar\psi\psi\right>=0$. Instead, if we
calculate the two-point function of the composite operator
$${\cal O}_{\tilde N}(x)=\prod_{i=1}^{\tilde N} \bar{\psi}_i(x)\psi_i(x),$$
we find
$$\left<{\cal O}_{\tilde N}(x){\cal O}_{\tilde N}(0)\right>\propto \ {\rm const.}\ .$$
We have thus identified an operator which has a non-zero
expectation value. As a consequence of the fermion antisymmetry,
if we perform a transformation under the group $U({\tilde
N})\times U({\tilde N})$ corresponding to matrices $U_+,U_-$, the
operator is multiplied  by $\det U_+/\det U_-$. The operator thus
is invariant under the group $SU({\tilde N})\times SU({\tilde
N})\times U(1)$. Its non-vanishing expectation value is the sign
of the spontaneous breaking of the remaining $U(1)$ chiral group.

\subsection The $U({\tilde N})$ Thirring model

We now consider the model \refs{\rThirringeps{--}\rDuHaMe}
$${\cal S}(\bar\psib,\psib)= - \int \d^d
x\left[\bar\psi\left(\sla{\partial}+m_0
\right) \psi - \ud g J_{\mu}J_{\mu}\right] ,\eqnd{\eactThir} $$
where
$$J_{\mu}=\bar\psib\gamma_{\mu}\cdot\psib\, . \eqnn $$
The special case ${\tilde N}=1$ corresponds to the simple Thirring
model. In two dimensions, it is then equivalent to a free massless
boson field theory (with mass term for fermions, one obtains the
sine--Gordon model). In order to bosonize the model in $d=2$ and
to study that large $N$ properties, one introduces an abelian
gauge field $A_\mu$ coupled to the current $J_\mu$:
$$\ud g  J_{\mu}J_{\mu} \longmapsto A_{\mu}^2/2g +iA_{\mu}J_{\mu}\, .
\eqnn $$
One then finds massive QED without a $F^2_{\mu\nu}$ gauge kinetic term:
$${\cal S}(A_{\mu},\bar\psi,\psi)= - \int \d^2
x\left[\bar\psi\left(\sla{\partial}+ i\Abar+ m_0 \right)\psi
-A_{\mu}^2 /2g\right] . \eqnd{\eactAmu} $$ If one integrates over
the fermions, the fermion determinant generates a  kinetic term
for the gauge field. For $m_0=0$, the situation is thus similar to
massless QED, except that the gauge field is now massive.

\vfill\eject
\def\r{\rm r}
\def\thetab{\theta}

\def\nub{\nu}
\def\l{m^2}
\def\lii{m^4}
\def\bphi{\bar\varphi}

\nref\rHohen{Dynamical models are reviewed from the RG point of
view in \lb P.C. Hohenberg and B.I. Halperin, {\it Rev. Mod.
Phys.} 49 (1977) 435.}

 \nref\rMart{The dynamic action associated with the
Langevin equation has been introduced in \lb P.C. Martin, E.D.
Siggia and H.A. Rose, {\it Phys. Rev.} A8 (1978) 423. }

\nref\rDomin{For an early discussion of the renormalization of
dynamic theories in a field theory language see for example \lb C.
De Dominicis and L. Peliti, {\it Phys. Rev.} B18 (1978) 353.}

 \nref\rWitt{
The relation between supersymmetry and dissipative Langevin or
Fokker--Planck equations have been shown in \rf E. Witten, {\it
Nucl. Phys.} B188 (1981) 513 and refs.
\refs{\rFeigel{--}\rEgo}.}\nref\rFeigel{ M.V. Feigel'man and A.M.
Tsvelik, {\it Sov. Phys.--JETP} 56 (1982) 823.}\nref\rNaka{ H.
Nakazato, M. Namiki, I. Ohba and K. Okano, {\it Prog. Theor.
Phys.} 70 (1983) 298}\nref\rEgo{ E. Egorian and S. Kalitzin, {\it
Phys. Lett.} 129B (1983) 320.}

 \nref\rZinnJ{
In the discussion of the dissipative dynamics we follow \lb J.
Zinn-Justin, {\it Nucl. Phys.} B275 [FS18] (1986) 135.}

 \nref\rHalper{
Th dynamic exponent $z$ at order $1/N$ is given in \lb B.I.
Halperin, P.C. Hohenberg and S.K. Ma, {\it Phys. Rev. Lett.} 29
(1972) 1548.}
 \nref\rBaus{The non-linear $\sigma$-model has been considered in
 \lb R.
 Bausch,
H.K. Janssen and Y. Yamazaki, {\it Z. Phys.} B37 (1980) 163.}

\section Dissipative dynamics in the large $N$ limit

We now study a dissipative stochastic dynamics described by a
Langevin equation, in the large $N$ limit. From the point of
critical phenomena, the dissipative Langevin equation describes
the simplest time evolution with prescribed equilibrium
distribution, and allows calculating relaxation or correlation
times or time-dependent correlation functions. We recall that the
correlation time diverges at a second order phase transition, a
phenomenon called critical slowing-down, and this leads to
universal long time evolution \rHohen. \par Purely dissipative
stochastic dynamics is a problem interesting in its own right, but
it will also serve as an introduction to the discussion of
supersymmetric models of section \label{\ssNSUSYsc}, since
correlation functions associated with the dissipative Langevin
equation can be calculated  from a functional integral with a
supersymmetric action. The corresponding algebraic structure
generalizes supersymmetric quantum mechanics. \sslbl\ssCrDY
\subsection Langevin equation in the large $N$ limit

A general dissipative  Langevin equation for a scalar field $\varphi$
is a stochastic differential equation of the form
$${\partial   \varphi (t,x)\over \partial t}=-{\Omega \over 2}{\delta{\cal A}
\over \delta \varphi (t,x)}+\nu(t,x), \eqnd\eqlandisp $$
where $ \nu(t,x)$ is a gaussian white noise,
$$ \left< \nu (t,x) \right> = 0\,,\qquad
\left< \nu (t,x)\nu (t',x' ) \right>  = \Omega
\,\delta  (t-t'  )\delta (x-x'  ),\eqnd\ebruitb $$
and the constant $\Omega $ characterizes the amplitude of the noise.
This equation generates a time-dependent field distribution, which converges
at large time towards an equilibrium distribution corresponding to the
functional measure $\e^{-{\cal A}(\varphi)}[\d\varphi]$  if it is normalizable.
\medskip
{\it $O(N)$ symmetric models in the large $N$ limit.}
We now consider a model where $\varphi$ is a $N$-component field and the
static action ${\cal A}(\varphi)$ has an $O(N)$ symmetry of the form \eactONgen:
$${\cal A}(\varphi)=\int\d^d x\left[\ud (\partial_\mu\varphi)^2
+NU(\varphi^2/N)\right] . \eqnd\eNAscalgen $$
The corresponding Langevin equation then reads
$$ \dot \varphi_i (t,x)=-\ud \Omega  \left[-\nabla _x^2 +2
U'( \varphi^2/N)\right]\varphi_i(t,x)+\nu _i(t,x)  \eqnd\eNlangev $$
with
$$ \left< \nu_i (t,x) \right> = 0\,,\qquad
\left< \nu_i (t,x)\nu_j (t',x' ) \right>  = \Omega
\,\delta  (t-t'  )\delta (x-x'  )\delta _{ij}. \eqnd\eNbruit  $$
Here, the components $\nu _i$ of the noise are independent variables
and in the large $N$ limit the central limit theorem applies to $O(N)$ scalar
functions of $\nu$. \par
As a boundary condition, we choose
$$\varphi(t=-\infty ,x)=0\,, $$
which ensures equilibrium at any finite time. \par
We then set
$$\rho (t,x)=\varphi^2(t,x)/N\,,\quad m^2=2 U'(\rho ),$$
and assume, as an ansatz, that $m^2$ goes to a constant in the large $N$ limit. The Langevin equation then becomes linear and can be solved. Introducing the
field Fourier components
$$\varphib(t,x)=\int\dd k\,\e^{ikx}\tilde\varphib(t,k),\quad \nub (t,x)=
\int\dd k\,\e^{ikx}\tilde\nub(t,k),$$
one finds
$$\tilde\varphib(t,k)=\int_{-\infty }^t \d\tau \,\e^{-\Omega (k^2+m^2)(t-\tau )/2}\tilde\nu (\tau ,k),$$
with
$$\left<\tilde\nu_i(t,k)\tilde\nu_j(t',k')\right>={1\over(2\pi)^d}\delta (t-t') \delta (k+k')\delta _{ij}\,.$$
The calculation of $\rho (t,x)$ then involves the quantity
with vanishing fluctuations for $N\to\infty $:
$$\sum_i \tilde\nu_i( \tau _1,k_1)\tilde\nu_i(\tau _2,k_2)
\sim {N\over(2\pi)^d} \delta (k_1+k_2)\delta ( \tau _1-\tau _2).$$
Therefore,
$$\rho (t,x)={1\over(2\pi)^d}\int \dd k\int_{-\infty }^t\d\tau\, \e^{-\Omega (k^2+m^2)(t-\tau ) }={1\over(2\pi)^d}\int {\dd k\over k^2+m^2}=\Omega _d(m),$$
a result that is consistent with the ansatz $m$ constant. One then recognizes the static saddle point equation \esaddleN{c}
in the symmetric phase. \par For the broken phase the ansatz or
boundary conditions have to be slightly modified. One verifies
that provided $m=0$, one can impose
$$\varphi(t=-\infty ,x)=\sigma \,,$$
where $\sigma $ is a constant. Then,
$$\tilde\varphib(t,k)=\int_{-\infty }^t \d\tau \,\e^{-\Omega  k^2 (t-\tau )/2}\tilde\nu (\tau ,k)+\sigma \delta (k),$$
and
$$\rho (t,x)=\sigma ^2/N+{1\over(2\pi)^d}\int \dd k\int_{-\infty }^t\d\tau\, \e^{-\Omega  k^2 (t-\tau ) }=\sigma ^2/N+ \Omega _d(0),$$
where one recognizes the static saddle point equation \esaddleN{c} in the broken phase. \par
However, we will not discuss  the problem of the large $N$ expansion in this formalism further, because we now introduce an alternative supersymmetric formalism.
\subsection Path integral solution: Supersymmetric formalism

In the case of the purely dissipative Langevin equation \eqlandisp\ with gaussian white noise, it can be shown that dynamic correlation functions can also be expressed in terms of a functional integral that generalizes supersymmetric quantum mechanics \refs{\rMart{--}\rWitt}.\par
One introduces a superfield  $\Phi$  function of two Grassmann coordinates
$\bar\theta,\theta$:
$$ \Phi  (t,x;\bar \theta ,\theta )=\varphi (t,x) +\theta \bar \psi(t,x)+ \psi (t,x)\bar \theta +\theta \bar \theta
\bphi (t,x)\,,  $$
and supersymmetric covariant derivatives
$$ {\rm \bar D} ={\partial \over \partial \bar\theta},\quad {\rm  D} = {\partial
\over \partial \theta} -\bar\theta{ \partial \over \partial
t},\eqnd{\egensusy}  $$
which satisfy the anticommutation relations
$$ {\rm D}^{2}= {\rm \bar D}^{2}
=0\,, \qquad  {\rm D} {\rm \bar D} + {\rm \bar D} {\rm D}  =-{ \partial
\over \partial  t} \, .  \eqnd\eantrela $$
The generating functional  of $\Phi$-field correlation functions
then is given by
$${\cal Z}(J)=\int[\d\Phi]\e^{-{\cal S}(\Phi)+J\cdot \Phi} $$
with
$$ {\cal S} (\Phi)= \int \d  \bar\theta\, \d  \theta\, \d  t
\left[{ 2 \over \Omega} \int \d ^{d}x\,\bar{\rm  D} \Phi{\rm  D} \Phi+{\cal
A} (\Phi)\right] . \eqnd{\eactLsup} $$
Here $J(x,\bar\theta,\theta)$ is a source for $\Phi$ field:
$$J\cdot \Phi\equiv \int\d t \,\d^d x\,\d\bar\theta\d\theta\,J(t,x,\bar\theta,\theta)
\Phi(t,x,\bar\theta,\theta).$$
Note that with our conventions for the $\theta$ integration measure
$$\delta^2(\theta-\theta')=(\theta-\theta')(\bar\theta-\bar\theta')\,.$$
\par
The generalized action ${\cal S} (\Phi)$ is supersymmetric.
The corresponding supersymmetry generators
 are
$$ {\rm Q}  ={\partial \over \partial\theta},  \qquad \bar {\rm Q}  ={\partial
\over \partial \bar\theta} + \theta{ \partial \over \partial t}\,.
\eqnd{\egensup} $$
Both {\it anticommute}\/ with ${\rm D}$ and ${\rm \bar D}$ and satisfy
$$ {\rm Q}^ 2  = \bar {\rm Q}^ 2
=0\,, \qquad  {\rm Q} \bar {\rm Q} + \bar {\rm Q} {\rm Q}  ={ \partial
\over \partial  t} \, .  \eqnd \erelant  $$
Let us verify, for instance, that
$ \bar {\rm Q} $ is the generator  of a symmetry. We perform
a variation of $\Phi$ of the form
$$ \delta\Phi  (t, \theta ,\bar\theta )= \bar\varepsilon\bar{\rm
Q}\Phi\,, \eqnd{\esuptr} $$
which in component form  reads
$$  \delta \varphi   = \psi \bar \varepsilon\, , \quad \delta
\psi   =0\,, \quad \delta \bar \psi = \left(\bphi-\dot\varphi\right)
\bar \varepsilon\,,  \quad  \delta \bphi  = \dot\psi\bar\varepsilon\, .
  \eqnd \esusytr   $$
The term  ${\cal A}$ is invariant because it does not  depend on
$t$ and $\bar\theta$  explicitly. For the remaining term, the additional property that $
\bar{\rm Q}  $ anticommutes with  ${\rm D}$ and $\bar{\rm D}$ has to be used:
$$\delta\left[\bar{\rm  D} \phi{\rm  D} \phi\right]= \bar{\rm  D}
\left[\bar\varepsilon\bar{\rm Q}\phi\right]{\rm  D} \phi +\bar{\rm  D} \phi{\rm
D} \left[\bar\varepsilon\bar{\rm Q}\phi\right]=\bar\varepsilon\bar{\rm Q}\left[
\bar{\rm  D} \phi{\rm  D} \phi\right].$$
The variation of the action density thus is a total derivative. A similar argument applies to $Q$. This proves that the action is supersymmetric. \par
This supersymmetry is directly related to the property that
the corresponding Fokker--Planck hamiltonian   is equivalent to a positive hamiltonian.
\medskip
{\it Static action.} The static action defines the equilibrium distribution.
In what follows we have in mind a static action of the form (this includes
the action \eNAscalgen)
$${\cal A}(\varphi)=\int\d^d x\left[\ud (\partial_\mu \varphi)^2 +\ud
m^2\varphi^2+ V(\varphi)\right]. $$ The propagator $\Delta $ is
the inverse of the kernel
$$K= -{2\over\Omega}[\bar {\rm D},{\rm D}]-\nabla_x ^2+m^2. $$
Introducing  Fourier components, frequency $\omega$
corresponding to time and momentum $k$ corresponding to space,
we can write this operator more explicitly. From
$$ [\bar {\rm D},{\rm D}]\delta ^2( \theta -\theta ')=2+  i\omega(\theta'-\theta)(\bar\theta+\bar\theta')  $$
one infers
$$\eqalign{K(\omega,{\bf k}, \theta',\theta)&=\left\{-{2\over\Omega}[\bar
{\rm D},{\rm D}]+k^2+m^2\right\} \delta^2(\theta-\theta') \cr
& =-{4\over\Omega}
\left[1+\ud i\omega(\theta'-\theta)(\bar\theta+\bar\theta')\right]
+(k^2+m^2) \delta^2(\theta-\theta'). \cr}$$
To obtain the propagator $\Delta $ in superspace, we note
$$\left([\bar {\rm D},{\rm D}]\right)^2 =-\left( 2\bar {\rm D} {\rm D}+i \omega \right)\left( 2 {\rm D}\bar {\rm D}+i \omega \right)  =   -\omega ^2\,.$$
Then,
$$\Delta ={\Omega \over2}{ [\bar {\rm D},{\rm D}]+  \Omega (k^2+m^2)/2\over
 \omega^2+
  \Omega^2  (k^2+m^2  )^2/4} $$
or, more explicitly,
$$\Delta  ( \omega ,{\bf k} , \thetab',\thetab  ) ={\Omega \left[ 1+
\ud i\omega \left(\theta' -\theta \right)
\left(\bar\theta +\bar\theta ' \right)+ \frac{1}{4} \Omega
\left(k^2+m^2\right)\delta^2(\bar\thetab' -\bar\thetab) \right] \over \omega^2+
 { \Omega^2 \over 4} \left(k^2+m^2 \right)^2}.
\eqnd\eprop $$
\subsection Ward--Takahashi (WT) identities and renormalization

{\it WT identities.}
The symmetry associated with the $\rm Q$ generator has a simple consequence,
correlation functions are invariant under a translation of the coordinate $
\theta  $. The transformation \esuptr\ has a slightly more complicated
form. Connected correlation functions $W^{ (n )} (t_i,x_i ,
\theta_i ,\bar\theta _i  t)$ and proper vertices
$\Gamma^{  n  )}  (t_i,x_i , \theta_i ,\bar\theta_i
 )$   satisfy  the WT identities  \sslbl\sssStfWT
$$ \bar{\rm Q} W^{ (n )} (t_i,x_i , \theta_i ,\bar\theta_i   )=0\,, \qquad  \bar{\rm Q}\Gamma^{ (n )} (t_i,x_i , \theta_i ,\bar\theta_i   )=0
\eqnd\eward $$
with
$$ \bar{\rm Q}\equiv\sum^{n}_{k=1} \left({\partial \over \partial
\bar\theta_{k}}+\theta_{k}{ \partial \over \partial t_k} \right). $$
After Fourier transformation over time, the operator $\bar{\rm Q}$
takes the form
$$\bar{\rm Q}= \sum^{n}_{k=1} \left({ \partial  \over \partial  \bar\theta
_{k}}-i\omega_{k}\theta_{k} \right).\eqnd\ewardft $$
One verifies  immediately that the propagator \eprop\ satisfies the identity
\eward.
Actually, the general solution can be written as
$$\widetilde W^{(n)}(\omega, {\bf k},\theta,\bar\theta)
=\exp\left[-{i\over4n}\sum_{k,l}(\theta_k-\theta_l)
(\omega_k-\omega_l)
(\bar\theta_k+\bar\theta_l)\right]F^{(n)}(\omega,
{\bf k},\theta,\bar\theta) ,$$ where the function $F^{(n)}(\omega,
{\bf k},\theta,\bar\theta)$ now is invariant under translations of
both $\theta$ and $\bar\theta$.
\medskip
{\it Example: a two-point function}. Let us explore the implications of WT identities for a two-point function. As the relations \eqns{\egensup,\erelant}
show, supersymmetry implies translation invariance on time and $\theta$.
Therefore, any two-point function $W^{(2)}$ can be written as
$$W^{(2)}= A ( t_1 -t_2 ) +  ( \theta_1 - \theta_2
 ) \left[  ( \bar\theta_1 + \bar\theta_2  )B ( t_1 -t_2  ) +
 ( \bar\theta_1 - \bar\theta_2  )C ( t_1 -t_2 )\right] .\eqnn $$
The WT identity \eward\ then implies
$$2B(t) = {\partial A \over \partial t}.\eqnn $$
The WT identity does not determine  the function $C$. An
additional constraint comes from causality. For the two-point
function, it implies that the coefficient of $\theta_1
\bar\theta_2$ vanishes for $t_1 < t_2$ and the coefficient of
$\theta_2 \bar\theta_1$ for $t_2 < t_1$. The last function is thus
determined, up to a possible distribution localized at $t_1 =
t_2$. One  finds
$$2C(t)= - \epsilon(t) {\partial A \over \partial t}, \eqnn $$
where $\epsilon(t)$ is the sign of $t$, and, therefore,
$$W^{(2)}= \left\lbrace 1+ \ud \left(\theta_1 -\theta_2 \right)
\left[ \bar\theta_1 +\bar\theta_2 -  (\bar\theta_1 -\bar\theta_2  )\epsilon
 (t_1-t_2 ) \right]{\partial \over \partial t_1 }
\right\rbrace  A ( t_1-t_2 ).\eqnd\etwopt $$
\medskip
{\it Renormalization.}
In the special case of the supersymmetric dynamical action \eactLsup, a
comparison between the two explicit quadratic terms in $\Phi$ of the action
yields the relation between dimensions \rZinnJ
$$ \left[ t \right] - \left[ \bar\theta \right] -
\left[\theta \right] = 0 \ \Rightarrow\left[ \d  t \right] +
\left[ \d  \bar\theta \right] + \left[ \d \theta
\right] =0\,. \eqnd{\edim} $$
(We recall that since integration and differentiation over anticommuting
variables are equivalent operations, the dimension of $ \d \theta
$ is $ -[ \theta] $.)  \par
Therefore, the term proportional to $ {\cal A}  (\Phi  ) $ in the
action has the same canonical  dimension as in the static case:
the power counting is thus the same and the dynamic theory is always
renormalizable in the same space dimension as the static theory. Note
that Eq.~\edim\ also implies
$$ 2[\Phi]  =d+ [t]\,,$$
an equation that relates the dimensions of field and time. \par
One then verifies that supersymmetry is preserved by renormalization
and that the most general supersymmetric renormalized action has the form
$$ {\cal S}_{\r}(\Phi)= \int \d  \bar\theta\, \d \theta\,
\d t \left[{ 2 \over \Omega} Z_\Omega \int \d ^{d}x\, {\rm \bar D}\Phi
{\rm D}\Phi +{\cal A}_{\r}(\Phi) \right] , \eqnd\eactssr $$
where $Z_\Omega$ is the renormalization of the parameter $\Omega $, and thus also of the
scale of time. \par
The renormalized Langevin equation thus remains dissipative; the drift
force derives from an action.
\subsection $O(N)$ symmetric models in the large $N$ limit: supersymmetric formalism

We now consider again the  $O(N)$ symmetric Langevin equation \eNlangev\  for an $N$-component field $\varphi$,  corresponding to the static action  \eNAscalgen:
$${\cal A}(\varphi)=\int\d^d x\left[\ud (\partial_\mu\varphi)^2
+NU(\varphi^2/N)\right] . $$
We apply the usual strategy and introduce in the dynamic theory two
superfields $L$ and $R$, which have the form
$$\eqalign{L( \theta)&=l+\theta\bar\ell+\ell\bar\theta+\theta\bar\theta
\lambda\,, \cr
R( \theta)&=\rho+\theta\bar\sigma+\sigma\bar\theta +\theta\bar\theta s\,.
\cr}$$
We implement the condition $R=\Phi^2/N$ by an integral over $L$. The functional
integral takes the form
$${\cal Z}=\int[\d\Phi][\d R][\d L]\e^{-{\cal S}(\Phi,R,L)}  $$
with
$${\cal S}(\Phi,R,L)= \int\d t\d\bar\theta\d\theta\d^d x
\left[{2\over\Omega}{\rm \bar D}\Phi{\rm
D}\Phi+\ud(\partial_\mu\Phi)^2+NU(R) +\ud L(\Phi^2-NR)\right]. $$
We  integrate over $N-1$ components of the $\Phi$ field, keeping one
component $\Phi_1=\phi$ as a test-component. The large $N$ action then reads
$$\eqalignno{{\cal S}_N&=
\int\d t\d\bar\theta\d\theta\d^d x \left[{2\over\Omega}{\rm \bar
D}\phi{\rm D}\phi +\ud(\partial_\mu\phi)^2+NU(R) +\ud
L(\phi^2-NR)\right] \cr &\quad +\ud(N-1)\,{\rm
Str}\,\ln\left\{-2\Omega^{-1} [\bar {\rm D},{\rm D}] -\nabla^2_x
+L\right\}, \cr}$$ where ${\rm Str}$ means trace in the sense of
space, time  and Grassmann coordinates.
\par
At leading order at large $N$, the functional integral can be
calculated by the steepest descent method. The two first saddle
point equations, obtained by varying the superfields $\phi$ and
$R$, are \eqna\eStsad
$$\eqalignno{\left(-{2\over\Omega}[\bar {\rm D},{\rm D}]+L\right)\phi&=0\,,&\eStsad{a} \cr
L-2U'(R)&=0\,.&\eStsad{b} \cr} $$
The last saddle point equation, obtained by varying $L$, involves the
$\phi$ super-propagator $\Delta $:
$$\eqalignno{R-\phi^2/N&={1\over(2\pi)^{d+1}}\int\d\omega \,\d^d k
\, \Delta (\omega ,{\bf k} , \thetab,  \thetab),&\eStsad{c}\cr}$$
where $\omega $ and $\bf k$ refer to the
time and space Fourier components, respectively. \par
The super-propagator  can be calculated, for example, by solving
$$\left\{-2\Omega^{-1} [\bar {\rm D},{\rm D}] +k^2
+L\right\}\phi=J\,.$$ For Eq.~\eStsad{c},  the $\phi$-propagator
in presence of a  $L$ field is needed, but only  for $L$ of the
form
$$L=\l+\theta\bar\theta\lambda\,,$$
with $\l,\lambda$ constants. Then, setting
$$G=k^2+\l+2i\omega/\Omega\,$$
one finds
$$\Delta  ( \omega , {\bf k} , \thetab' ,\thetab  )
={4\Omega^{-1}+G\theta'\bar\theta'+G^*\theta\bar\theta
-\lambda\theta'\bar\theta'\theta\bar\theta \over
GG^*+4\lambda/\Omega} -{\theta'\bar\theta\over G}-
{\theta\bar\theta'\over G^*}\,.$$
At coinciding points $\theta=\theta'$ it reduces to
$$\Delta  (\omega , {\bf k} , \thetab ,\thetab
 )={4\Omega^{-1}+2(k^2+\l)\theta\bar\theta \over GG^*+4\lambda/\Omega}
-{2(k^2+\l)\theta\bar\theta \over GG^*}
\,. $$
After integration over $\omega$, one obtains 
$${1\over2\pi}\int\d\omega\,\Delta  (\omega , {\bf k} , \thetab ,\thetab
 ) \equiv \bar\Delta({\bf k},\theta)={1+\Omega(k^2+\l)\theta\bar\theta/2\over
\sqrt{(k^2+\l)^2+4\lambda/\Omega}}-\ud\Omega \theta\bar\theta\,. $$
Eq.~\eStsad{c} then becomes
$$R-\phi^2/N={1\over(2\pi)^{d}}\int\d^d k
\,\bar\Delta ({\bf k} , \thetab ).$$ All saddle point equations
reduce to the static equations for $\lambda=0$, and then $F=s=0$,
which implies that supersymmetry is preserved, and the ground
state energy vanishes. Then, no further analysis is necessary. Of
course, we know that supersymmetry is broken when the measure
$\e^{-{\cal A}(\varphi)}[\d\varphi]$ is not normalizable. But this
effect  cannot be seen at leading order in perturbation theory nor
in the large $N$ limit. \par Finally, note  that the
super-propagator formalism  simplifies dynamic $1/N$ calculations.
\smallskip
{\it The action density: saddle point equations in component form.} Alternatively, one can start from the action density   for constant scalar fields and vanishing Grassmann fields
$$\eqalignno{{\cal E}&=-{2\over \Omega }F^2+NsU'(\rho)+{1\over 2}\lambda (\varphi^2-N\rho)+{1\over2}\l(2F\varphi-Ns)\cr&\quad +
{N\Omega \over 4}\int{\d^d k \over (2\pi)^d}\left(\sqrt{(k^2+\l)^2+4\lambda /\Omega }-k^2-\l\right).&\eqnn\cr}$$
By differentiating $\cal E$ with respect to all parameters,
one recovers the saddle point equations in component form:
$$\eqalignno{F&=\Omega \l\varphi/4\,,\quad \l F+\lambda \varphi=0\,, &\eqnd\eLangsada \cr
\l&=2U'(\rho)\,,\quad \lambda =2sU''(\rho),& \eqnd\eLangsadb \cr}$$
and
\eqna\eLangsadx
$$ \eqalignno{\rho-\varphi^2/N&={1\over(2\pi)^d}\int{\d^d k \over \sqrt{(k^2+\l)^2+4\lambda /\Omega }}&\eLangsadx{a}\cr
s-2F\varphi/N&={\Omega \over 2}{1\over(2\pi)^d}\int \d^d k\left[ {(k^2+\l)\over \sqrt{(k^2+\l)^2+4\lambda /\Omega }}-1\right].&\eLangsadx{b}\cr}$$
Note that the same action density is obtained in a large $N$
variational calculation, starting from
$${\cal S}_0(\Phi,L) = \int\d t\d\bar\theta\d\theta\d^d x
\left[{2\over\Omega}{\rm \bar D}\Phi{\rm
D}\Phi+\ud(\partial_\mu\Phi)^2  +\ud L(\Phi-\Phi_0)^2)\right],$$
but then the two equations \eLangsadx{} are constraints, and only
$\Phi_0$ and $L$ are variational parameters. \par
Eliminating $F$ between the two equations \eLangsada, one finds
$$\varphi(\lambda +\Omega\lii/4)=0\,.\eqnn $$
This equation has two solutions: $\varphi=0$ which corresponds to
the $O(N)$ symmetric phase, $\lambda +\Omega\lii/4=0$ which
corresponds to a broken massless phase. \par Eliminating
$F,s,\rho$ from the action density  using the saddle point
equations, one obtains the ground state energy density
$$\eqalign{{1\over N}{\cal E}&=-{1\over2}\lambda \varphi^2- {\lambda
\over2(2\pi)^d}\int{\d^d k \over \sqrt{(k^2+\l)^2+4\lambda /\Omega
}}\cr\quad&+{ \Omega \over 4}\int{\d^d k \over
(2\pi)^d}\left(\sqrt{(k^2+\l)^2+4\lambda /\Omega }-k^2-\l\right).
\cr}$$ In the symmetric phase the derivative of $\cal E$ with
respect to $\lambda $ is
$${\partial {\cal E}\over \partial \lambda }={N\over \Omega }{\lambda
\over(2\pi)^d}\int{\d^d k \over \left[(k^2+\l)^2+4\lambda
/\Omega\right]^{3/2} }.$$ The minimum is at $\lambda =0$,  that is
at the supersymmetric point, where $\cal E$ vanishes. \par In the
broken symmetry phase $\lambda $ is non-positive. A short
calculation shows that again $\cal E$ is positive. The minimum is
reached for $\l=0$, and therefore $\lambda =0$. The minimum again
is supersymmetric and $\cal E$ then vanishes at the minimum,
independently of the value of $\varphi$.
\subsection Quartic potential

We now specialize to the  quartic potential
$$U(R)=\ud rR+{u\over 4!}R^2. $$
Then, the integral over $R$ can be performed, leading to a contribution
$$L=r+\frac{1}{6}uR\ \Rightarrow\ \delta{\cal S}_N=- 3N  (L-r)^2/2u
.$$ Note that for $u>0$ the usual static results are recovered, a
symmetric phase for $r>r_c $ and a broken symmetry phase
otherwise. For $u<0$ one finds the opposite situation, and there
is no sign that the situation is pathological form the static
point of view. The absence of an equilibrium state requires higher
order calculations.
\smallskip
{\it RG equations.}
The RG differential operator \eRGgamln\ acting on dynamic correlation functions takes at $T_c$  ($r=r_c$) the  form
$${\rm D}_{\rm RG}=  \Lambda { \partial \over \partial  \Lambda } +\beta(g){\partial
\over \partial g}+\eta_ \Omega (g)\Omega{ \partial \over \partial \Omega}-{n
\over 2}\eta (g) ,  \eqnn $$
where $g=u\Lambda ^{4-d}/N$ and $\eta_ \Omega(g)$ is a new independent RG function related to the renormalization constant $Z_\Omega $.\par
The solution of the RG equation for the two-point function $\widetilde W^{  (2  )}$ leads to the scaling form
$$\widetilde W^{(2)}  (p,\omega , \thetab  =0 )\sim  p^{-2+\eta -z} G^{  (2  )}  ( \omega /p^{z}   ).
\eqnn $$ The dynamic exponent $z$ which also characterizes, near
the critical point, the divergence of  the correlation time in the
scale of the correlation length, keeps at leading order its
classical value $z=2$.\par At order $1/N$  the $\left<LL\right>$
propagator is needed. In the symmetric phase $\varphi=0$, it is
given by
$$[\Delta_L]^{-1}=-{3\over u}\delta^2(\bar\theta-\bar\theta')+{1\over(2\pi)^{d+1}}
\int\d^d k\,\d \omega'' \,\Delta( \omega '',{\bf k}, \theta,\theta')\Delta(\omega- \omega '',{\bf p-k},\theta,\theta') .$$
In the infrared limit $\omega, {\bf k}\to 0$, it can be evaluated and used to
calculate the $\phi$ two-point function at order $1/N$.
The value of the dynamic exponent $z$ at order $1/N$ follows. It can be written as \rHalper
$$z=2+c\eta\,, \quad c=\left(4\over  4-d\right)\left\{ {dB(\ud d-1, \ud d-1)
\over 8\int_0^{1/2}\d x\,[x(2-x)]^{d/2-2}} -1\right\},$$
where $\eta$ has been given in section \sssfivNRT\ and $B(\alpha ,\beta )$ is
the mathematical $\beta $-function.
\medskip
{\it The dissipative non-linear $\sigma $-model.} Within the framework of the large $N$ expansion, we have shown that the results of the static $(\phib^2)^2$ could be reproduced by the non-linear $\sigma $-model. This result generalizes to the dynamic theory \rBaus.
In terms of the superfield $\Phi$, the functional integral takes a form
$${\cal Z}=\int[\d\Phi]\delta (\Phi^2-N)\exp\left[-{1\over T}\int\d t\d\bar\theta\d\theta\d^d x
\left({2\over\Omega}{\rm \bar D}\Phi{\rm
D}\Phi+\ud(\partial_\mu\Phi)^2\right)\right].$$
Therefore, the strategy is the same as in the static case. Since supersymmetry is not broken, the saddle point equations again reduce to the static equations.



\vfill\eject

\nref\rMosZinn{Parts of this section are based on: \lb M. Moshe
and J. Zinn-Justin, ~{\it Nucl. Phys.} B648 (2003) 131}

\nref\rSUSYN{On supersymmetric $O(N)$ quantum field theory at
large $N$ see also in: \lb T. Suzuki, {\it Phys. Rev.} D32 (1985)
1017 and refs. \refs{\rRydnell,\rSuzuki}.} \nref\rRydnell{ G.
Rydnell and R. Gundsmundottir, {\it Nucl. Phys.} B254 (1985) 593.}
\nref\rSuzuki{T. Suzuki and H. Yamamoto, {\it Prog. Theor. Phys.}
 75  (1986) 126. }

\nref\rBHM { The phase structure of the $O(N)$ symmetric,
supersymmetric model in $d=3$ was studied in \lb W.A. Bardeen, K.
Higashijima and M. Moshe, {\it Nucl. Phys.} B250 (1985) 437. }

\nref\Sal{Studies of  the scalar $O(N)\times O(N)$ model in
dimensions 3 and  $4 -\varepsilon$  are found in: \rf P.Salomonson
and B.S. Skagerstam {\it Phys. Lett.} 155B   (1985) 100 and
ref.~\rRabin.}\nref\rRabin{ E.~Rabinovici, B.~Saering and
W.A.~Bardeen, {\it Phys. Rev.} D36 (1987) 562.}

\nref\rEM{The $O(N)\times O(N)$ symmetric, supersymmetric model in
$d=3$ was studied in \lb O. Eyal and M. Moshe, {\it Phys. Lett. }
B178 (1986) 379. }

\nref\rWittSUSY{ The supersymmetric $O(N)$ non linear $\sigma $
model has been described in: \lb E. Witten, {\it Phys. Rev.} D16
(1977) 2991. }

\nref\rFreed { The study of general two-dimensional supersymmetric
non-linear $\sigma $ models has been initiated in: \lb D. Z.
Freedman, P.K. Townsend,  {\it Nucl. Phys.} B177 (1981) 282 }

\nref\rAlvar{ UV properties were first discussed in: \lb L.
Alvarez-Gaume, D. Z. Freedman, {\it Commun. Math. Phys.} 80 (1981)
443.}

 \nref\rMarcus{ The general four-loop $\beta $-function is given
 in:
 \lb
Marcus T. Grisaru, A.E.M. van de Ven, D. Zanon, {\it Nucl. Phys.}
B277 (1986) 409, {\it Phys. Lett.} B173 (1986) 423.}
\nref\rGrac{Critical exponents  of the supersymmetric non-linear
$\sigma$ model are calculated as $1/N$ expansions in:
\rf J.A. Gracey, {\it Nucl. Phys.} B348 (1991) 737; {\it ibidem}\/
B352 (1991) 183; {\it Phys. Lett.} B262 (1991) 49. }
 \nref\rFerre{For other
supersymmetric models see for instance \lb P.M. Ferreira, I. Jack,
D.R.T. Jones, {\it Phys. Lett.} B399 (1997) 258, hep-ph/9702304
and ref.~\rFerre.}\nref\rFerre{
P.M. Ferreira, I. Jack, D.R.T. Jones, C.G. North, {\it Nucl.
Phys.} B504 (1997) 108, hep-ph/9705328.}

\section Supersymmetric models in the large $N$ limit

We have already discussed scalar field theories and
self-interacting fermions in the large $N$ limit. We want now to
investigate how the results are affected by supersymmetry, and
what new properties emerge in this case~\rMosZinn.\par
Unfortunately, not many supersymmetric models  can be constructed
which can be studied by large $N$ techniques. We consider here two
such models which involve an $N$-component scalar superfield, in
three and two euclidean dimensions. First, we solve at large $N$ a
$( \Phi ^2)^2$ supersymmetric field theory. We then examine the
supersymmetric non-linear $\sigma$ model,  very much as we have
done in the non supersymmetric examples.  \sslbl\ssNSUSYsc \par
Both models are the simplest generalization of supersymmetric
quantum mechanics as it naturally appears, for example, in the
study of stochastic evolution equations of Langevin type (see
section \label{\ssCrDY}). \par Again the main issue will be the
phase structures of these models, and the possibility of
spontaneous symmetry breaking \refs{\rSUSYN}.

\subsection Supersymmetric scalar field in three dimensions

Apart from the interest in the general phase structure of the
supersymmetric $(  \Phi ^2)^2$ model, it will be of interest here
to study the spontaneous breaking of scale invariance that occurs
in this model at large $N$ . As we will be seen below, in a
certain region of the parameter space, there is only spontaneous
breaking of scale invariance and no explicit breaking. Though one
may expect this as a result of non-renormalization of the coupling
constant, in fact this happens also in the non-supersymmetric case
(see appendix \label{\ssNscaleinv}). When the coupling constant
that binds bosonic and fermionic $O(N)$ quanta is tuned to a value
at which $O(N)$ singlet massless bound states are created. The
resulting massless Goldstone particles appear as a supersymmetric
multiplet of a dilaton and dilatino.


\medskip
\noindent {\it Conventions and notation: Supersymmetry and
Majorana spinors in $d=3$}. Since the properties of Majorana
spinors in three euclidean space dimensions may not be universally
known, we briefly recall some of them and explain our notation. In
three dimensions the spin group is $SU(2)$. Then, a spinor
transforms like
$$\psi_U=U\psi\,, \quad U\in SU(2).$$
The role of Dirac $\gamma$ matrices is played by the Pauli $\sigma$ matrices,
$\gamma_\mu\equiv \sigma_\mu$. Moreover, $\sigma_2$ is antisymmetric while
$\sigma_2\sigma_\mu$ is symmetric. This implies
$$\sigma_2\sigma_\mu\sigma_2=-{}^T\!\sigma_\mu \ \Rightarrow\ U^*=\sigma_2 U
\sigma_2\, .$$ A Majorana spinor corresponds to a neutral
fermion and has only two independent components $\psi_1,\psi_2$.
The conjugated spinor is defined by (${}^T$ means transposed)
$$\bar\psi={}^T\!\psi \sigma_2 \ \Leftrightarrow  \ \bar\psi  _\alpha =i  \epsilon  _{\alpha \beta }\psi _\beta\,,\eqnd\eMajo $$
($\epsilon _{\alpha \beta }=-\epsilon _{\beta \alpha }$, $\epsilon _{12}=1$) and thus $\bar\psi$ transforms like
$$[{}^T\!\psi \sigma_2]_U=[{}^T\!\psi \sigma_2] U^\dagger\,  .$$
In the same way, we define a spinor of Grassmann coordinates
$$\bar\theta={}^T\!\theta \sigma_2 \,.$$
Since the only non-vanishing product is $\theta_1\theta_2$, we have
$$\bar\theta_\alpha\theta_\beta=\ud\delta_{\alpha\beta}
\bar\theta\cdot\theta\,.$$
The scalar product of $\bar\theta$ and $\theta$ is
$$\bar\theta\cdot\theta=-2i\theta_1\theta_2\ \Rightarrow
\theta_\alpha\bar\theta_\beta=i\delta_{\alpha\beta}\theta_1\theta_2
\,.$$
 If $\bar \theta',\theta'$ is another pair of coordinates,
because $\sigma_2\sigma_\mu$ is symmetric,  one finds
$$\bar\theta \sigma_\mu \theta'={}^T\!\theta\sigma_2\sigma_\mu\theta'=
-\bar\theta' \sigma_\mu \theta\,, \eqnd\esigtetii $$
and for the same reason
$$\bar \theta\psi=\bar\psi\theta\,.$$
Other useful identities are
$$(\bar\theta\psi)^2=-\ud (\bar\theta\theta)(\bar\psi\psi),\quad
(\bar\theta\sla{p}\psi)^2=\ud p^2(\bar\psi\psi)(\bar\theta\theta). $$
It is convenient to integrate over $\theta_1,\theta_2$ with the measure
$$\d^2\theta\equiv {i\over2}\d\theta_2\d\theta_1\, .$$
Then,
$$\int\d^2\theta\,\bar\theta_\alpha\theta_\beta=\ud
\delta_{\alpha\beta}\,,\quad
\int\d^2\theta\,\bar\theta\cdot\theta=1\,.$$
With this convention the identity kernel $\delta^2(\theta'-\theta)$ in
$\theta$ space is
$$\delta^2(\theta'-\theta)=(\bar\theta'-\bar\theta)\cdot
(\theta'-\theta).
\eqnd\eIdtheta $$
\medskip
{\it Superfields and covariant derivatives.}
A superfield $\Phi(\theta)$ can be expanded in $\theta $:
$$\Phi(\theta)=\varphi+\bar\theta\psi+\ud\bar\theta
\theta F \,.\eqnd\PhiSupField$$
Again, although only two $\theta$ variables are independent, we define the
covariant derivatives ${\rm D}_\alpha$ and $\bar {\rm D}_\alpha$ ($\bar {\rm D}=\sigma_2 {\rm D}$)
$$ {\rm D}_\alpha \equiv
{\partial\over\partial\bar\theta_\alpha}-(\sla{\partial} \theta)_\alpha \,,\quad
\bar {\rm D}_\alpha \equiv{\partial \over\partial \theta_\alpha}
-(\bar\theta\sla{\partial})_\alpha \,. $$
Then the anticommutation relation is
$$\{{\rm D}_\alpha,\bar {\rm D}_\beta\}=-2
[\sla{\partial}]_{\alpha\beta}\,.$$
Also
$$\bar {\rm D}_\alpha {\rm D}_\alpha={\partial\over\partial\theta_\alpha}
{\partial\over\partial\bar\theta_\alpha}-(\bar\theta\sla{\partial})_\alpha
{\partial\over\partial\bar\theta_\alpha}
-{\partial\over\partial\theta_\alpha}(\sla{\partial}\theta)_\alpha
+\bar\theta_\alpha\theta_\alpha\partial^2\,.$$
Since the $\sigma_\mu$ are traceless, using the identity \esigtetii\ one
verifies that $\bar {\rm D} {\rm D}$ can also be written as
$$\bar {\rm D}_\alpha {\rm D}_\alpha={\partial\over\partial\theta_\alpha}
{\partial\over\partial\bar\theta_\alpha}-2(\bar\theta\sla{\partial})_\alpha
{\partial\over\partial\bar\theta_\alpha}
+\bar\theta_\alpha\theta_\alpha\partial^2\,,$$
and, therefore, in component form
$$\bar {\rm D}_\alpha {\rm D}_\alpha \Phi=2F-2\bar\theta\sla{\partial}\psi
 +\bar\theta\theta \partial^2\varphi\,.$$
Moreover,
$$\left(\bar {\rm D}_\alpha {\rm D}_\alpha\right)^2=4\partial ^2.$$
\medskip
{\it Supersymmetry generators and WT identities.} Supersymmetry is generated
by the operators
$$  Q_\alpha={\partial\over\partial\bar\theta_\alpha}+(\sla{\partial} \theta)_\alpha      \,,\quad \bar Q_\alpha ={\partial \over\partial \theta_\alpha}
+(\bar\theta\sla{\partial})_\alpha \,, $$
which anticommute with
${\rm D}_\alpha$ (and thus $\bar {\rm D}_\alpha$).
Then,
$$\{\bar Q_\alpha, Q_\beta\}=2[\sla{\partial}]_{\alpha\beta}\,.$$
Supersymmetry implies WT identities for correlation functions. The $n$-point
function $ \widetilde W^{(n)}(p_k,\theta_k)$ of Fourier components satisfies
$$Q_\alpha \widetilde W^{(n)}\equiv \left[\sum_k {\partial \over\bar \partial \theta^k_\alpha}
-i( \sla{p_k}\theta^k)_\alpha\right] \widetilde W^{(n)}(p,\theta)=0\,.$$
To solve this equation, we set
$$\widetilde W^{(n)}(p,\theta)=F^{(n)}(p,\theta)\exp\left[-{i\over2n}\sum_{jk}\bar\theta_j
(\sla{p_j}-\sla{p_k})\theta_k\right], \eqnd\eSUSYWTs $$ where
$F^{(n)}$ is a symmetric function in the exchange
$\{p_i,\theta_i\} \leftrightarrow \{p_j,\theta_j\}$. It then
satisfies
$$\sum_k {\partial \over\partial \theta^k_\alpha}
F^{(n)}(p,\theta)=0\,,$$
that is, is translation invariant in $\theta$ space.\par
In the case of the two-point function, this leads to the general form
$$\eqalignno{\widetilde W^{(2)}(p,\theta',\theta)
&=A(p^2)\left[1+C(p^2)\delta^2(\theta'-\theta)\right]
\e^{i \bar \theta\slam{p}\theta'} .&\eqnd\eSUSYiipt \cr
&=A(p^2)\left[1+C(p^2)
( \bar \theta'- \bar \theta) (\theta'-\theta)
+i \bar \theta\sla{p}\theta'-\frac{1}{4}p^2 \bar \theta\theta \bar \theta'\theta'
\right] .\cr }$$
Since the vertex functions $\Gamma^{(n)}$ satisfy the same WT identities, they take the same general form.
\medskip
{\it General $O(N)$ symmetric action.}
We now consider the $O(N)$ invariant action
$${\cal S}(\Phi)=\int\d^3 x\,\d^2\theta\left[\ud\bar {\rm D}\Phi\cdot
{\rm D}\Phi+NU(\Phi^2/N)\right], \eqnd\eSUSYact $$
where $\Phi$ is a $N$-component vector.\par
In component notation
$$\int\d^2\theta\,\bar {\rm D}\Phi {\rm D}\Phi=-\int\d^2\theta\,\Phi \bar {\rm D}_\alpha
{\rm D}_\alpha \Phi=-\bar\psi \sla{\partial}\psi
+(\partial_\mu\varphi)^2-F^2. \eqnd\esusykin $$
Then, since
$$\Phi^2=\varphi^2+2\varphi\bar\theta\psi-\ud(\bar\theta\theta)(\bar\psi\psi)
+F\varphi\bar\theta\theta\,,\eqnd\esuperFiii $$
quite generally
$$\int\d^2\theta\,{\cal U}(\Phi^2)=
{\cal U}'(\varphi^2)\left(-\ud\bar\psi\psi+F\varphi\right)
-{\cal U}''(\varphi^2)(\bar\psi\varphi)(\varphi\psi). $$
In the case of the free theory $U(R)\equiv \mu R$, the action in component
form is
$${\cal S}=\int\d^3 x\,\left[-\ud \bar\psi\sla{\partial}\psi+\ud
(\partial_\mu\varphi)^2 -\ud F^2+\mu\left(-\ud
\bar\psi\psi+F\varphi\right)\right]. $$
After integration over the auxiliary $F$ field, the action becomes
$${\cal S}=\int\d^3 x\,\left[-\ud \bar\psi\sla{\partial}\psi-\ud\mu
\bar\psi\psi +\ud (\partial_\mu\varphi)^2 +\ud \mu^2\varphi^2\right]. $$
For later purpose it is also convenient to notice that in momentum representation
$$[-\bar {\rm D} {\rm D}+2\mu]\delta^2(\theta'-\theta)= 2\left[-2+ \delta^2(\theta'-\theta)\mu  \right]\e^{-i\bar\theta\slam{k}\theta'} .\eqnd\eSUSYker $$
The free propagator can be written as
$$[- \bar {\rm D} {\rm D}+ 2\mu]^{-1}= {1\over4( k^2+\mu^2)}\left(  \bar {\rm D} {\rm D}+ 2\mu\right),$$
or more explicitly
$$[-  \bar {\rm D} {\rm D}+2 \mu]^{-1}\delta^2(\theta'-\theta)
= {1\over  k^2+\mu^2 }\left[1+\ud \mu \delta^2(\theta'-\theta)\right]
\e^{-i\bar\theta\slam{k}\theta'} .\eqnn $$
For a generic super-potential $U( R )$, we find
$$\eqalignno{ {\cal S}&=\int\d^3 x\,\left[-\ud \bar\psi\sla{\partial}\psi
+\ud (\partial_\mu\varphi)^2-\ud  U'(\varphi^2/N)\bar\psi\psi
-U''(\varphi^2/N)(\bar\psi\varphi)(\varphi\psi)/N \right.\cr&\quad\left.+\ud \varphi^2
U'{}^2(\varphi^2/N)\right].&\eqnd\eGenericAction \cr}$$ Note that the theory
violates parity symmetry. Actually, a space reflection is
equivalent to the change $U\mapsto -U$ (see definition \echirpar).
Therefore, theories with $\pm U$ have the same physical properties.\par
Finally, in the calculations that follow we assume,
when necessary, a supersymmetric Pauli--Villars regularization.
\subsection Large $N$ limit: superfield formulation

To study the large $N$ limit, we introduce a constraint on $\Phi^2/N=R$,
where $R$ now is a superfield, by integrating over another superfield $L$:
\sslbl \ssNSUSYsad
$${\cal Z}=\int[\d\Phi][\d R][\d L]\e^{-{\cal S}(\Phi,R,L)}, $$
where
$${\cal S}(\Phi,R,L)=\int\d^3x\,
\d^2\theta \left\{\ud\bar {\rm D}\Phi\cdot
{\rm D}\Phi+NU(R)  +  L(\theta)\left[\Phi^2(\theta)-NR(\theta)\right]\right\}.
\eqnd\eactLagm $$
We parameterize the scalar superfields $L$ and $R$ as
$$\eqalignno{L(\theta,x)&=M+\bar\theta\ell+\ud\bar\theta\theta \lambda\,,&\eqnd\eLsupField \cr
 R(\theta,x)&=\rho+\bar\theta\sigma+\ud\bar\theta\theta s\,.&\eqnd\eRsupField\cr}$$
The $\Phi$ integral is now gaussian. As usual we integrate over only
$N-1$ components and keep a test-component $\Phi_1\equiv \phi$. We find
$${\cal Z}=\int[\d\phi][\d R][\d L]\e^{-{\cal
S}_N(\phi,R,L)} \eqnn $$
with the large $N$ action
$$\eqalignno{{\cal S}_N&=\int\d^3 x\,\d^2\theta\left[\ud\bar {\rm D}\phi
{\rm D}\phi+NU(R)+ L \left(\phi^2-NR\right)\right] \cr
&\quad+\ud(N-1){\rm Str}\, \ln\left[-\bar {\rm D} {\rm D}+2L\right].&\eqnd\eactSUSYN
\cr}$$
The two first saddle point equations, obtained by varying $\phi$ and $R$,
are
\eqna\eSUSYsad
$$\eqalignno{2L\phi-\bar {\rm D} {\rm D}\phi&=0\,,&\eSUSYsad{a} \cr
L-U'(R)&=0\,.&\eSUSYsad{b} \cr} $$ The last saddle point
equation, obtained by varying $L$, involves the
super-propagator $\Delta $ of the $\phi$-field. For $N \gg 1$, it reads
$$\eqalignno{R-\phi^2/N& = \left<x,\theta \right| \left[-\bar {\rm D} {\rm D}+2L\right]^{-1}\left|x,\theta \right>=
{1\over (2\pi)^3} \int\d^3 k \,\Delta (  k,\theta ,\theta ).
&\eSUSYsad{c} \cr}$$
The super-propagator $\Delta$  is solution of the equation
$$\left(- \bar {\rm D} {\rm D}+2L(\theta)\right)\Delta(k,\theta,\theta')=
\delta^2(\theta'-\theta) . $$
It is here needed only for $\ell=0$ and $M,\lambda$ constants.
It can be obtained by solving
$$\left(- \bar {\rm D} {\rm D}+2L(\theta)\right)\Phi(\theta)=J(\theta). $$
In Fourier representation and
in terms of its components, the equation reads
$$2M\varphi-2F+2\bar\theta(-i\sla{k}+M)\psi+\bar\theta\theta
[(k^2+\lambda)\varphi+MF] =J(\theta). $$
>From its solution we
infer the form of the propagator
$$\Delta(k,\theta,\theta')={\left[1+\ud
M(\bar\theta\theta+\bar\theta'\theta') -\frac{1}{4}(\lambda+k^2)
\bar\theta\theta\bar\theta'\theta'\right]\over k^2+M^2+\lambda}
-{\bar\theta[i\sla{k}+M]\theta'\over k^2+M^2} \,.\eqnd\esupprop $$
Clearly, one reads in Eq.~\esupprop\ the $\langle\varphi (k)\varphi
(-k) \rangle $ propagator $(k^2+M^2+\lambda)^{-1}$ and the
$\langle\psibar (k)\psi (-k)\rangle$ propagator
$(i\sla{k}+M)/(k^2+M^2)$. The coefficients of
$(\bar\theta\theta+\bar\theta'\theta')$ and of $
(\bar\theta\theta\bar\theta'\theta')$ are the $\langle\varphi
(k)F(-k)\rangle$  and $\langle F(k) F(-k)\rangle $ propagators,
respectively. \par
At coinciding $\theta$ arguments, we obtain
$$\Delta(k,\theta,\theta)={1+ M\bar\theta\theta\over
k^2+M^2+\lambda}-{M \bar\theta \theta\over k^2+M^2}
\,. \eqnd\eSUSYpropcoin $$
Eq.~\eSUSYsad{c}  thus is
$$ R-\phi^2/N = {1\over (2\pi)^3} \int\d^3 k\left[{1+ M\bar\theta\theta \over
k^2+M^2+\lambda}-{M\bar\theta\theta\over k^2+M^2}\right].
 \eqnd\eSUSYsadc  $$
\medskip
{\it Saddle point equations in component form.}
It is now convenient to introduce a notation for the boson mass
$$m_\varphi\equiv m=\sqrt{M^2+\lambda }\,. \eqnn $$
Eq.~\eSUSYsad{a} implies
\eqna\eSUSYsada
$$\eqalignno{F-M\varphi&=0 \,, &\eSUSYsada {a}  \cr
\lambda\varphi+MF&=0\,. &  \eSUSYsada{b}      \cr}$$
Eliminating $F$ between the two equations, we find
$$\varphi m^2=0 \eqnd\eSUSYGold $$
and, thus,
 if the $O(N)$ symmetry is broken the boson mass $m_\varphi$ vanishes.\par
Then, Eq.~\eSUSYsad{b} yields
\eqna\eSUSYsadb
$$\eqalignno{M&=U'(\rho),  & \eSUSYsadb{a} \cr
\lambda=m^2-M^2&=sU''(\rho).&\eSUSYsadb{b} \cr} $$
Using Eq.~\esuperFiii, we write Eq.~\eSUSYsadc\ in component form
(a cut-off is implied) as \eqna\eSUSYii
$$\eqalignno{\rho-\varphi^2/N&={1\over(2\pi)^3}\int{\d^3 p\over p^2+ m^2}\,,
&\eSUSYii{a}\cr
s-2F\varphi/N &={2M\over(2\pi)^3}\int\d^3 p\left({1\over p^2+m^2}
-{1\over p^2+M^2}\right). &\eSUSYii{b}\cr} $$
Introducing the cut-off dependent constant
$$\rho_c={1\over(2\pi)^3}\int^\Lambda {\d^3 p\over p^2}=\Omega _3(0)\,,\eqnd\eRciiidef $$
(see Eqs.~\eqns{\etadepole{--}\eadef}) we rewrite these equations as
\eqna\eSUSYiv
$$\eqalignno{\rho-\varphi^2/N&=\rho_c-{1\over4\pi } m\,, &\eSUSYiv{a}\cr
s-2F\varphi/N &={1\over2\pi}M\left(|M|- m\right).
&\eSUSYiv{b}\cr} $$
Note that the change $U\mapsto -U$ here corresponds to
$$F\mapsto -F\,,\quad s\mapsto -s\,,\quad M\mapsto -M\,.$$
\medskip
{\it Action density.} Finally, we calculate the action  density
$\cal E$ corresponding to the action ${\cal S}_N$
(Eq.~\eactSUSYN), ${\cal E}={\cal S}_N/{\rm volume}$,  for
vanishing fermion fields. We use (see also Eq.~\esusykin)
$$ {1\over2} \int\d^2\theta\,\bar {\rm D}\phi {\rm D}\phi =  -{1\over2} F^2 \,,\quad \int\d^2\theta\,L\phi^2 =M F\varphi  +{1\over2}\lambda \varphi^2 \eqnn $$
and
$${\rm Str}\, \ln\left(-\bar {\rm D} {\rm D}+2L\right)=  \tr\ln(-\partial^2
+M^2+\lambda) - \tr\ln(\sla{\partial}+M).\eqnd\eNSUSYtr $$
Then,
$$\eqalignno{{\cal E}/N&=-\ud F^2/N+\ud s U'(\rho)+MF\varphi/N+\ud \lambda\varphi^2/N-\ud
Ms -\ud \lambda \rho \cr&\quad +\half\tr\ln(-\partial^2 +M^2+\lambda)
- \half\tr\ln(\sla{\partial}+M).&\eqnd\eSUSYV
}$$
In $d=3$, the $\tr\ln$ in Eq.~\eSUSYV\ is given by
$$    \half\tr\ln(-\partial^2
+M^2+\lambda) - \half \tr\ln(\sla{\partial}+M)  = \ud
\rho_c\lambda-{1\over12\pi}\left(m^3-|M|^3\right) .
\eqnd\eSupertraceLN  $$ The saddle point equations in component
form are then recovered from  derivatives of $\cal E$ with respect
to the various parameters. Using the saddle point equations
\eSUSYsad{a}, \eSUSYiv{} and \eSUSYsadb{a} to eliminate
$F,s,\rho$, one eliminates also the explicit dependence on the
super-potential $U$ and the expression is simplified into
$${\cal E}/N=\ud M^2\varphi^2/N+  {1\over 24\pi}(m-|M|)^2(m+2|M|)
.\eqnd\eSUSYground
$$
In this form we see that $\cal E$ is positive for all saddle
points, and, as a function of $m$, has an absolute minimum at
$m=|M|$, and thus $\lambda =0$, that is for a supersymmetric
ground state. \par Moreover, Eq.~\eSUSYGold\ implies $M\varphi=0$,
and thus ${\cal E}=0$. Therefore, if a supersymmetric solution
exists, it will have the lowest possible ground state energy  and
any non-supersymmetric solution will have a higher energy. \par
Since $M$, $m$, and $\varphi$ are related by the saddle point
equations, it remains to verify whether such a solution indeed
exists.\par In the supersymmetric situation the saddle point
equations reduce to $s=0$, $M\varphi=0$ and
$$\rho-\rho_c=\varphi^2/N-|M|/4\pi\,,\quad M=U'(\rho).$$
In the $O(N)$ symmetric phase $\varphi=0$ and $|M|=4\pi
(\rho_c-\rho)=|U'(\rho)|$. In the broken phase $U'(\rho)=0$ and
$\varphi^2/N=\rho-\rho_c$. We will show in the next section that
these conditions can be realized by a quadratic function $U(\rho)$
and then in both phases the ground state is supersymmetric and
$\cal E$   vanishes.
\subsection Variational calculations

For completeness, we present here the corresponding variational
calculations \rBHM\ and apply the arguments of section
\ssNVarfiv~to the action \eGenericAction. In terms of the two
parameters  \sslbl\ssSUSYNvar \eqna\eR
$$\eqalignno{\rho&=(1/N)\left<\varphi^2(x)\right>_0
={ \varphi^2 \over N} +{1\over(2\pi)^3}\int{\d^3 p\over
p^2+m_\varphi^2} ,&\eR{a}\cr \tilde \rho &=(1/N)\left< \bar\psi
(x)\psi (x)\right>_0 ={2m_\psi \over(2\pi)^3}\int{\d^3 p\over
p^2+m_\psi ^2 }\,, &\eR{b}\cr} $$
the variational energy density
can be written as
$$\eqalignno{{\cal E}_{\rm var}/N&=\ud \tilde \rho
\bigl(m_\psi -U'(\rho)\bigr)+\ud \rho U'{}^2(\rho)-\ud
m_\varphi^2(\rho-\varphi^2/N) \cr &\quad  +\half\tr\ln(-\partial^2
+ m_\varphi^2)- \half\tr\ln(\sla{\partial}+m_\psi
).&\eqnd\eVarEsusy }$$ In the following we choose $m_\psi \ge 0$
and $m_\varphi \ge 0$. \par
Since the variational energy is larger than or equal to the ground state energy, which in a supersymmetric theory is non-negative, it is sufficient to find values
of the three parameters $m_\psi,m_\varphi ,\varphi$ such that ${\cal E}_{\rm var}$ vanishes to prove that the ground state is supersymmetric.
Choosing $m_\psi=m_\varphi=M$ and using Eqs.~\eR{}, we find
$${\cal E}_{\rm var}/N=  U'{}^2(\rho ){\varphi^2\over 2N}+{1\over2(2\pi)^3}\int{\d^3 p\over
p^2+M^2}\left[U'(\rho )-M\right]^2. $$
This expression vanishes whenever $M=U'(\rho )$ together with $M\varphi=0$.
Together with Eq.~\eR{a}, we recover the three supersymmetric saddle point equations we discussed at the end
of section \ssNSUSYsad.
\par
A surprising  feature of the variational
energy density  is the appearance of a divergent contribution
$${\cal E}_{\rm var}/N=\ud \rho_c\bigl(M -U'(\rho)\bigr)^2+\ {\rm finite}\, .$$
Therefore, when the  equation $M =U'(\rho)$ is not
enforced, the variational energy is infinite with the cut-off. \par
Of course, we can also look for the minimum of ${\cal E}_{\rm var}$ by differentiating with respect to the three parameters.
Differentiating ${\cal E}_{\rm var}$ with
respect to $m_\psi $ and using the definition of $\tilde \rho$, we
obtain the fermion mass gap equation
$$m_\psi =U'(\rho).\eqnd\eFerMassGap $$
If this equation is taken into account the variational energy
density simplifies to
$$  {1\over N} {\cal E}_{\rm var}(\ma,\mpsi,\varphi ) = {\mpsi^2\over 2}
{\varphi^2 \over N} + {1\over 24\pi}(\ma-\mpsi)^2(\ma+2\mpsi).
 \eqnd\HartreeVTzeroC $$
 which is positive definite (recall that $m_\psi \ge 0$
and $m_\varphi \ge 0$) and vanishes when $m_\psi = m_\varphi$ and
$\mpsi\varphi = 0$, that is for a supersymmetric ground state with the
$O(N)$ symmetry either broken or unbroken. \par When the positive
fermion and the boson masses $m_\psi $, $m_\varphi$ are identified
with their value in terms of the parameters $M,m$,
$$m_\psi=|M| \quad {\rm and}
\quad m_\varphi=\sqrt{M^2+\lambda},\eqnd\FandBmasses$$ one
recognizes the expression \eSUSYground.  As we have shown this
expression has a unique  minimum $\ma=\mpsi$, which is
supersymmetric. Differentiating then with respect to $ \varphi$,
we find $\varphi m_\psi=0$, one solution $m_\psi=0$ corresponding
to $O(N)$ symmetry breaking, the other $\varphi=0$ to an $O(N)$
symmetric phase.

Differentiating Eq.~\eVarEsusy\ with respect to $m_ \varphi$ and
using the definition of $  \rho$, we obtain the boson mass gap
equation
$$m_\varphi^2=2\rho U'(\rho)U''(\rho)+U'{}^2(\rho)-\tilde \rho U''(\rho) \eqnd\eBosMassGap $$
or, using the value of $m_\psi $,
$$m_\varphi^2-m_\psi ^2=U''(\rho)\left(2m_\psi \rho-\tilde \rho\right).$$
Clearly,   the supersymmetric solution satisfies both gap
equations. In the combination $2m_\psi \rho-\tilde \rho$, we
recognize the parameter $s$ as in Eq.~\eSUSYii{b} ($F$ being taken
from Eq.~\eSUSYsada{a}), and thus the equation becomes
Eq.~\eSUSYsadb{b}:  $\lambda =sU''(\rho)$.\par

Eqs.~\eqns{\eFerMassGap,\eBosMassGap}  have a clear
Schwinger--Dyson  diagrammatic interpretation for a $U(\rho)=\mu
\rho + \half u \rho^2 $ potential (see Eq.~\eGenericAction).
Namely, \eqna\eGaps
$$\eqalignno{m_\psi &=\mu + {u\over N}\left<\varphi^2(x)\right>_0\,,
&\eGaps{a}\cr m_\varphi^2 &= \mu^2 + 4 {\mu u \over
N}\left<\varphi^2(x)\right>_0 + 3 {u^2\over
N^2}\left<\varphi^2(x)\right>_0^2 - {u \over N }\left< \bar\psi
(x)\psi (x)\right>_0 . \cr & &\eGaps{b}  }
$$


%
\subsection The $\Phi^4$ super-potential in $d=3$: phase structure

We now consider the special example
$$U(R)=\mu R+\ud uR^2\ \Rightarrow\ U'(R)=\mu+ uR\,.$$
The dimensions of the $\theta$ variables and the field $\Phi$ are
$$[\theta]=-\ud\,,\quad[\Phi]=\ud\ \Rightarrow \ [u]=0\,.$$
Power counting thus tells us that the model is renormalizable in
three dimensions. Prior to a more refined analysis, one expects
coupling constant and field renormalizations (with logarithmic
divergences) and a mass renormalization with linear divergences.
Using the solution \eSUSYiipt\ for the two-point function
$\Gamma^{(2)}$, one infers that the coefficient $A(p^2)$ has at
most a logarithmic divergence, which corresponds to the field
renormalization, while the coefficient $C(p^2)$ can have a linear
divergence which corresponds to the mass renormalization. \par
The invariance of physics under the change $U\mapsto -U$ was mentioned above and
seen in Eq.~\eGenericAction\ and the equations that followed. This
invariance will be reflected into the phase structure of the
model. For the quartic potential the equations \eSUSYsadb{} now
are
$$M=\mu+u \rho\,,\quad \lambda=us\,.\eqnd\eSUSYmpot $$
We introduce the critical value  of $\mu$,
$$\mu_c=-u \rho_c  \, . \eqnn $$
Taking into account equations  \eSUSYmpot, one finds that the equations \eSUSYiv{} can now be written as
\eqna\eSUSYivquartic
$$\eqalignno{ M&=\mu-\mu_c + u\varphi^2 /N- {u \over{4\pi}}
\sqrt{M^2+\lambda}\,,&\eSUSYivquartic{a}\cr \lambda &=2u M\varphi^2/N +
{u \over{2\pi}}M\left( |M|-\sqrt{M^2+\lambda} \right)\,.
&\eSUSYivquartic{b}\cr}$$
 Eqs.~\eSUSYivquartic{} relate the fermion
mass $m_\psi=|M|$, the boson mass $m_\varphi = \sqrt{M^2+\lambda}$
and the classical field $\varphi $. The phase structure of the
model is then described by the lowest energy solutions of these
equations in the $\{ \mu-\mu_c , u \}$ plane. Taking into account
the $U \to -U$ symmetry, mentioned above, one can restrict the discussion to $u>0$.
\par
We find, indeed, that supersymmetry is left unbroken
($\lambda = 0$) and the ground state energy ${\cal E}=0$ in each
quadrant in the $\{ \mu-\mu_c , u \}$ plane. This is consistent with Eqs.~\eSUSYivquartic{}
having a common solution with $\lambda =0$ (thus $m_\psi= m_\varphi = |M| $) and $M\varphi=0$.
They then reduce to
$$ M =\mu-\mu_c + u{\varphi^2 \over N} - {u \over{4\pi}}|M|\,,
\quad M\varphi  =0\, . \eqnd\eSUSYiii $$
\medskip
{\it The broken $O(N)$ symmetry phase.} The  $M=0$ solution
implies a spontaneously broken $O(N)$ symmetry, scalar and fermion
$O(N)$ quanta are massless and
$$\varphi^2 =-N(\mu-\mu_c)/u\,, \eqnd\ephi$$
which implies that this solution exists only for $ \mu<\mu_c$. The
solution exists in the fourth (and second) quadrant of the
$\{ \mu-\mu_c , u \}$ plane. Note that this yields the same  exponent $\beta=\ud $
as in the simple $(\varphib^2)^2$ field theory  (Eq.~\expbeta).
\smallskip
{\it The $O(N)$ symmetric phase.} We now choose the solution $\varphi=0$
of Eqs.~\eSUSYiii. Then the equation
$$M=\mu-\mu_c - (u/u_c)|M| \eqnd\eSymmetricPhase$$
yields the common mass $M$ for the fermions and bosons. In
Eq.~\eSymmetricPhase\ we have introduced the special value of the coupling $u$,
$$u_c= 4\pi  \ . \eqnd\egCritical$$
This equation splits into two equations, depending on the sign of $M$.
The first solution
$$M= M_+=(\mu-\mu_c)/(1+u/u_c) >  0\eqnd\eSUSYfivMscal $$
exists only for $\mu>\mu_c$ (first quadrant), as one would normally expect.
Note again that this corresponds to a correlation length exponent $\nu=1$, as in the ordinary $(\varphib^2)^2$ field theory (Eq.~\ecorlenb), though the form of the saddle point equations are different: in the supersymmetric theory this is the free field value.
Moreover, the exponent is independent of $u$, though the term proportional to $u$ is not negligible.  If one takes into account the leading correction coming from regularization (expansion \etadepolii), one finds
$$(1+u/u_c) M = \mu-\mu_c +u a(3)M^2/\Lambda \,.$$
Unlike what happens in the usual $\varphi^4$ field theory, no value of
$u$ can cancel the leading correction to the relation
\eSUSYfivMscal, and therefore no IR fixed point can be identified.
\par
The second solution
$$M=
M_-=(\mu-\mu_c)/(1-u/u_c) < 0 $$
is very peculiar. There are two different situations depending on the position of $u$ with respect to $u_c$: \par
(i) $u>u_c=4\pi$ and then $\mu>\mu_c$:  the solution is degenerate with another $O(N)$ symmetric solution $M_+$.\par
(ii) $u<u_c$ and then $\mu<\mu_c$: the solution is degenerate with a solution of broken $O(N)$ symmetry. \par
\midinsert \epsfxsize=13.5cm \epsfysize=11cm \pspoints=2.pt \vskip
-5.2cm \hskip2cm \epsfbox{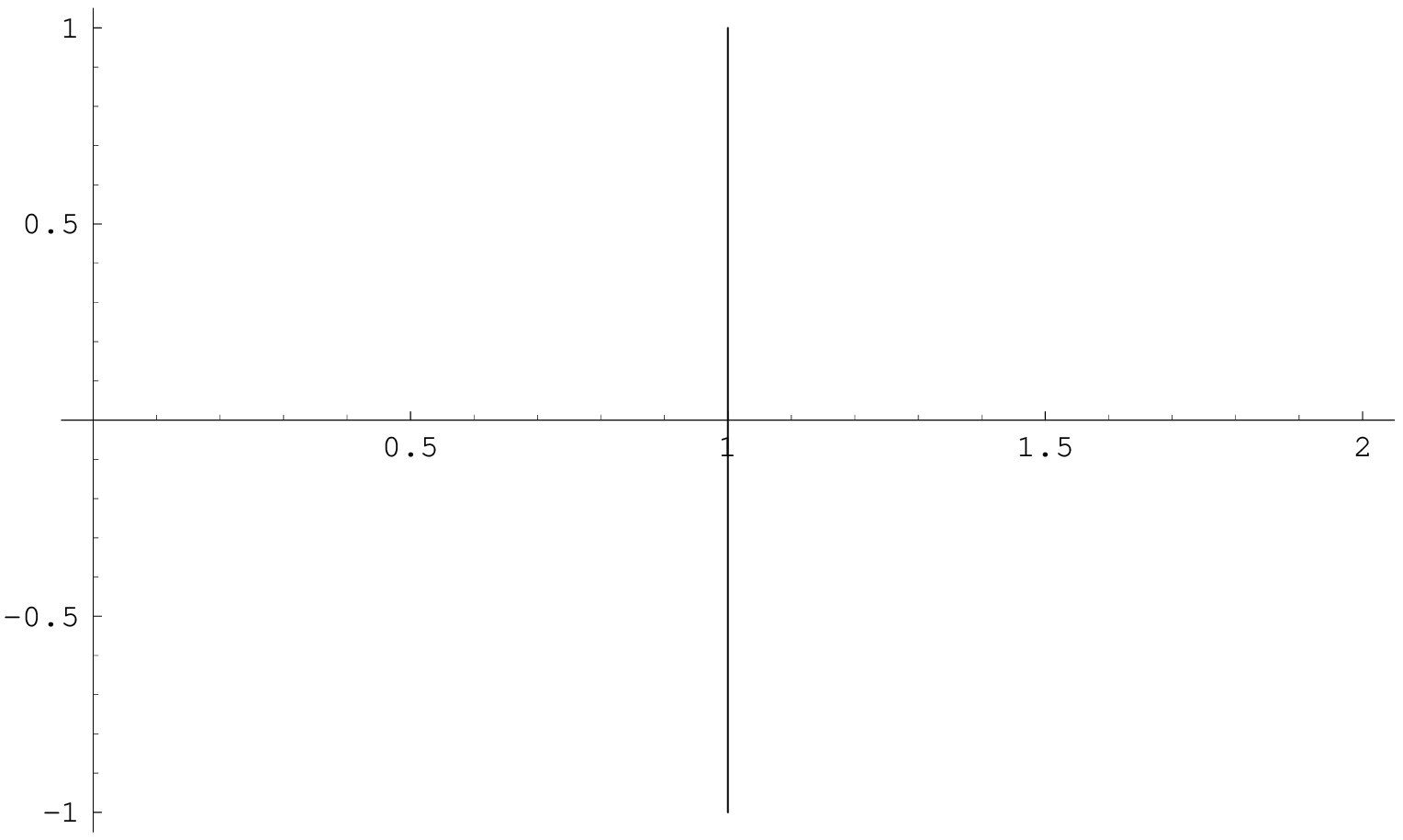}
\vskip -5.8cm
\hskip 5.cm $\varphi^2 = 0$ \vskip -.2cm \hskip 1.5cm
$~~~~\mu-\mu_c$ \vskip .2cm \hskip 3.5cm {\bf I} \hskip 3cm {\bf
II} \vskip .4cm \hskip 7.2cm $M_-$ \vskip .3cm \hskip 3.5cm $M_+$
\hskip 3cm $M_+$
\vskip 0.2cm
 \hskip 8.5cm $u/u_c$ \vskip .3cm \hskip 3.5cm $M_-$
\vskip .3cm \hskip 3.5cm {\bf IV} \hskip 3cm {\bf III} \vskip 1cm
\hskip 5cm $\varphi^2 \neq 0$ \vskip 5mm \figure{1mm}{Summary of
the phases of the model in the $ \{ \mu-\mu_c , u \} $ plane.
Here   $\ma=\mpsi=|M_\pm|=(\mu-\mu_c)/(u/u_c\pm 1)$. The
lines $u=u_c$ and $\mu-\mu_c=0$ are lines of first and second
order phase transitions.}
\figlbl\phases
\endinsert
 The phase structure is summarized
 in Fig. \label{\phases} in the first and
 second quadrant of the $ \{ \mu-\mu_c, u/u_c \} $
 plane where the following different phases appear:
\smallskip
{\bf Region I : ~~$\mu-\mu_c \geq 0 ~~,~~ {u/ u_c}\leq 1$}:

\noindent Here, there is only one $O(N)$ symmetric, supersymmetric
ground state with $\mpsi=\ma=M_+=(\mu-\mu_c)/(u/u_c+1)$ and
$\varphi^2=0$.
\smallskip

{\bf Region II : ~~$\mu-\mu_c \geq 0 ~~,~~ {u/ u_c}\geq 1$}:

\noindent There are two degenerate
$O(N)$ symmetric ($\varphi=0$) supersymmetric ground states with
masses $\mpsi=\ma=M_+=(\mu-\mu_c)/(u/u_c+1)$ and
$\mpsi=\ma=-M_-=(\mu-\mu_c)/(u/u_c-1)$.

\smallskip

{\bf Region III : ~~$\mu-\mu_c \leq 0 ~~,~~ {u/ u_c}\geq 1$}:

\noindent There is one supersymmetric ground state, it is an
ordered state with broken $O(N)$ symmetry ($\varphi^2 \neq 0$,
$\mpsi=\ma=0$).
\smallskip

{\bf Region IV : ~~$\mu-\mu_c \leq 0 ~~,~~ {u/ u_c}\leq 1$}:

\noindent There are two degenerate ground states: an $O(N)$
symmetric , supersymmetric ground states with masses
$\mpsi=\ma=m_-=(\mu-\mu_c)/(u/u_c-1)$ and $\varphi^2=0$. The
second ground state is a supersymmetric, broken $O(N)$ symmetry
state with $\mpsi=\ma=0$ and $\varphi^2 \neq 0$.

\medskip

{\it The action density.} To exhibit the phase structure in terms
of the variation of the  action density ${\cal E}$, we plot the
expression \eSUSYground, but use only the fermion gap equation in
Eq.~\eSUSYivquartic{a}, in such a way that $\cal E$ remains a
function of $\varphi$ and $\lambda $, or equivalently $\varphi$
and $m=\sqrt{M^2+\lambda }$:
$$  {1\over N} {\cal E}(m,\varphi)
 = \half M^2(m,\varphi ) {\varphi^2 \over N}
 +{1\over 24\pi} \left[m -\left|M(m,\varphi )\right|\right]^2
  \times \left(\ma + 2\left|M(m,\varphi)\right| \right) .
 \eqnd\HartreeVTzeroE $$
Two figures display the restriction of $\cal E$ to $\varphi=0$:
$$  {1\over N} {\cal E}(m,\varphi = 0)
=  {1\over 24\pi} \left[m - \left| \mu -\mu_c - (u/u_c)
m \right| \right]^2   \left( m +2\left| \mu -\mu_c
-(u/u_c)m \right| \right).   \eqnd\HartreeVTzeroF  $$
\midinsert \epsfxsize=17.5cm \epsfysize=14cm \pspoints=2.pt \vskip
-7cm \hskip 1cm \epsfbox{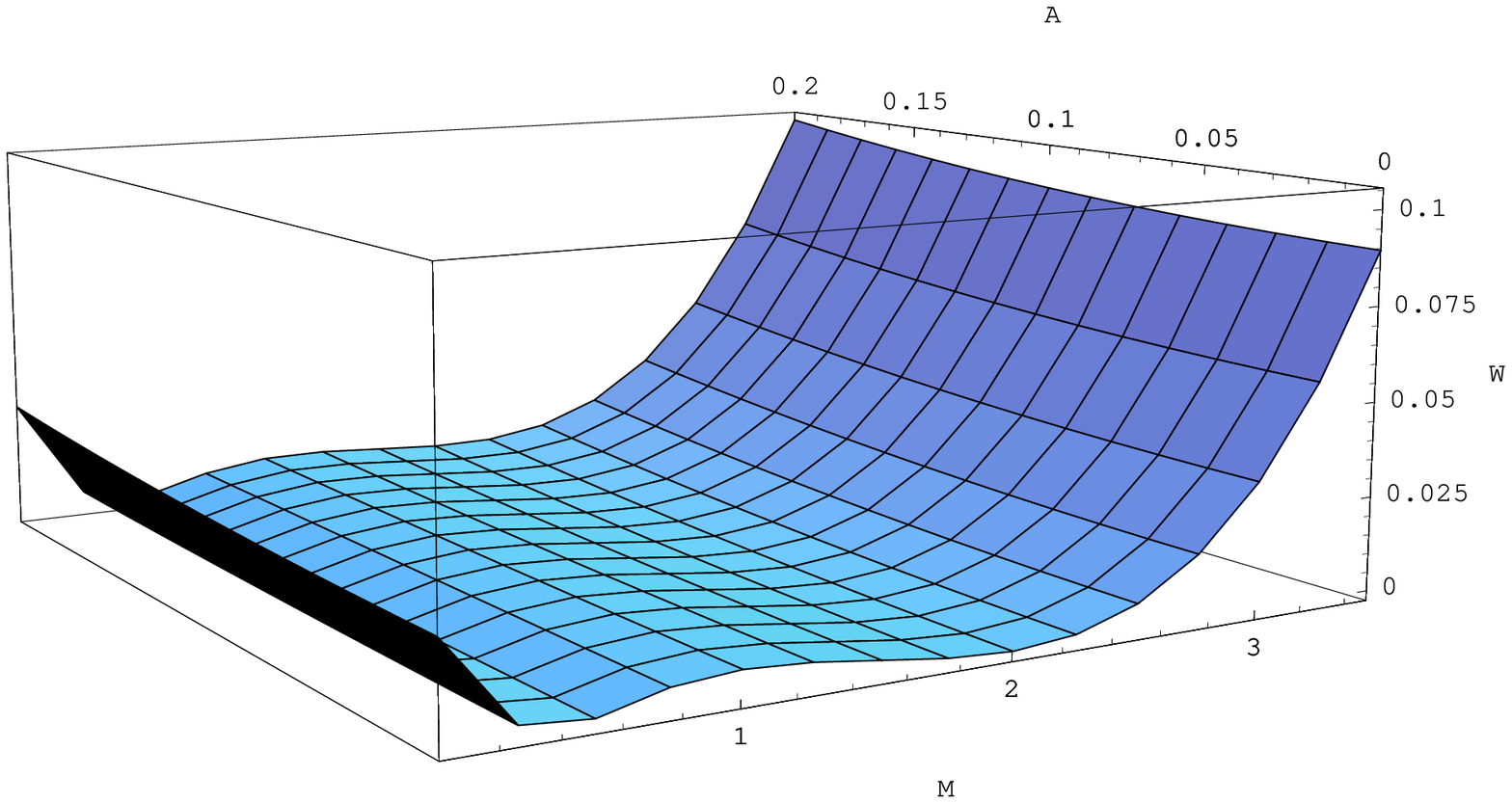} \figure{1mm}{{\bf Region II}
of Fig.~\phases: The energy density $W(m=\sqrt{M^2+\lambda
},\varphi ) \equiv {1\over N} {\cal E}(m,\varphi) $ as a function
of the boson mass ($m$) and $A$, where $A^2=\varphi^2/u_c$. Two
degenerate, $O(N)$ symmetric phases exist with  massive bosons
(and massive fermions). Here $\mu-\mu_c=1 $ (sets the mass scale)
\ $ u/u_c=1.5 $.} \figlbl\figTIIaaa
\endinsert

\midinsert \epsfxsize=13.5cm \epsfysize=10cm \pspoints=2.pt \vskip
-5cm \hskip 2cm \epsfbox{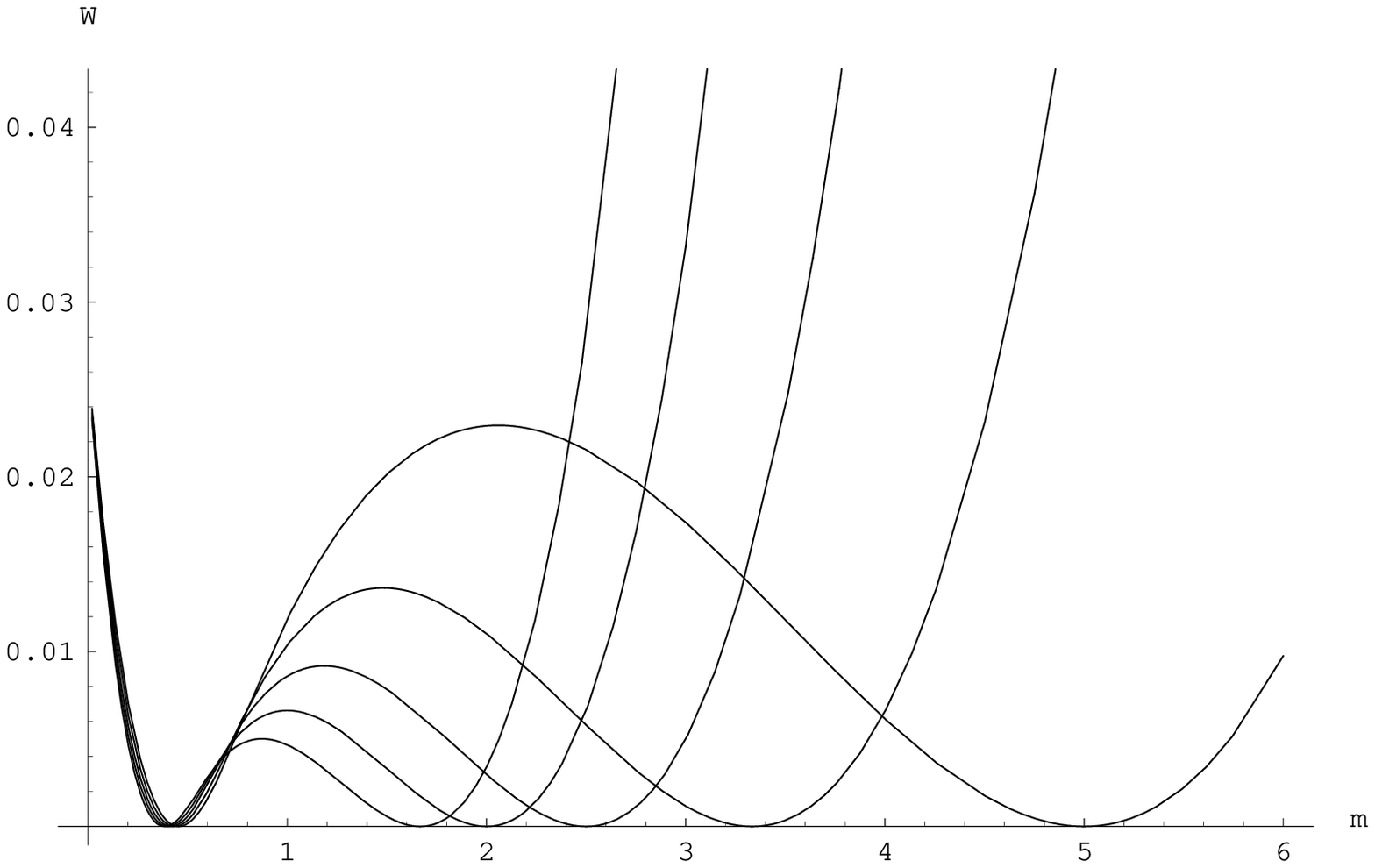} \figure{1mm}{The energy
density $W(m) \equiv {1\over N} {\cal E}(m,\varphi=0) $ from
Eq.~\HartreeVTzeroF\
in region {\bf II}
of Fig.~\phases. Here $\mu-\mu_c=1 $ (sets the mass scale) and
$u/u_c$= varies between 1.6 and 1.2. There are two degenerate
$O(N) $ symmetric SUSY vacua with $\varphi=0 $ at
$m_\psi=m_\varphi=M_+=(\mu-\mu_c)/(u/u_c+1)$ and at
$m_\psi=m_\varphi=-M_-=(\mu-\mu_c)/(u/u_c-1)$. } \figlbl\ZeroTone
\endinsert

\midinsert \epsfxsize=19cm \epsfysize=15cm \pspoints=2.pt \vskip
-7.5cm \hskip 1cm \epsfbox{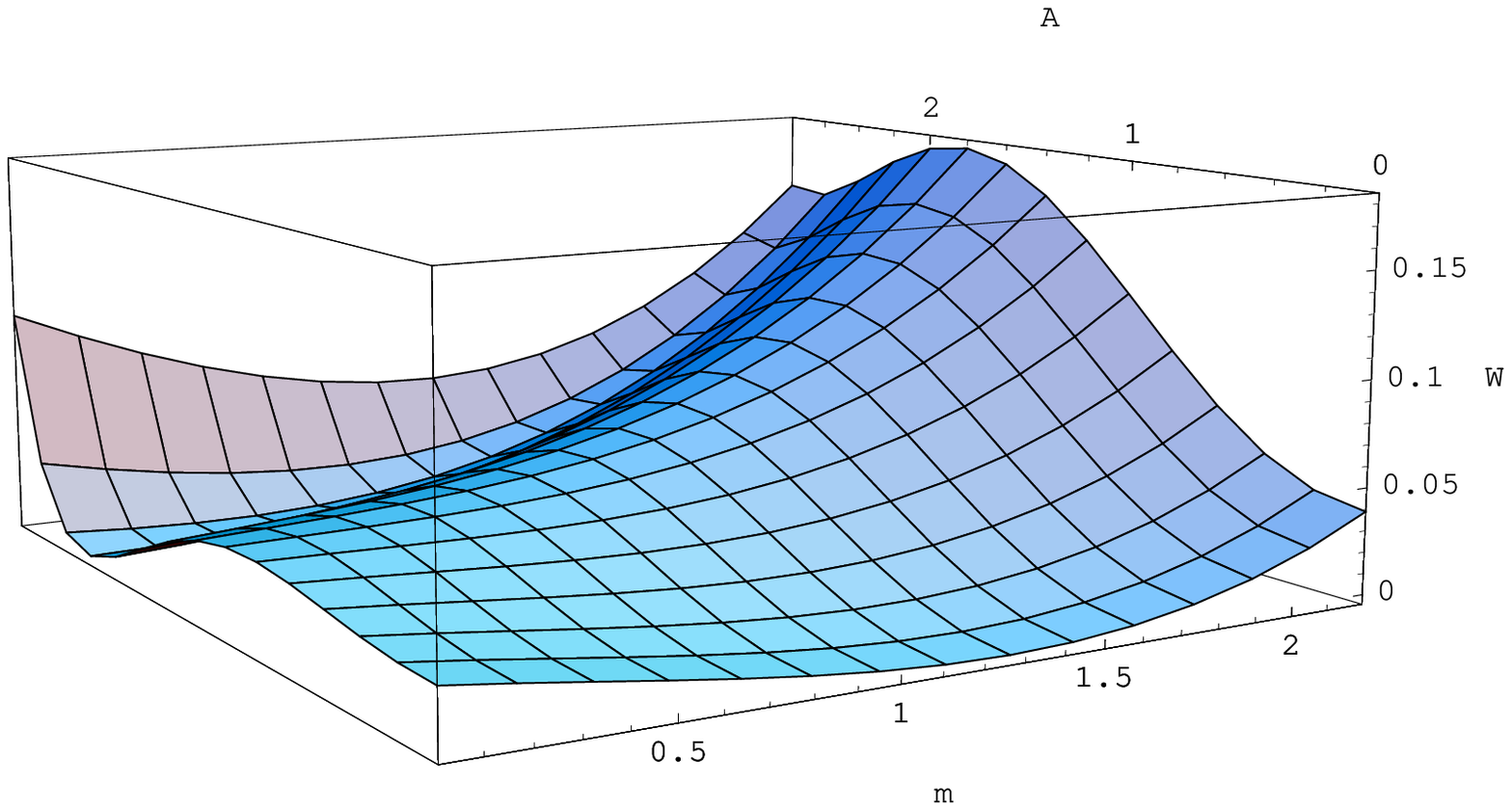} \figure{1mm}{{\bf
Region IV :} The  energy density $W(m,\varphi )={1\over N} {\cal
E}(m ,\varphi )$ as given in Eq.~\HartreeVTzeroE\ as a function of
the boson mass ($m$) and $A$, where $A^2=\varphi^2/u_c$. Here
$\mu-\mu_c=-1 ~,~ u/u_c=0.2 $. As seen here  there are two
distinct degenerate phases. One is an ordered phase ($\varphi \neq
0$) with a massless boson and fermion, the other is a symmetric
phase ($\varphi = 0$) with a massive ($m=|M_-|$) boson and
fermion.
 } \figlbl\figTIVaa
\endinsert


%

Several peculiar phase transitions can be easily traced now in
Eq.~\HartreeVTzeroE{}. First, one notes the phase transitions that
occur  when $\mu-\mu_c$ changes sign. When  $ 0 < u <u_c  $
and $\mu > \mu_c$, the system has a non-degenerate $O(N)$ symmetric
ground state with bosons and fermions of mass $M=M_+$. As $\mu -
\mu_c$ changes sign ($0 < u <u_c$ fixed), two degenerate ground
states appear. Either $M=0$ and $\varphi^2=-N(\mu-\mu_c)/u$ or the
system stays in an $O(N)$ symmetric ground state with a mass $|M_-|$
for the bosons and fermions. Similarly, when one goes from $\mu <
\mu_c$ to $\mu > \mu_c$ at $ u>u_c= 4\pi  $ the $O(N)$ symmetry is restored but there are two
degenerated ground states to choose from  $M=M_\pm$.

 In Fig.~\label{\figTIIaaa} and
 Fig.~\label{\ZeroTone}, $W(m=\sqrt{M^2+\lambda },\varphi) \equiv {1\over N}
{\cal E}(m, \varphi)$ from Eq.~\HartreeVTzeroF\ in region {\bf II}
($\mu-\mu_c \geq 0 ~,~ u/u_c \geq 1$) is plotted as a function of
$m$ and $\varphi$.  An unusual transition takes place when one
varies the coupling constant $u$. The transition from the
degenerate vacua at $u/u_c =1.4$ to a non-degenerate  ground state
at $u/u_c= 0.8$ is shown in Fig.~\label{\ZeroTtwo}  (from phase
{\bf II} to phase {\bf I}).

\midinsert \vskip -4.5cm \epsfxsize=15.5cm \epsfysize=10cm
\pspoints=2.pt \hskip 1.7cm \epsfbox{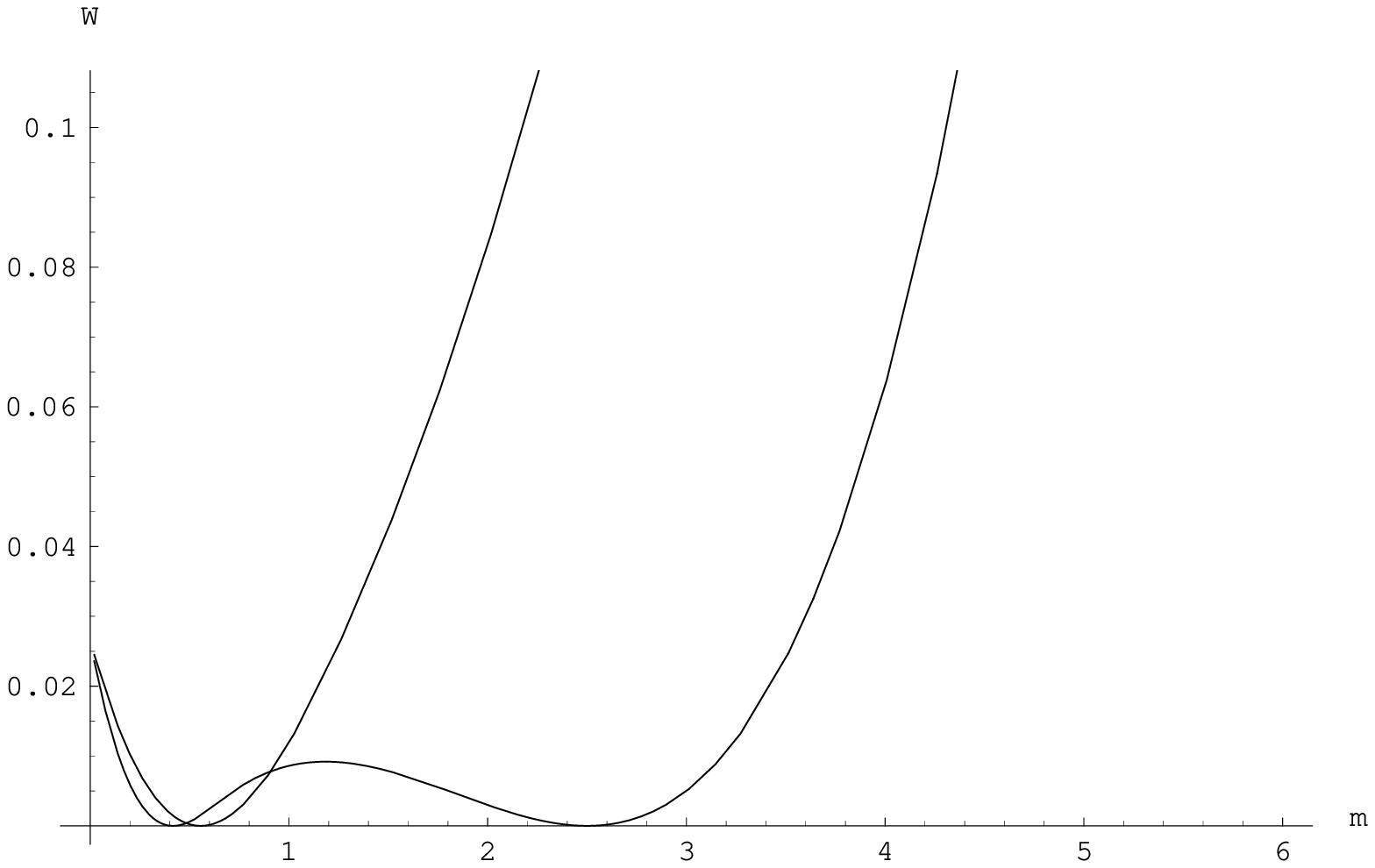} \figure{1mm}{The
energy density $W(m)={1\over N} {\cal E}(m,\varphi=0)$ as given in
Eq.~\HartreeVTzeroF.
Here $\mu - \mu_c=1 $ and $u/u_c = 1.2$ is changed to $u/u_c =
0.8$ (from region {\bf II} to region {\bf I} of Fig.~\phases).
There are two degenerate $O(N)$ symmetric SUSY vacua   at $u/u_c=
1.2$ with masses  $m_\psi=m_\varphi=|M_\pm|$ where
$|M_\pm|=\mu/(u/u_c \pm 1)$ while at $u/u_c= 0.8$ there is a
non-degenerate vacuum at $m_\psi=m_\varphi=M_+$.} \figlbl\ZeroTtwo

\endinsert

For positive $\mu-\mu_c$, we find two degenerate ground states if
$ u  > u_c $. As $u$ is lowered (at fixed $\mu-\mu_c$), the ground
state with mass $\ma=\mpsi=-M_-$ disappears ($|M_-/ M_+| \to
\infty$) and only the $O(N)$ symmetric phase remains with $M=M_+$
(Figs.~\label{\ZeroTone}  and \label{\ZeroTtwo}). Namely, suppose
we consider at $\{ u>u_c$, $\mu-\mu_c>0\}$ a physical system in a
state denoted by $A$ and defined by $\{\varphi^2=0, M=M_-\}$, such
a system will go into a state $B$ defined by $\{\varphi^2=0,
M=M_+\}$ when $u$ {\bf decreases} and passes the value $u=u_c=
4\pi $. Now, if we consider the reversed process; a physical
system at $\{ u < u_c , \mu-\mu_c> 0 \}$ that is initially in the
ground state $B$ and $u$ {\bf increases }and
 passes  $u=u_c$. There is no reason now for the
system to go through  the reversed transition from $B$ to $A$
since the $O(N)$ symmetric states with $M=M_-$ and $M=M_+$ are
degenerate and the fact that supersymmetry is preserved will avoid
that the energy of state $A$ will go below zero. These peculiar
phase transitions with the  ($|M_-/ M_+| \to \infty$) and with
``infinite hysteresis" in the $A \to B$ transitions are due to the
fact that supersymmetry is left unbroken in the leading order in
$1/N$. If supersymmetry would have been broken by some small
parameter, the lifted degeneracy would be, most probably,
translated into a slow transitions between the otherwise
degenerate ground states.

\smallskip

In section \label{\ssNSUSYT} we will study the transitions between
the different phases of ~Fig.~\phases~ as a function of the
temperature .

\medskip
{\it Special situation.} In general when $\mu=\mu_c$, the mass $M$ vanishes.
However, there is a special case when
$$u=u_c=4\pi  \,. $$
Then the value of $M$ is left undetermined. An accumulation point
of coexisting degenerate ground states exist in the phase
structure shown in Fig.~\phases. The case $\mu=\mu_c$ represents a
scale invariant theory where, however, the $O(N)$ fermionic and
bosonic quanta can have a non-vanishing mass $\ma=\mpsi=|M|$.
Since $u$ has not undergone any perturbative renormalization there is
no explicit breaking of scale invariance at this point. Thus, the
only scale invariance breaking comes from the solution of the gap
equation for $M$, which leaves, however, its numerical value
undetermined. We see a dimensional transmutation from the
dimensionless coupling $u$ that is fixed at a value of $u=u_c$
into an undetermined scale $M$. If $ M \neq 0$ the spontaneous
breaking of scale invariance will require the appearance of a
Goldstone boson at the point $ u= u_c$. The massless Goldstone
boson is associated here with the spontaneous breaking of scale
invariance. Moreover, since the ground state is supersymmetric, we
expect the appearance of a massless $O(N)$ singlet Goldstone boson
(a dilaton) and its massless fermionic partner (a ``dilatino'').
In order to see all this, we now calculate the $\langle LL \rangle$
propagator that will enable us to see these poles in the
appropriate four-point functions.\par
Finally, note that this analysis is only valid in the complete absence of cut-off corrections.
Otherwise if
$\mu=\mu_c$ and $u$ is tuned as $u=u_c - M_0/\Lambda$, where $M_0$
is an arbitrary mass scale, one finds for the $M_-$ solution:
$$M=- {M_0 \over a(3)u_c^2}\,.$$
\vskip .5cm
\subsection The $\langle LL\rangle$ propagator and massless
bound states of  fermions and bosons

{\it The $\langle LL \rangle$ propagator.}
We now calculate the $\langle LL\rangle$
propagator in the
symmetric phase. Then, the propagators of the fields $\phi$ and $L$ are
decoupled. In the quartic potential example the $R$ field can be
eliminated by gaussian integration. The relevant part of the $L$-action then
reads
$$-{N\over 2u}\int\d^3 x\,\d^2\theta(L-\mu)^2+{1\over2} (N-1) {\rm Str}\,
\ln\left(-\bar {\rm D} {\rm D}+2L\right) . $$
The calculation of the $\langle LL \rangle$ propagator  involves the
super-propagator \esupprop. For the inverse propagator  one
finds
$$\Delta_L^{-1}(p)=-{N\over u}\delta^2(\theta'-\theta)
-2N\int{\d^3 k\over(2\pi)^3}
\Delta(k,\theta,\theta')\Delta(p-k,\theta,\theta')  $$
with here (see Eq.~\eSUSYiipt)
$$\Delta(k,\theta,\theta')={1\over k^2+M^2}\left[1+\ud
M\delta^2(\theta'-\theta) \right]\e^{-i \bar\theta \slam{k}\theta'}\,. $$
Then,
$$\Delta(k,\theta,\theta')\Delta(p-k,\theta,\theta')
={[1+M \delta^2(\theta'-\theta)]\e^{-i \bar\theta \slam{p}\theta'}
\over (k^2+M^2)[(p+k)^2+M^2]}\,. $$
Notice the cancellation of the factor $\e^{-i \bar\theta \slam{k}\theta'}$
which renders the integral more convergent that one could naively expect.
The integral over $k$ then yields the three-dimensional bubble diagram
$$\eqalign{B(p)&={1\over(2\pi)^3}\int{\d^3 k\over  (k^2+M^2)[(p+k)^2+M^2]}\cr
&={1\over4\pi p}{\rm Arctan}(p/2|M|)\,. \cr}
$$
At leading order for $p$ small, we need only $B(0)=1/8\pi |M|$. Then,
$$\Delta_L^{-1}=
-{N\over4\pi |M|}\left[1+(M +4\pi |M|/u )
\delta^2(\theta'-\theta)\right]\e^{-i \bar\theta \slam{p}\theta'}.
$$
Comparing with the expression \eSUSYker, we conclude that for
$M>0$ small the $LL$ propagator corresponds to a super-particle of
mass $2M (1+4\pi/u ) $. For $M<0$ the mass is $2|M(1-4\pi/u)|$.
For $|u-u_c|$ small, it is a bound state and at the special point
$u=u_c$ the mass vanishes.\par
More generally, we find
$$\Delta_L(p)={2\over NB(p)}
{1\over p^2+m^2(p)}\left[1-\ud m(p)\delta^2(\theta'-\theta)\right]
\e^{-i \bar\theta \slam{p}\theta'} \eqnd\eLLprop$$
with
$$m(p)=2M+{1\over  u B(p)}\,.$$
We note that only a mass renormalization is required at leading order, a situation similar to the $\varphi^4$ scalar field theory.
As a consequence, dimensions of fields are not modified.\par

Clearly, the propagation of  the fields $M(x)$  and $\ell(x)$  (of
Eq.~\eLsupField), as indicated in Eq.~\eLLprop, when combined with
the $ L(\theta)\Phi^2$ interaction in Eq.~\eactLagm, namely,
$$  \int\d^2\theta\,L\Phi^2 =M(-\half\bar\psi\psi+F\varphi )
-\varphi\bar\ell \psi+\half\lambda \varphi^2 ,$$
describes the
bound states in the $\varphi\varphi$, $\psi\psi$ and $\psi\varphi$
scattering amplitudes. For example, in the supersymmetric ground
state case and with $\mu-\mu_c=0 $, the $\psi\varphi$ scattering
amplitude $T_{\psi\varphi,\psi\varphi}(p^2)$, in the limit $p^2
\to 0$ satisfies
$$T_{\psi\varphi,\psi\varphi}(p^2) \sim {2u\over
N}\left[1+ {u\over 4\pi}{M\over |M|} +{u\over 2\pi}{i\sla{p} \over
|M|} \right]^{-1}  \to -{4\pi i\over N}{|M|\over \sla{p}} \eqnn $$
for  $M<0$ and $u \to u_c$

One notes here that the fermionic massless bound state pole
appears when a non-zero solution  ($M$) to the gap equation exists
$ (m_\varphi=m_\psi=|M|) $ in the absence of any dimensional
parameters ($\mu-\mu_c=0$).  This happens  when the force between
the massive $\psi$ and $ \varphi$ quanta is determined by $u \to
u_c$. The massless $O(N)$ singlet fermionic bound state excitation
is associated with the spontaneous breaking of scale invariance.
Similarly, the bosonic partner of this massless bound state
excitation can be then seen, at the same value of the parameters,
in $\varphi\varphi$ and $\psi\psi$ scattering amplitudes as
  Eq.~\eLLprop~shows.
\par
At $\mu=\mu_c$ in the generic situation $M=0$, or for
$|p|\to\infty$, we find
$$B(p)={1\over 8p} $$
and, thus,
$$\Delta^{-1}_L(p)=-{N\over u}\delta^2(\theta'-\theta)
-{N\over4 p}\e^{-i \bar\theta \slam{p}\theta'}. $$
As a consequence,
the canonical dimension of the field $L$ is 1, as in
perturbation theory, and the interaction $L\Phi^2$ in Eq.~\eactLagm\
is renormalizable. The $L^2$ term thus is not negligible in the IR limit.
Nevertheless,  as we have noticed, the behaviour of physical quantities
does not depend on $u$. Therefore, we expect that, at least for generic values of $u$, we can describe the same physics with the supersymmetric non-linear $\sigma $-model.
\medskip
{\it $O(N)\times O(N)$ symmetric models.} The dynamics by which
scale invariance can be broken in a theory which has no trace
anomalies in perturbation theory has been studied also in
$O(N)\times O(N)$ symmetric models \refs{\Sal,\rEM}.
Here one finds that
spontaneous breaking of scale invariance is due to the breakdown
of the internal $O(N)\times O(N)$ symmetry or to non-perturbative mass
generation on a critical surface. The mass of the fermion
 and boson $O(N)$ quanta is arbitrary due to
the appearance of flat directions in the action density. Also the ratios
between the scales associated with breaking of the internal
$O(N)\times O(N)$ symmetry and scale
symmetries are arbitrary on the critical surface.
Massless bound states of fermions and bosons appear
due to the spontaneous breaking of scale invariance.
Here, again, in the large $N$ limit there is no explicit breaking of scale
invariance and the perturbative $\beta$ function vanishes.
The variational ground state energy, which is calculable in the
large $N$ limit, has flat directions which allow spontaneous breaking of
scale invariance by non-perturbative generation of mass scales.
This results in lines of fixed points in the
coupling constant plane which are
associated with dynamical scale symmetry
breaking. The interplay between internal symmetry, scale symmetry
and supersymmetry is reflected in a rich phase structure.
The novel issue in these models are phases in which
the breaking of scale invariance is directly related
to the breaking of the internal symmetry
$O(N)\times O(N) \rightarrow  O(N-1)\times O(N-1)$.

For the $O(N) \times O(N)$ symmetric potential $NU(\Phi _1^2/N ,\Phi_2^2/N ) $ where
$$ U( R _1,R _2) =  \mu  ( R _1+R _2)
 +\ud u   ( R _1 ^2 +  R _2^2 )
 + v R _1R _2\,,   \eqnn $$
one again finds in $d=3$, in the leading order in $1/N$, that  only a mass renormalization is needed. Correspondingly, the critical value of the parameter
$\mu$ now is $\mu_c=  -2 (u + v) \rho_c $.
  Here too, the interesting case is
$\mu = \mu_c$. As in the $O(N)$ symmetric model, one finds that
supersymmetry is left unbroken to leading order in $1/N$. The gap
equations are now
$$\left.\eqalign { m_1\left({{u}\over{4\pi}} + \sgn(\mpone) \right) + m_2\left({v\over{4\pi}}\right) &= u \varphi_1 ^2/N +v \varphi_2^2 /N   \,, \cr
m_1\left({v\over{4\pi}}\right) + m_2\left({{u }\over{4\pi}} +
\sgn(\mptwo)\right ) &= v{\varphi_1 }^2/N+ u{\varphi_2 }^2 /N\,,\cr}\right.\eqnn$$
where non-zero solutions are obtained on lines $v=v(u)$,
which is the condition for spontaneous breaking of scale
invariance. As expected, this is also the condition for the
appearance of massless dilaton and ``dilatino" in the spectrum.
One notes that scale invariance can be broken here as a
consequence of the breakdown of the internal $O(N) \times O(N)$
symmetry. Indeed, the massless dilaton and ``dilatino" appear
either as bound states of massive ``pions" and their supersymmetric
partners or, when the internal symmetry is broken, they are mixed
with the ``sigma" boson and fermion particles.
Thus, the non-zero scale that is responsible for the spontaneous
breaking of scale invariance may be set here also by $
\varphi^2~\neq~0$ rather than by $\mpsi=\ma~\neq~0$ only, as was
the case in the $O(N)$ symmetric model. On the lines $v = v(u)$,
one finds that $\cal E$ has flat directions in all dimensional
parameters.

\subsection  Dimensions $2\leq d \leq 3$

In $d$ space--time dimensions, $2\leq d \leq 3$, one can keep in the space of $\gamma $ matrices $\tr{\bf 1}=2$. From the point of view of power counting in perturbation theory the model is super-renormalizable since the dimension of $u$ now is $3-d$.\par
The calculations follow the same pattern discussed above.
Eqs.~\eSUSYii{} are replaced by (with the definition \etadepole)
\eqna\esadNSUSYnls
$$\eqalignno{\rho-\varphi^2/N&=\Omega _d(m), &\esadNSUSYnls{a}\cr
s-2F\varphi/N&=2M\left[\Omega _d(m)-\Omega _d(|M|)\right]. \cr}$$
Taking into account the other saddle point equations, and $\rho_c=\Omega _d(0)$,
one obtains a generalization of Eq.~\eSUSYground:
$$2 {\cal E}/N =  M^2\varphi^2/N+(M^2-m^2)\left[ \Omega _d(m)-\Omega _d(0)\right] +\int_0^\lambda \d\lambda ' \,
 \Omega _d\left(\sqrt{M^2+\lambda '}\right)  \eqnd \eDminEi $$
and, thus, from Eqs.~\etadepolii\ and \etadpolexp{a}:
$$ {\cal E}/N =\half M^2\varphi^2/N +K(d)\left[ \ud (m^2-M^2)m^{d-2}
-\left(m^d-|M|^{d }\right)/d\right],$$
 where
$$K(d)= -{ \Gamma(1-d/2)\over (4\pi)^{d/2}}.$$
Moreover, from Eq.~\eDminEi,
$${\partial {\cal E}\over\partial m }=- {N\over 2m}  (m^2-M^2)
 \Omega _d'(m),\eqnd \eDminE $$
which has the sign of $m-|M|$ because $\Omega _d'(m)$ is negative. The function, therefore, has a minimum  at $m=|M|$ for all values of $d$, and supersymmetry is maintained in the ground state for all $2\leq d \leq3$.\par
The critical exponents now are $\eta =0$, $\beta=1/2$, and for the mass in the unbroken phase one obtains
$$M=\mu -\mu _c-K(d)u |M|^{d-2}+O(M^2\Lambda ^{d-4}).$$
We see that the l.h.s.~is now negligible, the equation having a solution only for $(\mu -\mu _c)/u>0$. The non-perturbative value  $\nu =1/(d-2)$ of the correlation exponent follows.
One also verifies that the dimension of $L$ remains 1, and thus the $L^2$ term now is negligible in the IR limit, leading immediately to the non-linear $\sigma $-model. However, again one finds no IR fixed point, no value of $u$ cancels
the leading correction to scaling, and therefore the argument leading to the non-linear $\sigma $-model is not as solid as for the usual $(\varphi^2)^2$ field theory.
\medskip
{\it Two dimensions.}
It may be interesting to consider the same model in two
dimensions. The model now is super-renormalizable, but even at $\mu=0$ it is
not chiral invariant since a chiral transformation corresponds to a change
$U\mapsto -U$.\par
The expression \eDminE\  is cut-off independent and has a limit for $d=2$:
 $${\cal E}/N=\ud M^2\varphi^2/N +{1\over8\pi}
\left[m^2-M^2-2M^2\ln(m/M)\right]  ,\eqnd\eSUSYNEnii $$
an expression which again has an absolute minimum at $m=|M|$. Then,
a minimization with respect to $M$ and $\varphi$ yields $M\varphi=0$. At the
minimum $\cal E$ vanishes. \par
Taking into account $\lambda =0$ and Eq.~\esadNSUSYnls{a} in the $d=2$ limit, we obtain the gap equation
$$M=\mu+u\varphi^2/N+{u\over 2\pi}\left[\ln(\Lambda /|M|]+O(1)\right). $$
The solution $M^2=\lambda=0$, thus, is not
acceptable. The $O(N)$ symmetry
is never broken.
When the mass $M$ is small in the cut-off scale, the l.h.s.~is negligible
and
$$m=M\propto  \Lambda  \e^{2\pi\mu /u}. $$
This solution exists only when $-\mu/u$ is positive and large. Its behaviour
suggests immediately a relation with the non-linear $\sigma $-model, which we now briefly examine.

\subsection A supersymmetric non-linear $\sigma $ model at large $N$

We consider the supersymmetric non-linear $\sigma $ model in $d$
dimensions, $2\le d\le 3$. The action \refs{\rWittSUSY{--}\rFerre}
\sslbl\ssSUSYnls
$${\cal S}(\Phi)={1\over 2 \kappa }\int\d^d x\,\d^2\theta\, \bar {\rm D}\Phi\cdot
{\rm D}\Phi   \eqnd\eNLSM $$
involves an $N$-component   scalar superfield
$\Phi$,  which satisfies
$$\Phi\cdot\Phi = N \,.$$
This relation is implemented  by introducing  a superfield
$$L (x,\theta) = M(x) + \bar\theta\ell(x) +\half \bar\theta\theta
\lambda(x) ,$$
where $M(x)$, $\lambda(x)$ and $\ell(x)$ are the Lagrange multiplier fields,
and adding to the action
$${\cal S}_L ={1\over  \kappa }\int\d^d x\,\d^2\theta \, L (x,\theta)
\left[\Phi^2(x,\theta)-N \right] . \eqnd\eNLSML $$
The partition function is given by (${\cal S}(\Phi,L)={\cal S}+{\cal S}_L$):
$${\cal Z} = \int [d\Phi][\d L] \e^{ -  {\cal S  }(\Phi,L) }.$$
In terms of component fields, the total action reads
$$\eqalignno{ {\cal S}={1\over  \kappa }\int \d^dx \{ &\half { \varphi}(-\del^2 +
M^2){ \varphi}
-\half{ {\bar\psi}}(\delslash+M){ \psi} \cr
&+\ud \lambda
({ \varphi}^2- N  )
-\bar\ell({ {\bar\psi}}\cdot \varphi) \}.
&\eqnd\NLSMcomp }$$
As in the case of the $\Phi^4$ theory, we integrate out $N-1$ components
leaving out $\phi=\Phi_1$:
$${\cal Z} = \int [d\phi][\d L] \e^{ - {\cal S}_N(\phi,L) }$$
where
$$   {\cal S}_N(\phi,L)= {1\over\kappa } \int\d^d x\,\d^2\theta  \left[ \half\bar {\rm D}\phi\cdot
{\rm D}\phi+  L \left(\phi^2- N \right) \right]    + {{N-1}\over 2}
{\rm Str}\, \ln( -{\bar {\rm D}}{\rm D} +2 L )  .   \eqnd\eNLSMeff   $$
By varying the effective action with respect to $\phi$, one finds the
saddle point equation
$${\bar {\rm D}}{\rm D}\phi-2L \phi =0\,, \eqnd\eSadNLSMa $$
which implies for constant $\varphi$ and $\psi=0$:
$$ F-M\varphi=0 \qquad {\rm and} \qquad MF+\lambda\varphi = 0 \,.
\eqnd\eSadNLSMaa$$
When the large $N$ action is varied with respect to $L $ and using the
expression in \esupprop~for the $\phi$ propagator, one finds ($N\gg 1$)
$${1\over \kappa}-{\phi^2 \over N}
=  \int{\d^d k\over (2\pi)^d}\left[{1+ M\bar\theta\theta \over
k^2+M^2+\lambda}-{M\bar\theta\theta\over k^2+M^2}\right].
\eqnd\eSadNLSMb$$
We now introduce the boson mass parameter $m=\sqrt{M^2+\lambda }$.
 Eq.~\eSadNLSMb~in  component form (for  $\psi=0$) then reads
\eqna\eSadNLSMbb
$$\eqalignno{1-{\varphi^2\over N}&=
\kappa \,\Omega _d (m ),
&\eSadNLSMbb{a}\cr
M{\varphi^2\over N\kappa } &=M \left[\Omega _d(|M|)- \Omega _d\left( m\right)
\right]. &\eSadNLSMbb{b}\cr} $$
\medskip
{\it Dimension $d=3$.} Introducing the critical (cut-off dependent) value
 $$ \kappa_c = 1/\Omega _3(0)\,,$$
we can write
Eqs.~\eSadNLSMbb{} now as
\eqna\eSadNLSMbbb
$$\eqalignno{{1\over \kappa }-{1\over \kappa_c }-{\varphi^2\over N\kappa } &=-
{m\over4\pi } &\eSadNLSMbbb{a}\cr
M{\varphi^2\over N\kappa } &={1\over4\pi}M\left(m-|M|\right).
&\eSadNLSMbbb{b}\cr} $$
The calculation of the ground state energy density
of the non-linear $\sigma $ model follows  similar steps
as in the $(\Phi^2)^2$ model:
$$  {\cal E} /N={ 1\over 2N\kappa } m^2\varphi^2  - \ud(m^2-M^2)\left( {1\over\kappa } -{1\over\kappa_c}\right)
-{1\over12\pi}\left(m^3-|M|^3\right)
   \eqnd\eNLSMenergy $$
and, taking into account the saddle point equation \eSadNLSMbb{a},
 $$ {\cal E} /N={ 1\over 2N\kappa } M^2\varphi^2 +  {1\over 24\pi}(m-|M|)^2(m+2|M|)
,\eqnd\eSUSYnlsground
$$
an expression identical to \eSUSYground, up to the normalization of $\varphi$.
Again, if a supersymmetric solution exists it has the lowest energy.
We thus look for supersymmetric solutions.\par
In the $O(N)$ symmetric phase ($\varphi = 0$)
Eq.~\eSadNLSMbbb{b} is satisfied while
Eq.~\eSadNLSMbbb{a} yields
$$m=|M|=4\pi\left({1\over \kappa _c}-{1\over \kappa }\right).$$
This phase exists for   $\kappa \ge \kappa _c$.\par
In the broken phase $m=M=0$, and Eq.~\eSadNLSMbbb{b} is again satisfied.
Eq.~\eSadNLSMbbb{a} then  yields
$$\varphi^2/N=1-\kappa /\kappa _c \,, \eqnn $$
which is the solution for $  \kappa \le \kappa _c$.\par
Since we have found solutions for all values of $\kappa $, we conclude that
the ground state is always supersymmetric.
At $\kappa _c$, a  transition occurs between a massless phase with broken $O(N)$ symmetry  ($\varphi^2 \neq 0$)  and a symmetric phase with massive fermions and bosons of equal mass.
The supersymmetric non-linear $\sigma $ model and the usual non-linear $\sigma $-model studied in section \ssLTsN, thus, have analogous phase structures, which are less surprising than the structure of $(\Phi^2)^2$ field theory. \par
\medskip
{\it Renormalization group.} From the point of view of power counting, the model is analogous to its non-supersymmetric version (section \ssLTsN), and RG equations take the same form with two RG functions $\beta (\kappa )$ and $\zeta (\kappa )$. Both have been calculated to four loop order.\par
At leading  order for $N\to\infty $, one finds the same critical exponents as in the supersymmetric $(\Phi^2)^2$ field theory or the usual non-linear $\sigma $ model: $\eta=0$ and, in generic dimension $d$: $\nu=1/(d-2)$.
The $\beta $-function has at this order the same form as in the non-supersymmetric model:
$$\beta (\kappa )=(d-2)\kappa (1-\kappa /\kappa _c)+O(1/N).$$
In the supersymmetric model the critical exponents $\nu$ and $\eta$ have been calculated
to order $1/N^2$ and $1/N^3$, respectively. To order $1/N$ one finds
$$\eta= {X_1(d)\over N}+O(1/N^2)\,, \quad \nu={1\over d-2}+O(1/N^2),$$
where $X_1(d)$ is the constant \eXone, $X_1(3)=8/\pi^2$.
\medskip
{\it Dimension $d=2$.}   The phase structure of the supersymmetric
non-linear $\sigma$ model in two dimensions is rather simple, and again analogous to the structure of the usual non-linear $\sigma $-model.
Eq.~\eSadNLSMbbb{a} immediately implies that $M^2+\lambda =0$ is not a solution
and thus the $O(N)$ symmetry remains unbroken $(\varphi^2=0$)
for all values of the coupling constant $\kappa $.
Then, with a suitable normalization of the cut-off $\Lambda $,
$$m=\Lambda \e^{-2\pi/\kappa }. \eqnd\enlsSUSYNii $$
Correspondingly, the action density becomes
$$  {\cal E} /N=   -  {\lambda\over2\kappa}
 -{1\over8\pi}\left[(M^2+\lambda)\ln(M^2+\lambda)
-\lambda\ln  \Lambda ^2
-M^2\ln M^2 -\lambda\right]  \eqnd\eNLSMenergyiicr $$
and, taking into account Eq.~\enlsSUSYNii,
$${\cal E} /N={1\over 8\pi}\left[m^2-M^2+2M^2\ln(M/m)\right].$$
We recognize the expression \eSUSYNEnii, and conclude in the same way
 that a  supersymmetric solution has the lowest energy. Supersymmetry remains unbroken, and the common boson and fermion mass is
$$M=m=\Lambda \e^{-2\pi/\kappa }. \eqnn $$
The model, thus, is UV asymptotically free and the condition $M\ll\Lambda $ implies that non-trivial physics is concentrated
near $\kappa =0$.

\vfill\eject
\def\r{{\rm r}}
\def\phib{\phi}
\def\varphib{\varphi}
\def\psib{\psi}

\nref\rKiLin{The study of finite temperature quantum field theory
has been initially motivated by the discussions of cosmological
problems \rf D.A. Kirznits {\it JETP Lett.} 15 (1972) 529 and
refs. \refs{\rKirzn,\rLinde}.}\nref\rKirzn{ D.A. Kirznits and A.D.
Linde, {\it Phys. Lett.} B42 (1972) 471; {\it Ann. Phys.} 101
(1976) 195.}\nref\rLinde{ A.D. Linde {\it Rep. Prog. Phys.} 42
(1979) 389.} \nref\rBWDJ{The possibility of phase transitions in
heavy ion collisions has generated additional interest \rf
 C.W. Bernard, {\it Phys. Rev.} D9 (1974) 3312  and refs. \refs{\rWein{--}\rGross}.}
 \nref\rWein{S. Weinberg, {\it
 Phys. Rev.} D9 (1974) 3357.}\nref\rDolan{  L. Dolan and R. Jackiw, {\it Phys. Rev.}
 D9 (1974) 3320.} \nref\rGross{D.J. Gross, R.D. Pisarski and L.G. Yaffe, {\it
 Rev. Mod. Phys.} 53 (1981) 43.}
\nref\rTreview{Among the various reviews and textbooks see, for
example,  \rf N.P. Landsman and C. van Weert, {\it Phys. Rep.} 145
(1987) 141 and refs. \refs{\rKapus{--}\rBlaizot}.}
\nref\rKapus{J.~I.~Kapusta, {\it Finite Temperature Field Theory},
Cambridge Univ. Press (Cambridge 1989).}\nref\rLeBel{ M. Le
Bellac, {\it Thermal Field Theory}, Cambridge Univ. Press
(Cambridge 1996).}\nref\rMeyer{ H. Meyer-Ortmanns, {\it Rev. Mod.
Phys.} 68 (1996) 473.}\nref\rBlaizot{ J.P. Blaizot, E. Iancu, {\it
Phys. Rep.} 359 (2002) 355, hep-ph/0101103.} \nref\rTRGeq{RG
equations in the context of finite temperature dimensional
reduction are discussed in \rf N.P. Landsman, {\it Nucl. Phys.}
B322  (1989) 498.} \nref\rZJTN{J. Zinn-Justin, SACLAY preprint
-SPH-T-00-055, hep-ph 0005272.} \nref\rTdimred{Early articles on
dimensional reduction include \rf P. Ginsparg, {\it Nucl. Phys.}
B170 (1980) 388 and refs.
\refs{\rAppel{--}\rKajan}.}\nref\rAppel{T. Appelquist and R.D.
Pisarski, {\it Phys. Rev.} D23 (1981) 2305.}\nref\rNadka{ S.
Nadkarni, {\it Phys. Rev.} D27 (1983) 917, {\it ibidem} D38 (1988)
3287.}\nref\rBraaten{ E. Braaten, {\it Phys. Rev. Lett.} 74 (1995)
2164.}\nref\rKajan{ K. Kajantie, K. Rummukainen and M.E.
Shaposhnikov, {\it Nucl. Phys.} B407 (1993) 356.} \nref\rCuSaYe{A.
Chubukov, S. Sachdev and J. Ye, {\it Phys. Rev.} B49 (1994)
11919.} \nref\rNTGN{Numerical simulations concerning the NJL $2+1$
model with $U(1)$ chiral symmetry are reported in \rf
 S.J. Hands, J.B. Kogut, C.G. Strouthos,
   {\it Phys. Lett.} B515 (2001) 407, hep-lat/0107004.
}

\nref\rDaMaRa{ The thermodynamics of the Gross--Neveu and and
Nambu--Jona-Lasinio models at all temperatures and densities at
$d=2$ for $N\to\infty$ is discussed and the existence of
instantons responsible of the symmetry restoration is demonstrated
in \rf R.F. Dashen, S.K. Ma and R. Rajaraman, {\it Phys. Rev.} D11
(1975) 1499.} \nref\rBCMPG{More recently a more complete analysis
has appeared in A. Barducci, R. Casalbuoni, M. Modugno and G.
Pettini, R. Gatto, {\it Phys. Rev.} D51 (1995) 3042.}

 \nref\rSharev{For a review see, for example,\rf
 M.E. Shaposhnikov, {\it Proc. Int. School of Subnuclear Phys.} (Erice
 1996), World Scientific, hep-ph/9610247.}
\nref\rTgauge{Finite temperature calculations in gauge theories are reported in \rf
 P. Arnold and C. Zhai, {\it Phys. Rev.} D51 (1995) 1906 and
 refs.~\refs{\rFodor,\rArnold}.}\nref\rFodor{
Z. Fodor and A. Hebecker, {\it Nucl. Phys.} B432 (1994) 127.}
\nref\rArnold{ P. Arnold and O. Espinosa, {\it Phys. Rev.} D47
(1993) 3546.} \nref\rTSchwinger{The Schwinger model is solved in
\rf
 A.V. Smilga, {\it Phys. Lett.} B278 (1992) 371.}
\section Finite temperature field theory in the large $N$ limit

In this section, we first recall a few general properties of Statistical Quantum Field Theory (QFT) at equilibrium, that is QFT at finite temperature \refs{\rKiLin{--}\rTreview}. We argue that when the temperature varies, one generally observes a crossover from a zero temperature $d$ dimensional theory to an effective $d-1$ dimensional field theory at high temperature, a phenomenon called dimensional reduction. This corresponds also to a classical limit with a transition from a statistical quantum field theory to a statistical classical field theory. Note that high temperature here refers to an ultra-relativistic limit where the temperature, in energy unit, is much larger than the physical masses of particles, so that bosons can produce a classical field.  In perturbation theory it is often impossible to describe this crossover and therefore large $N$ techniques may again be useful.
 \sslbl\scFTQFT \par
We illustrate these ideas with several standard examples, $\phi^4$ field theory, the non-linear $\sigma $ model, the Gross--Neveu model and gauge theories.
\subsection Finite temperature QFT: general remarks

We first recall some general properties of QFT at thermal equilibrium in $(1,d-1)$ dimensions. We introduce the mode expansion of fields in the euclidean time variable, discuss the conditions under which statistical properties of finite temperature QFT in $(1,d-1)$ dimension can be described by an effective local classical statistical field theory in $d-1$ dimensions, and indicate how to construct it explicitly.
\smallskip
{\it Partition function.} Equilibrium properties of QFT
at finite temperature can be derived from the partition function ${\cal Z}=\tr\e^{-H/T}$, where $H$ is the quantum hamiltonian   and $T$ the
temperature (interesting physics related to the addition of
a chemical potential will not be discussed here).
 We will consider,
first,  a  euclidean action ${\cal S}(\phi)$ describing a scalar
boson field $\phi$. The partition function is given by the
functional integral
$${\cal Z}=\int[\d\phi]\exp\left[-{\cal S}(\phi)\right],\eqnd\eFTZp$$
where  ${\cal S}(\phi)$ is the integral of the euclidean
lagrangian density ${\cal L}(\phi)$:
$${\cal S}(\phi)=\int_0^{1/T}\d\tau\int\d^{d-1} x\,{\cal L}(\phi), $$
and the field $\phi$ satisfies {\it periodic boundary conditions}  in
the (euclidean or imaginary) time direction:
$$\phi(\tau=0,x)=\phi(\tau=1/T,x).$$
A QFT may also involve fermions. Fermion fields
$\psi(\tau,x)$, by contrast, satisfy anti-periodic boundary conditions:
$$ \psi(\tau=0,x)=-\psi(\tau=1/T,x).$$
\smallskip
{\it Mode expansion.} As a consequence of the finite temperature boundary conditions, fields have a
Fourier series expansion in the euclidean time direction with quantized
frequencies $\omega_n$ (also called Matsubara frequencies). For boson fields
$$\phib(t,x)=\sum_{n\in {\Bbb Z}}\e^{i\omega_n t}\phib_n(x) ,
\quad \omega_n=2n\pi T\,.
\eqnd\emodexp $$
In the case of fermions, anti-periodic boundary conditions lead to the expansion
$$\psib(t,x)=\sum_{\omega_n=(2n+1)\pi T}\e^{i\omega_n t}\psib_n
(x).\eqnd\efmodexp $$
\smallskip
{\it Remark.} The mode expansion \eqns{\emodexp} is well suited to simple situations where fields belong to a linear space. In the case of
non-linear $\sigma $ models or non-abelian gauge theories, the decomposition in modes leads to problems, because it conflicts with the geometric structure.
\medskip
{\it Classical statistical field theory and thermal mass.}  The quantum partition function \eFTZp\ can also be considered as the partition
function of a classical statistical field theory in $d$ dimensions. In this interpretation finite temperature for the quantum partition function  \eFTZp\ corresponds to a finite size $ \beta =1/T$ in one direction for
the classical partition function. The zero
temperature limit of the quantum theory corresponds to an infinite volume limit of the classical theory.
  \par
An important parameter then is the {\it thermal mass}\/ $m_T$, inverse of the correlation length $\xi_T $, which characterizes the decay of correlations in space directions. A cross-over is expected between a $d $-dimensional
behaviour when the correlation length $\xi_T$ is small compared with $
\beta $,  that is the mass scale $m_T$ is large compared with the
temperature $T$, to the $(d-1)$-dimensional behaviour when $m_T$ is small compared with $T$. Moreover, in this limit macroscopic properties and correlations for  momenta much
smaller than the temperature $T$ or distances much larger than $ \beta
$, can be described by an effective $(d-1)$-dimensional local field theory.
As we shall see, this corresponds to a classical limit, in the sense that classical fields replace quantum fields.
Within this framework, the temperature then plays the role of a large momentum cut-off.
 \par
The ratio $m_T/T$ can be expected to be small in several situations, at high temperature and near a finite temperature phase transition. Moreover, it is also small at low temperature in a third peculiar situation, when a symmetry is broken at zero temperature and no phase transition is possible at finite temperature.
\smallskip
{\it Finite temperature renormalization group.} From the classical
statistical interpretation one learns that general results
obtained in the study of finite size effects also  apply here.
Correlation functions satisfy the renormalization group (RG)
equations of the corresponding $d $-dimensional classical theory
\refs{\rTRGeq,\rZJTN}. Indeed, RG equations are  related to short
distance singularities  and are, therefore, insensitive to finite
sizes. A finite size  affects only  solutions of the RG equations,
because a new dimensionless  RG invariant parameter becomes
available, here for instance the ratio $ m_T/T$.  \par At high
temperature, the ratio $m/T$, where $m$ is the physical mass
scale, goes to zero. This does not automatically imply that
$m_T/T$ is small because  the parameter $m_T$ is identical with
$m$  only in the tree approximation. By solving the RG equations
in terms of the effective coupling at the temperature scale $T$,
one infers that if the effective coupling goes to zero at high
temperature, then the ratio $m_T/T$ really becomes small. Two
examples will be met: the first  corresponds to theories where the
free field theory is an IR fixed point, like $\phi^4_4$ scalar
field theory or QED${}_4$, the second corresponds to UV
asymptotically free field theories. Conversely, when a non-trivial
IR fixed point is present the ratio $m_T/T$ goes to a constant. At
high temperature, one then has  to rearrange the initial
perturbation theory  by adding and subtracting a mass term to
suppress fictitious perturbative large IR contributions.
\smallskip
{\it Zero mode and large distance behaviour.}
We consider first the example of a free scalar field
theory with the action
$${\cal S}(\phi)=\ud\int_0^{1/T}\d t\int\d^{d-1} x\left[ ( \partial _t \phi)^2+ (\nabla _x\phi )^2+m^2\phi^2\right]. $$
After introducing the mode expansion \emodexp\ into the action and integrating over time, one obtains a $(d-1)$-dimensional euclidean field theory with an infinite number of fields, the modes $\phi_n(x)$. The form of the action,
$${\cal S}(\phi)={1\over 2T}\int\d^{d-1} x\sum_n\left[|\nabla _x\phi_n (x)|^2+
(m^2+4\pi^2n^2 T^2)|\phi_n(x) |^2\right],$$ shows that $\phi_n$
has a mass $\sqrt{m^2+4\pi^2n^2T^2}$. Therefore, at high
temperature all modes become very massive except the zero-mode,
whose mass $m$ governs the decay of $\phi(t,x)$-field correlation
functions in space directions (here $m_T=m$). The large distance,
low momentum, physics can entirely described by an effective
$(d-1)$-dimensional field theory involving only the zero mode.
\par Note, however, that all
fermion modes  become very massive at high  temperature due to the
anti-periodic boundary conditions \efmodexp and generally
decouple.
\par In an interacting theory, if
$m_T$ remains much smaller than the temperature, one expects still
to be able to describe low momentum physics in terms of an
effective $(d-1)$-dimensional euclidean local field theory. The
thermal mass $m_T$ is the mass of the zero-mode in the effective
theory, and all other scalar and fermion modes have masses at
least of order $T$. The effective theory can thus be constructed
by integrating out perturbatively all non-zero modes and
performing a local expansion of the resulting effective action. As
mentioned above, in this framework the temperature $T$ acts as a
large momentum cut-off.
\medskip
{\it Dimensional reduction.}
We now briefly outline the construction of the $(d-1)$-dimensional effective theory in the example of a general scalar QFT, assuming that the mass $m_T$ of the zero-mode remains indeed much smaller than $T$ \refs{\rTdimred,\rSharev}.  We first separate the zero-mode $\varphi(x)$, setting
$$\phi( t,x)=\varphi(x)+\chi(t,x), \eqnd\eFTmodfc $$
where  $\chi$ the
sum of all other modes (Eq.~\emodexp):
$$\chi(t,x)=\sum_{n\ne 0}\e^{i\omega_n t}\phi_n
(x), \quad \omega_n=2n\pi T\,. \eqnn $$
The action ${\cal S}_T(\varphi) $ of the effective theory is then obtained by integrating over $\chi$:
$$\e^{-{\cal S}_T(\varphi)}=\int[\d\chi]\exp[-{\cal S}
(\varphi +\chi)]. \eqnd\eFTefact $$
\smallskip
{\it Leading order.} From the point of view of the $\chi$
integration, the tree approximation corresponds to setting
$\chi=0$ and one simply finds
$${\cal S}_T(\varphi)={1\over T} \int\d^{d-1} x\,{\cal L}(\varphi),\eqnn $$
an action that is obviously local.  \par We note that $T$ plays,
in this leading approximation, the formal role of $ \hbar$, and
the small $T$ expansion corresponds to a loop expansion.
 If the ratio $T/ \Lambda$, which is always assumed to be small,
 is the relevant expansion parameter, which means that the perturbative expansion
 is dominated by large momentum (UV) contributions, then the effective $(d-1)$
dimensional theory can still be studied by perturbative methods. This is
expected when the number $d-1$ of space dimensions is large and field
theories are non-renormalizable. However, another dimensionless
ratio can be found, $m/T$, which at high temperature
is small. This may be the relevant expansion parameter for theories that
are dominated by small momentum (IR) contributions,  a problem that
arises in low dimensions. Then, perturbation theory
is no longer possible or useful.   Actually, the relevant
parameter in the full effective theory  is $m_T/T$.
Therefore, the contributions to the mass of the zero-mode due to quantum
and thermal fluctuations need to be investigated.
\smallskip
{\it Loop corrections.} The integration over non-zero modes
generates non-local interactions.  To study long wave length
phenomena, one can, however, perform  a {\it local expansion}\/ of
the effective action, expansion that breaks down at momenta of
order $T$.  In general, higher order corrections coming from the
integration over non-zero modes generate terms that renormalize
the terms already present at leading order, and additional
interactions are suppressed by powers of $1/T$.  Exceptions are
provided by gauge theories where new low dimensional interactions
are generated as a consequence of the explicit breaking of the
$O(1,d-1)$ symmetry.
\smallskip
{\it Renormalization in the effective theory.} If the initial $(1,d-1)$ dimensional theory has been
renormalized at $T=0$, the complete theory is finite in the formal infinite cut-off
limit because finite size effects do not affect divergences. However, as a consequence of the zero-mode subtraction, cut-off
dependent terms may appear in the reduced $(d-1)$-dimensional action. These terms provide the necessary counter-terms that render the perturbative expansion of the effective field theory finite.
\def\Pib{ \Pi}
\subsection Scalar quantum field theory at finite temperature for $N$ large

General finite size effects and, therefore, as we have already
discussed, finite temperature physics involve crossover phenomena
between different dimensions when the temperature varies from zero
to infinity. This severely limits the applicability of
perturbation theory (even with RG improvement), in particular
because IR divergences are more severe in lower dimensions. Large
$N$ techniques, however, which are rather insensitive to changes
in the number of dimensions are, therefore, particularly well
suited to study a crossover situation \rZJTN.\par We first
consider again the self-interacting scalar field with $O(N)$
symmetric action, studied at zero temperature in section
\ssNbosgen. The field $\phib$ is an $N$-component vector and the
hamiltonian  \sslbl\ssNTphig
$$ {\cal H}(\Pib, \phib)={1\over2} \int \d ^{d-1} x \, \Pib^2(x)+\Sigma
(\phib) \eqnn $$
with
$$\Sigma (\phib)= \int \d ^{d-1} x \left\lbrace\ud
\left[ \nabla \phib (x) \right]^{2}+ NU\bigl(\phib^2(x) /N\bigr)
\right\rbrace . \eqnd\eactred $$ A cut-off $\Lambda$, as usual, is
implied that renders the QFT UV finite. The  quantum partition
function \eFTZp\ at finite temperature $T$,  $$ {\cal Z}= \int
\left[ \d \phib (x) \right] \exp \left[-{\cal S}(\phib)\right] ,
\eqnn $$ can then be expressed in terms of the action
$${\cal S}(\phib)=\int_0^{1/T} \d t\left[\int\d^{d-1} x\ud(\d_t\phib)^2
+\Sigma (\phib)\right], \eqnn $$
where the field $\phib$ satisfies periodic boundary conditions in the euclidean time direction.
 \par
Following Eq.~\eactONef, the action density at finite temperature
and large $N$ is
$${1\over N}{\cal F}(\rho,\sigma,m_T)=U(\rho)+\ud m_T^2 (\sigma^2-N\rho)+
{1\over 2\beta V_{d-1} }\tr\ln \left[ - \nabla^2  + m_T^2 \right]
,\eqnd\EBfiniteT
$$
where $V_{d-1}$ is the $d-1$ volume. The $\tr\ln$ contribution is now modified
by the boundary conditions:
$$ \eqalignno{{1\over V_{d-1}}\tr\ln(-\nabla^2 + m_T^2) &=
\int {\d^{d-1}k\over {(2\pi)^{d-1}} }\sum_ {n\in {\Bbb Z}}
 \ln(\omega_n^2+ k^2+m_T^2) \cr &=2  \int
{\d^{d-1}k\over {(2\pi)^{d-1}} }
\ln\left[2\sinh\bigl({\beta\omega(k)/2}\bigr)\right]&\eqnd\TRboson}$$
with $\omega(k)= \sqrt{k^2 +m^2_T}$. The result in Eq.~\TRboson\
is given up to a mass independent infinite constant. We have used
$\omega_n=2n{\pi/\beta}$ and the identity \eqns{\eFTgenidii}. \par
Differentiating  ${\cal F}$ in Eq.~\EBfiniteT \ with respect to
$\sigma $, $\rho $ and $m_T$, one obtains the saddle point
equations. The algebraic transformations which produce the large
$N$ effective action are clearly insensitive to boundary
conditions  and, therefore, only the specific form of the saddle
point equation in Eq.~\esaddleN{c} is modified. Eqs.~\esaddleN{}
become
$$m_T^2\sigma =0\,, \quad m_T^2=2U'(\rho)$$
 and
$$\rho-\sigma ^2/N =G_2(m_T,T ) \eqnd\esaddleNTci $$
 with
$$\eqalignno{G_2(m_T, T)&  ={T\over (2\pi)^{d-1}  }\sum_{n\in{\Bbb Z}}\int^\Lambda {\d^{d-1} k\over (2\pi n T )^2+k^2+m_T^2},\cr
 &=\int^\Lambda{\d^{d-1} k\over (2\pi)^{d-1}}{1\over \omega (k)}
 \left({1\over2} +{1\over \e^{\beta  \omega (k)}-1}\right). &\eqnd\eNfivTtp
\cr}$$
The quantum ($T=0 $) and thermal ($ T>0 $ finite) fluctuations are
clearly separated when the two terms in Eq.~\eNfivTtp\ are written
as
$$ \eqalignno{ G_2 & =[G_2]_{\rm quantum}  +
 [G_2]_{\rm thermal} \cr
&= {1\over (2\pi)^d}  \int^\Lambda  {\d^ d k\over
  k^2 + m_T^2} +{1\over  (2\pi)^{d-1}} \int {\d^{d-1} k\over \omega (k)} {1\over \e^{\beta  \omega (k)}-1}  . &\eqnd\eQuantTerm  \cr} $$
In particular, as expected, only the zero temperature contribution is cut-off dependent. For $T\to\infty $, the second contribution dominates
and at leading order is identical to the zero-mode contribution.  \par
Note that, alternatively, we could have used the finite size formalism
of section \ssNFSS\ and the corresponding Jacobi function \eJacobi.
\smallskip
{\it Symmetry breaking.} In the broken symmetry phase $\sigma \ne 0$ and thus $m_T=0$. Then,  $G_2$ can
be calculated explicitly:
$$G_2(0,T)=\rho _c+N_{d-1}\Gamma (d-2)\zeta (d-2) T ^{ d-2}, \eqnd\eNTGiilow $$
where $N_d$ is the  loop factor \etadpolexp{b} and $\zeta (s)$ Riemann's
$\zeta$-function. The function $\zeta (s)$ has a pole for $s=1$ and therefore, as expected, a phase transition for $T>0$ is possible only for $d>3$ when the result is IR finite.
The result has a simple
interpretation: at $d=3$, IR divergences come from  the contribution of the
zero-mode  in Eq.~\eNfivTtp\ and are those of a two-dimensional theory, where no
phase transition is possible. This is a direct example of the property of {\it dimensional reduction}\/ $d \mapsto d-1$.  \par
The parameter $\rho$ is the minimum of $U$ and thus takes its $T=0$ value. Finally, the expectation value $\sigma $ is given by
$$\sigma ^2/N=\rho(T=0) -\rho _c-N_{d-1}\Gamma (d-2)\zeta (d-2) T ^{ d-2}.$$
The field expectation value $\sigma $ decreases with the temperature, which
implies that the symmetry is broken at finite temperature only if it is already broken at zero temperature.
The critical temperature $T_c$ can be expressed, for $N\to\infty $,
in terms of the zero temperature expectation value:
$$T_c\propto \left[\sigma (T=0)\right]^{2/(d-2)}. \eqnn $$
A RG analysis actually shows that, in general,  $T_c$ can be related to the crossover scale between Goldstone and critical behaviour (section \label{\ssNTfiv}).
\smallskip
{\it The symmetric phase.} One now finds
$$\rho -\rho _c=\Omega_d(m_T)-\Omega_d(0)+T^{d-2}f_d(m_T^2/T^2 ),\eqnd\eNTphigsym $$
where $\Omega _d$ is defined by Eq.~\etadepole\ and
$$\eqalignno{f_d(z)&=N_{d-1}\int_0^\infty {x^{d-2}\d x \over \sqrt{x^2+z}}{1\over
\exp[\sqrt{x^2+z}]-1} \cr
&=N_{d-1}\int_{\sqrt{z}}^\infty(y^2-z)^{(d-3)/2}{\d y\over \e^y-1}\,
.&\eqnd\eThermfz\cr} $$
In particular,
$$f_d(0)=N_{d-1}\Gamma (d-2)\zeta (d-2),\quad
f'_d(0)=-\ud(d-3)N_{d-1}\Gamma (d-4) \zeta (d-4)
.\eqnd\eTfzzero $$
The other equation is $m_T^2=U'(\rho)$.
In the special limit of a critical potential at $T=0$ (the massless theory), which satisfies $U'(\rho_c)=0$, the equation  can be expanded as
$$m_T^2\sim (\rho -\rho _c)U''(\rho _c)$$
if $U''(\rho _c)$ does not vanish. Then,
$$ m_T^2/U''(\rho_c)\sim \Omega_d(m_T)-\Omega_d(0)+T ^{ d-2}f_d(m_T^2/T^2 ).\eqnd\eNTUgenmTcrit $$
For $d>4$, this equation implies
$$m_T/T\propto (T/\Lambda )^{(d-4)/2}\ll 1 \eqnd \eNTmTfivv$$
and, thus, dimensional reduction is justified. \par
For $d=4$, the conclusion is the same because
 $$\left(m_T \over T\right)^2\sim {2 \pi^2\over3 \ln (\Lambda /T)}\ll 1\,. \eqnd\eNTmTfiviv $$
Instead, for $d<4$, the l.h.s.\ is negligible and, therefore, $m_T$ is proportional to $T $.
These results have a simple interpretation from the RG point of view in the framework of the $(\phib^2)^2$ field theory.
\subsection The $(\phib^2)^2$ field theory at finite temperature

We now specialize to the $(\phib^2)^2$ field theory,
that is (section \ssfivNi) \sslbl\ssNTfiv
$$U(\rho )=\ud r \rho +{ u\over 4!}\rho^2\,.$$
We first summarize what can be learned from a simple RG analysis,
and then solve the large $N$ saddle point equations in this case. \par
We define the quantity $r_c(u)$, which has the form of a mass renormalization, as the value of $r$ at which the physical
mass $m$ of the field $\phib$
vanishes at $T=0$. At $T=0$, $r=r_c$, a  transition occurs between a symmetric phase ($r>r_c$) and
a broken symmetry phase ($r<r_c$). We recall that a QFT is meaningful only if
the physical mass $m$ is much smaller than the cut-off $\Lambda$. This implies
either (the famous {\it fine tuning}\/ problem) $|r-r_c| \ll \Lambda^2  $
or, for $N> 1$,  $r<r_c$ which corresponds to a spontaneously broken symmetry with massless Goldstone modes. The latter situation will be examined in
section \label{\ssFTnls} within the more suitable formalism of the non-linear
$\sigma $-model.  \par
It is also convenient to introduce a dimensionless coupling
$$ \lambda = u \Lambda^{d-4}/N  \,, \eqnd\eFTuLg $$
where later $N\lambda $ will be assumed to take generic (i.e.\ not very small) values. \par
\medskip
{\it RG at finite temperature.}
As we have already stressed, some useful information can be obtained from a RG analysis, which also explains the nature of some of the results
obtained in the large $N$ limit. \par
Correlation functions at finite temperature satisfy the RG equations
of the zero temperature QFT or the $d $-dimensional classical
field theory in infinite volume. The dimension $d=4$ is special, because then
the $\phi^4_4$ theory is just renormalizable.  One important quantity is the ratio $ m_T/T$,
where  $m_T$ governs the decay of correlations in space directions and is also, after dimensional reduction, the mass of the zero-mode in the effective theory.
\smallskip
{\it Higher dimensions}. For $d>4$, the theory is non-renormalizable, which
means that the gaussian fixed point $u=0$ is stable. The coupling constant
$u=N\lambda \Lambda^{4-d}$ is small in the physical domain, and perturbation theory is applicable. At zero temperature, the physical mass in the symmetric phase has the
scaling behaviour of a free or gaussian theory, $m\propto (r-r_c)^{1/2}$.
The leading corrections to the two-point function due to finite temperature
effects, are of order $u$. Therefore, in the symmetric phase, for dimensional
reasons,
$$m_T \propto ( r-r_c+{\rm const.}\ \lambda\Lambda^{4-d} T^{d-2})^{1/2}.$$
If at zero temperature the symmetry is broken ($r<r_c$),   a phase transition thus occurs
at a temperature $T_c$, which scales like
 $$T_c \propto \Lambda \left[(r_c-r)/\Lambda^2 \right]^{1/(d-2)} \gg (r_c-r)^{1/2} \,.$$
The critical temperature is large compared with the $T=0$ crossover mass scale \eNmcrossc.
 \par
At high temperature or in the massless theory ($r=r_c$), the thermal
mass behaves like
$$m_T  /T\propto ( T/\Lambda )^{(d-4)/2}\ll 1\,,$$
a behaviour consistent with the form \eNTmTfivv.
The property  $  m_T/T \ll 1 $ implies the validity of dimensional reduction.
\smallskip
{\it Dimension $d=4$.} The $(\phib^2)^2$ theory is just renormalizable and RG equations take the form
$$ \left[ \Lambda{  \partial \over  \partial \Lambda}
+\beta(\lambda){ \partial \over
 \partial \lambda}-{n \over 2}\eta
(\lambda)-\eta_{2}(\lambda)(r-r_c){ \partial \over  \partial r}
\right] \Gamma^{(n)}  (p_{i};r,\lambda,T,\Lambda
 )=0\,. \eqnd\eganTRG $$
The ratio $m_T/T=F( \Lambda /T, \lambda,r/T^2 )$  is dimensionless and
RG invariant, and thus  satisfies
$$ \left[ \Lambda{  \partial \over  \partial \Lambda}
+\beta(\lambda){ \partial \over
 \partial \lambda}-\eta_{2}(\lambda)(r-r_c){ \partial \over  \partial r}
\right] F=0\,.$$
The solution can be written as
$$  m_T/T=F(\Lambda /T, \lambda,  (r-r_c)/T^2)=F\bigl(\ell
\Lambda/T,\lambda(\ell ), ( r(\ell)-r_c)/T^2\bigr),\eqnd\eFTRGmass $$
where $\ell $ is a scale parameter, and $\lambda(\ell ), r(\ell )$ the
corresponding running parameters (or effective parameters at scale $\ell $):
$$\ell {\d \lambda(\ell )\over \d\ell }=\beta\bigl(\lambda(\ell )\bigr), \quad
\ell {\d r(\ell )\over \d\ell }=-\bigl(r(\ell )-r_c\bigr)\eta_2\bigl(\lambda(\ell
)\bigr).$$
The form of the RG $\beta$-function,
$$\beta(\lambda)={(N+8) \over 48\pi^2}\lambda^2+O\left(\lambda^{3}
\right), \eqnn $$
implies that the theory is IR free,  that is $\lambda(\ell )\to 0$ for $\ell
\to 0$. The effective coupling constant at the physical scale is
logarithmically small, implying logarithmic
deviations from naive scaling. To describe physics at the  scale
$T $, one has to
choose $\ell =T/\Lambda  \ll1$  and, thus,
$$\lambda(T/\Lambda  )\sim {48 \pi^2 \over (N+8)\ln(\Lambda
/T)}. \eqnd\eFTgrun $$
Therefore, RG improved perturbation theory can be used to derive the effective action of the reduced theory. In particular, for $r=r_c$ one recovers the behaviour \eNTmTfiviv:
$$\left(m_T\over T\right)^2\sim{N+2\over72} \lambda(T/\Lambda  )\sim {2\pi^2(N+2)\over 3(N+8)}{1\over\ln(\Lambda
/T)}.$$
\smallskip
{\it Dimension  $d=3$.}
The three-dimensional classical theory has an IR fixed point
$\lambda^*$. Then finite size scaling (Eq.~\eFTRGmass) predicts,
in the symmetric phase,
$$ m_T/T= f\bigl((r-r_c)/T^{1/\nu}\bigr),$$
where $\nu$ is the correlation exponent of the three-dimensional
system. Therefore, $m_T$ in general  remains of order $T$ at high
temperature.  \par The zero-mode plays a special role only if
there exist values such that the function $f$ is small (relative
to 1). This would happen near a phase transition, but in an
effective two-dimensional theory a phase transition is impossible
for $N>2$.
%
\medskip
{\it The large $N$ limit.} We now specialize the results of section
\ssNTphig, and verify consistency with the general RG analysis.
\smallskip
{\it Broken symmetry phase.} In the broken symmetry phase $m_T=0$,   $G_2$ is given by Eq.~\eNTGiilow:
$$G_2(0,T)=\rho _c+N_{d-1}\Gamma (d-2)\zeta (d-2) T ^{ d-2}. $$
We have already noted that a phase transition for $T>0$ is possible only for $d>3$ when the result is IR finite. \par
Then, using the last equation $U'(\rho )=0$ in the example of the
$(\phib^2)^2$  theory, one obtains
$$\sigma ^2/N= {6\over Nu}(r_c-r)- N_{d-1}\Gamma (d-2)\zeta (d-2)
T^{d-2} $$
and finds a critical temperature
$$T _c\propto \Lambda \bigl((r_c-r)/\Lambda ^2)^{1/(d-2)}.$$
The value of $T_c$ can be compared with $m_{\rm cr}$, the mass
scale at which a crossover between critical and Goldstone behaviours occurs. \par
For $d>4$,
$$T_c\propto \Lambda (m_{\rm cr}/\Lambda )^{2/(d-2)}\gg m_{\rm cr} $$
 (from Eq.~\eNmcrossc) and, thus, the temperature $T_c$ is large
 compared to the relevant mass scale.
\par For $d=4$, the explicit value is
$$ T _c=(72/Nu)^{1/2}(r_c-r)^{1/2} \propto m_{\rm cr}\sqrt{\ln(\Lambda /m_{\rm cr})},$$
where Eq.~\eNmcrossb\ has been used. In all cases the critical
temperature is large compared to the crossover mass.
\smallskip
{\it The symmetric phase.} The saddle point equations become
$$\eqalign{\rho -\rho _c&=\Omega_d(m_T)-\Omega_d(0)+T^{d-2}f_d(m_T^2/T^2 ), \cr
m_T^2&=(Nu/6)(\rho-\rho_c)+r-r_c\,. \cr}
$$
In the special limit $r=r_c$, which corresponds to the critical point
at zero temperature, Eq.~\eNTUgenmTcrit\ now reads
$$ (6/Nu)m_T^2=\Omega_d(m_T)-\Omega_d(0)+T ^{ d-2}f_d(m_T^2/T^2 ).$$
The behaviour of $m_T/T$ has already been discussed in section \ssNTphig.
The special dimension $d=4$ will be examined in more detail in  section \ssNvarfivT\ in the context of variational calculations, and the general large $N$ limit again in section \label{\ssFTnls}  from the point of view of the non-linear $\sigma $ model.

\subsection{Variational calculations in the $(\phib^2)^2 $ theory
}

To further compare variational calculations with large $N$ results, we
generalize the zero-temperature calculations of section \ssNVarfiv\ to a finite temperature system. The variational
principle is still based on the inequality \evarineq, but is
applied to a finite temperature system $T=1/\beta  $. For a system
of $N$ non-interacting massive bosons at temperature $T $ in $d-1$
space dimensions, the Helmholtz free energy density
 is given by \sslbl\ssNvarfivT
$$\eqalignno{ {\cal F}_0 =- {1\over V\beta  } \ln  {\cal Z}_0 &
=- {1\over V\beta } \ln \tr \e^{-\beta  H_0} \cr
& =N
  {1\over {(2\pi)^{d-1}} }\int \d^ {d-1}k \left( \half\omega(k) +
T\ln \left(1-\e^{-\beta \omega(k)}\right) \right).&
\eqnd\freeEzero \cr}$$
The inequality \evarTzero\ is replaced by
$$ {\cal F}\le  {\cal F}_{\rm var.}=  {\cal F}_0
+N\left<U\bigl(\phib^2(x)/N\bigr) -\ud m_T^2\bigl(
\phib(x)-\phib_0\bigr)^2/N\right>_0\,,\eqnn $$ where
$\left<\bullet\right>_0$ means average with respect to $\e^{-{\cal
S}_0}$ in a finite temperature, infinite space volume system, and
${\cal S}_0$ again is the trial free action:
$${\cal S}_0=\int \d^ d x \,
 \left({1\over2}{( \partial_{\mu} \phib)}^2 +
 {m_T^2 \over2} \bigl(\phib(x)-\phib_0\bigr)^2 ~\right)  , $$
$m_T^2$, $\phib_0^2 =\sigma ^2$ (in the notation of section \ssNTphig) being the two variational parameters. We now
evaluate the gaussian average in the large $N$ limit:
$$\left<U\bigl(\phib^2(x)/N\bigr)\right>_0\sim U\bigl(\left<\phib^2(x)/N\right>_0\bigr)
,$$
and introduce the parameter $\rho$, which is given by
$$\eqalignno{\rho=\left<\phib^2(x)/N\right>_0&=\phib_0^2/N+{T \over (2\pi)^{d-1}}
\sum_{n\in{\Bbb Z}}\int{\d^{d-1} k\over (2\pi n T )^2+k^2+m_T^2},\cr
 &= \sigma ^2/N+ G_2(m_T,T).&
\eqnd\esaddleNTc \cr}$$
Then, the variational free energy density becomes
$${\cal F}_{\rm var}= {\cal F}_0+ NU(\rho)-\ud m_T^2( N\rho- \sigma ^2) .\eqnd\FvarT $$
The relation
$${ \partial {\cal F}_0\over  \partial m_T^2}=\ud( N \rho-\sigma^2),$$
still holds, in such a way that
$$ { \partial {\cal F}_{\rm var.}\over  \partial m_T^2} =N{ \partial \rho \over
 \partial m_T^2}\left[U'(\rho)-\ud m_T^2\right].  \eqnd\eFvarTfiv $$
The condition of stationarity of ${\cal F}_{\rm var.}$ with respect
to variations of $m_T^2$, together with the equation defining $\rho$, are again
equivalent to the large $N$ saddle point equations.
 \medskip
{\it The $(\phib^2)^2$ field theory in dimension $d=4$.}
We now discuss the example of the $(\phib^2)^2$ field theory in four dimensions:
$$U(\rho)=\ud r \rho+{u\over 4!}\rho^2.$$ From Eqs.~\eqns{\eNTphigsym,\etadepoliv},
one finds
$$\eqalignno{ \rho(m_T^2)  &= \rho_c+ \sigma ^2/N-  {m_T^2 \over 8 \pi^2}
\ln(\Lambda /m_T) + { T^2\over 2 }f_4({m_T^2/T^2}),
 &\eqnd\phifourdimension \cr }$$
where $f_d(z)$ is given in Eq.~\eThermfz.
The variational free energy of the system
can be obtained from Eq.~\FvarT\ or by integrating Eq.~\eFvarTfiv:
$$\eqalignno{{1\over N}  {\cal F}_{\rm var}( \sigma ,m_T^2)  &= U(\rho)-\ud
 \int^{m_T^2}_0 \d s \ s{ \partial\rho(s)\over
 \partial s}\cr &= {1\over 32\pi^2} m_T^4 \ln( \Lambda/ m_T ) +
{1\over 4}T^4\int^\infty_{m_T^2/ T^2} \d z\, z  f_4'(z) \cr &+{u\over
24}\left\{  \sigma ^2/N +6 {r-r_c \over u} - {1\over 8\pi^2} m_T^2
\ln(\Lambda/ m_T) + {1\over 2}T^2 f_4({m_T^2/T^2}) \right\}^2 \cr
&-{3\over 2} {r^2 \over u}\,.&\eqnd\WfourDT  }$$
The renormalized theory with finite non-zero renormalized coupling $u_\r$
can be reached only with a negative $u$ in a metastable
state mentioned in section \ssNVarfiv. In the limit $\Lambda \to
\infty$,  $u \to 0^-$ and the renormalized free energy can be
calculated in terms of the renormalized
parameters $u_\r,m_T$ and using the gap equation. One finds
$$ \eqalignno{{1\over N}
{\cal F}_{\rm var}( \sigma ,m_T^2) + {3\over 2} {r^2 \over u} =
{m_T^2\over 2}(\sigma^2/N \pm 6{M^2/u_\r})
&- {r^2\over 4}\left\{ {1\over u_\r} + {1\over
16\pi^2}\ln\left({\e^ {3\over 2}M^2\over m_T^2 }\right)\right\}
\cr &-{1\over 4}T^4\int^\infty_{m_T^2/ T^2} \d z\, f_4(z).
&\eqnd\WfourDRe }
$$  
Here  $M$ is a normalization scale and the choice $\pm$ sign
depends on the sign of $M^2/ u_\r$. The basic instability that
originates from the negatively coupled theory is maintained also
in the renormalized expression in Eq.~\WfourDRe.
One finds that as the
temperature increases, the $O(N)$ symmetric metastable state that
 exists at low temperature becomes even less stable, and above a certain
critical temperature, disappears altogether. The theory  now has only the $O(N)$ broken symmetry phase that
we have found; it is unstable, as mentioned at the bottom of section
\ssNVarfiv, and thus  renders the theory inconsistent.

\subsection The non-linear $\sigma$ model  at finite temperature

We now discuss another, related, example: the non-linear $\sigma$
model because the presence of Goldstone modes introduces some new
features in the analysis. In the perturbative framework, due the non-linear character of the
group representation, one is confronted with difficulties which also
appear in non-abelian gauge theories. Moreover, the non-linear $\sigma $ model
and non-abelian theories share another property: both are UV asymptotically free
in the dimensions in which they are renormalizable.  \sslbl\ssFTnls \par
Finally, we recall that it has been proven in
section \label{\ssLTsN}, within the
framework of the $1/N$ expansion, that the non-linear $\sigma $ model is
equivalent to the $((\phib^2))^2$ field theory (at least for generic
$\phi^4$ coupling), both quantum field theories corresponding to
two different perturbative expansions of the same physical model.
\medskip
{\it The non-linear $\sigma $ model.}  The non-linear $\sigma $ model has been studied
at zero temperature in section \ssLTsN. It is an $O(N)$ symmetric
QFT, with an $N$-component scalar field
${\bf S}(t,x)$ which belongs to a sphere,  that is that satisfies the constraint ${\bf
S}^2(t,x)=1$. \par
The partition function of the non-linear $\sigma$ model can be written as
$${\cal Z}= \int \left[\d{\bf S}(t,x)\d\lambda(t,x)\right]
\exp\left[-{\cal S}({\bf S},\lambda)\right] \eqnn  $$
with
$$\eqalignno{{\cal S}({\bf S},\lambda) &= {1 \over 2g}\int_0^{1/T} \d t\,\d^{d-1}x \left[
\bigl( \partial_t {\bf S}(t,x) \bigr)^2\right.\cr&\quad\left.
+\bigl(\nabla{\bf S}(t,x) \bigr)^2  + \lambda(t,x) \bigl({\bf
S}^2(t,x) -1 \bigr) \right],& \eqnd\eactsigla \cr}$$ where the
$\lambda$ integration runs along the imaginary axis and enforces
the constraint ${\bf S}^2(x)=1$. The parameter $g$ is the coupling
constant of the quantum model as well as the temperature of the
corresponding $d $-dimensional classical theory. \par As we have
already explained,  a finite temperature $T$ leads, in the
corresponding  classical theory, to one finite size $ \beta =1/T$
with periodic boundary conditions.
 \smallskip
{\it Finite temperature saddle point equations.}
The non-linear $\sigma $ model has been discussed in the
large $N$ limit at zero temperature in section \ssLTsN\ with a slightly different notation ($T\to g$).
At finite temperature, the saddle point equation \emgNsig{a} remains
unchanged.  The saddle point equation \emgNsig{b} is modified because the frequencies in the time direction are quantized.  It can be expressed in terms of the function
$$ G_2(m_T, T)  ={T\over (2\pi)^{d-1}  }\sum_{n\in{\Bbb Z}}\int^\Lambda {\d^{d-1} k\over (2\pi n T )^2+k^2+m_T^2}  $$
defined earlier (Eq.~\eNfivTtp). In the symmetric phase $ \left<{\bf S}(T)\right>=0$, one finds
$$1= (N-1)\, g \, G_2(m_T,T)
. \eqnd\esaddNTf $$
Here, $m_T$ is the thermal mass and $\xi_T=m_T^{-1}$ has the meaning of a correlation length in  space
directions.  \par
In terms of the functions  \eqns{\etadepole, \eThermfz},
 Eq.~\esaddNTf\ (for $N$ large)  can be rewritten as
$${1 / Ng}=  \Omega_d(m_T ) +T^{d-2} f_d (m_T^2/T^2 ).  \eqnd\eNTnlssad $$
One recovers that a phase transition is possible only if $f_d(0)$ is finite,
which from Eq.~\eTfzzero\ implies $d>3$, a result that can be seen as a consequence of   dimensional reduction.
\medskip
{\it Dimension $d=2$}. We first examine the case $d=2$.
This corresponds to a situation where even at zero temperature  the
$O(N)$ symmetry remains always unbroken. In
the zero temperature QFT, or the infinite volume classical statistical
system, the continuum limit corresponds to $g\ll 1$ and the physical
mass $m$ then is given by  Eq.~\eNsigmii:
$$ 1/ Ng = \Omega _2(m) \ \Rightarrow \
m =\xi_0^{-1}\propto \Lambda \e^{-2\pi/N g}.$$
By subtracting this equation from Eq.~\eNTnlssad\ (the  finite
temperature gap equation), one finds
$$\ln(m_T/m )=\ln(\xi_0/\xi_T)=2\pi f_2 (m_T^2/T^2 ). \eqnd\eNTnlsgapii $$
High temperature corresponds to $ T/m   \gg 1 $ and thus one also
expects $ m_T \gg m  $. The integral \eThermfz\ then is dominated by
the contribution of the zero-mode and, therefore,
$${T\over m_T}  ={1\over \pi} \ln (m_T/m  )\sim  {1\over
\pi} \ln(T/m  ).\eqnd\eFTNnlsi $$
The logarithmic decrease of the ratio $m_T/T$ at high temperature  corresponds to the UV asymptotic freedom of the classical non-linear $\sigma
$ model in two dimensions.
\medskip
{\it Dimensions $d>2$.} In higher dimensions the system has a phase
transition for $T=0$ at a value $g_c$ of the coupling constant. The gap equation can then be rewritten as
 $${1\over Ng}-{1\over Ng_c}=\Omega_d(m_T )-\Omega_d(0)
+T^{d-2}f_d (m_T^2/ T^2 ).\eqnd\eFTnlsgap  $$
For $g>g_c$, the equation can also be expressed in terms of  the physical mass $m$ (Eq.~\emasNsig) as
$$\left[\Omega_d(m )-\Omega_d(m_T )\right]/T^{d-2}=f_d (m_T^2/
T^2 ).\eqnn $$
The behaviour of the system then depends
on the ratio $T/m $.
To obtain more explicit results, one has to distinguish between various possible dimensions.
\medskip
{\it Dimension $d=3$} \refs{\rGNDiiirev,\rCuSaYe}. The absence of phase transition in two dimensions prevents a phase transition at finite temperature. The
gap equation has a scaling form, as predicted by finite size RG arguments. A short calculation yields
$$f_3(s)=-{1\over
2\pi}\ln\left(1-\e^{-\sqrt{s}}\right) $$
and
$$\Omega_3(m_T)-\Omega_3(0)=-{m_T\over 4\pi}\,.$$
For $g>g_c$, after some simple algebra, the gap equation can be written as
$$2\sinh(m_T/2T)=\e^{m/2T} .$$
One verifies that for $m/T$ large (low temperature) $m_T\to m$, and at high
temperature $T\gg m$, $m_T$ becomes proportional to $T$:
$$m_T /T\sim 2\ln\bigl((1+\sqrt{5})/2\bigr).$$
For $g<g_c$,  that is when the symmetry is broken at zero temperature, one has to return to the general form
$$2\sinh(m_T/2T)=\exp\left[-{2\pi  \over
NT}\left({1\over g}-{1\over g_c}\right) \right]. \eqnn $$
One can also introduce the mass scale 
$$m_{\rm cr}(g)=  {1\over g}-{1\over g_c}  $$
(see Eq.~\eqns{\eNmcrossa}), which at zero temperature characterizes the crossover between
critical and Goldstone mode behaviours. Then,
 $$2\sinh(m_T/2T)=\e^{-2 \pi m_{\rm cr}/NT} .$$
For $g<g_c$, the zero-mode dominates if  the ratio  $m_T/T $  is small and thus if $ m_{\rm cr}/ T $ is large. This condition is  realized  for all temperatures  if $|g-g_c|$ is not small because then $m_{\rm cr}=O(\Lambda )\gg T$:
this is the situation of chiral perturbation theory, and corresponds to the deep  IR (perturbative) region where only Goldstone particles propagate.
It is also realized  in the critical domain $|g-g_c|$
small, if $T\ll m_{\rm cr}$, that is at low (but non-zero) temperature. Then,
$$ m_T \sim T \e^{-2 \pi m_{\rm cr}/NT} =T  \exp\left[-{2\pi  \over
NT}\left({1\over g}-{1\over g_c}\right) \right]. \eqnd\eFTnlsmLii$$
Note that, when the coupling constant $g$  or
the temperature $T$ go to zero, the mass $m_T$ has the exponential behaviour characteristic of the dimension 2.
This property  of dimensional reduction at low temperature
is somewhat surprising. It is in fact a precursor of the symmetry breaking at zero temperature.
\medskip
{\it Higher dimensions: the critical temperature.} For $d>3$, the quantity $f_d(0)$ is finite
and, therefore, a phase transition at finite temperature is possible, in agreement
with dimensional reduction and the property that a phase transition
is possible in dimensions larger than two (in the case of continuous symmetries). From Eq.~\eFTnlsgap\ one infers
$$T_c^{d-2}={1\over Nf_d(0)}  \left({1\over g}-{1\over g_c}\right).\eqnn $$
Since $f_d(0)$ is positive, this result confirms that a transition is possible only for $g<g_c$,  that is if at zero temperature the symmetry is broken. \par
However, this result is meaningful only if $T\ll \Lambda $ and thus only  for $|g-g_c|$ small.  Then, $T_c$ is proportional to the crossover mass scale $m_{\rm cr}(g)$ (Eq.~\eqns{\eNmcrossa}) between critical and Goldstone behaviours.
\smallskip
{\it Dimension $d=4$.}   Since $f_4(0)=1/12$  one finds
$$T_c^2={12\over N}\left({1\over g}-{1\over g_c}\right)={12 \over N} m_{\rm cr}^2.\eqnn $$

  \par
Another limit of interest is the high temperature or massless limit. For $m_T\ne 0$, one finds an additional cut-off dependence:
$$\Omega_4(m_T)-\Omega_4(0)\sim-{m_T^2\over 8\pi^2}\ln(\Lambda/m_T).$$
At $g=g_c$, one finds that $m_T/T $ decreases logarithmically with the cut-off. At leading order, using $f_4(0)=1/12$,  one obtains
$$(m_T/T )^2={2\pi^2\over 3\ln(\Lambda /T )}, $$
in agreement with Eq.~\eNTmTfiviv.
 \smallskip
{\it Dimension $d=5$.} From $f_5 (0)=\zeta(3)/4\pi^2$, one infers the critical temperature $T_c$:
$$T_c^3 \sim{4\pi^2\over N\zeta(3)}\left({1\over g}-{1\over g_c}\right) ={4\pi^2\over N\zeta(3)}m_{\rm cr}^3\,.$$
 \par
In the massless limit  $g=g_c$,
$$  (m_T/T) ^2\sim {\zeta(3)\over 4\pi^2}{T\over D_5(0)} \eqnd\eFTnlsmiv$$
with $D_5(0)\propto \Lambda $, a result consistent with the behaviour \eNTmTfivv.
\subsection The Gross--Neveu model at finite temperature

To gain some intuition about the role of fermions at finite temperature, we now
examine a simple model of self-interacting fermions, the Gross--Neveu (GN \refs{\rDaMaRa, \rBCMPG}.  The GN model is described in terms of a $U(\tilde N)$ symmetric action for a
set of $\tilde N$ massless Dirac fermions $\{\psi^i, \bar\psi^i \}$ (for details see section \ssNGNmod): \sslbl\ssGNNFT
$${\cal S} (\bar \psib, \psib )= -\int \d t\,\d^{d-1} x \left[
\bar\psib(t,x)\cdot \sla{ \partial} \psib(t,x) +\frac{1}{2N} G\left(\bar\psib(t,x)\cdot \psib(t,x) \right)^2
\right],\eqnd\eGNact $$
where $N=\tilde N\tr{\bf 1}$   is the total number of fermion components. \par
The GN model has in all dimensions a discrete symmetry  that prevents the addition of a mass term.
In even dimensions it corresponds to a discrete chiral symmetry, and in odd
dimensions to space reflection. Below, to simplify, we
will speak about chiral symmetry, irrespective of dimensions.  \par
The GN model is renormalizable in $d=2$ dimensions, where it is asymptotically
free and the chiral symmetry is always broken at zero temperature.  \par
It has been proven in section \sssGNYN\ that within the
$1/N$ expansion the GN model is equivalent to the GNY (Y for Yukawa)
model, at least for generic couplings: the GNY model has the same symmetry, but contains an elementary
scalar particle  coupled to fermions through a
Yukawa-like interaction, and is renormalizable in four dimensions. This
equivalence provides a simple interpretation to some of the results that
follow. \par
At finite temperature, due to the anti-periodic boundary conditions, fermions
have no zero modes. Therefore, limited insight about
the physics of the model at high  temperature can be gained from
perturbation theory; all fermions
are simply integrated out. Instead, we study here the GN model within the
framework of the $1/N$ expansion.   \par
Effects due to the addition of a chemical potential will
not be considered here. \par  
The integration over fermions,
leads to an effective non-local action for a periodic scalar field $\sigma $,
which has already been discussed in the zero temperature limit  in section
\sssGNYN:
$$ {\cal S}_N (\sigma )={N\over2 G}\int_0^{1/T} \d t\, \int
\d^{d-1} x \,
\sigma^2(t,x) -\tilde N\tr \ln\left(\sla{ \partial}+\sigma(\cdot) \right), \eqnd\eGNNactb $$
where $T  $ is the  temperature, and the
$\sigma $ field satisfies periodic boundary conditions in the euclidean time variable.
 \par
In the situations in which the thermal mass of the
$\sigma$ field is small compared with the temperature, one can perform a mode
expansion of the $\sigma $-field, integrate over the non-zero modes and obtain
an effective local $(d-1)$-dimensional action for the zero-mode.  It is important
to realize
that for $T>0$, since the reduced action is local and symmetric in $\sigma \mapsto
-\sigma $, it describes the physics of the Ising transition  with  {\it short range
interactions}, unlike what happens at zero temperature where the fermions are massless.
The question that then arises is the possibility of a breaking of  this
remaining reflexion symmetry.  \par
As we have seen,  a non-trivial perturbation theory is obtained by expanding
for large $N$. If a continuous order phase transition occurs at finite temperature,
IR divergences
generated by the $\sigma $ zero-mode will appear in the $1/N$
perturbation theory at the transition temperature $T_c$ for $d-1<4$.
Below $T_c$, in the same way as at zero temperature, the decay of $\sigma
$ correlation functions in  space directions is characterized by the
saddle point value $M_T$ of the field $\sigma (x)$. Above $T_c$, the correlation
length is also finite in contrast with the zero temperature situation. \par
One then finds two regimes, which have to be dealt with differently.
Near the transition temperature, $1/N$ perturbation theory
is not useful for $d<5$. Instead, one has to perform a mode expansion
of the $\sigma $-field and a local expansion of the dimensionally
reduced action for the $\sigma $ zero-mode. The effective field theory
relevant for long distance properties is of $\sigma ^4$ type (as in
the case of the Ising model) with coefficients depending on coupling
constant and temperature, which has to be studied by the usual RG methods.
Note that this implies the absence of phase transition for $d=2$ at finite temperature.
In the other regime where  the $\sigma $ correlation length is of
order $1/T$, all modes are similar and $1/N$ perturbation theory is directly applicable.

\subsection The gap equation

Calling $M_T$ the saddle point value of the field $\sigma (x)$, we
obtain the action density at finite temperature and large
$N$:
$${1\over N}{\cal F}(\rho,M_T) =  {M_T^2 \over 2G} - { T\over V_{d-1}\tr{\bf 1}}\tr \ln\bigl
(\sla{ \partial}+M_T \bigr),\eqnd\eTRfermionM $$
where $V_{d-1}$ is the $d-1$ dimension
volume and
$$ {1\over V_{d-1}}  \tr\ln(\sla{ \partial} + M_T)  =
  {1\over2} \tr{\bf 1}  \int {\d^{d-1}k\over {(2\pi)^{d-1}} }\sum_
{n\in {\Bbb Z}}
 \ln(\omega_n^2+ k^2+M_T^2)   $$
with $\omega_n=(2n+1){\pi T}$. The sum over frequencies  follows from the identity \eqns{\eFTgenidii}, and one obtains
 $$ { 1\over V_{d-1}\tr{\bf 1}} \tr\ln(\sla{ \partial} + M_T) =  \int {\d^{d-1}k\over {(2\pi)^{d-1}} }
\ln\left[2\cosh\bigl( \omega (k)/  2T\bigr)\right]
 \eqnd\eTRfermion $$
with $\omega(k)=\sqrt{k^2+M^2_T}$.
Alternatively, one could use
Schwinger's  representation  of the propagator and another
function of elliptic type
$$ \vartheta_1(s) =\sum_n\e^{-(n+1/2)^2\pi s}.\eqnd\eAellfer $$
The gap equation at finite temperature, obtained by differentiating
${\cal F}$ with respect to $M_T$, again splits
into two equations $M_T =0$ and
$$ 1 /G    = {\cal G}_2(M_T ,T)   \eqnd\eFTGNNsad $$
with
$${\cal G}_2(M_T ,T) =  \int^\Lambda{\d^{d-1} k\over (2\pi)^{d-1}}{1\over \omega(k)
}\left({1\over2} -{1\over \e^{ \omega(k)/T}+1}\right)  .
\eqnd  \eNGNTtp  $$
This is the fermion analogue
of Eqs.~\eqns{\esaddNTf,\eNfivTtp}  in which one recognizes again the sum of quantum and thermal contributions.
Note, however, that the function ${\cal G}_2(M_T ,T)$ has a regular expansion in
$M_T^2$ at $M_T=0$. \par
In terms of the function
$$\eqalignno{g_d(s)&= N_{d-1}\int_0^\infty {x^{d-2}\d x \over
\sqrt{x^2+s}}{1\over \exp[\sqrt{x^2+s}]+1}\cr
&=N_{d-1}\int_{\sqrt{s}}^\infty (y^2-s)^{(d-3)/2}{\d y\over \e^y +1}\,,
 &\eqnd\eThermfiis\cr}
 $$ where $N_d$ is  given in \etadpolexp{b},  the
gap equation can be rewritten as
$$  1 /  G  = \Omega_d(M_T)-T ^{d-2} g_d(  M_T ^2/T^2).  \eqnn  $$
If $d>2$, one can introduce the critical value $G_c$ where $M\equiv M_{T=0}$
vanishes at  zero temperature:
$$  {1 \over G}-{1\over G_c} =\Omega_d(M_T
)-\Omega_d(0)-T ^{d-2} g_d(  M_T ^2/T^2). \eqnd\eTGNNgapii $$
Finally, for $G>G_c$, the equation can be expressed in terms of the fermion physical mass $M\equiv m_\psi $ solution of Eq.~\eNGNmassgap,
$$ 1/G =\Omega_d(M) ,\eqnd\eGNNfms $$
and then reads
$$\Omega_d(M_T)- \Omega_d(M) =T ^{d-2} g_d(  M_T ^2/T^2). \eqnd\eTGNNgapiii $$
 \smallskip
{\it The $\sigma $ two-point function.} Since the correlation length $\xi_\sigma $
of the $\sigma $ zero-mode plays a crucial role, we also calculate the
$\sigma $ two-point function  $\Delta_\sigma (p)\equiv \Delta_\sigma
(p_0 =0,p)$ (see Eq.~\eGNsigprop). Note that in what follows we  use the notation $m_\sigma $ for the corresponding temperature-dependent
thermal mass:  $ m_\sigma(T)=1/\xi_\sigma (T)$, which is also the mass of the $\sigma $ field in the dimensionally reduced theory. \par
 For $M_T=0$, one finds
$${1\over N  \Delta_\sigma(p) } ={1 \over  G} -   {\cal G}_2(0,T)   + {T
\over 2 (2\pi)^{d-1} }  p^2 \sum_n \int^\Lambda{\d^{d-1} k \over \left(
k^2+\omega _n^2 \right) \left[(p+k)^2 +\omega _n^2  \right]} . \eqnn $$
For $d> 2$, the propagator can be expressed in terms of the constant $g_d(0)$:
$$ \eqalignno{{1\over N \Delta_\sigma(p) }&={1 \over  G} -{1\over G_c}+
T^{d-2}g_d(0) \cr& \quad   + {T \over 2 (2\pi)^{d-1} }
 p^2 \sum_n \int^\Lambda{\d^{d-1} k \over \left(
k^2+\omega _n^2 \right) \left[(p+k)^2 +\omega _n^2  \right]} .&\eqnd\eNTDsigsym\cr} $$
For $M_T\ne 0$, using the gap equation, one can write the propagator as
$${1\over N   \Delta _\sigma(p)}= { T\left(p^2+4M_T^2 \right)\over 2 (2\pi)^{d-1} }
\sum_n\int^\Lambda{\d^{d-1} k \over \left(
k^2+\omega _n^2+M_T^2 \right) \left[(p+k)^2+\omega _n^2 +M_T^2 \right]}.  \eqnd\eNTfergsig  $$
Therefore, when the symmetry is broken the mass $m_\sigma(T) =2 M_T$, generalizing the zero temperature result. \par
More detailed properties require distinguishing between dimensions.
\medskip 
{\it Local expansion.} When the $\sigma $ thermal mass or expectation value is small
compared with $T$, one can perform a mode expansion of the field $\sigma $, retaining only the
zero mode and then a local expansion of the action \eGNNactb,
and study it to all orders in the $1/N$ expansion.  In the reduced theory, $T$ plays the role of a large momentum cut-off.  \par
The first terms of the effective $(d-1)$-dimensional action are
$${\cal S}_{d-1} (\sigma )=N\int\d^{d-1} x \left[\ud Z_\sigma
( \partial_\mu \sigma
)^2+\ud r \sigma ^2 +{1\over 4!}u \sigma^4\right], \eqnd\eFTGNloc $$
where terms of order $\sigma^6$ and $ \partial ^2 \sigma ^4 $ and higher have
been neglected. The three parameters are given by
$$Z_\sigma =\ud   {\cal G}_4(0,T)/T\,, \quad r=\left[1/G-   {\cal G}_2(0,T)\right]/T \,,
\quad u=6   {\cal G}_4(0,T)/T\,, $$
where  ${\cal G}_4(m_T,T)$ can be calculated from
$${\cal G}_4(m_T,T)=-{ \partial \over  \partial m_T^2}{\cal G}_2(m_T,T).$$
For $d<4$, ${\cal G}_4(0,T)$ is finite and thus proportional to $T^{d-4} $.
For $d>2$, after the shift of the
coupling constant, one finds
$$rT  ={1\over G }-{1\over G_c  }-T^{ d-2}g_d(0)= g_d(0)\left(T_c^{d-2}-T^{d-2} \right) . \eqnn $$
 \par
As already explained, the properties of this model are those of the
critical $\phi^4$ theory and, for $d<5$, have to be studied by the usual non-perturbative techniques.
\subsection Phase structure for $N$ large

{\it Phase transition and critical temperature.}
One notes that $\Omega_d(M)-\Omega_d(M_T )$ is always
negative and, thus,  since $\Omega _d$ is a decreasing function, $|M|< |M_T|$. Therefore,  when the chiral symmetry is unbroken at zero temperature,
Eq.~\eTGNNgapiii\ has no solution, $M_T=0$ is the minimum  and the $\sigma\mapsto-\sigma$ symmetry is not
broken at $N=\infty$.   \par
We now assume a situation of symmetry breaking at zero temperature,
which implies $G>G_c$. For $d>2$,
$$g_d(0)=N_{d-1}\left(1-2^{3-d}\right)\Gamma (d-2)\zeta (d-2),$$
is finite  and, thus,  the gap equation \eTGNNgapiii\ has a solution up to a temperature $T_c$  where $M_T$ vanishes:
$$T_c=\left[{ 1\over g_d(0)}\left({1\over G_c}-{1\over
G}\right)\right]^{1/(d-2)}=\left[{\Omega_d(0) -\Omega_d(M)\over g_d(0)}\right]^{1/(d-2)}  .\eqnd\eGNNTTc $$
Moreover, the $p=0$ limit of the $\sigma $ propagator  \eNTDsigsym\ confirms that the symmetric phase is stable only if $T>T_c$.
Therefore, $T_c$ separates two
Ising-like phases, a low temperature  phase with symmetry breaking and a symmetric high
temperature phase. Since ${\cal G}_2(M_T,T)$ is a regular function of $M_T $
at $M_T=0$, one finds that, near $T_c$,
$$M_T^2\propto \Lambda ^{4-d}(T_c^{d-2}-T^{d-2})\propto \Lambda (T_c-T), $$
that is a quasi-gaussian or mean-field behaviour in all dimensions. \par
More specific results require considering various $d$ dimensions separately, first $d>2$, the case $d=2$ requiring a special analysis.
\smallskip
{\it High dimensions.} For $d>4$, the critical temperature scales like
$$T_c \propto M(\Lambda /M)^{(d-4)/(d-2)}\ \Rightarrow M\ll T_c \ll  \Lambda \,,$$
and, thus, $T_c$ is physical and large compared with the particle masses.
 \par
In the symmetric high temperature phase $T>T_c$, the $\sigma$ thermal mass ($\sigma $ inverse correlation length) behaves like
$$m_\sigma^2 \propto T^  2 (T/\Lambda )^{d-4}\left[ 1-(T_c/T)^{d-2}\right], $$
and thus is small with respect to $T$, justifying dimensional reduction.
For $T<T_c$ but $T\gg M$, one finds $|M_T^2-M^2|\ll T^2$ and the property remains true. \sslbl\ssNTGNNphase
\smallskip
{\it Dimension $d=4$.} In the high temperature symmetric phase, the thermal mass $m_\sigma $  is given by
$$m_\sigma^2 \propto {1\over \ln(\Lambda /T)}\left({1 \over  G}
-{1\over G_c}+  T^2g_4(0)\right). $$
The thermal  mass $m_\sigma $ is physical (i.e.~$m_\sigma \ll \Lambda $) only  for $|1/G-1/G_c|\ll \Lambda
^2$, that is in the critical domain of the zero temperature theory.
For $G>G_c$, one can introduce the critical temperature (Eq.~\eGNNTTc):
$$m_\sigma^2 \propto {1\over \ln(\Lambda /T)}(T^2-T_c^2).$$
Eventually, $m_\sigma $ vanishes as $(T-T_c)^{1/2}$ a typical mean-field or gaussian behaviour, and a
phase transition occurs.  Eq.~\eGNNTTc\ yields $T_c$ which,  expressed in terms of the
physical fermion mass $M $,  is given by
$$(T_c/M)^2\sim {3\over \pi^2}\ln(\Lambda /M).$$
The critical temperature is thus large compared with the
physical mass $M$.  \par
In the broken symmetry phase, for $T/M$ finite, the mass parameter $M_T$ remains
close to $M$
and one finds
$$\left(M_T\over M\right)^2=1- 8\pi^2  g_4(M^2/T^2)\left(T\over M\right)^2{1\over \ln (\Lambda /M)}\,.$$
\medskip
{\it Dimension $d=3$} \refs{\rGNDiiirev, \rNTGN}. In the symmetric phase
$$m_\sigma ^2\propto   {T \over  G} -{T\over G_c}+  T^2 g_3(0) . $$
The mass parameter $m_\sigma $ is physical only if $T(1/G-1/G_c)\ll \Lambda ^2$. At
the transition coupling constant
$G_c$, one finds as expected $m_\sigma \propto T$. \par
In the broken symmetry phase since at leading order $\Omega_d(M)-\Omega_d(0)=-M/4\pi$ and
$$g_3(s)={1\over 2\pi}\ln\left(1+\e^{-\sqrt{s}}\right),$$
the gap equation can be written as
$$2\cosh(M_T/2T)=\e^{M/2T},$$
an equation that has a scaling form expected for $d<4$ from the correspondence between GN and GNY models, and the existence
of an IR fixed point in the latter.
The critical temperature  is proportional to the fermion mass:
$$T_c/M={1\over 2\ln2}\,.$$
\medskip
{\it Dimension $d=2$} \refs{\rDaMaRa,\rBCMPG}. The situation $d=2$ is doubly special, since at zero temperature chiral
symmetry is always broken and at finite temperature the Ising symmetry is
never broken.  The GN model is renormalizable and UV free. For $N$ large
$$\beta(G)=-{1  \over 2\pi}G^2 . $$
All masses are  proportional to the RG invariant mass scale
$$\Lambda(G) \propto\Lambda \exp\left[-\int^G{\d G'\over \beta (G')}\right] .
$$
In particular, the fermion physical mass   has the form
$$M\propto \Lambda \e^{-2\pi/G}. $$
At finite temperature all masses, in the sense of inverse of the correlation length in the space direction, have a scaling property. For example, the
$\sigma $ mass has the form
$$m_\sigma/T=f(M/T). $$
For $T>M$, one can also express the scaling properties by introducing a temperature
dependent coupling constant $G_T$ defined by
$$\int_G^{G_T}{\d G'\over \beta(G')}=-\ln( \Lambda/T ).$$
At high temperature, $G_T$ decreases like
$$G_T  \sim {2\pi \over \ln( T/M)}.$$
One expects,  therefore, trivial high temperature physics with weakly interacting fermions.  \par
At high temperature  the mass parameter $m_\sigma $ is proportional to $T$
and, therefore, the zero-mode is not different from other modes. Eventually, it decreases
when $T$ approaches $T_c=M/\pi$. At leading order one finds
$$m_\sigma ^2\propto T^2 \ln(\pi T/M)  .\eqnd\eGNNTmii $$
This  result does not imply a phase transition since, for  $m_ \sigma /T\ll1 $, dimensional reduction is justified: the statistical system becomes essentially  one-dimensional  with short-range interactions and thus can have no phase transition. Due to fluctuations the correlation length $1/m_\sigma $ never diverges. \par
For $T<T_c$, the gap equation becomes
$$\ln(M/M_T)=2\pi g_2(M_T^2/T^2).$$
The function $g_2(s)$ is positive, which again implies $M_T<M$, goes to $ \infty$ for $s\to0$
and goes to $0$ for $s\to  \infty $.  At low temperature, $M_T/M$ converges to 1 exponentially in $M/T$.
At high temperature, $M_T/T$ goes to zero and
$$g_2(s)\sim -{1\over 4\pi}\ln s\,.$$
The equation implies $M\propto T$ and thus has no solution for $T\to\infty$, but instead  solutions at finite temperature, in agreement with Eq.~\eGNNTmii, which shows that $m_\sigma $ vanishes for some value $T_c\propto M$. The existence of non-trivial solutions to the gap equation here implies
only that the $\sigma $ potential has degenerate minima, but as a consequence of fluctuations
the expectation value of $\sigma $ nevertheless vanishes. \par
More precisely, the expansion \eFTGNloc\ can be applied to the $d=2$ example. One obtains a simple model in 1D quantum
mechanics: the quartic anharmonic oscillator. Straightforward considerations
show that the correlation length, inverse of the $\sigma $ mass parameter,
becomes small only when the coefficient of $\sigma^2$ is large and negative.
This happens only at low temperature where the two lowest eigenvalues of the
corresponding quantum hamiltonian are almost degenerate. Then, instantons  relate  the two classical minima of the potential and restore the symmetry.  Calculating the difference between the two lowest eigenvalues
of the  hamiltonian, one obtains a behaviour of the form
$$m_\sigma/T \propto (\ln M /T)^{5/4}\e^{-{\rm const.}(\ln M/T)^{3/2}}. $$
Again, the property that $m_\sigma /T$ is small at low temperature
is a precursor of the zero temperature phase transition.

\subsection Abelian gauge theories

The presence of gauge fields introduces some new features in finite temperature field theories, because the $O(d)$ space symmetry is explicitly broken. This affects the
mode decomposition of the gauge field and quantization. Therefore, we  discuss here only QED-like theories, that is abelian gauge fields coupled to fermions with $N$ flavours in $(1,d-1)$ dimensions, because the mode decomposition is gauge invariant  and the quantization simpler  as we now recall, before describing some explicit large $N$ calculations \rTgauge.
\medskip
{\it Mode expansion.} We first describe the effect of a mode
expansion on an abelian gauge field $A_\mu$. We set
$$A_\mu(t,x)=B_\mu(x)+Q_\mu(t,x), $$
where $B_\mu(x)$ is the zero-mode and, thus, $Q_\mu(t,x)$  satisfies
$$\int_0^\beta \d t\,Q_\mu(t,x)=0\,.$$
Gauge transformations then become
$$\delta A_\mu(t,x)= \partial_\mu \left[\varphi_1(t,x)+\varphi_2(t,x) \right] \eqnd\eNTAmugau $$
with
$$\delta Q_\mu(t,x)= \partial_\mu \varphi_1(t,x),\quad
\delta B_\mu(x)= \partial_\mu \varphi_2(t,x).$$
The constraints on $B_\mu$ and $Q_\mu$ then imply
$$\varphi_2(t,x)=F(x)+\Omega t\,,\quad \int_0^\beta \d t\, \varphi_1(t,x)={\rm const. }\
,\quad \varphi_1(0,x)=\varphi_1(\beta ,x), \eqnd\eNTBQgau $$
with $\Omega $ constant. \par
The interpretation of this result is simple, $Q_\mu$ transforms as a gauge field but the
gauge function has no zero-mode, $B_i$ transforms as a $(d-1)$-dimensional
gauge field, $B_0$ behaves like a $(d-1)$-dimensional massless scalar, since a translation of $B_0$ by a constant $\Omega$ leaves the action unchanged.
\medskip
{\it Quantization.} We may quantize by adding to the action a term
quadratic in $ \partial_\mu A_\mu$. Then,
$$  \partial_\mu A_\mu= \partial_t Q_0(t,x) + \partial_i B_i(x)
+ \partial_i Q_i(t,x ),$$
and, therefore,
$$\int_0^\beta \d t( \partial_\mu A_\mu)^2=\beta \bigl( \partial_i B_i(x)\bigr)^2
+\int_0^\beta \d t( \partial_\mu Q_\mu)^2.$$
An integration over the non-zero modes, then leads to a $(d-1)$-dimensional
gauge theory, quantized in the same covariant gauge, coupled to a neutral
massless scalar field.
\smallskip
{\it Massless QED.} If one now considers a massless QED-like theory
$${\cal S} (\bar \psib ,\psib,A _{\mu} ) = \int \d^d  x
\left[{\textstyle{ 1 \over 4e^2}}  F ^{2} _{\mu \nu}(x) -
\bar \psib (x) \cdot\left(\sla{ \partial} + i\Abar \right)
\psib(x) \right] , \eqnd\eactQEDN $$
quantized in the same covariant gauge,
one can integrate out the fermions because, due to
anti-periodic boundary conditions, they have no zero mode.  At one-loop order after fermion integration, the leading local corrections take the same form as in the free case, only coefficients are modified.  \par
Note, however, that one can change the fermion situation by introducing
a  chemical  potential term. \par
Gauge transformations
$$\psib(t,x)=\e^{i \varphi (t,x)}\psib'(t,x), $$
must now preserve the anti-periodicity of the charged fields, and thus be periodic. This implies
$$\varphi( \beta ,x)=\varphi (0,x) \pmod{ 2\pi} .$$
In terms of the decomposition $\varphi=\varphi_1+\varphi_2$, this condition implies that the parameter $\Omega $ in \eNTBQgau\ is quantized:
$$\Omega =2 \pi n T\,,\quad n\in{\Bbb Z}\,.$$
This affects the transformation of the scalar field $B_0$:
$$\delta B_0(x)=2 \pi n T\,$$
and, therefore, the thermodynamic potential is a periodic function of
$B_0$. One consequence of this quantization, then, is that
the field $B_0$, which is massless in the tree approximation, acquires a mass  from quantum corrections, as we verify below by explicit calculations.
\smallskip
{\it Temporal gauge.}
It is also instructive to quantize in the temporal gauge.  To change gauges, we make a gauge transformation with a periodic function
$$ \varphi(t,x)=\int_0^t \d t\, A_0(t',x)-{t\over \beta }\int_0^\beta  \d t\,
A_0(t',x) .$$
Then, $A_0$ is reduced to its zero-mode component, and
the other modes only appear in the gauge fixing function $ \partial_\mu
A_\mu$. Integration over these modes reduces it to $ \partial_i A_i$, where
now only the zero-modes of the other fields $A_i$ contribute.
For what concerns zero-modes, the situation is the same as before,
while the non-zero modes are now quantized in temporal gauge.
This means that one finds $d-1$ families of vector fields with masses $2\pi
n T$, $n\ne 0$, quantized in the unitary, and thus non-renormalizable gauge.
 \par
Let us confirm that the same result is obtained by  quantizing in the temporal gauge $A_0(t,x)=0$ directly:
$$ {\cal S} (\bar \psib ,\psib,A _{\mu}  ) = \int \d^{d-1}x\,\d t
\left[{\textstyle{ 1 \over 4e^2}}\left(2\dot A_i^2+  F ^{2} _{ij}(t,x)
\right)- \bar \psib (t,x) \cdot\left(\sla{ \partial} + i\Abar \right)
\psib(t,x) \right]. \eqnd\eactQEDAze $$
One has to remember that Gauss's law has to be imposed. Therefore,
a projector over gauge invariant states has to be introduced in the functional integral. This can be accomplished by imposing periodic, anti-periodic respectively, boundary conditions in the time direction up to a gauge
transformation:
$$\eqalign{A_i(\beta ,x)&=A_i(0,x)-\beta  \partial_i \varphi(x), \cr
\psib(\beta ,x)&=-\e^{i \beta  \varphi(x)}\psi(0,x),\cr
\bar\psib(\beta ,x)&=-\e^{i \beta  \varphi(x)}\bar\psi(0,x).\cr}
$$
We thus set
$$\eqalign{A_i(t,x)&=A_i'(t,x)-t \partial_i \varphi(x), \cr
\psib(t,x)&=\e^{i t \varphi(x)}\psi'(t,x) ,\cr
\bar\psib(t,x)&=\e^{-i t \varphi(x)}\bar\psi'(t,x) ,\cr} $$
where the fields $A'_i$, $\psi',\bar\psib'$ now are  periodic and anti-periodic, respectively. Two
modifications appear in the action:
$$\eqalign{\int\d x\d t\,( \partial_t A_i)^2\ &\mapsto\ \int\d x\d
t\,( \partial_t A_i)^2 + \beta \int\d x\,( \partial_i \varphi(x))^2 \cr
\int\d x\d t\,\bar \psib (t,x)\gamma_0  \partial_ t \psib(t,x) &
\mapsto \int\d x\d t\,\bar \psib (t,x)\gamma_0(  \partial_ t +i\varphi
(x))\psib(t,x) . \cr}$$
Therefore, $\varphi(x)$ is simply the zero-mode of the $A_0$
component. Not enforcing Gauss's law suppresses this additional mode. \par
The theory has a $(d-1)$-dimensional gauge invariance with the zero-mode
of $A_i(t,x)$ as a gauge field.
\medskip
{\it Fermion integration and large $N$ calculations.} We now consider the action \eactQEDN, and integrate over fermions, to render the $N$-dependence explicit:
$$ {\cal S}_N  (A _{\mu}  ) = \int \d^d x\,
{\textstyle{ 1 \over 4e^2}}  F ^{2} _{\mu \nu}(x) -\tilde N\tr\ln
\left(\sla{ \partial} + i\Abar \right) , \eqnn $$
where $\tilde N$ is the number of charged fermions, and below $N=\tilde N\tr{\bf 1}$. \par
The action density as a function of a constant field $\varphi\equiv A_0$ then is  given by
$${1\over N}{\cal F}(\varphi)=-  {T\over2} \sum_n{1\over (2\pi)^{d-1}}\int\d^{d-1} k\,\ln\left[k^2
+\bigl(\varphi+(2n+1)\pi T\bigr)^2\right]. $$
The sum over $n$  can be performed with the help of the identity \eqns{\eFTgenidii} and one obtains
$${1\over N}{\cal F}(\varphi)=- {T\over2}    \int {\d^{d-1} k\over (2\pi)^{d-1}}\,
\ln\bigl(\cosh(\beta k )+\cos(\beta\varphi) \bigr).\eqnd\eFTdetQED $$
One verifies that the difference ${\cal F}(\varphi)-{\cal F}(0)$   is UV finite
and has a scaling form $T^d f(\varphi/T)$.
This is not surprising since in the zero temperature  limit no
gauge field mass and quartic $\varphi$ potential are generated.  \par
The derivative
$${1\over N}{\cal F}'(\varphi)={1\over2}
{1\over(2\pi)^{d-1}}  \sinh(\beta \varphi)\int {\d^{d-1} k\over \cosh(\beta k)+\cos(\beta\varphi)},$$
is negative for $-\pi<\beta \varphi<0$ and positive for $0<\beta \varphi<\pi$.
The action density has a unique minimum at $\varphi=0$ in the interval
$-\pi<\beta \varphi<\pi$. \par
A special case is $d=2$ for which one finds \rTSchwinger
$$ {\cal F}'(\varphi)=\ud N \varphi/\pi \quad {\rm for}\
|\varphi|<\pi\quad{\rm and\ thus}\ {\cal
E}(\varphi)=\frac{1}{4} N \varphi^2/\pi\,. $$
 \par
Neglecting all $\varphi$ derivatives, one obtains a contribution to
the action ${\cal S}_T$:
$$-\tilde N\tr \ln \left(\sla{ \partial} + i\Bbar \right) \sim -{1\over2} N\int\d^{d-1} x
\int {\d^{d-1} k\over (2\pi)^{d-1}}\,
\ln\bigl(\cosh(\beta k )+\cos(\beta\varphi(x)) \bigr).$$
The coefficient $K_2(d)$ of $\ud\int\d x\,\varphi^2(x)$ follows:
$$\eqalign{ K_2(d)& =NN_{d-1}\Gamma (d-1)\left(1-d^{3-d}\right)\zeta (d-2)T^{d-3}\cr
&=N {8\over
(4\pi)^{d/2 }}\Gamma ( d/2 )\left(2^{d-3}
-1\right)\zeta(d-2)T^{d-3} . \cr}$$
\smallskip
{\it Discussion.} At leading order, the mass term is thus
proportional to $e T^{(d-2)/2}$. If $e$ is generic, that is of order 1
at the microscopic scale $1/\Lambda $, then $e\propto
\Lambda^{(4-d)/2}$ and the scalar mass $m_T$ is proportional to
$(\Lambda/T)^{(4-d)/2} T$. It is thus large with respect to the vector masses for $d<4$ and small for $d>4$.  \par
If one takes into account loop corrections, one finds for $d>4$ a finite coupling
constant renormalization $e \mapsto e_\r$, and the conclusion is not changed.
The zero-mode becomes massive but with a mass small compared with $T$, justifying  mode and local expansions.
 \par
For $d=4$,  QED is IR free,
$$\beta_{e^2}={\tilde N\over 6\pi^2} e^4 +O(e^6),$$
$e_\r$ has to be replaced by the effective coupling constant
$e(T/\Lambda)$, which is logarithmically small:
$$e^2(T/\Lambda)\sim {6\pi^2\over \tilde N\ln(\Lambda/T)}, $$
and the scalar mass thus is still small, although only logarithmically:
$$m^2_\varphi\propto {T^2\over \ln(\Lambda/T)}.$$
The separation between zero and non-zero modes remains justified. High temperature QED shares
some properties with high temperature $\phi^4$ field
theory, and a perturbative expansion for the same reason remains meaningful. \par
Note that if one is interested  in IR physics only, one can in a second
step integrate over the massive scalar field $\varphi$.
 \par
Finally for $d<4$, one finds an IR fixed point and, therefore,
one expects that in massless QED $m_T$ becomes proportional to $T$
and comparable to all other modes, in particular, to all gauge field
non-zero modes that become massive vector fields.

\vfill\eject

\nref\rGirard{ Early discussions and controversy on the breakdown
of supersymmetry at finite temperature are found in \rf  L.
Girardelo, M.T. Grisaru, P. Salomonson, {\it Nucl. Phys.} B178
(1981) 331 and in \rDas.}
\nref\rDas{ A. Das and M. Kaku, {\it
Phys.Rev} D18 (1978) 4540; A. Das {\it Physics A } 158 (1989) 1.}

\nref\rSenjan{ Restoration of broken internal symmetries were
discussed in \rf A. Riotto and G. Senjanovi\'c, {\it Phys. Rev. Lett.}
79 (1997) 349 [arXiv:hep-ph/9702319]  and references therein. }

\nref\rFotop{The thermodynamics of supersymmetric gauge theories
was studied in  \rf A. Fotopoulos and T.R. Taylor, {\it Phys.
Rev.} D59 (1999) 061701 and references therein. }

\nref\rEspin{The minimal supersymmetric model at finite
temperatures was discussed in \rf J.R. Espinosa, {\it Nucl. Phys.}
B475 (1996) 273 and in ref.~\rLaine.}
\nref\rLaine{ M. Laine, {\it
Nucl. Phys.} B481 (1996) 43, {\it Erratum ibid.} B548 (1999) 637.}

\section  $O(N)$ supersymmetric models  at finite temperature

In this section we would like to find out which of the peculiar
properties of the phase structure seen at $T=0$ in  section
\ssNSUSYsc\ are maintained at finite temperature and how the phase
transition occurs as the temperature, rather than the coupling
constant, is varied~\rMosZinn. \sslbl\ssNSUSYT

Supersymmetry is softly broken at finite temperature as bosons and
fermions behave differently when interacting with a heat bath.
Breakdown of supersymmetry at finite temperature has attracted
much attention since the early interest in supersymmetry and
involved much controversy on its consequences, appearance and
phase structure \refs{\rGirard,\rDas}. Other related issues such as
the restoration of broken internal symmetries at finite
temperature supersymmetric theories were also debated
\refs{\rSenjan}. More recently, there is a continued interest in
the thermodynamics of supersymmetric Yang-Mills theory
\refs{\rFotop} and temperature effects on the minimal
supersymmetric model \refs{\rEspin}.

\subsection {The free energy at finite temperature}

Following the notation of section \ssNSUSYsc~to which we refer for
details, we now derive the free energy at finite temperature. The
partition function is given by \sslbl\ssNSUSYfreeE
$$ {\cal Z}=\int[\d\Phi][\d\rho][\d L] \e^{-{\cal S}(\Phi,L,R)}
\eqnd\SUSYaction$$
with
$${\cal S}(\Phi,L,R)=\int\d^3 x\,\d^2\theta\left\{\ud\bar {\rm D}\Phi\cdot
{\rm D}\Phi+NU(R)+L(\theta)\left[\Phi^2(\theta)-NR(\theta)\right]
\right\},$$ where $\Phi $, $L$, $R$ are $N$-component scalar
superfields, parametrized in the form Eq.~\PhiSupField:
$$ \Phi(\theta,x) =
\varphi + \bar \theta \psi + \ud  \bar \theta\theta  F   $$ and
(Eqs.~\eqns{\eLsupField,\eRsupField})
$$L(\theta,x)=M+\bar\theta\ell+\ud\bar\theta\theta \lambda\,,
\quad R(\theta,x)=\rho +
\bar\theta\sigma+\ud\bar\theta\theta s\,.\eqnd\LandRho$$
The Grassmann coordinate  $\theta$ is a two-component
Majorana spinor; $\varphi$, $M $ and $\rho$, are $N$ component
real scalar fields $\psi$, $\ell $ and $\sigma$ are $N$ component,
two-component Majorana spinors $F$, $ \lambda $ and $s$ are $N$
component  auxiliary fields. $\D$ is the covariant derivative,
$\D=\partial /
\partial \bar \theta -\delslash ~\theta $, the integration
measure is $\d^2\theta={i\over 2}d\theta_2 d\theta_1$ and
$\bar\theta_\alpha\theta_\beta =
\half\delta_{\alpha\beta}\bar\theta\theta $.

Integrating out $N-1$ superfield components  of $\Phi$ and keeping
$\Phi_1\equiv \phi$ (the scalar component of the superfield $\phi$
is identified as $\varphi_1 \equiv \varphi$) one finds
$${\cal Z}=\int[\d\phi][\d R][\d L]\e^{-{\cal
S}_N(\phi,R,L)},\eqnd\Nintegration $$ where the  large $N$ action
is
$$\eqalignno{{\cal S}_N=\int\d^3 x\,\d^2\theta &\left[\half\bar \D\phi
\D\phi+NU(R)+ L \left(\phi^2-N R\right)\right] \cr &\quad+
\half(N-1){\rm Str}\, \ln\left[-\bar D
D+2L\right].&\eqnd\LargeNaction \cr}$$ The two saddle point
equations~\eSUSYsad{} are not changed. Only the third equation
\eSUSYsadc\ is affected  by the boundary conditions due to finite
temperature: \eqna\saddle
$$\eqalignno{ 2L\phi-\bar D D\phi&=0\,,&\saddle{a} \cr
  L-U'(R)&=0\,,&\saddle{b} \cr
 R-\phi^2/N&= {1\over N}\tr\Delta(k,\theta,\theta), &\saddle{c}\cr
} $$
$\Delta(k,\theta,\theta)$ is given by \eSUSYpropcoin:
$$
\Delta(k,\theta,\theta)= {1\over k^2+M_T^2+\lambda}+
\bar\theta\theta M_T \left( {1\over k^2+M_T^2+\lambda} - {1\over
k^2+M_T^2} \right). \eqnd\SUSYpropagator$$where $M_T$ is the
expectation value of $M(x)$ at finite temperature $T$. When
written in components, Eq.~\saddle{a} implies \eqna\saddleComponA
$$\eqalignno{F-M_T\varphi&=0 \  , &\saddleComponA {a} \cr
 \lambda\varphi+M_TF&=0 \,,&\saddleComponA{b} \cr} $$
from which  the Goldstone condition $\varphi(\lambda+M_T^2)=0 $
follows.\par Eq.~\saddle{b} implies:
$$M_T=U'(\rho)\ ,\quad  \lambda=sU''(\rho).\eqnd\saddleComponB $$
When calculating the trace in Eq.~\saddle{c}, one has to take into
account that bosons at finite temperature satisfy periodic, and
fermions anti-periodic boundary conditions. Then, combining
Eq.~\eNfivTtp~with the $\theta =0$ part of the finite temperature
Eq.~\saddle{c}, we obtain
$$ \rho-\varphi^2/N = \int {\d^2k\over (2\pi)^2 }{1\over \omega_\varphi (k)}
 \left({1\over2} +{1\over \e^{  \omega_\varphi (k)/T}-1}\right) $$
with
$$\omega_\varphi (k)=\sqrt{M_T^2+\lambda +k^2}. $$
It is convenient here to introduce the boson thermal mass
$$m_T=\sqrt{M_T^2+\lambda}\,.\eqnd\eSUSYbostherm $$
The thermal mass $m_T$ characterizes the decay of correlation
functions in space directions. Note that, on the other hand, $M_T$
does not characterize the decay of fermion correlations, because
fermions have no zero mode, the relevant parameter being
$\sqrt{M_T^2+\pi^2T^2}$. Then,
$$\rho-\varphi^2/N   =\rho_c
  -{ T\over 2\pi}\ln\bigl(2\sinh(m_T/2T)\bigr) \eqnd \esadSUSYNTa $$
where  $\rho_c$ has been defined in Eq.~\eRciiidef.

In the same way, the $\bar\theta \theta $ part of Eq.~\saddle{c}
combined with Eq.~\eNGNTtp~yields
$$\eqalign{s-2F\varphi /N&=2M_T \int {\d^2k\over (2\pi)^2  } \left\{ {1\over 2 \omega _\varphi(k)}
\left({1\over 2}+{1\over \e^{ \omega_\varphi(k)/T}-1}\right)
\right. \cr&\quad \left.
 -{1\over 2 \omega _\psi(k)}\left({1\over2} -{1\over \e^{ \omega_\psi (k)/T}+1}\right)\right\} \cr}$$
with $\omega _\psi=\sqrt{M_T^2+k^2}$. Integrating we obtain
$$ s-2F\varphi /N =  {  T M_T\over  \pi}\left[
 \ln\bigl(2\cosh( M_T/ 2)\bigr)-\ln\bigl(2\sinh(m_T/2T)\bigr)\right].
\eqnd  \esadSUSYNTb $$

The action density at finite temperature ${\cal F}$, to leading
order in $1/N$, is given by ($\beta =1/T$)
$${\cal F}={\cal S}_N/{V_2 \beta}\ , \eqnd\freeEnergyT$$
where ${\cal S}_N$ is given in Eq.~\LargeNaction~at constant
fields and $V_2$ is the two dimensional volume.

The supertrace term in Eq.~\LargeNaction~can be calculated at finite
temperature by  using the expressions
Eqs.~\TRboson~and~\eTRfermion:
$$\eqalignno{ {1\over V_2  } {\rm Str}\, &\ln\left [-\bar D D+2L\right] =
{1\over V_2  }\tr\ln(-\partial^2 +M_T^2+\lambda) -{1\over V_2  }
\tr\ln(\sla{\partial}+M_T) \cr
  &=2  \int {\d^2k\over {(2\pi)^2} }
  \ln[2\sinh( \beta \omega _\varphi /2 ) ]
  -  2 \int {\d^2k\over {(2\pi)^2} }
  \ln[2\cosh( \beta  \omega _\psi/2)]    \cr
&={1 \over T}\rho_c (m_T^2-M_T^2) -{ 1\over 6\pi T}\left(
m_T^3-|M_T|^3\right) \cr &\quad +2\int {\d^2k\over  (2\pi)^2
}\left\{
  \ln[ 1-\e^{- \beta \omega _\varphi  } ]-  \ln[1+\e^{- \beta  \omega _\psi}]
  \right\}.
  &\eqnd\SuperTraceT \cr} $$
The explicitly subtracted part reduces for $M_T=M$ to the $T= 0$ result.
The other contributions to ${\cal E}$ are the same (up to the change
$M\mapsto M_T$) as in Eq.~\eSUSYV\ and therefore
one finds
$$\eqalignno{ {1\over N}{\cal F}&=
   -  {F^2\over2 N} +
M_T{F\varphi\over N} +   \lambda{\varphi^2\over 2N}  + \half s
(U'(\rho)-M_T)   -{1\over 12\pi}\left(  m_T^3 - |M_T|^3\right)\cr
&\quad + \half\lambda (\rho_c-\rho) +T\int {\d^2k\over  (2\pi)^2
}\left\{
  \ln[ 1-\e^{- \beta \omega _\varphi  } ]-  \ln[1+\e^{- \beta  \omega _\psi}]
  \right\}  \,.
  &\eqnd\freeEnergyTiii \cr }$$
In the limit $T=0$, the free energy in Eq.~\freeEnergyTiii~reduces
to the action density of  Eq.~\eSUSYV,   ${\cal F}(T=0)\equiv{\cal
E}$.\par

Eq.~\freeEnergyTiii\ can be rewritten by using  Eq.~\esadSUSYNTa,
which is also obtained by setting to zero the $ \del/ \del\lambda
$ derivative of Eq.~\freeEnergyTiii, as well as $M_T-U'(\rho)=0$
from Eq.~\saddleComponB~(or equivalently, setting to zero the $
\del/ \del s $ derivative of Eq.~\freeEnergyTiii). One finds
$$\eqalignno{&{1\over N} {\cal F} = \half M_T^2 {\varphi^2\over N} + {1\over 24 \pi}(
m_T - |M_T|)^2(m_T +2|M_T|) \cr
&\quad+{T\lambda\over 4\pi
}\ln(1-\e^{-m_T/T})+T \int {\d^2k\over
{(2\pi)^2} } \{ \ln(1-\e^{-\beta \omega_\varphi} )
 - \ln(1 + \e^{-\beta \omega_\psi} ) \}.\hskip12mm
&\eqnd\FreeEnergy }$$
Eq.~\FreeEnergy~is the finite energy version of
Eq.~\HartreeVTzeroC.
 \par
Inserting Eqs.~\esadSUSYNTa~and \esadSUSYNTb~(with $F=M_T\varphi$)
into Eqs.~\saddleComponB~with $U( R )=\mu R  +\half u  R^2$, one
finds ($\mu_c=-u\rho_c$)
 $$\eqalignno{  M_T &=
  \mu-\mu_c   + u {\varphi^2\over N}
 -{u\over 2\pi}T\ln\left(2\sinh\left(\ud  m_T/T \right)\right) , &
 \eqnd\FermionMassi  \cr
{m_T^2-M_T^2 \over uM_T}&={2\varphi^2 \over N}     - {   T  \over  \pi}\left[
\ln\left(2\sinh\left(\ud  m_T/T \right)\right)-\ln\left(2\cosh
\left(\ud  |M_T|/T \right)\right) \right],
\hskip10mm & \eqnd  \esadSUSYNTla\cr} $$
 which are the   gap equations for  $T\neq 0$. \par
%
\medskip
{\it Solutions to saddle point equations:}
 One first notes that the r.h.s.~of the gap Eq. \FermionMassi\
 diverges when $m_T\to 0$.
This phenomenon has been already discussed in the example of the
scalar field theory. A finite temperature system in three
dimensions has the property of a statistical system in two
dimensions. Spontaneous breaking of a continuous symmetry is
impossible in two dimensions due to the  IR behaviour of a system
with potential massless Goldstone particles. Therefore, the $O(N)$
symmetry is never broken,  $\varphi=0$, and thus
\eqna\eNTSUSYgapii
 $$\eqalignno{ { M_T \over T} &=
 { \mu-\mu_c\over T}     -{u\over 2\pi} \ln\left(2\sinh\left(\ud   m_T/T
 \right)\right) , &
 \eqnd\eNTSUSYgapii{a} \cr
{M_T^2-m_T^2\over uTM_T}&=    {   1  \over  \pi}\left[
\ln\left(2\sinh\left(\ud  m_T/T \right)\right)-\ln\left(2\cosh \left(\ud
M_T /T \right)\right) \right].
\hskip2mm & \eqnd  \eNTSUSYgapii{b}
\cr} $$
Note that while the expression of $\cal F$ is complicated, its derivative with
respect to $m_T^2 $ remains simple
$${1\over N}{\partial  {\cal F} \over \partial m_T^2 }={(m_T^2-M_T^2)
\over 16\pi m_T  \tanh(m_T/2T)}.
\eqnn $$
Therefore, the minimum still occurs at $m_T=|M_T|$,  but
one verifies that $m_T=|M_T|$ is not a solution to the saddle
point equations. We have only found a lower bound on the free energy density  $\cal F$ as a function of $M_T$:
$$ {\cal F}=NT \int {\d^2k\over (2\pi)^2 }\ln\tanh\left(\ud\beta \sqrt{M_T^2+k^2} \right).
\eqnd\eSUSYTFlow $$
Its derivative with respect to $M_T$ is
$${\partial {\cal F}\over \partial M_T}=-N{T M_T\over 2\pi}\ln\tanh(|M_T|/2T).$$
Therefore the derivative has the sign of $M_T$. The lower bound  has a limit for $M_T\to 0$, which is thus an absolute lower bound:
$${\cal F}=-{7\over 8\pi}\zeta (3)NT^3\,.\eqnd\eSUSYFlow $$
Similarly, the  derivative of $\cal F$ with respect to $ M_T $ at $m_T$ fixed is
$$\eqalignno{{\partial {\cal F}\over\partial  M_T }&=   { M_T T\over 2\pi}
\left[
\ln\bigl(2\cosh(M_T/2T)\bigr)-\ln\bigl(2\sinh(m_T/2T)\bigr)\right] \cr
&={m_T^2-M_T^2 \over 2u}\,. \cr}$$

 We now have to find the solutions to the saddle point
 equations and compare their free energies.\par
Note a first solution in the regime $T\to0$, $\mu<\mu_c$ with $|m_T |\ll T$ and $|M_T\ll T$
Then, from \FermionMassi, we find the boson thermal mass
$$m_T \sim T\e^{-2\pi(\mu_c-\mu)/uT}\,.$$
Eq.~\esadSUSYNTla~yields the other mass parameter
$$M_T \sim {m_T^2 \over \mu_c-\mu}\,.$$
Since both $m_T$ and $M_T$ are very small, the free energy is very
close to the lower bound \eSUSYFlow. The solution found here is a
precursor of the zero temperature phase transition, and
corresponds to $\varphi\ne 0$ in region III and IV of
Fig.~\phases. ~Other solutions  exist in this regime but they
converge, up to exponential corrections, to the finite masses of
the $T=0$ spectrum, and thus their free energy is much larger from
the lower bound \eSUSYTFlow.\par It is expected that even for $T$
larger and $\mu\ge \mu_c$ the continuation of the small mass
solution remains the solution with lowest energy.\par

Variational calculations will produce here similar results to
those obtained from the path integral at $N \to \infty$.
We know from the general discussion at $T=0$ in section
\ssSUSYNvar\ that one of the gap equations $M_T=U'(\rho)$ is
immediately implied by demanding that the free energy density in
the variational calculations remains finite for large cut-off
$\Lambda $. Therefore, the variational free energy in Eq.~\eVarEsusy\
 (with $\rho$ and $\tilde \rho$  of Eqs.~\eR a ~and
 \eR b~replaced by their thermal expressions) takes the form  identical
to Eq.~\FreeEnergy.

$m_T$ and $M_T$ being related, in what follows,  we take as the
independent variable $m\equiv m_T$ and, using Eq.~\FermionMassi\
for $M_T=M_T(m,\varphi,\mu,u,T)$, ~we discuss
$$W(m,T)={1\over N}{\cal F}
(m_T=m,M_T(m,\varphi,\mu,u,T),T).\eqnd\eWvar$$
Since, as described above, at $T\neq 0$ there is no breaking of
 $O(N)$ symmetry, we will discuss mainly $\varphi=0$ in
Eq.~\eWvar.





 For $T=0$, Eq.~\eWvar\  results in
$ {1\over N}{\cal F}(m,\varphi^2, T=0) ={1\over N}{\cal
E}(m,\varphi^2)$ which has been analyzed in section \ssNSUSYsc. In
particular, see the phase structure in Fig.\phases~ and the
interesting degeneracy found in Regions II and IV and exhibited in
Figs.\ZeroTone-- \ZeroTtwo.

Rather than changing the value of $\mu-\mu_c$ and the coupling $u$
as done in section \ssNSUSYsc, we are interested here to see
the temperature dependence while the parameters $\mu-\mu_c$ and
$u$ are held fixed.

At finite temperature, the effective field theory describes  the
interactions of the fermions and bosons with the heat bath. This
interaction acts like a source of soft breaking of supersymmetry.
The short distance behavior is cured at finite temperature in a
similar way it happens at $T=0$. We will see now that the general
properties of the transitions as a function of the temperature $T$
have a similar character as the transitions seen at $T=0$ when the
coupling constant was varied.

As seen in Fig.~\phases, at $T=0$ there are two regions in the $\{
\mu-\mu_c , u \}$ space where the vacua are degenerate.
These are (region {\bf II :}
$\mu-\mu_c \geq 0$ and $u \geq u_c$ and region {\bf IV : }
$\mu-\mu_c \leq 0$ and $u \leq u_c$). Fig.~\figTIIaaa ~shows the
ground state  energy ($T=0$) in region {\bf II} and Fig.~\figTIVaa ~shows the  ground state energy  in region
{\bf IV}.

We will discuss first region {\bf (II)}  $\mu-\mu_c \geq 0$
and $u \geq u_c$:

\noindent Here, as $T$ increases the ground state with the smaller
mass ($\ma\equiv m =m_+$) has a lower free energy than the heavier
one ($\ma \equiv m=m_- > m_+$) due to a higher entropy
contribution,
(Fig.~\label{\figTIIa}).


Peculiar transitions can occur in this system. If the system was
initially at $T=0$ ~in the ground state with a boson mass
$\ma=m_-$, it will eventually go, as $T$ is increased, into the
state with mass $\ma=m_+$, namely, into the lower mass ground
state. This is due to the entropy negative contribution to the
free energy forcing the system to favors the lowest mass state. On
the other hand, a system that started at a high temperature in the
state of low mass $\ma=m_+$ will stay in this state as the system
is cooled and will never roll back into the $m=m_-$ high mass
phase. Favoring the lowest  mass phase as the temperature
increases is a general effect that will occur in any physical
system that is initially (at $T=0$) mass degenerated. Here,
supersymmetry imply that the $m_+$ and $m_-$ vacua are at the same
energy $E=0$ at $T=0$.

\midinsert \epsfxsize=14.5cm \epsfysize=8cm \pspoints=2.pt \vskip
-5.5cm \hskip 2cm \epsfbox{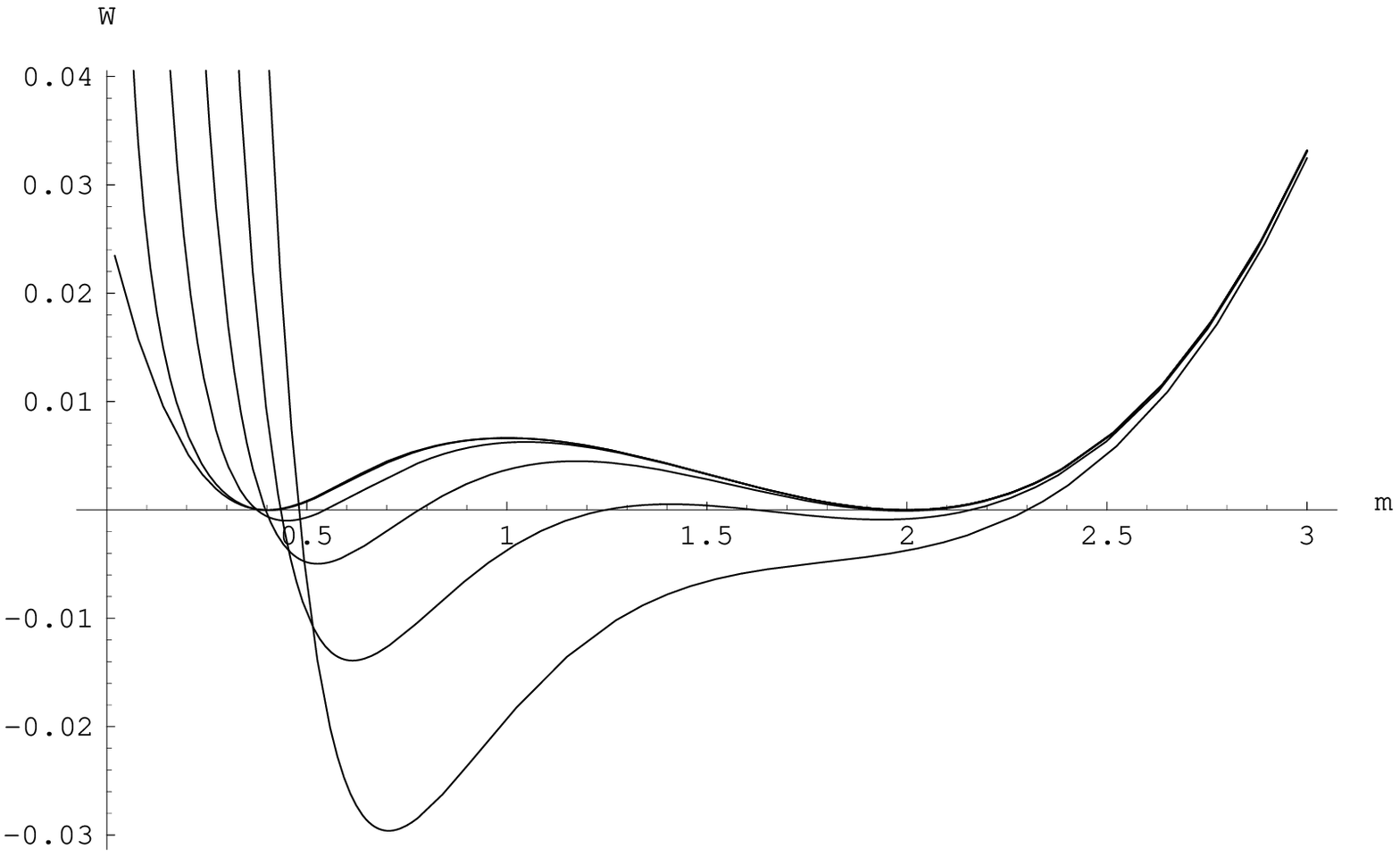} \figure{1mm}{The
 free energy $W(m,\varphi,T)={1\over N} {\cal
F}(m\equiv\ma,\varphi,T)$ as given in Eq.~ \FreeEnergy\ at
$\varphi=0$ as a function of the boson mass ($m$) at different
temperatures. Here $\mu-\mu_c=1 $ (this sets the mass scale),  $
u/u_c=1.5 $ and $T$ varies between $T=0 - 0.5$. At $T = 0$ two
degenerate $O(N)$ symmetric phases exist  with a light $m=m_+$ and
heavier $m=m_- > m_+$ massive boson (and fermion). As the
temperature increases the light mass phase is stronger affected as
its entropy increases faster and becomes the only ground state.}
\figlbl\figTIIa
\endinsert

\noindent Region {\bf (IV)}  $\mu \leq 0$ and $u \leq u_c$: Here
one finds at $T=0$ two distinct degenerate phases. One is an
ordered phase ($\varphi \neq 0$) with a zero boson and fermion
mass, the other is a symmetric phase ($\varphi = 0$) with a
massive ($m=m_-$) boson and fermion. (as seen in
Fig.~\label{\figTIVaa}).

As mentioned above at $T\neq 0$
the ordered phase with the broken
$O(N)$ symmetry,
 ($m=0 , \varphi^2 \neq 0$) disappears
into $\varphi^2 = 0$ symmetric phase and a very small mass ground
state $\ma=m\geq 0$ is created (as seen in Fig.~\label{\figTIVbb}
and in Fig.~\label{\figTIVa}).
 \midinsert
\epsfxsize=13cm
\epsfysize=10cm \pspoints=2.pt \vskip -4.5cm\hskip 2.5cm
\epsfbox{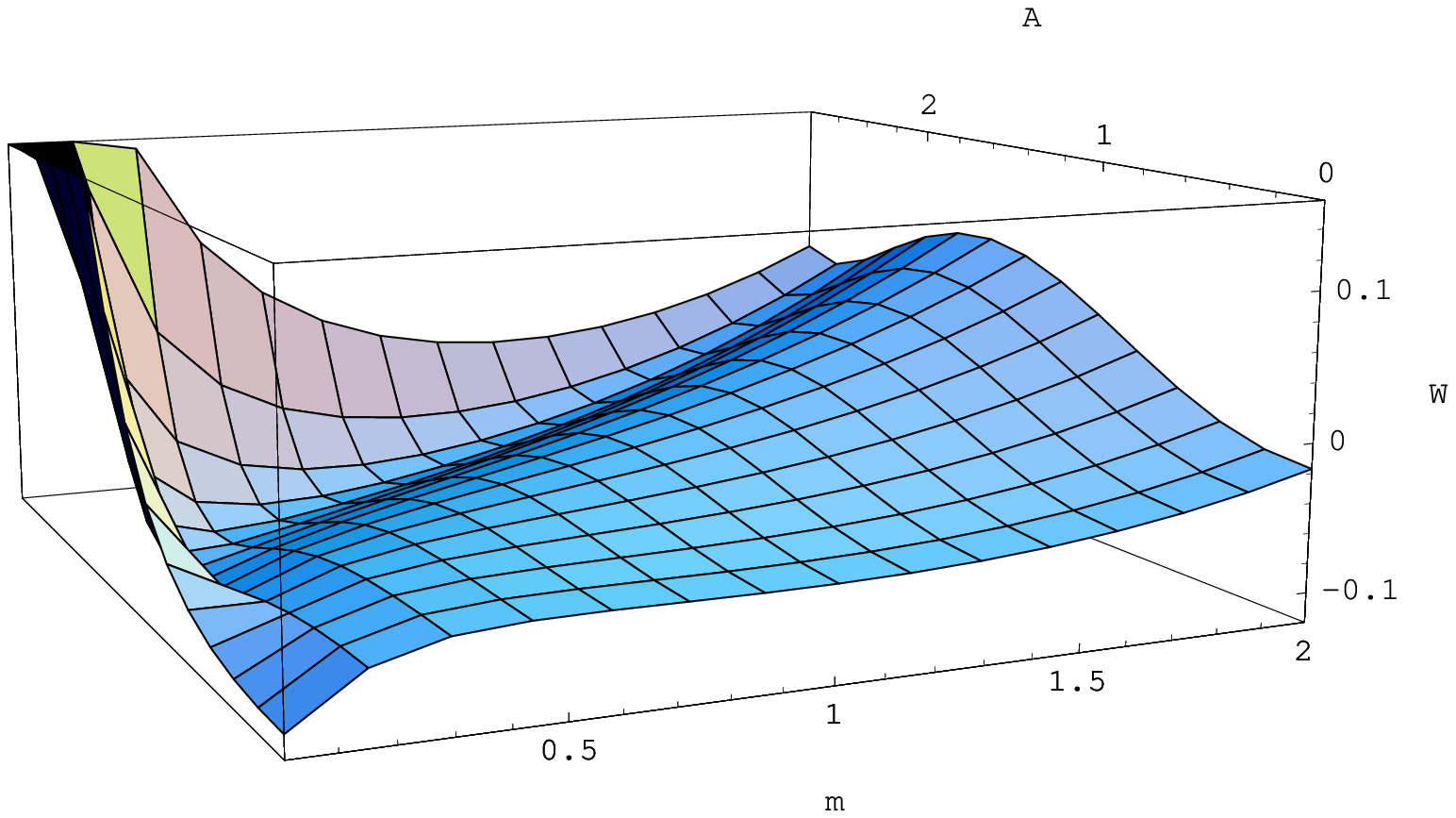} \figure{1mm}{Same as Fig.~\figTIVaa ~in
section~\ssNSUSYsc ~but the temperature has been increased from $T
= 0$ (in Fig.~\figTIVaa) to $T=0.7$ (here).
$W(m,\varphi,T)={1\over N} {\cal F}(m\equiv\ma,\varphi,T)$ as
given in
 Eq.~ \FreeEnergy\ as a function of the
boson mass ($m$)  and $A$, where $A^2=\varphi^2/u_c$.
 Here $\mu-\mu_c=-1 $ and
$ u/u_c=0.2 $. A non-degenerate $O(N)$ symmetric ground state
($\varphi = 0$) appears with a very small boson mass (the non-zero
mass is not seen here due to the limited resolution of the plot).
} \figlbl\figTIVbb
\endinsert

Eventually, also the other $O(N)$ symmetric vacua ($m=m_-,
\varphi^2 = 0$) goes  into the small mass $O(N)$ symmetric ground
state.

\midinsert \epsfxsize=13cm \epsfysize=11cm \pspoints=2.pt \vskip
-5cm \hskip 2.5cm \epsfbox{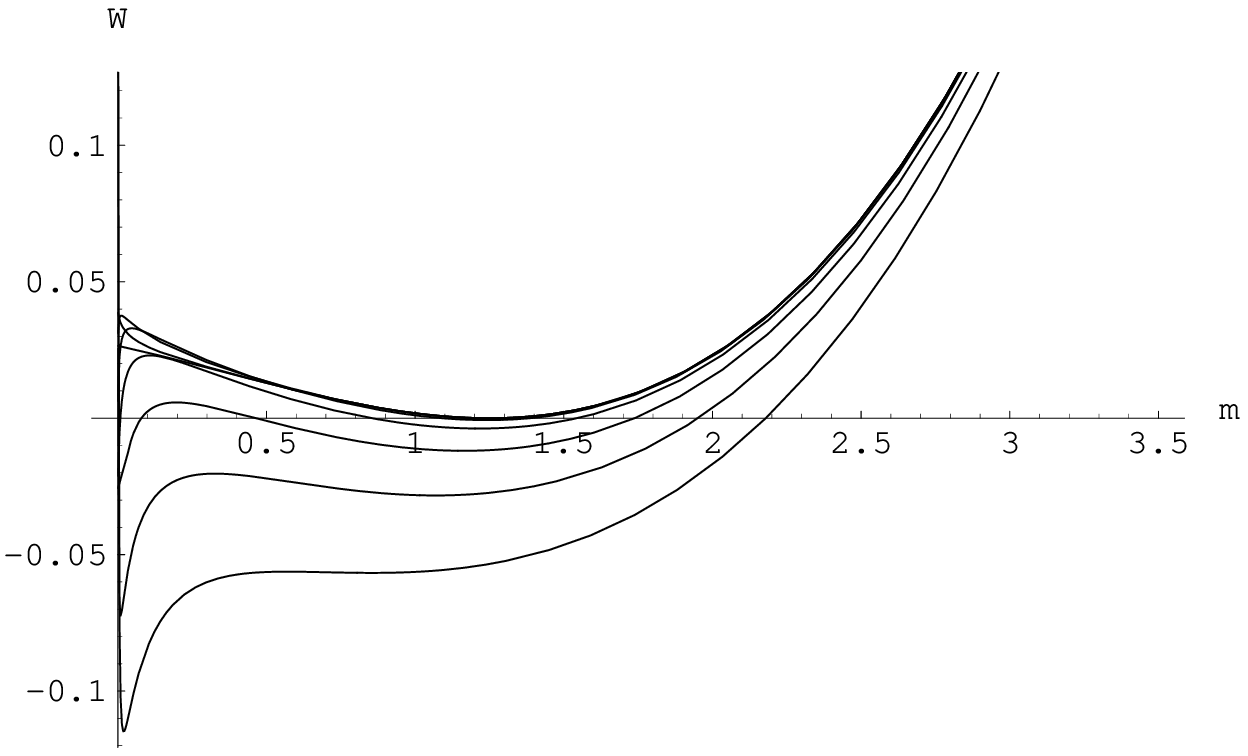} \figure{1mm}{This figure
displays the effect of increasing the temperature from $T = 0$ in
Fig.~\figTIVaa~to $T$ that varies in the range $ 0 \le T\le  0.7$
(Fig.~\figTIVbb ~has $T=0.7$). The  free energy
$W(m,\varphi,T)={1\over N}{\cal F}(m\equiv\ma,\varphi,T)$ as given
in Eq.~ \FreeEnergy\ at $\varphi=0$ is plotted as a function of
the boson mass ($m$)  at different temperatures. Here
$\mu-\mu_c=-1 $, $ u/u_c=0.2 $. At $T = 0$ there are two
degenerate phases: An $O(N)$ symmetric phase, shown here, with a
massive ($m=m_-$) boson and fermion  and an ordered phase
($\varphi \neq 0$) with massless particles (both phases are shown
in Fig.~\figTIVaa).  At finite temperatures the $O(N)$ symmetry is
restored (see Fig.~\figTIVbb) and a small mass ground state
appears, the heavy mass state decays into the small mass ground
state as seen here.} \figlbl\figTIVa
\endinsert

\noindent As in {\bf II}, a system that was initially, at $T=0$
in the $O(N)$ symmetric phase ($\ma=m_- , \varphi^2 = 0$) will
decay into the smaller mass state when the temperature. But upon
cooling the system in the small mass phase will roll into the
ordered phase ($\ma=m=0 , \varphi^2 \neq 0$) at $T=0$. The system
will never roll back into the symmetric phase ($\ma=m_- , \varphi
= 0$).

\subsection The supersymmetric $O(N)$
non-linear $\sigma $ model at finite temperature

The supersymmetric non-linear $\sigma $ model we consider here has already
been discussed at zero temperature in section \ssSUSYnls\ to which
we refer for details.\sslbl\ssNlsSUSYT
\par
The partition function of the $O(N)$ supersymmetric non-linear $\sigma $ model
in $d $ dimensions is given by
$${\cal Z}=\int[\d\Phi][\d L] \e^{-{\cal S}(\Phi,L)} \eqnd\NLSMpartition$$
with
$${\cal S}(\Phi)={1\over 2 \kappa }\int\d^d x\,\d^2\theta\, \bar D\Phi\cdot
D\Phi +L(\Phi^2-N)  \,, \eqnd\eNLSMT $$
where $\kappa$ is a constant. The scalar superfield
$\Phi(x,\theta)$  is an $N$-component vector:
$$ \Phi(x,\theta) =   \varphi + \bar \theta  \psi +
\ud \bar \theta\theta   F \,,\eqnd{\evecPhi} $$ and the scalar
superfield $L(x,\theta)$ is given by
$$L(x,\theta) = M(x) + \bar\theta \ell(x) +\half \bar\theta\theta
\lambda(x) .\eqnd\L$$ After integrating out $N-1$ superfield
components of $\Phi$, leaving out $\phi=\Phi_1$, one obtains
$${\cal Z} = \int [d\phi][\d L] \e^{ - {\cal S}_N(\phi,L) }$$
where
$$   {\cal S}_N(\phi,L)= {1\over\kappa } \int\d^d x\,\d^2\theta  \left[ \half\bar D\phi\cdot
D\phi+  L \left(\phi^2- N \right) \right]    + {{N-1}\over 2} {\rm
Str}\, \ln( -{\bar D}D +2 L )  .   \eqnd\eNLSTact   $$ Varying the
action $     {\cal S}_N(\phi,L)$
 by varying the superfields
$\phi$ and $L$,   one obtains Eq.~\eSadNLSMa\
and the generalization of Eq.~\eSadNLSMb:
\eqna\eSUSYNTnls
$$\eqalignno{ 2L\phi-\bar D D\phi &=0 \,,&\eSUSYNTnls{a} \cr
 {N\over \kappa}-{\phi^2 \over N}&=\tr\Delta(k,\theta,\theta) .&\eSUSYNTnls{b} \cr}
 $$
The first equation is equivalent to $F=M_T\varphi$ and $\lambda \varphi+FM_T=0$
while the second equation can be compared with Eq.~\saddle{c}, and thus
leads to

(for $N\gg 1$ and $\psi=0$)
\eqna\NLSMComponC
$$\eqalignno{ {1\over \kappa}-{\varphi^2\over N} &=\int {\d^dk\over (2\pi)^d }{1\over \omega_\varphi (k)}
 \left({1\over2} +{1\over \e^{  \omega_\varphi (k)/T}-1}\right), &\NLSMComponC{a}\cr
-{2F\varphi  \over N}&=2M_T \int {\d^dk\over (2\pi)^d  } \left\{ {1\over 2 \omega _\varphi(k)}
\left({1\over 2}+{1\over \e^{ \omega_\varphi(k)/T}-1}\right)
\right. \cr&\quad \left.
 -{1\over 2 \omega _\psi(k)}\left({1\over2} -{1\over \e^{ \omega_\psi (k)/T}+1}\right)\right\}  & \NLSMComponC{b}\cr} $$
with $\omega_\varphi (k)=\sqrt{m_T^2  +k^2}$ and  $\omega_ \psi (k)=\sqrt{M_T^2  +k^2} $.
\medskip
{\it Dimension $d=3$.} As it has been discussed already, the $d=3$
finite temperature theory is analogous from the point of phase
transitions to a two-dimensional theory. Therefore the $O(N)$
symmetry remains unbroken and $\varphi=0$. We thus write the two
gap equations only  in this limit

\eqna\eNLSMsadTi
$$\eqalignno{
 {1\over \kappa _c}  - {1\over \kappa} &=
 {  T\over 2
\pi}\ln\bigl(2\sinh(m_T/2T)\bigr)&\eNLSMsadTi{a}
\cr
0&= {  T M_T\over 2 \pi}\left[\ln\bigl(2\sinh(m_T/2T)\bigr)-
\ln\bigl(2\cosh(M_T/2T)\bigr)\right].
&\eNLSMsadTi{b} } $$
The calculation of the  energy density to leading order for $N\to\infty $
of the non-linear $\sigma $ model at finite temperature follows the similar
steps of the $(\Phi^2)^2$ field theory.
The   free energy at finite temperature is given by
$${\cal F}=T {\cal S}_N / V_2\,, \eqnd\NLSMFEnergyT$$
where ${\cal S}_N$ is given by Eq.~\eNLSTact~at constant fields and
$V_2$ is the two dimensional volume.
The expression has been calculated in section \ssSUSYnls\ but here
the supertrace term in Eq.~\eNLSTact~has to be replaced by the
finite temperature form as
it appears in Eq.~\freeEnergyTiii.
The free energy then is given by
$$\eqalignno{{1\over N} {\cal F}&=      \ud(M_T^2-m_T^2)\left( {1\over\kappa } -{1\over\kappa_c}\right)
-{1\over12\pi}\left(m_T^3-|M_T|^3\right) \cr\quad &
+T\int {\d^2k\over  (2\pi)^2  }\left\{
  \ln[ 1-\e^{- \beta \omega _\varphi  } ]-  \ln[1+\e^{- \beta  \omega _\psi}]\right\}.
&\eqnd\NLSMFreeEnergyTiii \cr}$$
The  free energy in Eq.~\NLSMFreeEnergyTiii~can be compared
with Eq.~\eNLSMenergy\ which gives the ground state  energy ${\cal F}$ at $T=0$.
\par
In section \ssSUSYnls\ we have verified that after using
the zero temperature limit of Eq.~\eNLSMsadTi{a} the energy densities
of the $(\Phi^2)^2$ theory and the non-linear $\sigma $ model become identical, up to a possible rescaling of the field.
It is now clear that the same mechanism works here. Using Eq.~\eNLSMsadTi{a} one indeed finds the expression \FreeEnergy\ (for $\varphi=0$):
$$\eqalignno{&{1\over N} {\cal F} =  {1\over 24 \pi}(
m_T - |M_T|)^2(m_T +2|M_T|) +{T \over 4\pi
}(m_T^2-M_T^2)\ln(1-\e^{-m_T/T})\cr
&\quad+T \int {\d^2k\over
{(2\pi)^2} } \{ \ln(1-\e^{-\beta \omega_\varphi} )
 - \ln(1 + \e^{-\beta \omega_\psi} ) \}.\hskip12mm
&\eqnn\cr} $$
\medskip
{\it Solutions.}
The derivative of $\cal F$ with respect to $|M_T|$ at $m_T$ fixed is
$${\partial {\cal F}\over\partial |M_T|}={|M_T|T\over 2\pi}
\left[
\ln\bigl(2\cosh(M_T/2T)\bigr)-\ln\bigl(2\sinh(m_T/2T)\bigr)\right].$$
It is convenient to introduce the notation
$$X(\kappa ,T)=\exp\left[{2\pi \over T}\left({1\over \kappa_c}-{1\over \kappa }\right)\right].$$
Then the behaviour of $\cal F$ depends on the position of $X$ with respect
to $2$:
$$X=2\ \Leftrightarrow\ T={2\pi\over \ln 2}\left({1\over \kappa_c}-{1\over \kappa }\right), $$
an equation that has a solution only for $\kappa >\kappa _c$.\par
 For $X<2$, the derivative vanishes at $M_T=0$ and is positive
for all $|M_T|>0$. Instead for $X>2$, the derivative vanishes both at $M_T=0$
and
$$|M_T|=2T\ln\left[\ud(X+\sqrt{X^2-4})\right],\eqnd\eSUSYNTnlsMT $$
which are the two solutions of Eq.~\eNLSMsadTi{a}. The minimum of $\cal F$
is located at the second value \eSUSYNTnlsMT. \par
We find therefore an interesting non-analytic behaviour:
$$\cases{M_T=0 &for $X<2$, \cr
M_T=2T\ln\left[\ud(X+\sqrt{X^2-4})\right] & for $X>2$ .\cr}$$
Eq.~\eNLSMsadTi{a} then yields directly the boson thermal mass
$$m_T=2T\ln\left[\ud(X+\sqrt{X^2+4})\right] .$$
For $\kappa <\kappa _c$ and $T\to 0$, we find the asymptotic behaviour
$$m_T\sim  TX(\kappa ,T),$$
which is exponentially small, and $M_T=0$.\par
For $\kappa >\kappa _c$ and $T\to 0$, both $m_T$ and $M_T$ converge
toward the finite $T=0$ limit with
exponentially small corrections.\par
In the opposite high temperature limit $T\gg | 1/\kappa -1/\kappa _c|$,
we find that $m_T$ is asymptotically proportional to $T$:
$$m_T\sim 2T\ln\bigl((1+\sqrt{5})/2\bigr),$$
while  $M_T=0$.\par
It is not clear whether such a result can survive $1/N$ corrections.
 \medskip
{\it Dimension $d=2$.} In generic dimensions $2\le d\le 3$ a phase
transition is not even possible at finite temperature  and $\varphi=0$
(in two dimensions it is even impossible at zero temperature).
The gap equations take the form
$$\eqalignno{{1\over \kappa }&=\Omega _d(m_T)+T^{d-2}f_d(m_T^2/T^2), &\eqnn \cr
0&=M_T\left[\Omega_d(|M_T|)-T^{d-2}g_d(M_T^2/T^2)-
\Omega _d(m)-T^{d-2}f_d(m_T^2/T^2)\right] ,
\hskip6mm &\eqnn \cr}$$
where $f_d$ and $g_d$ are defined in Eqs.~\eThermfz~and \eThermfiis~respectively.
\par
For $d=2$ it is then convenient to introduce the physical mass $m$ solution of
$${1\over \kappa }=\Omega _2(m). $$
The equations can then be rewritten as
$$\Omega _2(m)=\Omega _2(m_T)+  f_2(m_T^2/T^2)  \eqnn $$
and either $M_T=0$ or
 $$ \Omega_2(|M_T|)- g_2(M_T^2/T^2)-\Omega _2(m_T)- f_2(m_T^2/T^2) =0\,.\eqnn $$
The first equation, which determines $m_T$, is identical to the
Eq. \eNTnlsgapii~already  obtained  for the usual non-linear
$\sigma $ model. At low temperature $m_T=m$ up to exponentially
small corrections. At high temperature
$${T\over m_T}\sim{1\over \pi}\ln(m_T/m)\sim {1\over \pi}\ln( T/m),$$
a consequence of the domination of the zero mode and the UV
asymptotic freedom of the non-linear $\sigma $-model.\par
Combining both gap equations we find
$$ \Omega_2(|M_T|)- g_2(M_T^2/T^2)=\Omega _2(m),$$
which is identical to Eq.~\eTGNNgapiii.\par
The analysis of section \ssNTGNNphase\ for $d=2$ has shown that for $T$ large
the equation has no solution and thus $M_T=0$, but it has a solution for $T$ small.
The situation therefore is similar to what has
been encountered in three dimensions. Again an analysis
of $LL$ propagator and $1/N$ corrections is required to
understand whether this result survives beyond the large $N$ limit.

\vfill\eject

\def\lfrac#1#2{{\displaystyle{#1\over#2}}}
\section Weakly interacting Bose gas and large $N$ techniques


\nref\LY{T.D.  Lee and C.N.  Yang, {\it Phys.  Rev.}   112 (1957)
1419.}

\nref\huang{K. Huang, in {\it Stud.  Stat.  Mech}, {\bf II}, J. de
Boer and G.E.  Uhlenbeck, eds.  (North Holland Publ., Amsterdam, 1964), 1.}

\nref\toyoda{T.~Toyoda, {\it Ann. Phys.} (NY)   141  (1982)  154.}

\nref\stoof{H.T.C.  Stoof, {\it Phys.  Rev.}  A45 (1992)  8398.}
\nref\rBijl{M. Bijlsma and H.T.C.  Stoof, {\it Phys.  Rev.}  A54
(1996) 5085.}

\nref\GCL{P. Gr\"uter, D. Ceperley, and F. Lalo\"e, {\it Phys.  Rev.
Lett.} 79 (1997) 3549.}

\nref\laloe{M. Holzmann, P. Gr\"uter, and F. Lalo\"e, cond-mat/9809356, {\it
Euro.  Phys.  J.} B  10 (1999) 739.}

\nref\club{G. Baym, J.-P.  Blaizot, M. Holzmann, F. Lalo\"e, and D.
Vautherin, {\it Phys.  Rev.  Lett.} 83 (1999) 1703, cond-mat/9905430.}
\nref\rBBZJ{G. Baym, J.-P.  Blaizot and J. Zinn-Justin, {\it Euro. Phys. Lett.}
49 (2000) 150.}

\nref\rPA{P. Arnold, B. Tomasik,
{\it Phys. Rev.} A62 (2000) 063604, cond-mat/0005197.}



\def\cite{\refs}

\nref\rarnold{
P. Arnold, G.D. Moore, {\it Phys. Rev. Lett.} 87 (2001) 120401,
cond-mat/0103228.}\nref\rarnoldB{
 P. Arnold, G.D. Moore, B. Tomasik, {\it
Phys. Rev.} A65 (2002) 013606, cond-mat/0107124.}
\nref\rKPS{V.A.
Kashurnikov, N. Prokof'ev and B. Svistunov, {\it Phys.  Rev. Lett.}
87 (2001) 120402,
 cond-mat/0103149.}
\nref\rGBHV{G. Baym, J.-P. Blaizot, M. Holzmann, F. Laloe, D.
Vautherin; cond-mat/0107129, to appear in EJP B.}\nref\rHolz{ M.
Holzmann, G. Baym, J.-P. Blaizot, F. Laloe,
 {\it Phys. Rev. Lett.} 87 (2001)  120403, cond-mat/0103595.}

 \nref\rKPR{J.-L. Kneur, M. B. Pinto and R. O. Ramos,
cond-mat/0207295.}


%

As it is well known a free Bose gas undergoes at a low but finite temperature a transition called Bose--Einstein condensation. The condensed low temperature phase is characterized by a macroscopic occupation of the one-particle ground state. \sslbl\scBEC \par
The effect of a weak repulsive two-body interaction on the transition temperature of a dilute Bose gas at fixed density has been controversial for a long time
\refs{\LY
-\laloe}. In ref.~\refs{\club} it has first been argued theoretically
 that the transition temperature $T_c$ increases linearly with the strength of the interaction, parametrized in terms of the scattering length $a$. However, the coefficient cannot be obtained from perturbation theory.  A simple self-consistent approximation was thus used to derive an explicit estimate. \par
A better understanding of the physics of the weakly interacting Bose gas came from the recognition that the universal properties of the system under study, like the helium superfluid transition, can be described by a particular example of the general $N$ vector model, for $N$ =2 \refs{\rBBZJ}.
Renormalization group (RG) arguments then allowed to prove the existence, besides the universal IR behaviour common with the superfluid transition, of a universal large momentum (UV) behaviour peculiar to systems with a very weak two-body interaction. The long wavelength properties of the weakly interacting Bose gas can be described by the three-dimensional, super-renormalizable, euclidean  $(\phib^2)^2$ field theory. A more direct and general RG derivation of the linear behaviour of the shift of the critical temperature and the universality of the coefficient followed.
\par
However, the calculation of the coefficient remained a non-perturbative problem. A possible method to calculate the temperature shift is to generalize the problem to arbitrary $N$.  This generalization makes new tools available; in particular, the coefficient of $\Delta T_c/T_c$ can be calculated by carrying out an expansion in $1/N$.  The leading contribution to $\Delta T_c/T_c$
already requires  a $1/N$ calculation \refs{\rBBZJ}, as we show explicitly below.
The result happens to be independent of $N$, for non-trivial reasons.  The
calculation involves subtle technical points, which are most easily dealt with by dimensional regularization.  More recently, a $1/N^2$ calculation
has been performed that yields the $1/N$ correction to $ \Delta
T_c/T_c$ \refs{\rPA}. The relative correction for $N=2$ is about
$26\%$, a  correction which is smaller than what is typically found in the calculation of critical exponents. \par
%
\subsection{Quantum field theory and Bose--Einstein condensation}

We consider a system of identical non-relativistic bosons of mass $m$,
at temperature $T=1/\beta $ close to the critical temperature $T_c$.  When the two-body potential $V_2$ is
short-range  and one is interested only in long wavelength
 phenomena, one can approximate the potential by a $\delta$-function
pseudo-potential,
$$V_2(x-y)=G\, \delta ^{(d)}(x-y), $$
regularized at short distance because such a potential is singular for $d>1$.
\par
 The partition function \sslbl\ssQSDPBdel
$${\cal Z}(\beta ,\mu )=\tr\e^{-\beta (H-\mu N)},$$
where $H$ is the hamiltonian in Fock space, $N$ the
particle number operator (which commutes with $H$), and $\mu$ the chemical potential, can then be expressed as a functional integral
$${\cal Z}=\int[\d\varphi\d\bar\varphi]\exp[-{\cal S}(\varphi,\bar\varphi)],
\eqnd\eZBE $$
where ${\cal S}(\varphi,\bar\varphi)$ is a  non-relativistic {\it local}\/ action:
$$\eqalignno{{\cal S}(\varphi,\bar\varphi)&=\int_0^\beta \d{t}\int\d^d{x}\left[
 \bar\varphi(x,t)\left(-{\partial\over\partial t}-{\hbar^2\over2m}\nabla^2 -\mu
\right)\varphi(x,t) \right. \cr &\quad \left.
+{G\over2} \bigl(\bar\varphi(x,t) \varphi(x,t)\bigr)^2\right],&\eqnd\eactBEdel \cr}$$
and the fields satisfy periodic boundary conditions in the euclidean
time direction,
$$\varphi(x,\beta =1/T)=\varphi(x,0) \quad \bar\varphi(x,\beta =1/T)=\bar\varphi(x,0).$$
The strength $G$ of the interaction must be positive,  corresponding to a repulsive interaction, for the boson system to be stable. \par
 It is customary to parametrize the strength $G$ of the pair
potential in terms of the scattering length $a$. In three
dimensions $a= m G/4\pi\hbar^2$.    Furthermore, we assume that
$a\ll \lambda$, where
$$
  \lambda=\hbar\sqrt{   2\pi / m k_BT}
$$
is the thermal wavelength.  (In the following we set $k_B=\hbar=
1$, and write simply $\lambda^2=2\pi/mT$.) \par Note that, unlike
what would happen for a fluid, in a very dilute system even though
the pair-potential is weak, the $N$-body potentials, $N>2$, are
even much smaller and can thus be totally  neglected.
\par
To compute the effects of the interactions on the transition temperature,
we write the equation of state, the relation between particle number density, temperature and chemical potential. The particle number density can be expressed as a sum of the single particle Green's
function over Matsubara frequencies $\omega_\nu=2\pi \nu T$:
$$
n = {T\over \Omega } {\partial \ln {\cal Z}\over \partial \mu}=
\left<\bar\varphi(x,t) \varphi(x,t)\right>=
T\int\lfrac{\d^d k}{(2\pi)^d} \sum_{\nu\in{\Bbb Z}} \tilde G^{(2)}( \omega_\nu,k)\eqnd{\ematsuz}
$$
($\Omega $ is the space volume).
Above the transition, the single particle  Green's function  can
be parametrized as
$$
\left[\tilde G^{(2)}( \omega ,k)\right]^{-1} =- i\omega  - \mu + \lfrac{k^2}{2m}+ \Sigma( \omega ,k).
\eqnd{\egf}
$$
The Bose--Einstein condensation temperature is determined by the point where $ [\tilde G^{(2)}(0, 0)]^{-1}=0$, that is where $\Sigma(0,0) = \mu$.  At this point,
$$
\left[\tilde G^{(2)}( \omega,k )\right]^{-1}  =-i \omega +\lfrac{k^2}{2m} - \left[\Sigma( \omega,k )-\Sigma(0,0)\right].
\eqnd{\egfi}
$$
At $(T_c,\mu_c)$ the Fourier transform of the two-point correlation
function at zero frequency diverges at zero momentum, and so does the correlation length. \par
In the absence of interactions,
$$
 \sum_{\nu\in{\Bbb Z}} \tilde G^{(2)}(k,\omega_\nu) ={\beta \over
\e^{\beta (k^2/2m-\mu)}-1}, $$
$\mu=0$ at the transition, and
$$ n=\zeta(d/2)/\lambda_c^d\, , \eqnd\eBeTcideal
$$
where
$$\lambda_c^2=2\pi/mT_c^0\,,$$
 and $T_c^0$ is the condensation temperature of the ideal gas.

In the presence of weak interactions, the temperature of the Bose--Einstein
condensation becomes the critical temperature of the interacting model, and is
shifted by the interactions. From the theory of critical phenomena we know
that the variation of the critical temperature in systems with dimension $d$
below four depends primarily on contributions from the small momenta or large
distance (which we refer to as the infrared, or IR) region.  This property,
which we later verify explicitly for $d=3$, simplifies the problem, since to
leading order the IR properties are only sensitive to the $\omega_\nu=0$ component.

It follows from the relations \eqns{\egfi,\eBeTcideal} that at leading order in the dilute limit, where only the
$\omega_\nu=0$ Matsubara frequency contributes, the shift in the transition
temperature at fixed density, $\Delta T_c = T_c - T_c^0$, can be related to
the change $\Delta n$ in the density at fixed $T_c$ by \refs {\club}
$$ \lfrac{\Delta T_c}{T_c} =-\lfrac{2}{d}\lfrac{\Delta n}{n},\eqnd{\edelta}
$$
where
$$
\Delta n={4\pi \over\lambda^2}N_d \int_0^\infty {\rm d}k\,
k^{d-1}\left(\lfrac{1}{k^2+{\cal M}(k)}-\lfrac{1}{k^2}\right),
$$
$N_d$ is the usual loop factor \etadpolexp{b} ($N_3=1/2\pi^2$), and
$$
{\cal M}(k)\equiv 2m\left[\Sigma(k,0)-\Sigma(0,0)\right].
$$

Restricting  the fields $ \varphi$ and $\bar \varphi $ to their zero Matsubara frequency components, corresponds to take a classical field limit.
Equivalently,  the time dependence of the fields  $\varphi(x,t),\bar\varphi(x,t)$
in \eactBEdel\ can be neglected.
It is then
convenient to rescale the field $ \varphi$ in order to introduce  the
field theory normalizations used elsewhere in this review, and to parametrize it in terms of two real fields
$\phi_1,\phi_2$:
$$\varphi(t,x)\sim \sqrt{mT}\bigl(\phi_1(x)+i\phi_2(x)\bigr).$$
  The partition function
then reads
$$
 {\cal Z}= \int \left[ \d \phib (x) \right] \exp \left[-{\cal S}(\phib)\right]
 , \eqnd{\eeONpart}
$$
where now ${\cal S}(\phib)={\cal H}/T$ is given by
$$
{\cal S}   \phib  )= \int \left\lbrace{ 1 \over 2} \left[
\partial_{\mu} \phib (x) \right]^2+{1 \over 2}r
\phib^2 (x)+{u \over 4!}  \left[ \phib^2(x) \right]^2 \right\rbrace \d^{d}x\,,
\eqnd\eeactfivOii
$$
with $r=-2m \mu$, and for $d=3$:
$$
u=96 \pi^2\lfrac{a}{\lambda^2} = 12 m^2 T G.\eqnd\euaGrel
$$
where $G$ is the strength of the pair potential in expression \eactBEdel.  In Eq.~\eeactfivOii\ we have kept the dimension
$d$ of the spatial integration arbitrary in order to use dimensional regularization later. The two-point correlation
function $\tilde G^{(2)}(p)$ is related to the two-point vertex
function $\tilde\Gamma^{(2)}(p)$ of the classical statistical
field theory by
$$
\tilde G^{(2)}( \omega =0,p)=\lfrac{2mT}{ \tilde\Gamma^{(2)}(p)}.
$$

The model described by the euclidean action \eeactfivOii\ reduces to the
ordinary $O(2)$ symmetric $(\phib^2)^2$ field theory, which indeed describes the universal properties of the superfluid Helium transition.  As it stands this
field theory suffers from more UV divergences than the
original theory, the higher frequency modes providing a large momentum
cutoff $\sim \sqrt{m T}\sim 1/\lambda$.  This cutoff may be restored when
needed, for instance, by replacing the propagator by the regularized propagator:
$$
{2mT \over  k^2}\rightarrow {1\over {\rm e}^{k^2/2mT}-1}\,.
$$
In fact, as we show later, since the shift of the critical temperature is
dominated by long distance properties it is independent of the precise
cutoff procedure, that is universal.

    A second effect of the non-zero frequency modes is to renormalize the effective coefficients of the euclidean action.  This problem can be explored
by returning to the functional integral representation \eactBEdel\ of the complete quantum
theory and integrating over the non-zero modes perturbatively.  The
corrections generated are of higher order in $a$ and can thus be neglected.

    Because the interactions are weak, one may imagine calculating the change in the transition temperature by perturbation theory.  However, the
perturbative expansion for a critical theory does not exist for any fixed
dimension $d<4$; IR divergences prevent a straightforward calculation.  If one
introduces an infrared cutoff $k_c$ to regulate the momentum integrals, one
finds that perturbation theory breaks down when $k_c \sim a/\lambda^2$, all
terms being then of the same order of magnitude.  To discuss this problem in
more detail, we now generalize the model to $N$ component fields with an $O(N)$ symmetric hamiltonian.

\subsection{The $N$-vector model. Renormalization group}

We consider the $O(N)$ symmetric generalization \eactON\ of the model
corresponding to the euclidean action \eeactfivOii:
$$
{\cal S} ( \phib )= \int \left\lbrace{ 1 \over 2} \left[
\partial_{\mu} \phib (x) \right]^2+{1 \over 2}r
\phib^2 (x)+{1 \over 4!}{u\over N} \left[ \phib^2(x) \right]^2 \right\rbrace \d^{d}x\,,
\eqnd\eeactfivON
$$
where the field $\phib(x)$  now has $N$ components (note the change in the normalization of the interaction strength $u\mapsto u/N$).
The advantage of this generalization is that it provides
us with a tool, the large $N$ expansion, which allows calculating directly at
the critical point.

The goal is to obtain the leading order non-trivial contribution at
criticality (in the massless theory) to
$$
n=2mT \sum_{i=1}^N  \left<\phi_i^2 \right> \equiv {2mT \,
N\,}\rho
$$
with
$$
\rho= \int\lfrac{\d^d k}{(2\pi)^d}\,\lfrac{1}{\tilde\Gamma^{(2)}(k)}\,.\eqnd{\eeONrho}
$$
Here, $\delta_{ij}/{\tilde\Gamma^{(2)}(k)}$ is the connected two-point correlation
function.

We first recover, by a simple renormalization group analysis, the result of
\refs {\club} that the change in $\rho$ due to the interaction is linear in the
coupling constant.  We introduce a large momentum cutoff $\Lambda
\propto\sqrt{mT}\sim 1/\lambda$, and the dimensionless coupling constant
$$
g=\Lambda^{d-4}{u\over N} \propto \left(a\over
\lambda\right)^{d-2}\,. \eqnd{\ecoupling}
$$
At $T_c$ the inverse two-point function in momentum
space satisfies the renormalization group equation \refs {\rbook}
$$
\left(\Lambda{\partial\over\partial \Lambda}+\beta(g){\partial \over\partial
g}-\eta(g)\right)\tilde\Gamma^{(2)}(p,\Lambda,g)=0\,.
$$
This equation combined with dimensional analysis implies that the two-point
function has the general form
$$
\tilde\Gamma^{(2)}(p,\Lambda,g)=p^2 Z(g)F\bigl(p/\Lambda(g)\bigr).
\eqnd{\ef}
$$
On dimensional grounds $\Lambda(g)$ is proportional to $\Lambda$, with
$$\eqalignno{
\beta(g){\partial \ln Z(g)\over \partial g}&= \eta(g), &
\eqnd\ebeta \cr
\beta(g){\partial \ln \Lambda (g)\over \partial g}&= -1 \,. &
\eqnd\elam \cr}
$$
Since
$$\beta(g)=-(4-d) g +(N+8)g^2/48\pi^2+{\cal O}(g^3),$$
$\beta(g)$ is
of order $g$ for small $g$ in $d<4$; similarly
$$\eta(g)=(N+2)g^2/(72(8\pi^2)^2)+{\cal O}(g^3).$$
  Therefore,
$$
 Z(g)=\exp\int_0^g{\eta(g') \over \beta(g')}\d g'=1+{\cal O}(g^2).$$
Thus to leading order $Z(g)$ =1.  The function $\Lambda(g)$ is then obtained by
integrating Eq.~\elam:
$$
\Lambda(g)=g^{1/(4-d)}\Lambda\exp\left[-\int_0^g\d g'\left({1\over
\beta(g')} +{1\over (4-d)g'}\right)\right].
$$
In the generic situation (like in a fluid) $g=O(1)$ , the scale $\Lambda(g)$ is indistinguishable from
the cut-off scale $\Lambda $. Universal behaviour
exists only for momenta $|p|\ll \Lambda $ where
$$\tilde\Gamma^{(2)}(p)\propto p^{2-\eta} \quad {\rm for}\  p\ll \Lambda(g)=O(\Lambda ) .
\eqnd{\eIRscaling}
$$
However, for $g\ll 1$ one finds
$$\Lambda(g) \sim g^{1/(4-d)}\Lambda \ll \Lambda\,.$$
There exists therefore an intermediate scale $\Lambda^{-1}(g)$ between the IR and the microscopic scales.
The scale $\Lambda(g)$ corresponds to a crossover separating a universal long-distance regime governed by the non-trivial zero $g^*$ of the $\beta$-function, from a
universal short distance regime governed by the gaussian fixed point $g=0$,
where
$$
\tilde\Gamma^{(2)}(p)\propto p^2 \qquad \Lambda(g) \ll p\ll \Lambda\,.
$$
In a generic situation $g$ is
of order unity and the universal
large momentum region is absent. \par
Since $g$  is equal to $a/\lambda$ (see Eq.~\ecoupling) which is $\ll 1$, this
condition is satisfied in the present situation. Moreover, unlike what is seen in ordinary critical phenomena, no $(\phib^2)^3$ term is here present, because this would correspond to three-body interactions, which for such dilute systems are even smaller.
\par
In the language of the running or effective coupling constant, the
situation can be described as follows: because the initial
coupling constant at the microscopic scale is very close to the (unstable) gaussian fixed point value, it first moves very slowly away from the fixed point. For a dilatation $\lambda $ such that $g(\lambda )$
all irrelevant couplings are already negligible, justifying the existence of a universal small distance, large momentum regime,
whose physics here is the physics of the Bose--Einstein condensation.
Then, very rapidly, $g(\lambda )$ moves toward the IR fixed point $g^*$, where the universal IR behaviour of the superfluid transition is seen.

We now show that with this condition, $\Delta T_c\propto \Lambda(g)$.
First, from the $g=0$ limit we infer that $F(\infty)=1$ (see Eq.~\ef.)
The function $F(p)$ behaves for large $p$ as
$$
F(p)=1+ O (p^{2d-8}),
$$
up to $\ln p$ factors, as can be verified directly from perturbation theory ($g=0$ is the UV fixed point).
Therefore, the first correction to the density \eeONrho\ is convergent at
large momentum and independent of the cutoff procedure, that is universal,
$$
\delta\rho= \int\lfrac{\d^d p}{(2\pi)^d} {1\over p^2}\left(
{1\over F(p/\Lambda(g))} -{ 1}\right).
$$
Similarly, the IR scaling result (Eq.~\eIRscaling) implies that this
integral is IR convergent.  Setting $p=\Lambda(g)k$, we then find the general
form
$$\delta\rho=[\Lambda(g)]^{d-2}\int {\d^d k \over(2\pi)^d} {1\over
k^2}\left( {1\over F(k)} - 1\right), \eqnd{\edeltarho} $$
the $g$ dependence is entirely contained in $\Lambda(g)$.  For $g$ small
we conclude
$$\lfrac{\delta\rho}{\rho}\propto[\Lambda (g)/\Lambda ]^{d-2}\propto
g^{(d-2)/(4-d)}. $$
In the physical dimension 3 we recover
$$
\lfrac{\delta\rho}{\rho}\propto g \propto an^{1/3}\,,
$$
in agreement with \refs {\club}.  It is important to note that both the
perturbative large momentum region and the non-perturbative IR region
contribute to the integral in Eq.~\edeltarho.  Therefore, we cannot
evaluate it from a perturbative calculation of the function $F(p)$.  However,
as we now show, we can calculate $\delta\rho$ in the form of an $1/N$ expansion.
\par
Finally, the result for generic dimension $d$ shows that the linear behaviour found for $d=3$, that could lead to believe
that the result is in some way perturbative, is just  accidental.
\subsection{The large $N$ expansion at order $1/N$}

We now use the techniques of large $N$ expansion explained in section \scfivN, where the large $N$ limit is taken
at $Nu$ fixed.  To leading order the critical two-point function has simply
its free field form.  However, a non-trivial correction is generated at order
$1/N$; one finds the inverse two-point function \eONpropi\ \refs {\rFAAAH,\rbook},
$$\tilde
\Gamma^{(2)}(p)= p^2 +{2\over N }\int
 \lfrac{\d^{d}q}{(2\pi)^{d}}{1 \over(6/u)+B_\Lambda (q,0)}\left({1 \over
(p+q)^2} -{1 \over q^2}\right) + O(1/ N^2),
\eqnd\eeONpropi
$$
where $B_\Lambda(p,0) $ is the one-loop function defined by Eq.~\ebullecrit:
$$
B_\Lambda (q,0)=\int^\Lambda  \lfrac{\d^{d}k}{(2\pi)^{d}}\lfrac{1}{k^2(k+q)^2}\mathop{=}_{q\to0}b(d)q^{d-4} -{6\over N g^*}\Lambda^{d-4}
+{\cal O}(1/\Lambda^2),
$$
and the large $N$ value of $g^*$ has been used. We recall (Eq.~\econstb)
$$
b(d)=-{\pi
\over\sin(\pi d/2)}  {\Gamma^2 (d/ 2) \over \Gamma (d-1)} N_d\,.
 $$

For $d=3$ one finds $b(3)=1/ 8$.\par
Note that in the large $N$ limit, the chemical potential $\mu$ is
proportional to $1/N$.

    In the large $N$ limit, the $\beta$-function takes the simple form
$\beta(g)=(d-4) g (1-g/g^*)$.  Therefore, the leading
cutoff-dependent correction to $B_\Lambda(q,0)$ combines with
$6/u$ to yield $(6/u)(1-g/g^*)$,
as expected from renormalization group arguments. This cut-off
dependent correction, however, can be neglected because $g\ll
g^*$. \par
We evaluate
$$
\delta\rho=-{2\over N }\int\lfrac{\d^d p}{(2\pi)^{2d}}{1 \over p^4}
{\d^{d}q \over(6/u)+b(d)q^{d-4}}\left({1 \over
(p+q)^2} -{1 \over q^2}\right)
\eqnd{\ero}
$$
by keeping the dimension $d$ generic and using dimensional regularization.
The integral over $p$ is
$$\eqalign{
 \int\lfrac{\d^d p}{(2\pi)^d}{1 \over p^4}{1\over(p+q)^2}&=
{1\over(4\pi)^{d/2}}{\Gamma(3-d/2)\Gamma(d/2-1)\Gamma(d/2-2)
\over \Gamma(d-3)}q^{d-6}\,. \cr
&={1\over(4\pi)^{d/2}}{\Gamma(d/2-1)\over\Gamma(d-3)}{\pi\over\sin \pi d/2}\,
q^{d-6}\,.\cr}
$$
Note that the singularity at $d=3$, which would apparently entail the
vanishing of the integral, is cancelled in the subsequent $q$ integral, which
reduces to
$$ \eqalign{
\int\lfrac{\d^{d}q}{(2\pi)^{d}}\, {q^{d-6}
\over(6/u)+b(d)q^{d-4}} &=
{N_d \over 4-d} \quad {\pi\over\sin\left(\pi(d-2)/(4-d)\right)} \cr
&\quad \times b^{(2d-6)/(4-d)}\left(6\over u\right)^{(2-d)/(4-d)}.\cr}
$$
In the $d=3$ limit the two integrations in Eq.~\ero\ yield
$(1/32\pi^2)(u/6)$. As expected, $\delta\rho\propto u$:
$$
\delta\rho=-Ku/N \,,\quad K={1\over96\pi^2} \,,
$$
or in terms of the original parameters  $(u/N)=96\pi^2 a/ \lambda^2
$,
$$\delta\rho=-\lfrac{a}{\lambda^2}. \eqnd{\edelrhoa}
$$
Using this result in Eq.~\edelta, we finally obtain the change in
the transition temperature \refs{\rBBZJ}:
$$\eqalignno{
\lfrac{\Delta T_c}{T_c} &=
  \lfrac{8\pi}{3\zeta(3/2)}\lfrac{a}{\lambda} \cr
&=  \lfrac{8\pi}{3\zeta(3/2)^{4/3}} a n^{1/3}
 = 2.33\, a n^{1/3}. & \eqnd{\eresult} \cr}
$$

    Note that although the final result does not depend on $N$ and, therefore,
replacing $N$ by two is easy, the result is only valid for $N$ large.

\medskip {\it Conclusion.}
The properties of the weakly
interacting Bose gas remain dominated by the UV fixed point of the
renormalization
group equations up to very large length scales; this is why we can still
refer to the
Bose--Einstein condensation when discussing the phase transition of the
dilute interacting
Bose gas.  Renormalization group arguments also  confirm
directly that the
shift of the transition temperature at fixed density is proportional to the
dimensionless
combination
$an^{1/3}$ for weak interactions.  This result is non-perturbative, and the
proportionality
coefficient cannot be obtained from perturbation theory.  However, a non-perturbative method, the large $N$ expansion,  allows a
systematic calculation of this coefficient as a power series in $1/N$, where
eventually one has to set $N=2$.The explicit calculation of  the
leading order contribution has been given in this section.  The first correction is formally of order $1/N$
multiplied by a function of the product $aN$ which is fixed in the large $N$
limit.  Because the final result in three dimensions is linear in $a$, the
$1/N$ factor somewhat surprisingly cancels, and the result is independent of
$N$.  As mentioned earlier the $1/N$ correction has now been calculated
\rPA, and yields a rather low $26\%$ correction for $N=2$.
Moreover, the value found agrees within a factor 2 with the most recent numerical estimates \refs{\rarnold{--}\rKPR}.


\vfill\eject

\nref\rEMNZJ{The following sections are taken mainly from \rf G.
Eyal, M. Moshe, S. Nishigaki and J. Zinn-Justin, {\it Nucl. Phys.}
B470 (1996) 369, hep-th/9601080.} \nref\rmatrix{For a review on
matrix models and double scaling limit see \rf
 P. Di Francesco, P. Ginsparg and J. Zinn-Justin,
{\it Phys. Rep.} 254 (1995) 1.} \nref\rNscaling{Previous
references on the double scaling limit in vector models include
\rf A. Anderson, R.C. Myers and V. Perival, {\it Phys. Lett.} B254
(1991) 89, {\it Nucl. Phys.} B360 (1991) 463. and
refs.~\refs{\rNishi{--}\rYoneya}.} \nref\rNish{S. Nishigaki and T.
Yoneya, {\it Nucl. Phys.} B348 (1991) 787.} \nref\rDiVe{ P. Di
Vecchia, M. Kato and N. Ohta, {\it Nucl. Phys.} B357 (1991)
495.}\nref\rZinnJ{ J. Zinn-Justin, {\it Phys. Lett.} B257 (1991)
335.}\nref\rDiVecc{ P. Di Vecchia, M. Kato and N. Ohta, {\it Int.
J. Mod. Phys.}A7, (1992)1391.}\nref\rYoneya{T. Yoneya, {\it Prog.
Theo. Phys. Suppl.} 92, 14 (1992).} \nref\rmatrixDi{The $d=1$
matrix problem is discussed in \rf P. Ginsparg and J. Zinn-Justin,
{\it Phys. Lett.} B240 (1990) 333 and
refs.~\refs{\rBrezin{--}\rGrossM}.}\nref\rBrezin{E. Br\'ezin, V. A.
Kazakov, and Al. B. Zamolodchikov, {\it Nucl. Phys.} B338 (1990)
673.}\nref\rPa{ G. Parisi, {\it Phys. Lett.} B238 (1990) 209, 213;
{\it Europhys. Lett.} 11 (1990) 595.}\nref\rGrossM{D. J. Gross and
N. Miljkovic, {\it Phys. Lett.} B238 (1990) 217.}

\nref\rBBFS{Early large $N$ calculation in zero and one dimensions
are reported in \rf A.J.~Bray, {\it J.~Stat.~Phys.} 11 (1974) 29;
{\it ibidem} {\it J. Phys.} A7 (1974) 2144 and ref.~\rFerrelS.
}\nref\rFerrelS{R.A. Ferrel and D. Scalapino, {\it Phys. Rev.} A9
(1974) 846.}
\nref\rNtricrit{References on tricritical behaviour and massless
dilaton include \rf W.A.~Bardeen, M.~Moshe, M.~Bander, {\it
Phys.~Rev.~Lett.} 52 (1984) 1188 and
refs.~\refs{\rDavidKN{--}\rSchnitzer}.}\nref\rDavidKN{ F.~David,
D.A.~Kessler and H.~Neuberger, {\it Phys.~Rev.~Lett.} 53 (1984)
2071, {\it Nucl.~Phys.} B257 [FS14] (1985) 695.}\nref\rKessler{
D.A. Kessler and H. Neuberger, {\it Phys.~Lett.} 157B (1985)
416.}\nref\rDiVecMM{ P. Di Vecchia and M. Moshe, {\it
Phys.~Lett.~}B300 (1993) 49.}\nref\rSchnitzer{ H.J.~Schnitzer,
{\it Mod.~Phys.~Lett.~}A7 (1992) 2449.}

\section{Multicritical points and double scaling limit}

We use now the general formalism presented in section \ssNbosgen\
to discuss various additional issues concerning the general
$N$-vector model with one scalar field, in the large $N$ limit
\rEMNZJ. One obvious application concerns multi-critical points.
Of particular interest are the subtleties involved in the
stability of the phase structure at critical dimensions. The
example of the tricritical $(\phib^2)^3$ theory will illustrate
explicitly, however, the limitations of the large $N$
method.\sslbl\ssdblescal\par Another issue involves the so-called
{\it double scaling limit}. Statistical mechanical properties of
random surfaces as well as randomly branched polymers can be
analyzed within the framework of large $N$ expansion. In the same
manner in which matrix models in their double scaling limit
provide representations  of dynamically triangulated random
surfaces summed on different topologies \rmatrix, $O(N)$ symmetric
vector models represent discretized branched polymers in this
limit \refs{\rNscaling{--}\rYoneya}, where $N\to\infty$ and the
coupling constant $g \to g_{c}$ in a correlated manner. The
surfaces in the case of matrix models, and the randomly branched
polymers in the case of vector models are classified by the
different topologies of their Feynman graphs and thus by powers of
$1/N$. Though matrix theories attract most  attention, a detailed
understanding of these theories exists only for dimensions  $d
\leq 1$ \rmatrixDi. On the other hand, in  many cases, the $O(N)$
vector models can be successfully studied  also  in dimensions $d
> 1$, and thus, provide us with intuition for the search for a
possible description of quantum field theory in terms of extended
objects in four dimensions, which is a long lasting problem in
elementary particle theory.

The phase structure of $O(N)$ vector quantum field theories at $ N
\to \infty $ is generally well understood, there are, however,
certain cases where it is still unclear which of the features
survives at finite $N$, and to what extent. One such problem is
the multicritical behaviour of $O(N)$ models {\it at critical
dimensions}. Here, one finds that in the $N \to \infty$ limit,
there exists a non-trivial UV fixed point, scale invariance is
spontaneously broken, and the one parameter family of ground
states contains a massive vector and a massless bound state, a
Goldstone boson--dilaton. However, since it is unclear whether
this structure is likely to survive for finite $N$, one would like
to know whether it is possible to construct a local field theory
of a massless dilaton via the double scaling limit, where all
orders in $1/N$ contribute. The double scaling limit is viewed as
the limit at which the attraction between the $O(N)$ vector quanta
reaches a value at $g \to g_c$, at which a massless bound state is
formed in the $N \to \infty$ limit, while the mass of the vector
particle stays finite. In this limit, powers of $1/N$ are
compensated by IR singularities and thus all orders in $1/N$
contribute.

The special case of field theory in two dimension is discussed in
section \label{\stwodim}. In higher dimensions a new phenomenon
arises: the possibility of a spontaneous breaking of the $O(N)$
symmetry of the model, associated to the Goldstone phenomenon.
Before discussing a possible double scaling limit, the critical
and multicritical points of the $O(N)$ vector model are
re-examined in section \label{\smulticr}. In particular, a certain
sign ambiguity that appears in the expansion of the gap equation
is noted, and related to the existence of the IR fixed point in
dimensions $2<d<4$ discussed in section \label{\sssEGRN}.  In
section \label{\ssNtricritical}, the  interesting physical example
of the tricritical point in three dimension is discussed. Some
insight in the problem is obtained there from  variational
calculations.

\par
In section  \label{\sboundst}, we discuss the subtleties and
conditions for the existence of an $O(N)$ singlet massless bound
state along with a small mass  $O(N)$ vector particle excitation.
It is pointed out that the correct massless effective field theory
is obtained after the massive $O(N)$ scalar is integrated out.
Section \label{\sdouble} is  devoted to the double scaling limit
with a particular emphasis on this limit in theories at their
critical dimensions.
\subsection{The 2D  $O(N)$ symmetric field theory in the double scaling
limit}

We first re-examine and summarize the results for the $O(N)$
symmetric field theory with a potential $NU(\phib^2/N)$, where
$\phib$ is $N$-component field, in the large $N$ limit in two
dimensions. Indeed, there are no phase transitions in two
dimensions and we therefore discuss this case separately. The
action is\sslbl\stwodim
$${\cal S}(\phib)= N\int\d^2 x \left\{\half \left[ \partial_{\mu} \phib (x)
\right]^{2} +U\left(\phib^2/N\right) \right\} ,\eqnd{\eactONgii}$$
where an implicit cut-off $\Lambda$ is always assumed below. Whenever the
explicit dependence in the cut-off will be relevant, we shall assume a
Pauli--Villars's type regularization replacing the propagator by a regularized form
\epropreg.\par
 As  in section \ssNbosgen, one introduces two fields
$\rho(x)$ and $\lambda(x)$ and uses the identity \egeniden.
The large $N$  action obtained by integrating over the field $\phib$ is then
$${\cal S}_N=N\int\d^2 x \left[U(\rho)-\half \lambda
\rho \right]+\half N \tr\ln(-\nabla^2+\lambda ).\eqnd\eactefNgii $$
The integral is  evaluated for $N$ large by the steepest descent method.
The saddle point value $\lambda$ is the
$\phib$-field mass squared, and thus we set in general $\lambda=m^2$.  Since in two dimensions
there is no phase transition, $\left<\phib\right>=0$, the three
saddle point equations \esaddleN{} reduce to two:
 \eqna\esaddNgenii
$$\eqalignno{
U'(\rho_s)&=\half m^2\,,&\esaddNgenii{a}\cr
\rho_s&= \Omega _2(m)\,, & \esaddNgenii{b} \cr}$$
where  the function $\Omega_d(m)$ is defined by Eq.~\etadepole.\par
For $m\ll\Lambda$, one finds
$$\Omega_2(m)={1\over2\pi}\ln(\Lambda/m)+{1\over4\pi}\ln (8\pi K)
+O(m^2/\Lambda^2) , $$
where $K$ is a regularization dependent constant.
\par
As was discussed in the case of quantum mechanics in ref.~\rEMNZJ,
a critical point is characterized by the vanishing at zero
momentum of the determinant of second derivatives of the action at
the saddle point. The mass-matrix has then a zero eigenvalue
which, in field theory, corresponds to the appearance of a new
massless excitation other than $\phib$ (this implies also that the
forces between $\phib$ quanta are attractive, a serious problem in
the case of bosons). \par
 In order to obtain the effective action
for this scalar massless mode, we must integrate over one of the
fields. In the field theory case the resulting effective action
can no longer be written in a local form. In order to discuss the
order of the critical point we only need, however, the action for
space independent fields, and thus, for example, we can eliminate
$\lambda$ using the $\lambda$ saddle point equation. The action
density   ${\cal E} (\rho)$ (Eq.~\eNEner) can then be written as
$${1\over N}{\cal E} (\rho)=U(\rho)-{1\over2}\int^{\lambda(\rho)}\d \lambda' \,\lambda'
{\del \over \del \lambda'} \Omega_d(\sqrt{\lambda'}),\eqnd\eWeffrho$$
where at leading order for $\Lambda$ large
$$\lambda(\rho)=8\pi K \Lambda^2\e^{-4\pi\rho}. $$
The expression  \eWeffrho\ is
valid  for any $d$ and will be used  also in section \sdouble.
Here, it reduces to
$${1\over N}{\cal E}(\rho)=U(\rho)+K\Lambda^2\e^{-4\pi\rho}=U(\rho)+\frac{1}{8\pi}m^2
\e^{-4\pi(\rho-\rho_s)}.$$
A multicritical point is defined by the condition
$$ {\cal E}(\rho)- {\cal E}(\rho_s) =O\left((\rho-\rho_s)^n\right).\eqnd\eWcrit $$
This implies the conditions
$$U^{(k)}(\rho_s)=\half (-4\pi)^{k-1}
m^2\quad \ \ {\rm for}\ \ \ 1\le k\le n-1\,.$$ Note that the
coefficients $U^{(k)}(\rho_s)$ are the coupling constants
renormalized at leading order for $N$ large. If $U(\rho)$ is a
polynomial of degree $n-1$ (the minimal polynomial model), the
multicritical condition in Eq.~\eWcrit\ determines the critical
values of renormalized coupling constants as well as $\rho_s$.
\par
When the fields are space-dependent it is, instead,
 simpler to eliminate $\rho$
because the corresponding field equation
$$U'\bigl(\rho(x)\bigr)=\half \lambda(x)  \eqnd\esadroge $$
is local. This equation can be solved by expanding $\rho(x)-\rho_s$ in a power
series in $\lambda(x)-m^2$:
$$\rho(x)-\rho_s= {1\over2 U''(\rho_s)}\bigl(\lambda(x)-m^2\bigr)
+O\left((\lambda-m^2)^2\right) . \eqnd\esolrola $$
The resulting action for
the field $\lambda(x)$ remains non-local but because, as we
shall see,  adding powers of $\lambda$
as well as adding derivatives make
terms less relevant, only the few first terms of a local expansion of the effective action are important. \par
If in the local expansion of the determinant we keep only the two
first terms, we obtain an action containing at leading order a kinetic
term proportional to $(\del_\mu\lambda)^2$ and the interaction
$(\lambda(x)-m^2)^n$:
$${\cal S}_N(\lambda) \sim N\int\d^2 x\left[{1\over96\pi
m^4}(\del_\mu\lambda)^2 + \frac{1}{n!}S_n \bigl(\lambda(x)-m^2)^n\right],$$
where the neglected terms are of order $(\lambda-m^2)^{n+1}$, $\lambda\del^4
\lambda$, and $\lambda^2\del^2\lambda$ and
$$S_n={1\over N}{\cal E}^{(n)}(\rho_s)[2U''(\rho_s)]^{-n}={1\over N}{\cal E}^{(n)}(\rho_s)(-4\pi m^2)^{-n} .$$
Moreover,  we note that, together with the cut-off $\Lambda$, $m$
now also acts as a cut-off in the local expansion. \par
To eliminate the $N$ dependence in the action we have, as in the
example of quantum mechanics \rEMNZJ, to rescale both the field
$\lambda-m^2$ and space:
$$\lambda(x)-m^2=\sqrt{48\pi}m^2 N^{-1/2}\varphi(x)\,,\quad x\mapsto
N^{(n-2)/4}x \,.
\eqnd\elamrescal $$
We obtain
$${\cal S}_N(\varphi)
\sim\int\d^2x\left[\half(\partial_\mu\varphi)^2+\frac{1}{n!} g_n
\varphi^n\right] .$$
In the minimal model, where the polynomial $U(\rho)$ has exactly degree $n-1$,
we find $g_n=6(48\pi)^{(n-2)/2}m^2$.

As anticipated, we observe that derivatives and powers of
$\varphi$ are affected by negative powers of $N$, justifying a
local expansion. However, we also note that the cut-offs
($\Lambda$  or the mass $m$) are now also multiplied by
$N^{(n-2)/4}$. Therefore, the large $N$ limit becomes also a large
cut-off limit.
\medskip
{\it Double scaling limit.} The existence of a double scaling
limit relies on the existence of IR singularities due to the
massless or small mass bound state which can compensate the $1/N$
factors appearing in the large $N$ perturbation theory. We refer
the reader to the examples in ref.~\rEMNZJ\ of a simple integral
$(d=0)$ and a quantum mechanical $(d=1)$ example of the double
scaling limit.
\par
 We now add to the action relevant perturbations
$$\delta_k U=v_k(\rho(x)-\rho_s)^k,  \quad 1\le k\le n-2\,,$$
proportional to $\int\d^2x(\lambda-m^2)^k$:
$$\delta_k {\cal S}_N(\lambda)=N S_k \int\d^2x \bigl(\lambda(x)-m^2\bigr)^k ,$$
where the coefficients $S_k$ are functions of the coefficients
$v_k$. After the rescaling of Eq.~\elamrescal, one finds
$$\delta_k {\cal S}_N(\varphi)=\frac{1}{k!} g_k
N^{(n-k)/2}\int\d^2x\,\varphi^k(x)  \quad \ \ \ 1\le k\le n-2\,.
$$
However, unlike quantum mechanics \rEMNZJ, it is not sufficient to
scale the coefficients $g_k$ with the power $N^{(k-n)/2}$ in order
to obtain a finite scaling limit. Indeed, perturbation theory is
affected by UV divergences, and we have just noticed that the
cut-off diverges with $N$. In two dimensions the nature of
divergences is very simple: it is entirely due to the
self-contractions of the interaction terms and only one divergent
integral appears
$$\left<\varphi^2(x)\right>={1\over4\pi^2}\int {\d^2 q\over q^2+\mu^2}\,,$$
where $\mu$ is the small mass of the bound state, required as an IR cut-off to
define perturbatively the double scaling limit.
We can then extract the $N$ dependence:
$$\left<\varphi^2(x)\right>={1\over8\pi}(n-2)\ln N+O(1) .$$
Therefore, the coefficients $S_k$ have also to cancel these UV divergences,
and thus have a logarithmic dependence in $N$ superposed to the natural
power obtained from power counting arguments. In general, for any potential,
(Eq.~\eqns{\enormorderInv})
$$V(\varphi)=:V(\varphi):+\left[\sum_{k=1}{1\over2^k
k!}\left<\varphi^2\right>^k \left(\partial\over\partial \varphi\right)^{2k}
\right]:V(\varphi):\,, \eqnd\eNVnormorder $$
where $:V(\varphi):$ is the potential from which self-contractions have been
subtracted (it has been normal-ordered). For example, for $n=3$
$$\varphi^3(x)=:\varphi^3(x):+3\left<\varphi^2\right> \varphi(x) ,$$
and thus the double scaling limit is obtained with
$$N g_1+{1\over16\pi}\ln N g_3  \ \ \ \ \ {\rm held \ \ fixed \ \ as}
\ \ N \to \infty \ .$$ For the example $n=4$ ~(
$\varphi^4(x)=:\varphi^4(x):+6\left<\varphi^2\right>
\varphi(x)^2-3\left<\varphi^2\right>^2$ ) one finds that the
double scaling limit is obtained when
$$ g_1 N^{3/2}\quad {\rm and}\quad N g_2
+{g_4\over8\pi}\ln N \ \ \ {\rm are  \ held \  fixed  \ as}\ \ \
N\to \infty .$$
\subsection{The $O(N)$ symmetric model in higher dimensions: phase transitions}

In higher dimensions, a phase transition  associated with  the spontaneous breaking of the $O(N)$ symmetry is possible. In a first part we thus study the $O(N)$
symmetric $NU(\phib^2/N)$ field theory, in the large $N$ limit in
order  to explore the possible phase transitions and identify the
corresponding multicritical points. \par
 The action is\sslbl\smulticr
$${\cal S}(\phib)=  \int\d^d x \left\{\half \left[ \partial_{\mu} \phib (x)
\right]^{2} +NU\left(\phib^2/N\right) \right\} ,\eqnd{\eactONg}$$
where an implicit cut-off $\Lambda$ is again assumed. \par
Following the strategy of section \ssNbosgen, to which we refer for detail,
we again introduce two auxiliary fields $\lambda (x)$, $\rho (x)=\phib^2(x)/N$,
integrate over $N-1$ components of $\phib$,  and obtain the large $N$ action
$${\cal S}_N=N\int\d^d x \left[\half \left(\partial_\mu\sigma\right)^2+
U(\rho)+\half \lambda\left(\sigma^2/N- \rho\right)\right]+\half
(N-1)\tr\ln(-\nabla^2+\lambda ),\eqnd{\eactefNg}$$
with $\sigma (x)\equiv \phib_1(x)$.
\medskip
{\it The saddle point equations: the $O(N)$ critical point.}
The large $N$ saddle point equations are given by Eqs.~\esaddleN{}:
 \eqna\esaddNgen
$$\eqalignno{m^2\sigma&=0\,, & \esaddNgen{a} \cr
U'(\rho)&=\half m^2\,,&\esaddNgen{b}\cr
\sigma^2/N&=\rho-\Omega _d(m)\,. & \esaddNgen{c} \cr}$$
In the ordered phase $\sigma\ne0$ and thus $m$ vanishes.
Eq.~\esaddNgen{c} has a solution only for $\rho>\rho_c$,
$$\rho_c={1\over (2\pi)^d }\int^\Lambda {\d^d k \over k^2}\ \ \Rightarrow\
\sigma=\sqrt{\rho-\rho_c}\,.$$
Eq.~\esaddNgen{b} which reduces to $U'(\rho)=0$ then yields the critical temperature. Setting
$$U(\rho)=V(\rho)+\half r\rho ,$$ one finds
$$r_c=-2 V'(\rho_c).$$
In order to find the magnetization critical exponent $\beta$, we
need the relation between the $r$ and $\rho$ near the critical
point.
\par
 In the disordered phase, $\sigma=0$, Eq.~\esaddNgen{c}
relates $\rho$ to the $\phib$-field mass $m$. For $m\ll\Lambda$,
$\rho$ approaches $\rho_c$, and the relation becomes
(Eq.~\etadepolii)
$$\rho-\rho_c =-K(d)
m^{d-2}+a(d)m^2\Lambda^{d-4}
+O\left(m^4\Lambda^{d-6}, m^d \Lambda ^{-2}\right). \eqnd\edmum  $$
The constant $K(d)$ is universal (Eq.~\etadpolexp{a}).
The constant  $a(d)$, which also appears in Eq.~\etadepolii, on the other hand,
depends on the cut-off procedure (Eq.~\eadef). \par
For $2<d<4$, (the situation we  assume below except when stated otherwise)
the universal non-analytic part in $m^{d-2}$ dominates.\par
For $d=4$, Eq.~\edmum\ becomes
$$\rho-\rho_c=\frac{1}{8\pi^2} m^2\left(\ln m/\Lambda+\ {\rm const.}\right),$$
and for $d>4$ the analytic contribution dominates and
$$\rho-\rho_c\sim a(d)m^2\Lambda^{d-4}.$$
\medskip
{\it Critical point.} In a generic situation $V''(\rho_c)=U''(\rho_c)$ does
not vanish, a situation we have examined in section \ssNUfige. We  find in the low temperature phase
$$t=r-r_c\sim -2 U''(\rho_c)(\rho-\rho_c) \ \Rightarrow\ \beta=\half\,.
\eqnd\erminrc$$
This is the case of an ordinary critical point. Stability implies
$U''(\rho_c)>0$ so that $t<0$.\par
At high temperature, in the disordered phase, the $\phib$-field mass $m$ is
given by $2V'(\rho)+ r=m^2$ and thus, using \edmum, at leading order
$$t\sim 2U''(\rho_c)K(d)m^{d-2}.$$
Of course, the
simplest realization of this situation is to take $U(\rho)$ quadratic, and we
recover the $(\phib^2)^2$ field theory.
\medskip
{\it Multicritical points.} A new situation arises if we can
adjust the parameters of the potential in such a way that
$U''(\rho_c)=0$. This can be achieved only if the potential $U$ is
at least cubic. We then expect a tricritical behaviour \rNtricrit.
Higher critical points can be obtained when more derivatives
vanish. We shall examine the general case though, from the point
of view of real phase transitions, higher order critical points
are not especially interesting. Indeed, for continuous symmetries
phase transitions occur only for $d>2$ and quasi-gaussian
behaviour is then obtained for all dimensions $d\ge 3$. The
analysis will, however, be useful in the study of  double scaling
limit. \par
Assuming that the first non-vanishing derivative is
$U^{(n)}(\rho_c)$, we expand further Eq.~\esaddNgen{b}. In the
ordered low temperature phase, we now find
$$t=-{2\over (n-1)!}U^{(n)}(\rho_c)(\rho-\rho_c)^{n-1},\ \Rightarrow\
\sigma\propto (-t)^\beta,\quad \beta={1\over2(n-1)}\,  ,  \eqnd\ecritbeta $$
which is the magnetic exponent  obtained in the mean field approximation for such a multicritical point.
We have in addition the condition $U^{(n)}(\rho_c)>0$.\par
In the high temperature phase, instead,
$$m^2=t+ (-1)^{n-1}{2\over (n-1)!}U^{(n)}(\rho_c)K^{n-1}(d)m^{(n-1)(d-2)}.
\eqnd\emsq$$
For $d>2n/(n-1)$, for $m$ small the equation reduces to $m^2=t$, which yields a simple gaussian behaviour, as expected above the upper-critical dimension. \par
For $d<2n/(n-1)$, we find a peculiar phenomenon, the term in the r.h.s.\ is
always dominant, but depending on the parity of $n$ the equation has solutions
for $t>0$ or $t<0$. For $n$ even, $t$ is positive and we find
$$m\propto t^\nu,\qquad \nu={1\over(n-1)(d-2)},\eqnd\emoft$$
which is a non gaussian behaviour below the critical dimension.
However, for $n$ odd (this includes the tricritical point), $t$ must be
negative,
in such a way that we have now two competing solutions at low temperature.
We have to find out which one is stable. We verify below that only the
ordered phase is stable, so that the correlation length of the $\phib$-field
in the high temperature phase remains always finite. Although these dimensions
do not correspond to physical situations because $d<3$, the result is
peculiar and inconsistent with the $\varepsilon$-expansion.
\par
For $d=2n/(n-1)$, we find a gaussian behaviour without logarithmic
corrections, provided the condition
$$U^{(n)}(\rho_c)< \Omega _c\,,\quad \Omega _c=\ud(n-1)! \left[K(2n/(n-1))\right]^{1-n},\qquad
K(3)=1/(4\pi),\eqnd\etriccond$$
is met. In particular, we will see that the special point
$$ U^{(n)}(\rho_c) =\Omega _c
\eqnd\eendtric$$
 has several peculiarities. \par
We examine, in section \label{\ssNtricritical}, in more detail,  the most interesting example: the tricritical point.
 \medskip
{\it Discussion.} In the tree approximation, the dominant configuration
is given by the minimum of the function $U(\rho)\propto\rho^n$. For $n$ odd,
the function is not bounded from below, but $\rho=0$
is the minimum because by
definition $\rho\ge0$. Here, however, we are in the situation where $U(\rho)
\sim (\rho-\rho_c)^n$ with $\rho_c$ is positive. Thus, this extremum of the
potential is likely to be unstable for $n$ odd. To check the global
stability requires further work. This difficulty shed immediately some doubts about the possibility of studying  such multicritical
points  by the large $N$ method.
\par
Another point to notice concerns renormalization group: the $n=2$ example is
peculiar in the sense that the large $N$ limit exhibits a non-trivial IR fixed
point. For higher values of $n$, no coupling renormalization arises in the
large $N$ limit and only the gaussian fixed point is found. We are in a
situation quite
similar to usual perturbation theory, the $\beta$ function can only be
calculated perturbatively in $1/N$ and below the upper-critical dimension the IR fixed point is outside the $1/N$ perturbative regime.
\medskip
{\it Local stability and the mass matrix.}
The matrix of the general second partial derivatives of the action \eactefNg\
is
$$N\pmatrix{p^2+m^2 & 0 & \sigma \cr
0 & U''(\rho) & -\half \cr
\sigma & -\half & -\frac{1}{2}B_\Lambda(p,m)  \cr}  ,  \eqnd\ematrix$$
where $B_\Lambda(p,m)$ is defined in \ediagbul.\par
We are in position to study the local stability of the critical points.
Since the integration contour for $\lambda=m^2$ should be parallel to the
imaginary axis, a necessary condition for stability is that the determinant
remains negative.
\smallskip
{\it The disordered phase.} Then $\sigma=0$ and thus we have only to study
the $2\times2$ matrix $\bf M$ of the $\rho,m^2$  subspace. Its determinant
must remain negative, which implies
$$\det{\bf M}<0\ \Leftrightarrow\
2U''(\rho)B_\Lambda(p,m)+1>0\,.\eqnd\estable $$
For Pauli--Villars's type regularization, the function
$B_\Lambda(p,m)$ is decreasing
so that this condition is implied by the condition at zero momentum
$$\det{\bf M}<0\ \Leftarrow\ 2U''(\rho)B_\Lambda(0,m)+1>0\,.$$
For $m$ small, we use Eq.~\eBLamze\ and
at leading order the condition becomes
$$ K(d)(d-2)m^{d-4}U''(\rho)+1>0\,.$$
This condition is always satisfied by a normal critical point since $U''(\rho_c)>0$.
For a multicritical point, and taking into account Eq.~\edmum, one finds
$$(-1)^n {d-2\over (n-2)!}K^{n-1}(d)m^{n(d-2)-d}U^{(n)}(\rho_c)+1>0\,.
\eqnd\multicr $$
We obtain a result consistent with our previous analysis. For $n$ even it is
always satisfied; for $n$ odd, it is always satisfied above the critical
dimension and never below. At the upper-critical dimension we find a condition
on the value of $U^{(n)}(\rho_c)$, which we recognize to be identical to
condition \etriccond\ because then $2/(n-1)=d-2$.
\smallskip
{\it The ordered phase.} Now $m^2=0$ and the determinant $\Delta$ of the
complete matrix is
$$-\Delta>0\ \Leftrightarrow\
2U''(\rho)B_\Lambda(p,0)p^2+p^2+4U''(\rho)\sigma^2>0\,.\eqnd\eorder$$
We recognize a sum of positive quantities, and the condition is always
satisfied. Therefore, in the case where there is a competition with a disordered saddle point, only the ordered one can be stable.
%
\subsection{The tricritical point: variational analysis}

The most interesting physical example is the tricritical point in
three dimensions. Some insight in the problem can be obtained from
the variational calculations of section \ssNVarfiv.
Eq.~\esaddleNcii\ becomes\sslbl\ssNtricritical
$$\rho -\rho _c={\sigma ^2 \over N} -{m \over 4\pi}\,.$$
The variational action density for large $N$ then reads
$${\cal E}_{\rm var.}/N\mathop{\sim}_{N\to \infty }    -{m^3\over 12\pi}+
 U(\rho)-\ud m^2 (\rho-\rho _c) \,,$$
The polynomial of minimal degree near a tricritical point can be parametrized as
$$U(\rho)={1\over3!}U'''(\rho _c) (\rho-\rho _c)^3+{1\over2}t (\rho-\rho _c).$$
In the disordered phase, after elimination of $\rho -\rho _c$,
the variational action density becomes
$$ {\cal E}_{\rm var.}/N\mathop{\sim}_{N\to \infty } {m^3\over24\pi}\left(1-
  U'''(\rho _c)/ \Omega _c\right) -{1\over8\pi}m t \,.$$
For $U'''(\rho _c)/ \Omega _c>1$,  the action is unbounded from below
and, in the absence of other interactions, $m$ is of the order of the cut-off,
independently of the value of $t$, in such that way that the transition has disappeared.  For $U'''(\rho _c)/ \Omega _c<1$, instead, and $t>0$, one finds a minimum such that
$$m^2=\left(1-  U'''(\rho _c)/ \Omega _c\right)^{-1/2} t^{1/2},$$
in agreement with Eq.~\emsq. For $t<0$ the minimum corresponds to $m=0$.
\par
It is also easy to understand from the same arguments what happens for general $d$. The variational action density reads (Eq.~\edmum)
$$ {\cal E}_{\rm var.}/N\mathop{\sim}_{N\to \infty } -{1\over3!}U'''(\rho _c) K^3(d)
m^{3(d-2)}+{d-2\over 2d} K(d)m^{d}-{1\over2}K(d)t m^{d-2}.$$
In the formal situation $d<3$ and for $t=0$, the negative term proportional to $m^{3(d-2)}$ dominates for $m$ small. The minimum is obtained for
$$m^{2(3-d)}=K^2(d)  U'''(\rho _c),$$
that is for a mass of the order of the cut-off. Of course, when $m$ is of the order
of the cut-off, several expressions based on a small $m$ expansion are no longer valid.
For $t>0$, Eq.~\emsq\ now has a different interpretation: it has a solution because $m$ again is of the order
of the cut-off and  $m^2$ is larger than $m^{2(d-2)}$. Therefore, no particle propagates in the high temperature phase. For $t<0$, by contrast, the term in $tm^{d-2}$ dominates for $m$ small
and ${\cal E}_{\rm var.}$ first increases. The solution of Eq.~\emsq\ yields a
small value for $m$, but it corresponds to a maximum. Thus, for all values of
$t\ll \Lambda ^2$, the contribution proportional to $t$ is negligible at the
minimum of the action density and $m$ remains of the order of the cut-off.
This situation is clearly pathological and invalidates the large $N$ expansion. \par
By contrast for $d>3$, the term proportional to $U'''(\rho _c)$ is negligible,
the minimum is given by Eq.~\emsq, and the quasi-gaussian (or mean-field) theory applies.
\par
In the ordered phase, for $d=3$,
$$ {\cal E}_{\rm var.}/N\mathop{\sim}_{N\to \infty }
  {1\over3!}U'''(\rho _c){\sigma ^6\over N^3}+{t\over2}{\sigma ^2\over N}\,.$$
For $t>0$, the minimum corresponds to $\sigma =0$. For $t<0$,
the minimum is given by
$$\sigma ^2/N=[-t /U'''(\rho _c)]^{1/2}.$$
Therefore, in contrast of what is seen in $d<3$, the variational analysis confirms
for $U'''(\rho _c)/ \Omega _c<1$ a non-pathological situation.
\smallskip
{\it Dimension three: the end-point $U'''(\rho _c)=\Omega _c$.}
In the disordered phase the action density reduces to one term. It is interesting to add the first correction for $m$ small
$$ {\cal E}_{\rm var.}/N\mathop{\sim}_{N\to \infty }  -{1\over8\pi}m t
+{a(3)\over 4}{m^4 \over \Lambda }\,.$$
The results now depend on
the sign of $a(3)$. A comment concerning the non-universal
constant $a(d)$, given in \eadef\ , is here in order because,
while its absolute value is irrelevant, its sign plays a role in
the discussion of multicritical points. The relevance of this sign
to the RG properties of the large $N$ limit of the simple
$(\phib^2)^2$ field theories has already mentioned (section
\sssEGRN). For the simplest Pauli--Villars's type regularization,
 $D(k^2) $ is an increasing function and thus $a(d)$ is
finite and positive in dimensions $2 < d < 4$, but this clearly is
not a universal feature.\par
 If $a(3)$ is positive, for $t>0$ the
action density has a minimum at a non-trivial value of $m$:
$$ m\sim \left(t\Lambda \over 8\pi a(3)\right)^{1/3}.$$
For $t$ small, $m$ is small, justifying the analysis, but has a different scaling behaviour, $\nu =1/3$, a property one would expect from an IR unstable fixed point.
For $t<0$, the minimum occurs at $m=0$, as one expects for a phase transition.\par
If $a(3)$ is negative, the action density, in this small $m$ approximation, is not bounded from below and the situation is pathological.
\par
The variational analysis has clarified some peculiarities of the large $N$ limit. However, it does not allow to investigate whether the results survive at $N$ finite.
This is here specially relevant because, for example, the  point $U'''(\rho _c)=\Omega _c$ is at the boundary of a pathological region in parameter space, and its peculiar properties
are expected to be especially sensitive to $1/N$ corrections.
To calculate them, one has to return to the steespest descent method.
\subsection The scalar bound state

In this section, we study the limit of stability in the disordered phase
($\sigma=0$). This is a problem which only arises when $n$ is odd, the first case being provided by the tricritical point.\sslbl\sboundst\par
The mass-matrix has then a zero eigenvalue, which corresponds to the appearance
of a new massless excitation other than $\phib$. Let us denote by $\bf M$ the
$\rho,m^2$ $2\times2$ submatrix. Then,
$$\det{\bf M}=0\ \Leftrightarrow\ 2U''(\rho)B_\Lambda(0,m)+1=0\,.$$
In the two-space the corresponding eigenvector has components
$(\half,U''(\rho))$.
\medskip
{\it The small mass region.}
In the small $m$ limit, the equation can be rewritten in terms of the
constant $K(d)$ defined in \etadepolii\ as
$$ (d-2)K(d)m^{d-4}U''(\rho)+1=0\,.\eqnd\ephitwoa $$
Eq.~\ephitwoa\ tells us that $U''(\rho)$ must be small. We are thus
close to a multicritical point. Using the result of the stability analysis,
we obtain
$$(-1)^{n-1}{d-2\over (n-2)!}K^{n-1}(d)m^{n(d-2)-d}U^{(n)}(\rho_c)=1\,.
\eqnd\estabil$$
We immediately notice that this equation has solutions only for $n(d-2)=d$,
that is at the critical dimension. The compatibility then fixes the value of
$U^{(n)}(\rho_c)$. We  find again the point \eendtric, $U^{(n)}(\rho_c)=\Vc$.
If we take into account the leading correction to the small $m$ behaviour
we  find,  instead,
$$U^{(n)}(\rho_c)\Vc^{-1}-1\sim(2n-3){a(d)\over K(d)}\left({m\over\Lambda}
\right)^{4-d}.\eqnd\estabil  $$
This means that when $a(d)>0$, there exists a small region
$U^{(n)}(\rho_c)>\Vc$ where the vector
field is massive with a small mass $m$ and the bound-state massless. The value $\Vc$ is a fixed point value. 
\medskip
{\it The scalar field at small mass.} We now extend the analysis to a
situation where the scalar field has a small but non-vanishing mass $M$ and
$m$ is still small. The goal is in particular to explore the neighbourhood of
the special point \eendtric.
Then, the vanishing of the determinant of $\bf M$ implies
$$1+2U''(\rho)B_\Lambda(iM,m)=0\,.\eqnd\emassless$$
Because  $M$ and $m$ are small, this equation still implies
that $\rho$ is close to a point $\rho_c$ where $U''(\rho)$ vanishes.
Since reality imposes $M<2m$, it is easy to verify that this equation has also solutions for only the critical dimension. Then,
$$U^{(n)}(\rho_c)f(m/M)=\Vc\,,\eqnd\eOmegaC$$
where we have set
$$f(z)=\int_0^1\d x\left[1+(x^2-1)/(4z^2)\right]^{d/2-2},\qquad \half<z\,.
\eqnd\efofz$$
In   dimension three, it reduces to
$$f(z)=z\ln\left({2z+1\over 2z-1}\right).$$
$f(z)$ is a decreasing function which diverges for $z=\half$ because $d\le3$.
Thus, we find solutions in the whole region $0<U^{(n)}(\rho_c)<\Omega_c$,
that is, when the multicritical point is locally stable. \par
Evaluating the propagator near the pole, we find the matrix
$${\bf \Delta}={2\over G^2}\left[N\left.{\d B_\Lambda(p,m)\over \d
p^2}\right \vert_{p^2=-M^2}\right]^{-1}{1\over p^2+M^2}\pmatrix{1 &G
\cr G  &  G^2 \cr},\eqnd\edeltapole$$
where we have set
$$ G={2(-K)^{n-2} {\cal E}^{(n)} \over N (n-2)! } m^{4-d}\,. $$ 
For $m/M$ fixed, the residue goes to zero with $m$ as
$m^{d-2}$ because the derivative of $B$ is of the order of $m^{d-6}$.
Thus the bound state decouples on the multicritical line.
%

\subsection Stability and double scaling limit

In order to discuss more thoroughly the stability issue and the double scaling
limit, we
now construct the effective action for the scalar bound state. We consider
first only the massless case. We  need the action only in the IR limit, and in
this limit we can integrate out the vector field and the second massive eigenmode.\sslbl\sdouble
\medskip
{\it Integration over the massive modes.}
As we have already explained in section \stwodim, we can integrate over one of
the fields, the second being fixed, and we need  the result only at leading
order. Therefore, we replace in the functional integral
$$\e^{\cal Z}=\int [\d\rho\d\lambda]\exp\left[-\frac{1}{2}N\tr \ln(-\nabla ^2 + \lambda)
+N\int \d^dx\left(- U(\rho) +\half \rho \lambda\right)\right] \eqnd\eZ$$
one of the fields by the solution of the field equation. It is useful to discuss the effective potential of the
massless mode first. This requires calculating the action only for constant fields.
It is then simpler to eliminate $\lambda$. We
assume in this section that $m$ is small (the vector propagates).
For $\lambda\ll\Lambda$, the $\lambda$-equation reads ($d<4$)
$$\rho-\rho_c=-K(d)\lambda^{(d-2)/2}.  \eqnd\erholambda $$
It follows that the resulting action density $ {\cal E}(\rho)$, obtained from
Eq.~\eWeffrho, is
$${1\over N}{\cal E}(\rho)=U(\rho)+{d-2\over 2d (K(d))^{2/(d-2)}}(\rho_c-\rho)^{d/(d-2)}.
\eqnd\effUrho$$
In the sense of the double scaling limit, the criticality conditions are
$$ {\cal E}(\rho)=O\bigl((\rho-\rho_s)^n\bigr).$$
It follows
$$U^{(k)}(\rho_s)=- { K^{1-k}(d)\over2}{\Gamma\bigl(k-d/(d-2)\bigr)\over
\Gamma\bigl(-2/(d-2)\bigr)}m^{d-k(d-2)}\,,\quad 1\le k\le n-1\, .$$
For the potential $U$ of minimal degree, we find
$${1\over N}{\cal E}(\rho)\sim { K^{1-n}(d)\over 2 n!}
{\Gamma\bigl(n-d/(d-2)\bigr)\over
\Gamma\bigl(-2/(d-2)\bigr)}m^{d-n(d-2)}(\rho-\rho_s)^n .$$
\medskip
{\it The double scaling limit.}
We recall here that quite generally one verifies that a non-trivial double
scaling limit may exist only if the resulting field theory of the massless
mode is
super-renormalizable, that is below its upper-critical dimension $d=2n/(n-2)$,
because perturbation theory has to be IR divergent. Equivalently, to
eliminate $N$ from the critical theory, one has to rescale
$$\rho-\rho_s\propto N^{-2\theta}\varphi\,,\quad x\mapsto x N^{(n-2)\theta}
\quad {\rm with}\quad 1/\theta=2n-d(n-2),$$
where $\theta$ has to be positive.
\par
We now specialize to dimension three, since $d<3$ has already been
examined, and the expressions above are valid only for $d<4$.
The normal critical point ($n=3$), which leads to a $\varphi^3$ field theory,
and can be obtained for a quadratic potential $U(\rho)$  (the $(\phib^2)^2$ field theory)
has been discussed in section \ssfivNi. We thus concentrate on the next critical
point $n=4$ where the minimal potential has degree 3.
\medskip 
{\it The $d=3$ tricritical point.}
The action density  then becomes
$${1\over N}{\cal E}(\rho)= U(\rho)+ \frac{8\pi^2}{3}(\rho_c-\rho)^3. \eqnd\effWthreeD$$
If the potential $U(\rho)$ has degree larger than 3, we obtain
after a local expansion and a rescaling of fields,
$$\rho-\rho_s= -{ 1\over 32\pi^2 \rho_c}
(\lambda-m^2)\propto \varphi/N\,,\quad x\mapsto Nx\,,\eqnd\rescal$$
a simple super-renormalizable $\varphi^4(x)$ field theory.
If we insist,
on the other hand, that the initial theory should be renormalizable, then we remain with
only one candidate, the renormalizable $(\phib^2)^3$ field theory, also
relevant for the tricritical phase transition with $O(N)$ symmetry breaking.
Inspection of  $ {\cal E}(\rho)$ immediately shows a remarkable feature:
because the term added to $U(\rho)$ is itself a polynomial of degree 3,
the critical conditions lead to an action density $ {\cal E}(\varphi)$ that vanishes identically. This result reflects the property that
the two saddle point equations  ($\del {\cal S}/\del \rho = 0$, $
\del {\cal S}/\del \lambda = 0$ in Eqs.~\esaddNgen{}) are proportional
and thus have a continuous one-parameter family of solutions. This results in
a flat effective potential for $\varphi(x)$.
The effective action for $\varphi$ depends only on the derivatives
of $\varphi$, like in the $O(2)$ non-linear $\sigma$ model. \par
We conclude that no non-trivial double scaling limit can be obtained in this way.
In 3 dimensions with a $(\phib^2)^3$ interaction, we can generate at
most a normal critical point $n=3$, but then a simple $(\phib^2)^2$
field theory suffices.
\smallskip
{\it Remark.} The problem of the sign of $a(d)$ discussed in section \smulticr\ has an interesting appearance in $d= 3$ in the small $m^2$ region.
If one keeps the
extra term proportional to $a(d)$ in Eq.~\effUrho, one finds
$${1\over N}{\cal E}(\rho)=U(\rho)+{8\pi^2 \over 3}(\rho_c-\rho)^3
+{a(3)\over\Lambda} 4\pi^2 (\rho_c-\rho)^4.$$
Using now Eq.~\erholambda\ and, as mentioned in section \sboundst,
the fact that in the small $m^2$ region the potential is proportional to
$(\rho-\rho_c)^3$, we can solve for $m^2$.
Since $m^2>0$, the appearance of a phase with small mass depends
on the sign of $a(d)$. Clearly, this shows a
non-commutativity of the limits of $m^2/\Lambda^2 \to 0 $ and $ N \to \infty$.
The small  $m^2$ phase can be reached by a special tuning  and
cannot be reached with an improper sign of $a(d)$.
Calculated in this way, $m^2$ can be made proportional to the deviation  of
the  coefficient of $\rho^3$ in $U(\rho)$ from its critical  value $16\pi^2$.
\medskip
{\it A few concluding remarks.}
In this section, we studied of several subtleties in the phase structure of  $O(N)$ vector
models around multicritical points of odd and even orders.
One of the main topics is the understanding of the multicritical behaviour
of these models  at their critical dimensions and the effective field theory
of the $O(N)$-singlet bound state obtained in the $N \to \infty$,  $g \to g_c$
correlated limit. It was pointed out that the integration over massive $O(N)$ singlet modes is essential in order to extract the correct effective field theory of the small mass scalar excitation.
After performing this integration, it has been established here that
the double scaling limit of $(\phib^2)^K$ vector model
in its critical dimension $d=2K/(K-1)$
results in a theory of a free massless $O(N)$ singlet bound state.
This fact is a consequence of the existence of
flat directions at the scale invariant multicritical point in the effective action. In contrast to the case $d < 2K/(K-1)$
where IR singularities compensate powers of $1/N$ in the double scaling
limit,  at  $d=2K/(K-1)$ there is no such compensation and only a
non-interacting effective field theory  of the massless bound state is left.
\par
Another interesting issue in this study is the ambiguity of the
sign of $a(d)$.
The coefficient of $m^2\Lambda^{d-4}$ denoted by $a(d)$ in the expansion
of the gap equation in Eqs.~\esaddNgen{c} and \edmum\ seems to have a
surprisingly important role in the approach to the continuum limit ($\Lambda^2
\gg m^2$). The existence of an IR fixed point  at $g \sim O(N^{-1}),$ as seen
in the $\beta$ function for the unrenormalized coupling constant (section
\sssEGRN), depends on the sign of $a(d)$. Moreover, 
the existence of a phase with a small mass $m$ for the $O(N)$ vector quanta
and a massless $O(N)$ scalar depends also on the sign of
$a(d)$. It may very well be that the importance of
the sign of $a(d)$ is a mere reflection of the limited  coupling constant
space used to described the model.

\vfill\eject

\centerline{\bf Acknowledgements }

\noindent MM thanks the Saclay and Fermilab theory groups for
their warm hospitality while different parts of this work was
done.
\nref\rEMTen{ The structure of the energy momentum tensor in
renormalizable quantum field theories and a discussion of explicit
and spontaneous breaking of scale invariance can be found in \lb
C.G. Callan, S. Colman and R. Jackiw {\it Annals of Phys.} 59
(1970) 42;  S. Coleman and R. Jackiw {\it Annals of Phys.}  67
(1971) 552  and reference \refs{\rAdler}.}

\nref\rAdler{A general discussion on breaking of scale invariance
is found in  S.L. Adler, {\it Rev. Mod. Phys.} 54 (1982) 729.}

\vfill\eject
\listrefs

\appendix{}


\section Spontaneous breaking of scale invariance,
non-trivial fixed points

The subject of scale invariance breaking has a long history in
particle physics. Under very general conditions,
 a $d$-dimensional field theory that is scale
 invariant, is also invariant under the transformations
 of the  conformal group $SO(d+1,1)$ (for $d>2$).
 The scale and conformal current, $S_\mu$ and $K_{\mu \nu}$
are then constructed ~\refs{\rEMTen} from the ``improved"  energy
momentum tensor $\tilde T_{\mu \nu}$. The divergences of these
currents are proportional to the trace of the energy momentum
tensor ${\tilde T_\mu}^\mu$. They are: \sslbl\ssNscaleinv
$$ S^\mu~=~x_\lambda \tilde T^{\mu \lambda}~~~~~~
{\rm and} ~~~~~~K^{\mu \nu} ~=~x^2 \tilde T^{\mu \nu}~-~ 2 x^\mu
x^\lambda {\tilde T_\lambda}^\nu. \eqnd\scaleCurrent $$ There are
two reasonably well understood mechanisms for breaking these
symmetries in a theory which is scale and conformal invariant at
tree level (see, for example, Ref.~\refs{\rAdler}): \par

 {\bf (a)}  Spontaneous breaking of scale invariance of the Nambu--Jona-Lasinio type encountered in the BCS theory of
superconductivity.\par {\bf (b)} Explicit breaking  of scale
invariance, which is expressed at the quantum level by the anomaly
in the trace of the energy momentum tensor, as the result of
radiative corrections. \par In conventional quantum field theories
the two mechanisms occur simultaneously and the spontaneous
breaking of scale symmetry is not normally accompanied by the
appearance of a massless Nambu--Goldstone boson. Thus, for
example, we do not find a massless dilaton in QCD. An interesting
possibility exists, however, in case of theories with an hierarchy
of scales in which a light dilaton will appear as a
pseudo-Goldstone boson associated with the spontaneous breaking of
scale symmetry.
 \sslbl\SpnSclBrkg

Finite theories like $N=2$ and $N=4$ supersymmetric Yang--Mills
theories are especially interesting. In these theories there is no
trace anomaly
and thus mechanism {\bf b} mentioned above does not apply; but the
alternative dynamical generation of mass scales in these cases is
unclear.
Mechanisms {\bf a} and {\bf b} and related physical issues can be
also analyzed in solvable large $N$ vector theories, $O(N)$
symmetric scalar, and supersymmetric  theories. In particular, we
will mention in this appendix the case of spontaneous breaking of
scale invariance, unaccompanied by explicit breaking (namely, {\bf
a} without {\bf b}). This will be demonstrated in the leading
large $N$ analysis of the euclidean action of an $O(N)$ symmetric
quantum field theory in three dimensions.

We will recall first a few general points on scale invariance. The
scale current is conserved at tree level for a scale invariant
potential
$$V(\phi)=g\phi^n$$
if the trace of the improved energy momentum tensor vanishes,
namely
$$
\del_\mu S^\mu = \tilde T^\nu_{~\nu} = 0 \eqnd\eScaleCurrent$$
where, in general,
$$
\tilde T^{\mu\nu}={2\over \sqrt{-g}}{\delta\over \delta
g_{\mu\nu}}{\cal S} ( \phi ). \eqnd\TmunuDef$$
Here, $\cal S(\phi)$ is the action on a Riemannian manifold:
$$
{\cal S}(\phi)= \int \sqrt {-g}\d^dx
\ud\{g^{\alpha\beta}(x)\nabla_\alpha\phi(x)\nabla_\beta
\phi(x)-m^2\phi^2(x) - \xi R(x)\phi^2(x)\}, \eqnd\RiemanAction$$
$g
= \det \{g_{\alpha\beta}\}$, $\nabla_\alpha$ is the covariant
derivative (which reduces to $\del_\alpha$ for the scalar $\phi$)
and $R(x)$ is the curvature scalar (trace of the Ricci curvature
tensor). Varying with respect to $h_{\mu\nu}$ near flat space
$g_{\mu\nu}=\eta_{\mu\nu}+h_{\mu\nu}$, one then obtains from
Eq.~\TmunuDef:
$$\tilde T^{\mu\nu}=\del^\mu\phi\del^\nu\phi -
\eta^{\mu\nu}\left(\half (\del\phi)^2 - V(\phi) \right) +
\xi(\eta^{\mu\nu}\del^2 - \del^\mu\del^\nu)\phi^2.
\eqnd\energyMomentum$$
Using the equation of motion, the trace is
given by
$$\eqalignno{ \del_\mu S^\mu &= \tilde T^\nu_{~\nu} \cr
&=(\del\phi)^2\left[1-{d/2}+2 \xi (d-1) \right] + \phi\del^2\phi
\left[-{d/ n}+ 2 \xi (d-1)\right].  &\eqnd\eTrace }$$
So, in $d$ dimensions
$$ \xi ={1\over 4}\left({d-2\over d-1}\right)\quad {\rm
and }\quad n={2d\over d-2}\,. \eqnd\eKandn$$
 ($\xi$ is independent
of the potential, e.g.  $\xi = {1\over 6}$ in $d=4$ \refs{\rEMTen}
).

{}Following the variational method described in section
\ssNVarfiv, one defines the on-shell single particle state of
momentum $\vec p$ and mass $m$ to be used as a variational
parameter as $|\Psi_{\vec p}\rangle = a^\dagger (\vec p) |0\rangle
$ where we normalize by
$$ [a(\vec p_1), a^\dagger (\vec p_2)]=(2\pi)^{d-1}2\omega
\delta^{d-1}(\vec p_1-\vec p_2).$$ The energy is $\omega=(\vec
p^{~2}+m^2)^{1/2}$ and the field operators are
$$\phi(t,\vec x )=\int {d^{d-1}k\over (2\pi)^{d-1}2\omega}
\{\e^{ikx}a^\dagger(\vec k) + \e^{-ikx}a(\vec k) \}. \eqnd\ephiField
$$
 One finds from Eq.\energyMomentum
$$\left<\vec p_2\right|\tilde T^{\mu\nu} \left|\vec p_1\right>
= p_2^\mu p_1^\nu + p_1^\mu p_2^\nu -\eta^{\mu\nu} p_1\cdot p_2  -
2\xi(q^2 \eta^{\mu\nu}-q^\mu q^\nu) + \eta^{\mu\nu}\left<\vec
p_2\right|V(\phi)\left|\vec p_1\right> \eqnd\eEMtensor$$ where
$q^\mu=p_2^\mu-p_1^\mu$.

As an illustration, we will discuss the $d=3$, $ O(N)$ symmetric
model with the Euclidean action
$${\cal S}( \phi)= \int \d^3 x
\left[ \ud{(\partial_{\mu} \phi)}^2~ + V(\phi) \right] \eqnd\PhiSixAction  $$
and the potential
$$V(\phi) = { {\mu_0}^2 \over2} \phi^2  +
 {\lambda_0 \over 4N} {(\phi^2)}^2
 + {\eta_0 \over 6N^2}{(\phi^2)}^3.\eqnd\ePotentialPhiSix $$
Following section \ssNVarfiv, the variational ground state
energy density can be calculated from
$$\eqalignno{ {\cal E}_{\rm var.}/N\mathop{\to}_{N\to \infty } & {1\over 2 (2\pi)^d}
\int \d^d k\, \ln[(k^2+m^2 )/k^2] -\ud m^2 \Omega_3(m) +
 U(\left<\phib^2(x)/N\right>_0) \cr & = \int^m_0 s ds \Omega_3(s)
 -\ud m^2 \Omega_3(m) +U\bigl(\Omega_3(m)+{\sigma^2\over N}\bigr).&
\eqnd{\eEnergDensVar}\cr}$$
 The only divergence (see Eqs.~\eqns{\etadepole,\esaddleNc}) occurs in
$$\left< \phi^2 / N \right>_0
 = \Omega_3(m) + {\sigma^2\over N}= \int^\Lambda {\d^3 k\over
{(2\pi)}^{3 } } {1 \over k^2+m^2 } + {{\sigma}^2\over N} =
\Omega_3(0) - { |m|\over 4\pi}  + {{\sigma}^2\over N},
\eqnd\phiExpec $$ and the only renormalization needed in order to
obtain a finite ground state energy density ${\cal
E}(m^2,{\sigma}^2) $ is
$$\mu^2 = {\mu_0}^2 + \lambda_0 \Omega_3(0)
 + \eta_0 (\Omega_3(0))^2 , \quad\lambda = \lambda_0 + 2{\eta_0}\Omega_3(0)
\,,\quad \eta = \eta_0 \,.\eqnd{\renorCoupl}
$$
The ground state  energy density is then given, at leading order
for $ N\to\infty $, by
$$\eqalignno{ {1\over N}{\cal E}(m^2,{\sigma}^2) & =  {m^3 \over 24\pi}
+{\mu^2 \over 2} \left( {{\sigma}^2\over N} -{|m|\over 4\pi}
\right) +{\lambda\over 4} \left( {{\sigma}^2\over N} -{|m|\over
4\pi} \right)^2  \cr &\quad +{\eta\over 6} \left( {{\sigma}^2\over
N} -{|m|\over 4\pi} \right)^3. & \eqnd{\GrStatEnergy}\cr}
$$
In what follows, we discuss the region of parameter space where the
renormalized dimensional parameters are set to zero $(
\mu^2=0$, $\lambda=0 )$. We show  that the non-trivial
critical end-point $( \mu^2=0$, $\lambda=0$, $\eta=\eta_c=
(4\pi)^2 )$ can govern  the continuum limit of the theory. At this
point the mass $m$ of the quanta created by  $\phi (x)$, as found
from the gap equation, is non zero though all dimensional
parameters are set to zero. The appearance of a propagating mass $m$ is associated with spontaneous breakdown of scale invariance. This
is shown by the appearance of a massless bound state as $\eta \to
\eta_c$ and the vanishing of the trace of the energy momentum
tensor.
\smallskip
{\it Explicit calculation.} 
 The term $\eta^{\mu\nu}\langle\vec
p_2|V(\phi)|\vec p_1\rangle$ in Eq.~\eEMtensor ~contributes  to
$\langle\vec p_2|\tilde T^{\mu\nu} |\vec p_1\rangle$ at the tree
level  a term
$$\eqalignno{ &\mu_0^2
+{\lambda_0\over N}\left<\phi^2\right>_0 +{\eta_0\over
N^2}\left<\phi^2\right>_0^2 \cr &= \mu^2 + \lambda\left(
{{\sigma}^2\over N} -{|m|\over 4\pi}\right)+\eta \left(
{{\sigma}^2\over N} -{|m|\over 4\pi} \right)^2 .
&\eqnd\eGapEtaPhiSix }
$$ Using the gap equation, the contribution to $\langle\vec
p_2|\tilde T^{\mu\nu} |\vec p_1\rangle$ in Eq. \eGapEtaPhiSix~is
given, simply, by $\eta^{\mu\nu}m^2$. In the leading order in
an $1/N$ calculation of $\langle\vec p_2|\tilde T^{\mu\nu} |\vec
p_1\rangle$, one encounters an effective four-point vertex:
$${\bar\lambda \over N}\equiv {\lambda_0\over N}+
2 {\eta\over N^2}\left<\phi^2\right>_0= {\lambda\over N }- \eta
{|m|\over 2\pi N}\,. \eqnd\eEffectivCoupling$$ Summing all bubble
graphs, one finds (using the Euclidean action in
Eq.~\PhiSixAction)
$$\eqalignno{ \left<\vec p_2|\tilde T^{\mu\nu} |\vec p_1\right> &=
p_2^\mu p_1^\nu + p_1^\mu p_2^\nu -\eta^{\mu\nu} p_1\cdot p_2
-\eta^{\mu\nu} m^2 \cr &- \left[\frac{1}{ 4}(q^2
\eta^{\mu\nu}-q^\mu q^\nu) + \bar\lambda
I^{\mu\nu}(q)\right]\left({1\over 1+\bar\lambda B(q) }\right),
&\eqnd\eEMtensori }
$$
where
$$\eqalignno{ B(q)&={1 \over  (2\pi )^{3}} \int { \d
^{3}k \over \left(k^{2}+m^{2} \right) [ \left(q+k
\right)^{2}+m^{2} ]}\cr &={1\over 8\pi}\int_0^1 \d\alpha {1\over
\sqrt{\alpha(1-\alpha)q^2+m^2}}={1\over 4 \pi q}\arctan\left({q\over
2|m|}\right),
&\eqnd\ediagbuli}
$$
and
$$\eqalignno {I^{\mu\nu}(q)&=\int {\d^3k\over (2\pi)^3}{(q+k)^{\mu}k^\nu +k^\mu(q+k)^\nu
-\eta^{\mu\nu}(q+k)\cdot k -\eta^{\mu\nu}m^2\over
\left(k^{2}+m^{2} \right)[ \left(q+k \right)^{2}+m^{2} ] }  \cr &=
{q^\mu q^\nu - q^2 \eta^{\mu\nu} \over 4\pi } \int_0^1 \d\alpha
{\alpha(\alpha-1)\over \sqrt{m^2+\alpha(\alpha-1)q^2}
}\,.&\eqnd\eImunu }
$$
As mentioned above, we are interested in the theory when all
dimensional, renormalized parameters are set to zero and  the
dimensionless $\eta$ at its critical value:
$$\mu=\lambda=0 \quad {\rm and }\quad \eta=16\pi^2.$$

The first line in Eq.~\eEMtensori\ contributes $\half q^2-2m^2$ to
the trace of the energy momentum tensor. As a result of
spontaneous breaking of scale invariance, we will find that the
term in the second line in Eq.~\eEMtensori\ contributes $-\half
q^2 +2m^2$ and ensures a traceless energy momentum tensor  and a
conserved scale current. Indeed, one notices that the factor
$(1+\bar\lambda B(q) )^{-1}$ is exactly the term that appears in a
four-point function of the $\phi$ field. Thus, the appearance of a
massless bound state in this amplitude is directly associated with
the vanishing of the trace of $T^{\mu\nu}$ as expected.
\par

 In Eq.~\eEMtensori, the
 induced effective coupling ($\bar\lambda$  in Eq.~\eEffectivCoupling)  is
 $$ \bar\lambda = -8\pi |m| $$
 and thus $1+\bar\lambda B(q) = 0 $ at $q^2=0$. A
 massless $\phi-\phi$ bound state is created at this effective binding
 strength. Eq.~\eEMtensori\ can be written
 as
 $$\eqalignno{& \left< \vec p_2|\tilde T^{\mu\nu} |\vec p_1\right> =
p_2^\mu p_1^\nu + p_1^\mu p_2^\nu -\eta^{\mu\nu} p_1\cdot p_2
-\eta^{\mu\nu} m^2 +  {1\over 4}(q^\mu q^\nu - q^2 \eta^{\mu\nu}
)\cr &\quad\times \left[1- 8 m \int_0^1 \d\alpha
{\alpha(1-\alpha)\over \sqrt{m^2+\alpha(1-\alpha){q^2}} }\right]
\left[ 1-m\int_0^1 \d\alpha {1\over \sqrt{m^2+\alpha(1-\alpha) q^2
} }  \right]^{-1}. \cr & &\eqnd\eEMtensorii }
$$
 Indeed, due to the following simple identity
$$\eqalignno{&{1\over 4}\left[1- 8m\int_0^1 \d\alpha {\alpha(1-\alpha)\over
\sqrt{m^2+\alpha(1-\alpha) {q^2} }} \right]\cr &\quad=
\left({1\over4}-{m^2\over q^2}\right) \left( 1-m\int_0^1 \d\alpha
{1\over \sqrt{m^2+\alpha(1-\alpha) q^2}} \right) ,&\eqnd\eIdentity
\cr}
$$
as clearly seen now in Eq.~\eEMtensorii, the  term in the trace of
$\langle \vec p_2|\tilde T^{\mu\nu} |\vec p_1\rangle$ at tree
level is exactly canceled by the induced quantum correction term
due to the bound state massless Goldstone boson--dilaton pole.
that appears at $\eta=\eta_c$ when $\mu^2=\lambda=0$.

More can be seen on Eq.~\GrStatEnergy:   Clearly, at $\mu^2=\lambda=0$ one
finds ${\cal E }(m^2,{\sigma}^2) \rightarrow - \infty $ as $m
\rightarrow \infty$ unless $0<\eta<\eta_c= (4\pi)^2 $. One also
notices that it is the large negative renormalization of
$\phivec^2$ that makes the energy unbounded from below. In the
regularized theory positivity of $\phivec^2$ is maintained, but
the instability at $ \Lambda \rightarrow \infty $ is now reflected
in the solution of the gap equation. In general, $ m=\Lambda
f(\eta_0)$ where $f(\eta_0)$ is a function of the only
(unrenormalized) dimensionless coupling constant in the theory.
Thus, $m \rightarrow \infty$ as $\Lambda \rightarrow \infty$ {
unless} the theory has an UV fixed point exhibited here by a zero
of $f(\eta_0)$ at some $\eta=\eta_c$. In this case, as seen in
Eq.~\GrStatEnergy, the ground state energy density ${\cal
E}(m^2,{\sigma}^2)$ is flat, namely, is  $m$ independent  after
$\Lambda \to \infty $ has been removed. A cutoff independent,
finite physical mass can appear in the ground state spectrum.
Namely,
$$ m=\Lambda f(\eta_0) \rightarrow
{\rm ~finite~ value ~as~} \Lambda \rightarrow \infty
 \quad {\rm since }\quad \eta_0(\Lambda) \to {\eta_c}^- {\rm ~as~}
\Lambda \to \infty \,. \eqnd{\critEta} $$
One finds
$$ m=\Lambda f(\eta_0) = \Lambda {\pi \over 2}
\left( 1-\sqrt{{\eta_c \over \eta_0} }\right) .
\eqnd{\FinitMass}$$
This gives (from $\partial m/ \partial \Lambda=0 $) a
$\beta(\eta_0)$ function with a UV fixed point at
$\eta_0=\eta_c$. The bare coupling constant is then solved and
one finds
$$ \eta_0(\Lambda)=\eta_c +{\tilde \mu \over \Lambda}\quad
{\rm  and~ thus }\quad m={\pi \over 4 \eta_c}\tilde \mu\, ,
\eqnd{\runningEta} $$ where $\tilde \mu $ is a new normalization
scale. The above is just the manifestation of dimensional
transmutation at the non-trivial UV fixed point.
\smallskip
 {\it Summary :} Two facts should be
noted now:  {\bf (a)} In the massive phase described above, scale
invariance has been broken only spontaneously. As seen from
$\eta_0=\eta$, there is no explicit breaking of scale invariance
(at $\mu^2=\lambda=0$), and the perturbative $\beta (\eta)$
function vanishes in the large $N$ limit. Indeed, one finds that
the trace of the energy momentum tensor stays zero. A massless
dilaton --- the Goldstone boson associated with this breaking ---
appears in the ground state spectrum as a reflection of the
Goldstone realization of scale symmetry. ${\cal
E}(m^2,{\sigma}^2)$ is $m$ independent. {\bf (b) } The normal
ordering of $\phi^6$ induces a new $\bar\lambda \phi^4$
interaction, where $\bar \lambda=\lambda-\eta_0 [{m \over 2\pi}]$.
This new interaction guarantees the appearance of the dilaton pole
in the physical amplitudes as $\eta_0 \to \eta_c$ .

Finally, higher orders in $1/N$ introduce an explicit breaking of scale invariance and probably destabilize the finite $N$ scalar theory
in Eq.~\PhiSixAction.

\vfill\eject

\section One-loop diagrams

We present some technical details concerning one-loop calculations relevant to the gap equation
at leading order.
\subsection The regularized one-loop diagrams

We expand here for $m \ll \Lambda  $ the regularized  one-loop diagram \etadepole\ \sslbl\apiloop
$$\Omega_d(m) =\int{  \d^d k \over (2\pi)^d } \tilde\Delta _\Lambda (k)= {1\over (2\pi)^d }
 \int{  \d^d k \over m^2+ k^2D(k^2/\Lambda ^2)}=\Lambda^{d-2} \omega_d  (m/\Lambda) .\eqnd\eAptadepole $$
The function $D(z)$, $D(0)=1$, is  strictly positive for $z> 0$, analytic in
the neighbourhood of the real positive semi-axis, and increasing faster than
$z^{(d-2)/2}$ for $z\to+\infty $. \par
To extract the coefficients in the expansion we use the Mellin
transform and dimensional  continuation (see section \ssNplarge). We define
$$M(s)=\int_0^\infty \d z\, z^{-1-s} {1\over (2\pi)^d }
 \int{  \d^d k \over z+ k^2D(k^2 )}\,.$$
If the $k$ integral has a contribution of the form $z^\alpha $ in a small $z$ expansion, the Mellin transform has a pole at $s=\alpha $ and the coefficient can be identified. \par
We change variables $z \mapsto zk^2 D(k^2)$ and obtain
$$\eqalign{M(s)&=\int_0^\infty \d z\, {z^{-1-s} \over 1+z}
 \int{   \d^d k\over (2\pi)^d } \bigl( k^2D(k^2 )\bigr)^{-1-s} \cr
&=-{\pi \over \sin(\pi s)}
 \int {   \d^d k\over (2\pi)^d } \bigl( k^2D(k^2 )\bigr)^{-1-s}.\cr}$$
$$ M(s) =\int_0^\infty \d z\, {z^{-1-s} \over 1+z}
 \int{   \d^d k\over (2\pi)^d } \bigl( k^2D(k^2 )\bigr)^{-1-s}
 =-{\pi \over \sin(\pi s)}I_d(s)$$
with
$$I_d(s)= \int {   \d^d k\over (2\pi)^d } \bigl( k^2D(k^2 )\bigr)^{-1-s}=N_d
\int_0^\infty \d k\,k^{d-3-2s}\bigl(  D(k^2 )\bigr)^{-1-s}
. $$
We find two  series, the first  corresponding to the zeros of the sine function
with $s$ non-negative integer (the poles with $s<0$ corresponding to the
singularities in the large $z$ behaviour)
$$ M(s)\mathop{\sim}_{s\to n} (-1)^n  I_d(n){1\over n-s}. $$
Note that the analytic continuation of $I_n(d)$ for low dimensions is obtained by subtracting to $D^{-1-n}$ the first terms of its Taylor series at $k=0$.\par
The second corresponds to the divergence at $k=0$ of the $k$ integral $I(s)$. The residues are obtained by integrating near $k=0$ and expanding $D(k^2)$ in powers of $k^2$
 $$\eqalign{I_d(s)&\sim N_d
\int_0^1\d k\,k^{d-3-2s}\bigl(  D(k^2 )\bigr)^{-1-s} \cr
&\sim {N_d \over d-2-2s}-(1+d/2)D'(0){N_d \over d-2s}+\cdots .\cr} $$
In general the poles are at $s=\ud(d-2)+n $
with  $n$ non-negative integer. Therefore the small $z$ expansion
of $\omega _d( \sqrt{z})$ has the general form
$$\omega _d( \sqrt{z})=\sum_{n=0} \alpha _n(d)z^{(d-2)/2+n}+\beta _n(d)
z^n. $$

\smallskip
{\it Parametrization.}
We parametrize the expansion for  $z\to 0$ and $d>2$
as 
$$\omega_d(z)= \omega _d(0)-K(d)z^{d-2}+a(d)z^2 +O\left(z^{d}, z^4\right).
\eqnd\eAptadepolii $$
The first contribution  is proportional to $\rho _c$ (equation \eONrhoc)
$$ \omega _d(0)=
{1\over(2\pi)^{d }} \int{\dd k\over k^2 D(k^2)}. \eqnn $$
The constant $K(d)$ is independent of the cut-off procedure
because it involves only the leading behaviour of the integrand for $k\to 0$ and thus $D(0)=1$:
\eqna\eApintNcor
$$ \eqalignno{N_d &={2 \over(4\pi)^{d/2} \Gamma(d/2) } & \eApintNcor{a}
\cr  K(d)&=- { \pi \over2\sin(\pi d/2)}N_d=-{1\over
(4\pi)^{d/2}}\Gamma(1-d/2) \, , & \eApintNcor{b} \cr}
$$  where we have introduced   the usual loop factor
$N_d$.  \par
The constant $a(d)$, which characterizes the leading
correction in equation \etadepolii, depends explicitly on the regularization,
i.e.~the way large momenta are cut,
$$a(d)=\cases{\displaystyle - {1\over(2\pi)^{d }}  \int {\dd k \over k^4} {1\over D^2(k^2)}& for $d>4$ \cr
\displaystyle {1\over(2\pi)^{d }}  \int {\dd k \over k^4}\left(1-{1\over D^2(k^2)} \right) & for $d<4$ \cr}  . \eqnd\eApadef   $$
For $d=4$ the integrals diverge at $k=\infty $. For $\varepsilon=4-d\to 0$
$$ a(d)\mathop{\sim}_{\varepsilon=4-d\to 0} 1/ (8\pi^{2}\varepsilon). \eqnd \eApaespzero $$
A logarithmic contribution then appears in the expansion \eAptadepolii:
$$ \omega _d(z)- \omega _d(0)\sim {1\over8\pi^2}z^2\ln z \,. \eqnd \eAptadepoliv $$
\subsection Finite temperature

Let us add a few remarks concerning the calculation of Feynman diagrams.
General methods explained in the
framework of finite size scaling can also be used here, involving
Jacobi's elliptic functions. However, more specific techniques are also
available in finite temperature quantum field theory. The idea is the
following: in the mixed $(d-1)$-momentum, time
representation  the propagator is the two-point function $\Delta(t,p)$ of the
harmonic oscillator with frequency $\omega (p)=\sqrt{p^2+m^2}$ and time
interval $\beta=1/T$:
$${1\over p^2+\omega^2+m^2}\mapsto  \Delta(t,p)={1\over 2 \omega(p)}
{\cosh\left[(\beta/2-|t|)\omega (p)\right]\over \sinh\bigl(\beta\omega
(p)/2\bigr)}. $$
Summing over all frequencies is equivalent to set $t=0$. For the simple
one-loop diagram one finds
$${1\over 2\pi }\int{\d\omega \over \omega^2 +p^2+m^2}\mapsto {1\over 2
\omega(p)} {\cosh\bigl(\beta\omega (p)/2)\bigr)\over \sinh\bigl(\beta\omega
(p)/2\bigr)}, $$
This expression can be written in a way that separates quantum and thermal contributions:
$${1\over 2\omega(p)\tanh(\beta\omega(p) /2)}={1\over 2\omega
(p)}+{1\over \omega
(p)(\e^{ \beta\omega (p)}-1)}, $$
where the first term is the zero-temperature result, and the second term,
which involves the relativistic Bose statistical factor,  decreases
exponentially at large momentum.\par
Finally in the example of fermions or gauge theories we can use a more
general identity that can be proven by replacing the sum by a contour integral,
$$\sum_n{x\over (n+\nu)^2+x^2}=\pi {\sinh(2\pi x)\over \cosh(2\pi
x)-\cos(2\pi\nu)},\eqnd\eFTgenid $$
and thus after integration,
$$\ln\bigl(\cosh(2\pi x)-\cos(2\pi \nu)\bigr)-\ln 2=\lim_{N\to \infty}\sum_{n=-N}^N \left[\ln\bigl((n+\nu)^2+x^2\bigr) -2\ln(n+1/2)\right] .\eqnd\eFTgenidii $$

\subsection $\Gamma$, $\psi$, $\zeta$ functions: a few useful identities

Two useful identities on the $\Gamma$ function are \sslbl\appGPZid
$$\sqrt{\pi}\,\Gamma(2z)=2^{2z-1}\Gamma(z)\Gamma(z+1/2)\,,\quad
\Gamma(z)\Gamma(1-z)\sin(\pi z)=\pi\,.\eqnd\eGammadup $$
They translate into a relations for the $\psi(z)$ function,
$\psi(z)=\Gamma'(z)/\Gamma(z)$
$$2\psi(2z)=2\ln 2+\psi(z)+\psi(z+1/2),\quad \psi(z)-\psi(1-z)+\pi
/\tan(\pi z)=0\, \eqnn $$
where $\zeta(s)$ is the standard Riemann $\zeta$-function
$$\zeta(s)=\sum_{n \ge 1}{1\over n^s} \,. \eqnd\ezetadef $$
It is also useful  to remember the reflection formula
$$\zeta(s)\Gamma(s/2)=\pi^{s-1/2}\Gamma\bigl((1-s)/2\bigr)\zeta(1-s),
\eqnd\ezetaref  $$
which can be written in different forms using $\Gamma$-function
relations. Moreover
$$\sum_{n=1}{(-1)^n\over n^s}=(2^{1-s}-1)\zeta(s),\eqnd\ezetamin  $$
Finally
$$\eqalignno{\zeta(1+\varepsilon)&={1\over\varepsilon}-\psi(1)+O(\varepsilon),
&\eqnd\ezetaone \cr
\zeta(\varepsilon)&=-\ud(2\pi)^{\varepsilon}+O(\varepsilon^2). &
\eqnd\ezetanul \cr}$$
\section Gaussian measure and normal product

We assume that we are given some gaussian measure $\exp[-\ud
\sum_{ij} x_i A_{ij} x_j]$ for a finite or infinite number of
random variables $x_i$. In the following applications the index
$i$ will describe a parameter that can be either discrete (e.g.
group theoretic index) or continuous (e.g. space-time). We denote
the expectation values \sslbl\appnormprod
$$\left<x_{i_i}\ldots x_{i_\ell}\right>={\cal N}({\bf A})
\int\d x\, x_{i_i}\ldots x_{i_\ell}\exp[-\ud \sum_{ij} x_i A_{ij}
x_j],
$$
where the normalization ${\cal N}({\bf A})$ is such that
$\left<1\right>=1$. We denote by $\bf \Delta $ the inverse of the
symmetrix (or operator) $\bf A$ and, therefore,
$$\Delta _{ij}=\left<x_ix_j\right>.$$
We now define the normal ordering of a formal power series in the
variables $x_i$  as a linear map, with the following property: If
$V(x)$ is a homogeneous polynomial of degree $n$, the
normal-ordered polynomial $:V(x):$ is a polynomial of degree $n$
such that
$$V(x)-: V(x) :=O(|x|^{n-2}),$$
and
$$\left<: V(x) :x_{i_1}x_{i_2}\ldots x_{i_\ell}\right>=0 \quad
\ \ \ \ \forall \ell<n\,.$$
The normal ordering amounts to subtract from $V(x)$ all terms
corresponding to self-contractions in the sense of Wick's theorem.
\par
We now establish an explicit expression for the normal order of
any formal power series in the variables $x_i$. First, we consider
an exponential $\exp\left[\sum_i h_i x_i\right]$ and we generate
expectation values by a generating function $\exp\left[\sum_i g_i
x_i\right]$. We, therefore, calculate
$$\eqalign{{\cal Z}(g,h)&={\cal N}({\bf A})
\int\d x\, \exp\left[ -{1\over2} \sum_{ij} x_i A_{ij} x_j+\sum_i
(g_i+h_i)x_i\right]\cr &=\exp\left[{1\over2}\sum_{ij}
(g_i+h_i)\Delta _{ij}(g_j+h_j)\right].\cr } \eqnd\eGenFunc $$ The
suppression of the self-contractions is now easy; we divide by the
value for $g=0$. Indeed, then the remaining dependence in $h$, in
the exponential, is linear in $g$. A non vanishing result can only
be obtained by differentiating at least as many times with respect
to $g$ as to $h$ before taking the $g=h=0$ limit. Therefore,
$$ :\e^{{\bf  h} \cdot{\bf x}}:\,=\e^{-  {\bf h}\cdot{\bf \Delta}{\bf  h}/2} \e^{{\bf  h} \cdot{\bf x}}.$$
We now write any function as a Laplace transform
$$V(x)=\int\d h\, \tilde V(h)\e^{{\bf  h} \cdot{\bf x}}$$
and use the linearity of the normal order operation to get
$$:V(x):=\exp\left[-{ 1\over2}\sum_{ij}\Delta _{ij}{\partial \over\partial x_i}
{\partial \over\partial x_j}\right]V(x).\eqnn $$ Note that
analogous expressions can be derived by the same method for
complex or grassmannian gaussian measures.\par We now discuss a
few applications.
\medskip
{\it  Local polynomials in field theory.} In the exemple of a
gaussian functional measure over a field $\varphi$, $\bf \Delta $
is the propagator. If the field theory is invariant under space
translations the propagator is of the form $\Delta (x-y)$. If $V$
is a local function of a field of the form $V[\varphi(x)]$, the
normal order takes the form
$$ :V(\varphi):= \exp\left[-{ 1\over2} \left<\varphi^2(0)\right> \left( {  \partial \over  \partial \varphi   }\right)^2
 \right]V(\varphi ),\eqnd\enormorderfld $$
justifying equation \eNVnormorder.\par
By acting with $\partial/
\partial \varphi $ on both sides of the equation, we infer
$$  {  \partial \over  \partial \varphi   }
:V(\varphi):= :{  \partial \over  \partial \varphi   }  V(\varphi
):\ \ \ \ \ .$$ Namely, Eq.~\enormorderfld\ implies that the normal
ordering commutes with differentiation. Therefore, the relation
\enormorderfld\ can be inverted as
$$ V(\varphi )=:  \exp\left[ { 1\over2}
\left<\varphi^2(0)\right> \left( {  \partial \over  \partial \varphi   }\right)^2
 \right]  V(\varphi ): \ \ \ \ \ . \eqnd\enormorderInv $$
\smallskip
{\it $O(N)$ invariant theories.} With an $O(N)$ invariant gaussian
measure and $N$ component field $\varphi$ the two-point function
takes the form
 $$\Delta _{ij}(x)=\delta _{ij}\Delta (x)
 ={\delta _{ij}\over N}\left<\varphib(x)\cdot
\varphib(0)\right>.$$ The normal ordered polynomial is then given
by
$$ :V(\varphib):= \exp\left[-{ 1\over2N} \left<\varphib^2(0)\right>\sum_i \left( {  \partial \over  \partial \varphi_i   }\right)^2
 \right]V(\varphi ).\eqnn $$
In case of an $O(N)$ invariant local function of the form
$U[\varphib^2(x)/N]$,
$$\sum_i \left( {  \partial \over  \partial \varphi_i   }\right)^2U(\varphib^2 /N)=
2U'(\varphib^2 /N)+{4\over N^2}\varphib^2 U''(\varphib^2 /N), $$
and, thus, setting $\varphib^2 /N=\rho $:
$$:U(\rho ):=\exp\left\{-{ 1\over2N} \left<\varphib^2(0)\right>
\left[2{\partial \over\partial \rho }+{4\rho \over
N}\left({\partial \over\partial \rho
}\right)^2\right]\right\}U(\rho ). $$ The exponential of a first
order derivative is a translation operator. Moreover,
$\left<\varphib^2(0)\right>/N=\left<\rho \right>$. Therefore, the
normal-ordered polynomial can be rewritten
$$:U(\rho ):=\exp\left[ -{2\over N} \left< \rho \right>\rho
  \left({\partial \over\partial \rho }\right)^2\right] U\bigl(\rho-\left<\rho \right>\bigr ). \eqnn $$
In this way the normal ordered expression can be expanded in
powers of $1/N$. At leading order for $N\to\infty $:
$$:U(\rho ):\mathop{\sim}_{N\to \infty }
 U\bigl(\rho-\left<\rho \right>\bigr ). \eqnn $$
This relation,  when written in the form $:U(\rho+\left<\rho
\right> ):\mathop{\sim}
 U\bigl(\rho)$ immediately implies the
result \eNvarav.

\bye